\documentstyle[12pt,amsfonts,graphics]{thesis91Esub}
\title{A Quasilocal Hamiltonian for Gravity\\
       with\\
       Classical and Quantum Applications}
\author{Ivan S. N. Booth}
\discipline{Physics}
\degree{Doctor of Philosophy}

\newcommand{\R}{{\mathbb R}}

\newcommand{\be}{\begin{equation}}
\newcommand{\ee}{\end{equation}}
\newcommand{\bea}{\begin{eqnarray}}
\newcommand{\eea}{\end{eqnarray}}

\newcommand{\tal}{\tilde{\alpha}}

\newcommand{\cQ}{{\mathcal{Q}}}
\newcommand{\tQ}{{\tilde{\cQ}}}
\newcommand{\cB}{{\mathcal{B}}}
\newcommand{\cE}{{\mathcal{E}}}
\newcommand{\cI}{{\mathcal{I}}}
\newcommand{\cF}{{\mathcal{F}}}

\newcommand{\cG}{{\mathcal{G}}}
\newcommand{\cH}{{\mathcal{H}}}

\newcommand{\cR}{{\mathcal{R}}}
\newcommand{\cT}{{\mathcal{T}}}
\newcommand{\cA}{{\mathcal{A}}}  
\newcommand{\lt}{\frac{\Lambda}{3}}
\newcommand{\cM}{{\mathcal{M}}}

\newcommand{\ssg}{{\sqrt{\sigma}}}

\newcommand{\nn}{\nonumber}

\newcommand{\e}{{\varepsilon}}

\newcommand{\az}{{\alpha}}
\newcommand{\bz}{{\beta}}
\newcommand{\cz}{{\gamma}}

\newcommand{\dz}{{\delta}}
\newcommand{\ez}{{\epsilon}}
\newcommand{\kz}{{\kappa}}
\newcommand{\lz}{{\lambda}}
\newcommand{\mz}{{\mu}}
\newcommand{\nz}{{\nu}}

\newcommand{\bu}{{\bar{u}}}
\newcommand{\bn}{{\bar{n}}}
\newcommand{\bN}{{\bar{N}}}
\newcommand{\bV}{{\bar{V}}}
\newcommand{\bk}{{\bar{k}}}
\newcommand{\ba}{{\bar{a}}}
\newcommand{\bep}{{\bar{\varepsilon}}}
\newcommand{\bj}{{\bar{\jmath}}}
\newcommand{\bs}{{\bar{s}}}

\newcommand{\bPhi}{{\bar{\Phi}}}
\newcommand{\bE}{{\bar{E}}}
\newcommand{\bB}{{\bar{B}}}

\newcommand{\bT}{\bar{T}}
\newcommand{\bI}{\underline{I}}
\newcommand{\pa}{p_\alpha}
\newcommand{\pb}{p_\beta}

\begin{document}

\prepages

\maketitle

\sigpages

\begin{abstract}
I modify the quasilocal energy formalism of Brown and York into a 
purely Hamiltonian form. As part of the reformulation, I remove 
their restriction that the time evolution of the boundary
of the spacetime be orthogonal to the leaves of the time foliation.
Thus the new formulation allows an arbitrary evolution of the boundary
which physically corresponds to allowing general motions of the 
set of observers making up that boundary. 
I calculate the rate of change of the quasilocal energy in such
situations, show how it transforms with respect to boosts of the 
boundaries, and use the Lanczos-Israel thin shell formalism to 
reformulate it from an operational point of view. These steps are
performed both for pure gravity and gravity with attendant 
matter fields. I then apply the formalism to characterize naked black
holes and study their properties, investigate gravitational
tidal heating, and combine it with the path integral formulation of 
quantum gravity to analyze the creation of pairs of
charged and rotating black holes. I show that one must use complex
instantons to study this process though the probabilities
of creation remain real and consistent with the view that the
entropy of a black hole is the logarithm of the number of its quantum
states. 

\end{abstract}

\begin{acknowledgements}
I would like to thank Robert Mann, my supervisor, for all of his
ideas, support, and help. I have had many
useful conversations about quasilocal energy 
with other people as well and in particular with 
Eric Poisson, Richard Epp, and Jolien Creighton.
A question from Eric led me to the thin shell work 
and the Hamiltonian part of his Physics 789 course
greatly influenced my presentation of chapters 
3 and 4. Jolien suggested the tidal heating
calculation and collaborated with me on the paper
where those results first appeared.
The Natural Sciences and Engineering Research Council of 
Canada (NSERC), the Ontario provincial government through
their Ontario Graduate Scholarships in Science and Technology
(OGS-ST) programme, and the University of Waterloo
have all provided financial support. Finally, I thank 
Valeri Frolov, John Wainwright, Ray MacLenaghan, Peter Hoffman,
and Eric and Robb (again), who were my examining committee and made
many constructive criticisms that have been incorporated into this final version of the thesis.

\end{acknowledgements}

\tableofcontents

\listoftables

\listoffigures

\mainbody

\chapter{Introduction}

The conservation of energy is one
of the most fundamental ideas in all of physics. 
As a principle, its history dates back three hundred years
to Gottfried Wilhelm Leibniz whose philosophy of nature led
him to propose that kinetic energy (which he called {\it vis viva} 
or ``living force'') is conserved in an isolated system.
This notion gained currency as the eighteenth century 
progressed and gradually widened to include various types
of potential energy so that by the end of the 
century, the notion of conservation of {\it vis viva} was 
essentially equivalent to the conservation of total mechanical energy.

At the turn of the nineteenth century however, there was a large gap
between the prevalent idea of energy conservation and how that
concept is understood today. Most conspicuously, 
people didn't realize that
heat was a form of kinetic energy and instead believed that it was
an independently conserved immaterial fluid called ``calor''. In fact
Sadi Carnot developed his theory of heat engines based on that concept
and it was believed that the power of steam engines originated
from flows of calor from high to low temperature just as water wheels
are powered by the flow of water. It wasn't until the middle of the 
nineteenth century that James Prescott Joule performed his 
decisive experiment to demonstrate that heat was
also a form of energy and so paved the way for the modern
formulation of energy conservation as expressed in the 
1850's by W.\ J.\ Macquorn 
Rankine's definitive statement that ``\dots the sum of the
actual and potential energies in the universe is unchangeable \dots''.
His notion of actual energy was identical to kinetic
energy and in writing those words he was fully aware that heat is a 
form of kinetic energy. 

With the recognition that heat is energy, a theory was required to 
explain under what circumstances work
could become heat and vice versa. Recasting
Carnot's theory of heat engines in the light of the new ideas,
scientists such as Rudolf Clausius and William Thomson 
(Lord Kelvin) developed thermodynamics to meet this need. 
As part of this science and to explain why a given amount of heat 
cannot be fully transformed into work, they developed the notion 
of entropy and the second law of thermodynamics which says that
the entropy of a isolated system can never decrease. Then, 
to try to explain the macroscopic and general  
laws of thermodynamics at a microscopic and 
mechanical level, physicists such as 
Clausius, James Clerk Maxwell, and Ludwig Boltzman
created statistical mechanics in the second half of the 
nineteenth century.

Now apart from emphasizing the central role that the notion of 
energy has played in physics, 
what makes the preceding bit of history relevant
to this thesis is that the classical statistical mechanics that 
was developed could never fully explain the reality
revealed by experience and experiment. For example,
with its ideas of equipartition of energy this statistical
mechanics could not successfully predict the low temperature
heat capacities of an ideal gas, or much more dramatically,
explain why the heat in a closed container doesn't all shift into
ultra-high frequency radiation (the ``ultraviolet catastrophe'').
Problems such as these ultimately led to the conception 
and birth of quantum mechanics\footnote{A more complete 
discussion of the development of all of these ideas can be found 
in any history of physics. See for example \cite{gamow}
or \cite{harman}.}.

Finally, momentarily leaving aside the thermodynamics, no 
discussion of energy is complete without a mention of
Emmy No\"{e}ther's celebrated theorem which states that
the conserved quantities of a physical system are in one-to-one
correspondence with the transformations which leave 
the value of its Hamiltonian (or Lagrangian) invariant. 
In particular, energy is conserved if and only if the 
Hamiltonian exhibits a time translation symmetry and angular momentum
is conserved if and only if there is a 
rotation symmetry. Mathematically this result
is straightforward and indeed almost trivial, but from a physical
point of view its influence on theoretical physics since it was
introduced early in the last century has been profound. Not 
surprisingly it will show up in my discussion of gravitational energy
in the following chapters. 
 
\section{Gravity, energy, and thermodynamics}
Today, the situation in gravitational physics is in some ways analogous
to that of physics in general near the end of the nineteenth century. 
For almost thirty 
years physics has had a theory of black hole thermodynamics. It 
originated in the early 1970s with Bekenstein's recognition that if the
temperature of a black hole is proportional to its surface gravity and
its entropy is proportional to the surface area of its horizon, then the
laws of black hole mechanics are laws of thermodynamics \cite{bek1,bek2}. The first of these speculations was confirmed
by Hawking's discovery that a black hole emits radiation as a perfect black body with temperature proportional to its surface gravity
\cite{hawkNat} and the second supported by calculations which 
used the Euclidean path integral formulation of gravity
(proposed by Gibbons and Hawking in \cite{OrigPathInt})
to predict that a black hole has an entropy
equal to one quarter of its surface area.

Thus, the classical laws of thermodynamics were extended 
to black holes with semiclassical calculations to bolster 
their interpretation and application.
What was missing was a full theory of quantum gravity that could 
generate a statistical mechanics to explain them at a microscopic level.
Now, the difficulties in constructing such a theory  
need not concern us here but the important point is that any 
successful candidate must have those laws as one of its
predictions. Indeed in the quest for a theory of quantum gravity, the
laws of black hole mechanics are one of the few clues to its final form. 
Recently the two leading candidates, string theory and canonical 
quantum gravity, have passed muster and predicted
the entropy/area relationship (see for 
example \cite{string} and \cite{canon} respectively) 
but the issue is by no means fully
resolved. 

As such, a proper formulation of the laws of black hole thermodynamics
remains of great interest and that is one of the reasons why a good
definition of energy is important in general relativity. To someone
who is not directly involved in the field it would probably come as
a surprise that such a definition doesn't already exist. Afterall, 
I have just finished emphasizing how central is the role of energy in
physics and general relativity has been part of that science
for over 80 years. Thus, there has been no shortage of time in which 
to investigate how energy fits into the theory. 
What is more, the energy contained in the other
fields of physics is well-understood. In general, all aspects of the
energy content of a non-gravitational field may be described by a 
four-dimensional 
stress-energy tensor $T_{\az \bz}$. Roughly speaking, at any point in 
spacetime, the time-time components of this tensor define the
field's energy density, 
the time-space components describe the momentum carried by the
field, and the space-space components describe
the stresses associated with the field. Indeed, such 
stress-energy tensors play a central role in determining the 
general relativistic curvature of spacetime according to 
Einstein's field equations which say that 
\be
G_{\az \bz} = 8 \pi T_{\az \bz},
\ee
where the Einstein tensor $G_{\az \bz}$ describes the curvature of 
spacetime. Of course, in general relativity gravity is curvature so 
the equations roughly say that matter curves spacetime and so 
creates gravitational fields\footnote{I advisedly use the word
roughly here since this split is not so clear as it might first 
appear. Namely, since $G_{\az \bz}$ defines the geometry of spacetime
it defines the background in which the matter dwells. So, this set
of equations is much more complicated than those of, say,
electromagnetism where electric charge determines the electric 
field over an immutable background space.}.

Seeing these field equations one gets the first inklings that
there might be a problem in defining energy in general relativity.
Gravity plays the dual role of being a field and determining
the spacetime in which it and all other fields live, so it
seems likely that there could be problems in isolating its
energy. Still and all, it seems possible that
a stress-energy tensor could be conjured from somewhere.
Such hopes are dashed by the equivalence principle. 
Recall that this states that there 
is no way for an observer making measurements entirely at a single
point in spacetime to distinguish between her own acceleration
and the effects of a gravitational field. 
Therefore there is no invariant way for a single observer to assign 
a ``strength'' to the gravitational field at a point and by extension 
no way to assign it an energy density. Thus there is apparently 
no way to define a purely local energy for gravity. An extended
discussion of this point can be found in section 20.4 of 
reference \cite{MTW}.

How then are the laws of black hole mechanics defined if there is
no way to define energy in a spacetime? Well, the answer is that 
the prohibition against a purely local definition of energy 
does not extend to the total energy of a spacetime. At least
for asymptotically flat spacetimes, there are well known 
and accepted measures of the total energy such as
the ADM \cite{ADM} or Trautman-Bondi-Sachs \cite{bondi} masses and
it is usually one of these measures of energy that is used in 
the traditional formulations of black hole mechanics. 
However, this is not really a satisfactory way of proceeding since 
the thermodynamic
system of interest is supposed to be the black hole itself 
rather than the entire, often infinite, spacetime of which it is a part. 
As an example consider a black hole spacetime which also contains
a sprinkling of regular stars all situated many light years away from the hole and each other. Then, no one would
argue that the stars should be considered as integral parts of the
black hole system, yet
the ADM energy would include the masses of those stars. 
Quasilocal definitions of gravitational energy attempt
to meet this concern by defining the energy of just a part
of the full spacetime while not attempting to fully localize
the energy in a stress-energy tensor\footnote{An alternate
view on this point can be found in reference 
\cite{nesterchen} which argues that any quasilocal energy
is equivalent to a stress-energy pseudo-tensor that fully
localizes the energy.}.

\section{Quasilocal energy}
\label{QLEsect}
A quasilocal definition of energy is a procedure that associates an
energy with each closed and spacelike two-surface in a spacetime.  Though there are many definitions
of quasilocal energy in the literature 
(see for example \cite{BY1, shayward, nester99,epp} and
references contained those papers) a large subset of them can be 
characterized as Hamiltonian approaches. That is, they start with a
Hamiltonian functional for finite three-surfaces in a spacetime
which will generate the Einstein equations in the usual Hamiltonian
way. Then the energy of the finite region is taken to be the 
value of that Hamiltonian evaluated thereon. 
Usually all bulk terms of the functional are proportional to constraints
and so its numerical value evaluated on-shell\footnote{That is 
for solutions to the field equations.} 
is a functional on the boundary two-surface only. 
Given this property, the energy can't really be said to be
associated with the three-surface but instead is a property
of the two-boundary alone. Any number of three-surfaces
could be associated with that boundary, but which one
actually is is completely irrelevant to the final evaluation. 
This property is in accord with the equivalence principle
prohibition against a point-by-point localization of the
energy. Since the energy can't even be associated with a
particular three-volume, it certainly can't be assigned
to individual points. 

In some ways these definitions can be thought of as analogous 
to the Gauss law for electric charge.
Just as that rule defines the electric charge contained by a 
closed two-surface from measurements of the electric field
made at the surface, the quasilocal energies define the
energy ``contained'' by a two-surface based on measurements
of the gravitational field made at the surface. 

One of the main aims of this thesis is to extend and generalize
the popular Hamilton-Jacobi definition of quasilocal energy that was
originally proposed by Brown and York \cite{BY1}. Advantages of this
definition include its appealing geometric form (discussed in some
detail in chapter \ref{gravChapter}) and  its natural interface
with the path integral formulation of quantum gravity which 
allows one to do gravitational thermodynamics (briefly
discussed in the next section and chapter \ref{PC}, and in
more detail in references 
\cite{BY2, BCM, jolien}). Further, in common with other 
definitions of quasilocal energy, it can be shown to 
behave in ways that one would expect an energy to behave.
For example it is additive, negative for binding energies, and 
in the appropriate limits (and spacetimes)
it reduces to such total measures of energy as the 
ADM energy \cite{BY1}, the 
Trautman-Bondi-Sachs energy \cite{BYnull}, and the Abbot-Deser energy \cite{BCM}. In the small sphere limit in the presence of 
matter, it can be shown to recover intuitive notions of matter
energy density  \cite{BYsmallsphere}.

The first part of this thesis
reformulates the Brown-York definition into a pure Hamiltonian form
and removes the slight dependence on the spanning three-surface 
from which their energy suffered. That is, I 
modify the Hamiltonian they proposed, show that it 
does indeed generate the correct field equations in the 
usual Hamiltonian way, and
further show that its numerical value depends only on the 
values of fields at the bounding two-surface in a way that
doesn't care about what three-volume it contains.  
From there I show how the value of the Hamiltonian does depend
on the motion of the observers measuring it, allow for the inclusion 
of Maxwell and dilaton fields, and show how the energy can be
defined from an operational point of view. 

Moving away from the mathematical formalism I get 
my hands dirty and try to develop an intuitive feel
for the quasilocal energy by examining the
distribution of energy in the standard static and 
spherically symmetric spacetimes. I then investigate
naked black holes and calculate the energy 
flows that occur during gravitational tidal heating. 

\section{Path integrals, thermodynamics, and quantum tunnelling}

The last part of the thesis deals with a quantum 
application of the Hamiltonian work.
As noted above, the quasilocal formalism of Brown and York
naturally combines with the path integral formulation 
of quantum gravity and thereby gives some insights into
gravitational thermodynamics. Recall that in general,
path integral versions of quantum mechanics calculate
the probability that a quantum system passes from an initial
state $X_1$ to a final state $X_2$ by considering all conceivable
``paths'' that the system can take between the two states (not
just those that satisfy classical equations of motion). The 
action of each of those paths can be computed using a classical
Lagrangian action functional and then, up to a normalization factor,
the probability amplitude that the system takes a specific path
is $e^{-i I}$ where $I$ is the action of the 
path. Then, the sum of all of these probability amplitudes is the 
probability amplitude that the system will pass to the final state
$X_2$. There are a myriad of unsolved problems involved in 
rigorously defining these integrals, but nevertheless
history has shown that many physical insights can be gained 
through their judicious use. 

The problems of mathematical rigor are even more serious 
for path integral gravity than for regular quantum mechanics, but
all the same its usefulness as a conceptual and provisional
computational tool remains. In particular, as is usual with path
integrals, one can use it to study thermodynamics by viewing  
the ``paths'' as members of a thermodynamic ensemble and so
reinterpret the path integrals in terms of partition functions.

The connection with the quasilocal formalism arises because
the classical behaviour of a system is not sufficient to 
specify the action functional that should be used to assign
the probability amplitude to each path. However, it turns out that
the choice of an action functional also corresponds to a choice
of restrictions on the ensemble of paths considered. The 
Brown-York formalism provides a convenient way to see
those restrictions from a thermodynamic perspective. 
With this insight one can associate each action functional
with a specific thermodynamic partition function (for example 
grand canonical, canonical, or microcanonical) as was
first discussed in reference \cite{BY2}.

That said, one can also use the path integrals in their original
form to estimate the probability that a quantum event will occur.
In this case, one must recognize that the action functional still
places restrictions on the physical properties of the paths 
considered and so should be chosen to conserve essential 
physical properties (such as the angular
momentum or electric charge of a spacetime)
through the quantum transition. 

As a specific application, 
in recent years there has been a considerable
interest in black hole pair production. Inspired by the well
understood particle pair production of quantum field theory
(for example $ 2 \gamma \rightarrow e^+ + e^-$), theorists
have investigated the corresponding phenomenon for black holes
and studied the possibility that a spacetime with a source of 
excess energy will quantum tunnel into a spacetime containing
a pair of black holes. The earliest  
investigations considered pair creation due to background electromagnetic fields \cite{em1,em2,em3,hhr} 
but since then have been extended to include
pair creation due to cosmological vacuum energy
\cite{robbross,othercosmo}, cosmic strings \cite{str1,str2,str3}, and 
domain walls \cite{dom1,dom2,dom3}. In all cases the chance of such 
an event happening has been found to be extremely small, but 
perhaps an equally interesting outcome of the calculations has been
the evidence that they have provided that black hole entropy does indeed
correspond to the number of quantum states of the hole.

In the last part of the thesis I show how the pair creation
results can be extended to include pairs of rotating black holes, which
were not considered in the above referenced papers. This is quite an 
involved process which starts with the identification of classical 
solutions to the Einstein equations that properly describe pairs of
black holes in the appropriate context. From there instantons
are constructed from the classical solutions that will be used 
to approximate the path integrals and it is seen that requirements
of regularity restrict the possible physical parameters 
of the created spacetime. The Brown-York 
formalism is used to choose the correct action for use in each
situation and finally, with all of the preparation completed,
I calculate and interpret the creation probabilities.

\section{Overview}
With these ideas in mind I now outline the rest of the thesis.
As its name implies, chapter \ref{Setup} establishes the background
for the work that follows.
Much of it is a review of well-known ideas and results but
it will serve to refresh these ideas for the reader who is not
especially familiar with this area of general relativity 
and establish notation and 
sign conventions. Since I will be working with a Hamiltonian 
formulation of general relativity, section \ref{geometry} explains
how a spacetime may be foliated into ``instants'' of time and how
a vector field is set up to define the ``flow of time'' from instant to
instant. I focus on a finite region of that spacetime and discuss
its boundaries and the fields on those boundaries in some detail as 
well as give a physical interpretation of the boundaries as being 
defined by the history of a closed two-surface of observers. 
Extending the spacetime foliation to the timelike boundary, I 
foliate it with closed two-surfaces which define the observers' 
notion of simultaneity. The quasilocal energy will be defined for these
surfaces. 

Section \ref{appMatter} reviews the field
equations for fields of interest to this thesis. 
Specifically they are gravity, electromagnetism, and a dilaton field
where a coupling exists between the dilaton and Maxwell fields. 
First examining these
from a covariant four-dimensional perspective, I then review
how they become constraint and evolution equations if they 
are projected into the leaves of the time foliation. I discuss
how one-half of the Maxwell equations are implied by the assumption that
a gauge potential exists, a simple fact that will have larger
consequences later on, and discuss duality for these three
fields.

With this theoretical stage set, chapter \ref{gravChapter} begins
the main work of the thesis. Starting with a modified
Einstein-Hilbert action for gravity, section \ref{gravLag} briefly
reviews how it variation produces the standard field equations for
gravity. The action (proposed by Geoff Hayward in ref.\ \cite{hayward})
differs from the classical Einstein-Hilbert
action in that it is formulated for a finite region of spacetime bounded
by a combination of spacelike and timelike hypersurfaces and
disagrees with the one used by Brown and York in that
it allows for those boundaries to be non-orthogonal.

From that action, subsection \ref{Hamform} derives a 
Hamiltonian functional defined on the slices of the 
time-foliation of the spacetime. This Hamiltonian 
differs from the Brown-York Hamiltonian in that it does not 
restrict the time foliation to be orthogonal to the 
timelike boundary. It is noted that even though
the functional is defined for a finite region of a spatial 
three-surface, its actual on-shell
{\it numerical} value depends only
on the fields at the boundary of that surface and how they are evolving
in time. It is indifferent to what three-surface it bounds. This
means that a quasilocal energy derived from the Hamiltonian really 
is a functional of the boundary two-surface only; a fact which is
crucial for its proper definition  
since there is no natural way to uniquely 
associate a spanning three-surface with that boundary (or for that
matter even guarantee that such a surface exists). Thus, this 
approach differs markedly from that taken by Hawking and Hunter
in ref.~\cite{hhunter} which required reference terms
to remove the dependence of their Hamiltonian on the intersection
angle between the foliation surfaces and timelike boundary. 

With the proposed
Hamiltonian functional in hand, subsection \ref{HamVar}
confirms that it really is a properly defined Hamiltonian
(this is the first time that this has been explicitly demonstrated) and 
shows that, as would be expected, the calculated variation of the
Hamiltonian is in accord with the full variation of the action
functional as considered in such papers as \cite{BY1, nopaper}.

Section \ref{energySect} presents a definition of quasilocal
energy in terms of the quasilocal Hamiltonian of the previous section.
Its exact form is dependent on the time four-vector that determines the
evolution of the boundary observers and it is seen that if that vector
field is a Killing vector field for the induced metric on the 
boundary of the spacetime, then the 
quasilocal energy is a conserved quantity. A special case
of the general quasilocal energy where the observers are
stationary relative to the foliation hypersurfaces and measure
proper time is considered and geometrical interpretations of 
that energy are discussed.
Transformation laws for the quasilocal quantities with respect 
to boosts of the measuring observers are derived and investigated
in section \ref{gravtransform}. These laws
are shown to be Lorentz-like and
a comparison is made with corresponding laws from special relativity.

Next, section \ref{RefTerm} defines reference terms for the quasilocal
energy of the previous sections. These terms are necessary because 
without them the quasilocal energy of a spherical region of flat
space is non-zero and actually diverges as the radius of such a 
sphere goes to infinity. Within the quasilocal formalism there
is quite a lot of freedom to define these reference terms and
the choice of a particular one is essentially a choice of where
to set the zero-level of the energy. I examine three
choices of reference terms, starting with the original 
Brown-York term which is defined by embedding the
instantaneous two-surface of observers into a reference
three-space. A well recognized problem with this term is
that it is not always defined
and I point out that it also runs into problems for boosted
observers in flat space. From there I discuss an alternate proposal 
involving embedding the instantaneous two-surface of observers into
a four-dimensional reference space (discussed in \cite{nopaper, naked}
and from a different perspective in \cite{epp}). It is more likely to
exist than the three-dimensional proposal but
unfortunately is not uniquely defined. Finally,
I briefly comment on the so-called intrinsic reference terms that
have recently been inspired by the AdS-CFT correspondence.

The chapter ends with section
\ref{thinshell} which investigates 
the close relationship between the quasilocal energy and the thin
shell formalism of Israel \cite{thinshell}. 
I show that there is an exact correspondence
between the mathematics of the quasilocal energy and the thin shell
formalism. This means that for a two-surface of observers 
with a specified time evolution in a given spacetime, the quasilocal
energy with the two-into-four reference terms discussed above 
is defined if and only if that
two-surface could be replaced with a thin shell of matter so that
outside of the shell the spacetime would be unchanged while inside
it would be isometric to a part of the reference space. The
quasilocal energy measured by the observers is exactly equal to 
the total {\it matter stress-energy} of the thin shell. This 
equivalence means that one can reinterpret the (modified)
Brown-York quasilocal
energy from an operational point of view. That is, the quasilocal
energy ``contained'' by a two-surface could be defined as the 
matter stress-energy required to reproduce that spacetime outside
of a matter shell that is isometric to the two-surface and embedded in the reference space.

Though similar in length to the previous one, chapter 
\ref{matterChapter} can be summarized quite a bit more 
quickly since it covers much the same ground except that 
this time Maxwell and dilaton matter fields are included in the
mix. 
It starts in section \ref{matLag} 
with a review of the Lagrangian action whose variation will
generate all of the field equations. From there, subsection 
\ref{FormMatHam} derives a Hamiltonian functional from that
Lagrangian action and it is noted that the assumption that
a single gauge potential exists over the region being studied
implies not only two of the Maxwell equations, but also that
no magnetic charge can exist in that region. Subsection 
\ref{matterVariation} checks that the proposed functional
really is a proper Hamiltonian and again compares the
variation of the Hamiltonian with previously 
calculated action variations \cite{jolien, jolienrobb, naked}.
Section \ref{MatHamProps} reviews conserved charges, 
reference terms, transformation laws, 
and the thin shell correspondence when 
the matter fields are included along with the gravitational
field. Finally, in recognition of the fact that the formalism 
as constituted cannot handle magnetic charges, 
section \ref{EMdual} uses
duality to define an action and Hamiltonian that can handle
those charges, though in doing so it loses the ability to 
deal with electric charges.

Chapter \ref{quasiExamples} applies the work of the previous
two chapters to several spacetimes both to gain 
insight into the quasilocal energy and to 
demonstrate its utility. Section \ref{sss} is targeted
mainly towards the first goal as it examines Schwarzschild
and Reissner-Nordstr\"{o}m spacetimes. It starts with 
static and spherically symmetric 
sets of observers and shows that the quasilocal
energies that they measure are physically reasonable though not 
entirely in accord with intuition. Interestingly it is seen
that the definition of the quasilocal energy derived for 
a purely gravitational field appears to also include contributions
from matter fields. The extra terms generated by matter fields
are gauge dependent and are not directly related to the physical
configuration of the system but instead seem to give the potential
energy for the system to exist relative to an (almost) arbitrarily
set gauge potential. The next section considers
radially boosted observers which the original 
Brown-York formalism couldn't easily handle and so
demonstrates the nonorthogonal formalism. Finally
a spherical set of ``z-boosted'' observers is considered
for Schwarzschild space and interesting but
slightly enigmatic results are obtained.

Section \ref{nbh} applies the formalism to study naked black holes
(first discussed by Horowitz and Ross in \cite{naked1, naked2}). 
These are massive, near-extreme, Maxwell-dilaton black holes
that are characterized by how different sets of observers feel
gravitational tidal forces close to the event horizon.
Specifically, static observers measure relatively small transverse 
tidal forces while those who are infalling on radial geodesics
measure huge (though not divergent) forces. Though at first thought
this might not seem to be especially surprising, it should be kept 
in mind that equivalent observers near to similar Reissner-Nordstr\"{o}m
black holes all measure small tidal forces irrespective of their 
radial motion. I calculate the 
quasilocal energies measured by corresponding spherical sets
of observers and find that the static ones measure a very
large quasilocal energy while the infalling ones measure
it to be extremely small. I explain all of these measurements
in terms of the geometry of the naked spacetimes. 

The final classical application is found in section 
\ref{tidal} where I apply the formalism to calculate 
energy flows during gravitational tidal heating. The
prototypical example of this in our own solar system 
is found in Jupiter and its moon Io, where the
gravitational tidal forces 
provide the energy that powers Io's volcanism. 
I successfully reproduce the results of energy 
flow calculations that in the past have been 
done with Newtonian and stress-energy pseudo-tensor \cite{purdue:1999}
methods. The calculation is cleaner than the pseudo-tensor
calculations and has the added advantage of providing a 
simple geometric interpretation of gauge ambiguities in the
energy flow. Thus, this section can be viewed both as
an application of the quasilocal formalism and as an 
additional check on its physical relevance.

Chapter \ref{PC} contains a quantum mechanical 
application of the formalism as it applies it to 
the pair production of charged and rotating black hole
pairs in a cosmological background. It begins with a
brief review of the Euclidean path integral formulation of
quantum gravity in section \ref{formalism}.
Section \ref{KNdSsect} examines the classical 
spacetimes that describe pairs of charged and rotating
black holes in a cosmological background. It starts with the
generalized C-metric of Plebanski and Demianski \cite{PlebDem}
which can be interpreted as describing a pair of charged
and rotating black holes accelerating away from each other 
in a cosmological background, and then shows that matching 
the acceleration of the holes to that of the rest of the universe
(as is demanded by conservation of energy) reduces this metric
to the Kerr-Newman-deSitter (KNdS) metric. Thus, those will
be the class of spacetimes that I aim to create and so I examine
them in some detail, working out the allowed range of their physical
parameters and examining limiting cases. Traditionally it has
been asserted that only spacetimes in full thermodynamic
equilibrium can be created by quantum tunnelling processes,
so I finish off
by considering the various KNdS spacetimes from this point
of view. I show that three limiting cases of the 
KNdS spacetime, the cold limit (which corresponds 
to a pair of extreme black holes), the Nariai
limit (where the outer black hole and cosmological
horizon become coincident), and an ultracold limit
(the overlap of the cold and Nariai limits) are in thermal 
(but not full thermodynamic) 
equilibrium as are a class of non-extreme black holes that are
dubbed lukewarm (where
the outer black hole and cosmological horizons simply
have the same temperature).
If the rotation parameter is
set to zero, then these reduce to equivalent cases
considered in extant non-rotating calculations.

Section \ref{InsCons} assembles instanton solutions 
to mediate the creation of each of the classes of KNdS
spacetimes that are in thermal equilibrium. These
instantons have complex metrics, in contrast with the
usual ones used to create non-rotating pairs of black holes.
I show that this is a necessary feature of the instanton metrics if
one requires that these solutions match onto the corresponding
Lorentzian ones along a spacelike hypersurface. Finally I
construct the actual instantons treating the non-degenerate two horizon 
(lukewarm and Nariai), non-degenerate single horizon 
(cold, ultracold I), and zero horizon (ultracold II) cases separately. 
I note that only thermal (rather than full thermodynamic) equilibrium 
is required to construct a smooth instanton. 

Section \ref{ActPick} examines the instantons to identify their
essential features and then uses the Brown-York formalism
to select the actions that will preserve those characteristics
during the quantum tunnelling. Then with that action in hand 
section \ref{evAct} evaluates those actions for the instantons
and so finds the probability for a pair creation event to occur.
It is shown that the probability of pure deSitter space tunnelling
into a spacetime containing a pair of black holes with opposite
spins and electric/magnetic charges is proportional
to $e^{-\Sigma \cA_i /4}$ where $\Sigma \cA_i$ is the sum 
of the areas of the non-degenerate horizons in the created
spacetime. This is in accord with the corresponding results
for non-rotating holes \cite{robbross} as well as an interpretation
of black hole entropy as the logarithm of the number of quantum 
states of the hole. Section \ref{CompExt} shows how the methods
that I have used compare with the procedures 
that other people have used. 

Finally, chapter \ref{discuss} attempts to summarize the results 
of the thesis, put them into some perspective, and discusses 
future work related to the topics of this thesis.

\chapter{Set-up}
\label{Setup}
This chapter sets the stage on which the rest of the thesis
will play. In the first section I define a quasilocal region of space,
show how it may be foliated, and define a variety of geometric
quantities that will be used extensively in the rest of the thesis. 
The second section reviews the interlocking field equations for 
gravity, electromagnetism, and the dilaton field in their full 
four-dimensional and projected three-dimensional forms. It further discusses electromagnetic potentials and the duality inherent in the
electromagnetic and dilaton fields. Much of this chapter is
a review of well known facts but it serves to establish notation
and cast some of these ideas in a new light. 

\section{The geometric background}
\label{geometry}

\begin{figure}
\rotatebox{270}{\resizebox{9cm}{!}{\includegraphics{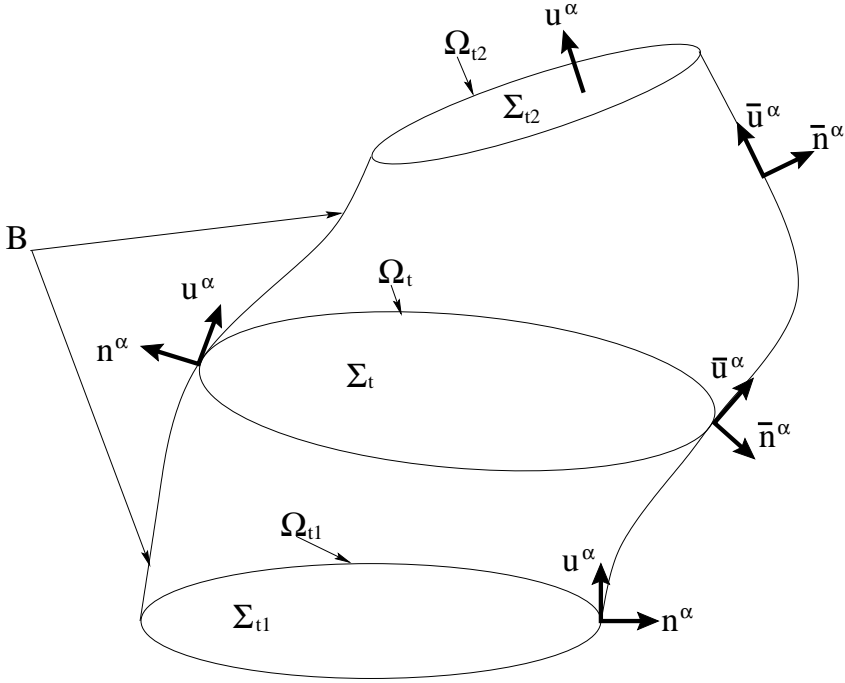}}}
\caption[The quasilocal region]{A three-dimensional schematic of
the Lorentzian region $M$, 
assorted normal vector fields, and 
typical elements of the foliation.}
\label{Mfig} 
\end{figure}

Let $M$ be a compact and topologically trivial region of a 
four-dimensional spacetime
$\cM$. It is specified to be the region bounded 
by two spacelike surfaces $\Sigma_{t1}$ and $\Sigma_{t2}$
(each homeomorphic to $\R^3$)
and a timelike surface $B$ (homeomorphic to 
$\R \times S^2$). Such a region is depicted schematically
in figure \ref{Mfig}\footnote{In 
later chapters $M$ and $B$ will sometimes be taken to
have more complicated topologies. The extensions to those cases
will be straightforward 
so for simplicity I now consider only the simplest
case.}. Let $M$ be foliated with a set of three-dimensional 
spacelike surfaces $\{ \Sigma_t \}$, labelled by a time
coordinate $t$, such that $\Sigma_{t1}$ and $\Sigma_{t2}$ are 
leaves of the foliation. This then induces a corresponding foliation
of $B$ by spacelike two-surfaces $\{ \Omega_t \equiv \Sigma_t \cap B \}$
each with topology $S^2$. Finally, in association with the foliations
define a smooth timelike vector field $T^\az$ such that 
$T^\az \partial_\az t = 1$ and is tangent to $B$. 
These conditions are not sufficient to uniquely
specify $T^\az$, so there is a certain arbitrariness in its definition. 

Intuitively one can think of $B$ as defining the history
of a set of observers and each $\Sigma_t$ as defining an ``instant'' in
time. Then each $\Omega_t$ defines an ``instantaneous'' configuration of 
those observers and in the regular way, I 
will say that $\Omega_{t1}$ ``happens'' before $\Omega_{t2}$ 
if $t_1 < t_2$. Further $T^\az$ can be thought of as the
(unnormalized) four-velocity of the observers and so $B$ can
be viewed as the history of a set of observers who 
started out in the configuration $\Omega_{t1}$ and then 
evolved through time with $T^\az$ as their four-velocity.
Because $T^\az$ isn't normalized the time $t$ doesn't correspond
to proper time. The freedom in the
definition of $T^\az$ corresponds to how individual observers can
evolve differently while leaving their evolution as a 
set invariant. 

Note that while a $\Sigma_t$ foliation surface 
uniquely specifies a corresponding $\Omega_t$, the
converse isn't true. Any number of $\Sigma_t$ foliations
can be defined that are compatible with a given $\Omega_t$ foliation.
In fact in spite of the way that the foliations have been set up in this
section, a main goal of this thesis is to show that only the foliation
of $B$ is important. The foliation of the rest of the spacetime is
irrelevant, basically because there are no observers in the interior of
$B$ to define it. The only observers are thought of as residing on the
boundary $B$. 

Up to this point no real use has been made of a metric. Terms
like spacelike and timelike have been used for clarity but everything
could equally well have been formulated in terms of a manifold
without metric. Now however, I'll introduce a signature $+2$ metric
field $g_{\az \bz}$ over $M$. With this metric one can define a (unit
normalized) forward-pointing timelike vector field $u^\az$ normal to 
the $\Sigma_t$ surfaces 
as well as induce a spatial metric field $h_{\az \bz} 
\equiv g_{\az \bz} + u_{\az} u_{\bz}$ on those surfaces. 
Then, one can project $T^\az$ into its components perpendicular and
parallel $\Sigma_t$. Namely
\be
T^\az = N u^\az + V^\az,
\ee
where $N$ and $V^\az$ are called the lapse function 
and shift vector
field respectively and $V^\az u_\az = 0$. Conversely one can define the
spacetime metric in terms of the spatial metric, lapse, shift, and time
vector field by 
\be
g^{\az \bz} \equiv 
h^{\az \bz}-\frac{1}{N^2}(T^\az - V^\az)(T^\bz - V^\bz).
\ee

Define unit normal vector fields for the various hypersurfaces. Already
$u^\az$ has been defined as the future-pointing timelike unit normal
vector field to the $\Sigma_t$ surfaces. Similarly, define $\bu^\az$ as
the future-pointing timelike unit normal vector field to the surfaces
$\Omega_t$ in the timelike hypersurface $B$. 
The spacelike outward-pointing unit
normal vector field to $\Omega_t$ that is 
perpendicular to $u^\az$ (and thus 
in the tangent bundle $T \Sigma_t$) is $n^\az$ and the corresponding 
normal vector field to $\Omega_t$ perpendicular to $\bu^\az$ is
$\bn^\az$ which is also the outward-pointing unit 
normal vector field to $B$.

Next define the scalar field $\eta = \bu^\az n_\az = -u^\az \bn_\az$
over $B$.  If $\eta = 0$ everywhere, then the foliation surfaces are
orthogonal to the boundary $B$ (the case dealt with in refs.
\cite{BY1,hhorowitz}) and the barred vector fields are equal to 
their unbarred counterparts. If $\eta \neq 0$ then $\bu^\az$ and
$\bn^\az$ may be written in terms of $u^\az$ and $n^\az$ 
(or vice versa) as,
\bea
\label{nutrans}
\bn^\az =\frac{1}{\lambda} n^\az + \eta u^\az &\mbox{ and }& \bu^\az 
= \frac{1}{\lambda} 
u^\az + \eta n^\az, \label{nou} \\
&\mbox{or,}& \nn \\
n^\az =\frac{1}{\lambda} \bn^\az - \eta \bu^\az &\mbox{ and}& u^\az 
= \frac{1}{\lambda} 
\bu^\az - \eta \bn^\az, \label{onu}
\eea
where $\lambda^2 \equiv \frac{1}{1 + \eta^2}$. 
$\eta$ and $\lambda$ may also be written without direct 
reference to the barred normals. To do that first define
\be
v_\vdash \equiv (V^\az n_\az)/N, \label{v}
\ee
which is the three-velocity in the direction $n^\az$ of an object 
with four-velocity $T^\az$ as measured by an observer 
with four-velocity $u^\az$. Then,
\bea
\eta \equiv v_\vdash/\sqrt{1-v_\vdash^2} & \mbox{ and} &
\lambda = \sqrt{1-v_\vdash^2}. \label{veta}
\eea
These quantities then begin to look like the terms that 
appear in special relativistic Lorentz transforms. This correspondence
will be explored in some detail in section \ref{gravtransform}.

On the surface $B$ one may write, 
\bea
\label{boundaryTime}
T^\az = \bN \bu^\az + \bV^\az,
\eea
where 
\bea
\bN \equiv \lambda N = \sqrt{N^2 - [V^\az n_\az]^2} 
& \mbox{ and} & \bV^\az
\equiv \sigma^\az_{\ \bz} V^\bz = V^\az - [V^\bz n_\bz] n^\az
\label{bNbV}
\eea
are respectively labelled the boundary lapse and the boundary shift. 
This split is possible because $T^\az$ on $B$ has been restricted to lie
in the tangent bundle $TB$. Equivalently, $B$ is the history of the
observers $\Omega_t$ and $T^\az$ is their four-velocity, so naturally
$T^\az \in TB$ on $B$. In any case $T^\az \bn_\az = 0$. 

Next consider the metrics induced on the hypersurfaces by the spacetime
metric $g_{\az \bz}$. Just as $h_{\az \bz} \equiv g_{\az \bz} + u_\az
u_\bz$ is the metric induced on the $\Sigma_t$ surfaces,
the other metrics may also be written with respect to the normals. 
$\gamma_{\az \bz} \equiv
g_{\az \bz} - \bn_\az \bn_\bz$ is the metric induced on $B$ 
and $\sigma_{\az \bz} \equiv h_{\az \bz} - n_\az n_\bz 
= \gamma_{\az \bz} + \bu_\az \bu_\bz$ is the metric induced on
$\Omega_t$. Raising one index of these metrics defines  
projection operators into the corresponding surfaces. These have 
the expected properties: $h^\az_{\ \bz} u^\bz =
\gamma^\az_{\ \bz} \bn^\bz = \sigma^\az_{\ \bz} u^\bz = 
\sigma^\az_{\ \bz} n^\bz = \sigma^\az_{\ \bz} \bu^\bz = 
\sigma^\az_{\ \bz} \bn^\bz = 0$, and 
$h^\az_{\ \bz} h^\bz_{\ \cz} = h^\az_{\ \cz}$,
$\gamma^\az_{\ \bz} \gamma^\bz_{\ \cz} = \gamma^\az_{\ \cz}$, and
$\sigma^\az_{\ \bz} \sigma^\bz_{\ \cz} = \sigma^\az_{\ \cz}$. 

Let $\epsilon^{\az \bz \cz \dz}$ be the four-dimensional Levi-Cevita
tensor defined over $M$. Then, fix the orientation of the corresponding 
Levi-Cevita tensors on $\Sigma_t$, $B$, and $\Omega_t$ by setting
\bea
\epsilon_{\Sigma}^{\bz \cz \dz} &\equiv& 
u_\az \epsilon^{\az \bz \cz \dz}, \label{eps} \\
\epsilon_B^{\az \cz \dz} &\equiv& 
\bn_\bz \epsilon^{\az \bz \cz \dz}, \mbox{ and} \nn \\
\epsilon_{\Omega}^{\cz \dz} &\equiv& 
u_\az n_\bz \epsilon^{\az \bz \cz \dz}
= \bu_\az \bn_\bz \epsilon^{\az \bz \cz \dz}. \nn
\eea
Often, where it won't cause confusion, I drop the subscripts to 
get a slightly tidier notation.

Coordinate invariant integrals on $M$ and the various hypersurfaces
are defined in terms of tensor densities (relative
tensors of weight one). Thus, the rest of this thesis should really
be formulated in terms of tensor densities rather than tensors to 
maximize its aesthetics and remove any appearance of coordinate 
dependence (similar work is formulated in that way in 
\cite{jolien,wald1}). For ease of reading however, it is more convenient
to principally
stick with tensors and a coordinate system over the region $M$. 
The final results will come out the same. 

That said, assume that one can 
define a coordinate system $\{r, \theta, \phi\}$ on 
$\Sigma_{t1}$ such that $\Omega_t$ is surface of constant $r$.
Continuously extend it to the other $\Sigma_t$ surfaces so that
$\{t,r,\theta,\phi \}$ is a coordinate system over $M$ and 
$B$ is a constant $r$ surface. Then, if 
$\hat{\epsilon}^{\az \bz \cz \dz}$, $\hat{\epsilon}_{\az \bz \cz \dz}$, 
$\hat{\epsilon}_{\Sigma}^{\az \bz \cz}$, 
$\hat{\epsilon}^{\Sigma}_{\az \bz \cz}$, 
$\hat{\epsilon}_{B}^{\az \bz \cz}$, $\hat{\epsilon}^{B}_{\az \bz \cz}$, 
$\hat{\epsilon}_{\Omega}^{\az \bz}$, and
$\hat{\epsilon}^{\Omega}_{\az \bz}$ are the Levi-Cevita symbols
(relative tensors of weights $\pm 1$) in the spaces $M,\Sigma_t, B, 
\mbox { and }\Omega_t$ respectively with orientations chosen
to match those of the corresponding tensors, the determinants of the
coordinate representations of the metrics are the scalar
functions $g$, $h$, $\gamma$, and $\sigma$ that satisfy the relations
\bea
-g \hat{\epsilon}_{\az \bz \cz \dz} &=& \hat{\epsilon}^{\kz \lz \mz \nz}
g_{\az \kz} g_{\bz \lz} g_{\cz \mz} g_{\dz \nz} \label{dets} \\
h \hat{\epsilon}^{\Sigma}_{\az \bz \cz} 
&=& \hat{\epsilon}_{\Sigma}^{\kz \lz \mz}
h_{\az \kz} h_{\bz \lz} h_{\cz \mz}, \nn \\
- \gamma \hat{\epsilon}^B_{\az \bz \cz} &=& \hat{\epsilon}_B^{\kz \lz \mz}
\gamma_{\az \kz} \gamma_{\bz \lz} \gamma_{\cz \mz}, \mbox{ and } \nn \\
\sigma \hat{\epsilon}^\Omega_{\az \bz} &=& 
\hat{\epsilon}_\Omega^{\kz \lz}
\sigma_{\az \kz} \sigma_{\bz \lz}. \nn
\eea
Combining these relations with equations (\ref{eps}) it 
is straightforward to show $-g = N^2 h$ and $-\gamma = \bN^2 \sigma$.

Define the following extrinsic curvatures.  Taking $\nabla_\az$ as
the covariant derivative on $\cM$ compatible with $g_{\az \bz}$, the
extrinsic curvature of $\Sigma_t$ in $\cM$ is $ K_{\az \bz} \equiv -
h^\cz_{\ \az} h^\dz_{\ \bz} \nabla_\cz u_\dz = - \frac{1}{2} \pounds_u
h_{\az \bz}$, where $\pounds_u$ is the Lie derivative in the direction
$u^\az$. The extrinsic curvature of $B$ in $\cM$ is 
$\Theta_{\az \bz} = -
\gamma^\cz_{\ \az} \gamma^\dz_{\ \bz} \nabla_\cz \bn_\dz$ while the
extrinsic curvature of $\Omega_t$ in $\Sigma_t$ is $k_{\az \bz} \equiv -
\sigma^\cz_{\ \az} \sigma^\dz_{\ \bz} \nabla_\cz n_\dz$. Contracting
each of these with the appropriate metric define 
$K \equiv h^{\az \bz}
K_{\az \bz}$, $\Theta \equiv \gamma^{\az \bz} \Theta_{\az \bz}$, and $k
\equiv \sigma^{\az \bz} k_{\az \bz}$. The addition of an overbar to
any quantity will indicate that it is defined with respect to $\bu^\az$
and/or $\bn^\az$ rather than $u^\az$ and $n^\az$ -- for example,
$\bk \equiv - \sigma^{\az \bz} \nabla_\az \bn_\bz$. 

Further, define the following intrinsic quantities over $\cM$ and
$\Sigma_t$. In $\cM$, the Ricci tensor, Ricci scalar, and Einstein
tensor are $\cR_{\az \bz}$, $\cR$, and $G_{\az \bz}$ respectively.
$D_\az$ is the covariant derivative on $\Sigma_t$ 
compatible with $h_{\az \bz}$, 
and $d_\az$ is the covariant derivative on $\Omega_t$ compatible with 
$\sigma_{\az \bz}$. $R_{\az \bz}$ and $R$ are the 
Ricci tensor and scalar intrinsic to the $\Sigma_t$ hypersurfaces. 
The sign convention for the Riemann tensor is such that 
$\nabla_\az \nabla_\bz \omega_\cz - \nabla_\bz \nabla_\az \omega_\cz
= \cR_{\az \bz \cz}^{\ \ \ \ \: \dz} \omega_\dz$ for a covariant
vector field $\omega_\az$. 

Finally, from the preceding it is clear that tensors defined over
$M$ will usually be written with Greek indices. However, in cases
where these tensors can defined entirely in the tangent and cotangent
bundles of the surfaces $\Sigma_t$ they will often be written with
Latin indices instead.

\section{Field equations}
\label{appMatter}
With the stage set, I now consider the fields that are the players in
this spacetime.

\subsection{The 4D field equations}
\label{4Dfield}
Consider spacetimes with a cosmological constant
$\Lambda$, a massless scalar field $\phi$ (the dilaton), and a Maxwell
field $F_{\az \bz}$. In units where $c$, $\hbar$, and $G$ are unity, 
the field equations are:
\bea
\frac{1}{2}
\epsilon^{\az \bz \cz \dz} \nabla_\bz F_{\cz \dz} &=& 0, \label{dF} \\
\nabla_\bz (e^{-2a\phi} F^{\az \bz} ) &=& 0, \label{nabF} \\
\nabla^\az \nabla_\az \phi + \frac{1}{2} a e^{-2 a \phi} F_{\az \bz} 
F^{\az \bz} &=& 0, \mbox{ and} \label{dil}\\
G_{\az \bz} + \Lambda g_{\az \bz} - \kappa T_{\az \bz} &=& 0 \label{Ein},
\eea
where $a$ is the coupling constant between the scalar and Maxwell fields, $\kappa \equiv 8 \pi$ (it would take on a less trivial value 
in other systems of units) and 
\be
\label{Tab}
T_{\az \bz} \equiv  \frac{1}{4 \pi} \left( [\nabla_\az \phi][\nabla_\bz \phi] 
- \frac{2}{\kappa} [\nabla^\cz \phi][\nabla_\cz \phi] g_{\az \bz} 
+ e^{-2a\phi}[F_{\az \cz} F_\bz^{\ \cz} - \frac{1}{4} g_{\az \bz} F_{\cz
\dz}F^{\cz \dz}]  \right)
\ee
is the stress-energy tensor associated with the matter. From the
field equations it is clear that there are no 
EM or dilaton charges or currents in the region under consideration. 
I work with the sign convention that $\Lambda$ is positive for 
deSitter space.

The first equation implies that, at least locally, it
is possible to define a vector potential $A_\az$ such that 
$F_{\az \bz} = \partial_\az A_\bz - \partial_\bz A_\az$. Conversely
if one takes the vector potential as a pre-existing field and $F_{\az \bz}$ as a quantity derived from it, then equation (\ref{dF})
automatically holds (it simply expresses the identity $d(dA)=0$
for any differential form $A$). This is a common viewpoint, and
from chapter \ref{matterChapter} onwards,
$A_\az$ will be taken as the primary field and so equation (\ref{dF})
will be that identity. The other equations of motion will then 
be derived from the variational principle.

\subsection{The 3D field equations}
\label{3dfieldeq}
Much of this thesis works with Hamiltonians and as such it will be 
useful to know how these field equations project down into the 
three-dimensional spatial hypersurfaces $\Sigma_t$. First define
(dilaton modified) three-dimensional electric and magnetic vector 
fields in the usual way. That is,
\bea
E_\az &\equiv& e^{-2a\phi} F_{\az \bz} u^\bz \mbox{ and}\\
B_\az &\equiv& \frac{1}{2} u_\dz 
\epsilon^{\dz \az \bz \cz} F_{\bz \cz}.
\eea
Conversely $F_{\az \bz}$ and may be rewritten
in terms of the electric and magnetic field 
three-vectors as
\bea
F_{\az \bz} &=& e^{2a\phi}(u_\az E_\bz - u_\bz E_\az) 
+ u^\dz \epsilon_{\dz \az \bz \cz} B^\cz
\label{Fab}.
\eea
While these are the most commonly seen definitions of the electric
and magnetic fields, for much of the following 
it will be more convenient to work with the related vector densities 
on the $\Sigma_t$ hypersurfaces, defined by
\be
\cE^\az \equiv - \frac{2 \sqrt{h}}{\kappa} E^\az \mbox{ and } 
\cB^\az \equiv \frac{2 \sqrt{h}}{\kappa} B^\az.
\ee
With respect to these vector field densities, the Maxwell
equations can be projected into components perpendicular and 
parallel to the $\Sigma_t$ hypersurfaces using the identities
\bea
\frac{\sqrt{h}}{\kappa} u_\az \epsilon^{\az \bz \cz \dz} \nabla_\bz
F_{\cz \dz} &=& D_\bz \cB^\bz, \\
\frac{N \sqrt{h}}{\kappa} h^\mu_\az \epsilon^{\az \bz \cz \dz} 
\nabla_\bz F_{\cz \dz} &=&
h^\mu_\bz {\pounds_u} [N \cB^\bz] -  
h^\mu_\bz u_\az \ez^{\az \bz \cz \dz} D_\cz \left[ N e^{2a \phi}
\cE_\dz \right], \\
\frac{2 \sqrt{h}}{\kappa} u_\az \nabla_\bz (e^{-2a\phi} F^{\az \bz}) 
&=& D_\bz \cE^\bz, \mbox{ and}\\
\frac{2 N \sqrt{h}}{\kappa} h^\mu_\az  \nabla_\bz (e^{-2a\phi} F^{\az \bz}) &=&
h^\mu_\bz {\pounds_u} [N \cE^\bz] + 
h^\mu_\bz u_\az \ez^{\az \bz \cz \dz} D_\cz \left[ N e^{-2a \phi}
\cB_\dz \right].
\eea
If these equations were written 
with respect to vector fields instead of vector field densities
they would include unaesthetic extrinsic curvature terms. 

Then, for time derivatives defined as the Lie derivative with 
respect to the time vector
$T^\az$ rather than $u^\az$ and using Latin indices to emphasize 
that all quantities are defined exclusively in the hypersurface,
the three-dimensional Maxwell equations are
\bea
D_b \cB^b &=& 0, \label{DB}\\
h^b_\bz \pounds_T \cB^\bz &=& \ez^{bcd} D_c \left[ N e^{-2a\phi} \cE_d
\right] + \pounds_V \cB^b, \label{cB}\\
D_b \cE^b &=& 0, \mbox{ and}\label{DE}\\
h^b_\bz \pounds_T \cE^\bz &=& - \ez^{bcd} D_c \left[ N e^{2a\phi} \cB_d
\right] + \pounds_V \cE^b, \label{cE}
\eea
where $\ez^{bcd} = u_\az \ez^{\az b c d}$ as 
was defined in the previous section. 

Next consider the dilaton equation (\ref{dil}). It takes its simplest
three-dimensional form written in terms of the scalar density
\be
\wp \equiv \frac{2 \sqrt{h}}{\kappa} \pounds_u \phi. 
\ee
Then
\bea
&& 2 \frac{N \sqrt{h}}{\kappa} \left( \nabla^\az \nabla_\az \phi + 
\frac{a}{2} e^{-2a\phi} F^{\az \bz} F_{\az \bz} \right) \\
&&= - {\pounds_u (N {\wp})} + \frac{2 \sqrt{h}}{\kappa} D^b [N D_b \phi]
+ a \frac{N \kappa}{2 \sqrt{h}} \left( e^{-2a\phi} \cB^b \cB_b 
- e^{2 a \phi} \cE^b \cE_b  \right), \nn
\eea
or equivalently the time rate of change of $\wp$ is
\be
{ \pounds_T \wp }
= \frac{2 \sqrt{h}}{\kappa} D^c [N D_c \phi] 
+ a \frac{N \kappa}{2 \sqrt{h}} \left( e^{-2a\phi} \cB^b \cB_b 
- e^{2 a \phi} \cE^b \cE_b \right) + \pounds_V \wp, \label{wpT}
\ee
where again I've used Latin indices to emphasize the three-dimensional
nature of the equation.

Finally consider the projections of Einstein's equation (\ref{Ein}).
There are three: time-time, time-space, and space-space. Again it
is most convenient to work with a tensor density, namely
\be
P^{\az \bz} \equiv \frac{\sqrt{h}}{2 \kappa} 
\left( K h^{\az \bz} -  K^{\az \bz} \right), \label{Pab}
\ee
which is contracted as $P = h_{\az \bz} P^{\az \bz}$. 
Ignoring the matter terms, the equations project as
\bea
\cH &\equiv& - \frac{\sqrt{h}}{\kappa} (G_{\alpha \beta} 
+ \Lambda g_{\az \bz}) u^\alpha  u^\beta \\
&=& -\frac{\sqrt{h}}{2 \kappa} 
\left( R - 2 \Lambda \right) + \frac{2 \kappa}{\sqrt{h}}
\left(P^{\az \bz} P_{\az \bz} - \frac{1}{2} P^2 \right), \nn \\
\cH_\bz &\equiv& \frac{\sqrt{h}}{\kappa} h_\bz^\cz 
(G_{\cz \dz} + \Lambda g_{\cz \dz} ) u^\dz  \\
&=&  -2 D_\bz P^{\ \bz}_\az, \hspace{.5cm} \mbox{ and} \nn \\
\cH^{\az \bz} &\equiv&
 \frac{N \sqrt{h}}{2 \kappa} h^{\az \cz} h^{\bz \dz} 
( G_{\cz \dz} + \Lambda g_{\cz \dz} ) \\
& = & 
{ h^\az_\cz h^\bz_\dz \pounds_u (N P^{\cz \dz}) } + \frac{N \sqrt{h}}{2 
\kappa} \left( R^{\az \bz} - \frac{1}{2} R h^{\az \bz} + \Lambda h^{ab} \right) \\
&& - \frac{\sqrt{h}}{2 \kappa} \left( D^\az D^\bz N - h^{\az \bz} 
D^\cz D_\cz N \right)
- \frac{N \kappa}{\sqrt{h}} \left( P^{\cz \dz} P_{\cz \dz} - 
\frac{1}{2} P^2 \right) h^{\az \bz} \nn \\
&& + 4\frac{N \kappa}{\sqrt{h}}
\left(P^{\cz (\az} P^{\ \bz)}_\cz - \frac{1}{2} P P^{\az \bz} \right). 
\nn
\eea
Next, the matter terms project as
\bea
\cH' &\equiv& \sqrt{h} T_{\az \bz} u^\alpha u^\beta \nn \\
&=& \frac{\sqrt{h}}{\kappa} [D^\cz \phi][D_\cz \phi] + 
\frac{\kappa}{4 \sqrt{h}} \wp^2 + \frac{\kappa}{4 \sqrt{h}}
\left( e^{2a\phi} \cE^\az \cE_\az + e^{-2a\phi} \cB^\az \cB_\az \right),
\\
\cH'_\bz &\equiv& - \sqrt{h} h_\bz^\cz T_{\cz \dz} u^\dz =
\wp D_\bz \phi + \frac{\kappa}{2 \sqrt{h}} \ez_{\bz \cz \dz} 
\cE^\cz \cB^\dz, \hspace{.5cm} \mbox{ and}\\
\cH'^{\az \bz} &\equiv& - \frac{N \sqrt{h}}{2} h^{\az \cz} h^{\bz \dz} 
T_{\cz \dz} \nn \\
&=& - \frac{N \sqrt{h}}{\kappa} \left( [D^\az \phi][D^\bz \phi] - \frac{1}{2} h^{\az \bz} [D^\cz \phi][D_\cz \phi] \right)
- \frac{N \kappa}{8 \sqrt{h}} \wp^2 h^{\az \bz} \\
&& - \frac{N \kappa}{4 \sqrt{h}}
\left( e^{2a\phi} [\cE^\az \cE^\bz- \frac{1}{2}\cE^\cz \cE_\cz h^{\az \bz}] \right) 
- \frac{N \kappa}{4 \sqrt{h}}
\left( e^{-2a\phi} [\cB^\az \cB^\bz- \frac{1}{2}\cB^\cz \cB_\cz h^{\az \bz}] \right). \nn
\eea
Again the right-hand sides of the equations are composed entirely
of three-surface terms and so could be written with Latin indices.

Combining the projections, the Einstein equations may be rewritten as
two constraints and a time evolution equation. They are
\bea
\cH^m &\equiv& \cH + \cH' = 0, \label{C1} \\
\cH^m_b &\equiv& \cH_b + \cH'_b = 0, \mbox{ and} \label{C2} \\
h^a_\cz h^b_\dz \pounds_T P^{\cz \dz} 
& = & - \frac{\sqrt{h}}{2 \kappa} \left( N ^{(3)}G^{ab} + \Lambda h^{ab} 
- \left[ D^a D^b N - h^{ab} D_c D^c N \right] \right) \label{C3} \\
&& + \frac{N \kappa}{\sqrt{h}} \left(
[ P^{cd} P_{cd} - \frac{1}{2} P^2 ] h^{ab} 
- 4 [ P^{c(a}P_c^{\ b)} - \frac{1}{2} P P^{ab} ] \right) \nn \\
&& + \pounds_V P^{ab} - \cH'^{ab}, \nn
\eea
where $^{(3)}G_{ab} \equiv R_{ab} - (1/2) R h_{ab}$. 

\subsection{3D electromagnetic potentials}
As noted at the end of section \ref{4Dfield}, either by assumption or
equation (\ref{DB}) there (locally) exists an electromagnetic 
vector potential $A_\az$ such that $F_{\az \bz} = \partial_\az A_\bz
- \partial_\bz A_\az$. Here I examine how that four-dimensional
potential breaks up into the regular Coulomb potential and a 
three-dimensional vector potential. To wit, define
\bea
\Phi \equiv - A_\az u^\az  & \mbox{ and} & 
\tilde{A}_\az \equiv h_\az^\bz A_\bz.
\eea
Then,
\bea
E_\bz &=& - e^{-2a\phi} \left( \frac{1}{N} D_\bz (N\Phi) + 
h_\bz^\cz \pounds_u \tilde{A}_\cz \right) \mbox{ and} \label{Epot} \\
B^\bz &=&  u_\az \epsilon^{\az \bz \cz \dz}
D_\cz \tilde{A}_\dz \label{Bpot}.
\eea
Strictly on the hypersurface and with respect Lie derivatives in 
the $T^\az$ direction, these become a time evolution equation for
the three-vector potential and a definition of 
$\cB^\az$ in terms of the vector potential respectively. That is,
\bea
h_b^\cz \pounds_T
\tilde{A}_\cz &=& \frac{N \kappa}{2 \sqrt{h}} e^{2a \phi}
\cE_b + \pounds_V \tilde{A}_a - D_a [N \Phi], \label{Et} \mbox{ and } \\
\cB^b &=& \frac{2 \sqrt{h}}{\kappa} \ez^{bcd} D_c \tilde{A}_d.
\eea

\subsection{Duality}
\label{dualreview}

Defining the dual $\star F_{\az \bz} = \frac{1}{2} e^{-2 a \phi} 
\epsilon_{\az \bz}^{\ \ \ \cz \dz} F_{\cz \dz}$ of $F_{\az \bz}$,
the four-dimensional field equations (\ref{dF}, \ref{nabF}, \ref{dil},
and \ref{Ein}) may be written as 
\bea
\nabla_\bz (e^{2a\phi} \star \! \! F^{\az \bz} ) &=& 0, \label{nabstarF} \\
- \frac{1}{2}
\epsilon^{\az \bz \cz \dz} \nabla_\bz \star \! \! F_{\cz \dz} &=& 0, 
\label{dstarF} \\
\nabla^\az \nabla_\az \phi - \frac{1}{2} a e^{2 a \phi} \star \! \! 
F_{\az \bz} \star \! \! F^{\az \bz} &=& 0, \mbox{ and} \\
G_{\az \bz} + \Lambda g_{\az \bz} - 8 \pi T_{\az \bz} &=& 0,
\eea
respectively where this time the stress energy is
\be 
T_{\az \bz} =  \frac{1}{4 \pi} \left( [\nabla_\az \phi][\nabla_\bz \phi] 
- \frac{1}{2} [\nabla^\cz \phi][\nabla_\cz \phi] g_{\az \bz} 
+ e^{2a\phi}[ \star F_{\az \cz} \star \! \! F_\bz^{\ \cz} 
- \frac{1}{4} g_{\az \bz} \star \! \! F_{\cz \dz} \star \! \! F^{\cz \dz}]  
\right).
\ee 
Thus, the four-dimensional field equations as a set are invariant
under the full duality
transformation $( \phi \rightarrow - \phi, F_{\az \bz} \rightarrow \star
F_{\az \bz})$. Note however that this time it is equation 
(\ref{dstarF}) that is 
equivalent to a statement that there (locally) exists a 
potential one form. It implies that there exists an 
$A^\star_\az$ such that $\star F_{\az \bz} = 
\partial_\az A^\star_\bz - \partial_\bz A^\star_\az$.

Next,
in terms of $\star F_{\az \bz}$ the electric and magnetic vector
fields may be written as
\bea
E^\bz &=& 
 \frac{1}{2} u_\az \epsilon^{\az \bz \cz \dz} \! \! 
\star \! \! F_{\cz \dz} \mbox{ and} \\
B_\az &=& -e^{2a\phi} \! \star \! \! 
F_{\az \bz} u^\bz, 
\eea 
or equivalently
\bea
\star F_{\az \bz} &=& -e^{-2a\phi}(u_\az B_\bz - u_\bz B_\az) 
+ u^\dz \epsilon_{\dz \az \bz \cz} E^\cz \label{Fstarab}.
\eea
Thus, with respect to these fields, the duality transform
$F_{\az \bz} \rightarrow \star F_{\az \bz}$ becomes $E_\az
\rightarrow - B_\az$ and $B_\az \rightarrow E_\az$. Combining 
these two with $\phi \rightarrow - \phi$ and the corresponding
$\wp \rightarrow -\wp$, it is a simple exercise
to check that the set of three-dimensional field equations for gravity,
electromagnetism, and the dilaton field are also unchanged. 

Decompose $A^\star_\az$ in the same way as $A_\az$. That is define
\bea
\Phi^\star = - A^\star_\az u^\az &\mbox{and}& \tilde{A}^\star_\az = h_\az^\bz A^\star_\bz.
\eea  
In terms of these potentials the electric and
magnetic fields may be written as
\bea
E^\bz &=& u_\az \epsilon^{\az \bz \cz \dz} 
D_\cz \tilde{A}^\star_\dz \mbox{ and} \\
B_\az &=& e^{2a\phi} \left( \frac{1}{N} D_\az (N \Phi^\star)  
+ \pounds_u \tilde{A}^\star_\az \right),
\eea
which alternately 
may be viewed as a definition of $\cE^a$ in terms of 
$A^\star_a$ and a time evolution equation for $A^\star_a$. Namely,
\bea
h_b^\cz \pounds_T \tilde{A}^\star_\cz 
&=& \frac{N \kappa}{2 \sqrt{h}} e^{- 2a \phi}
\cB_b + \pounds_V \tilde{A}^\star_a - D_a [N \Phi], \mbox{ and } \\
\cE^b &=& - \frac{2 \sqrt{h}}{\kappa} \ez^{bcd} D_c \tilde{A}^\star_d.
\eea

\chapter{A quasilocal Hamiltonian for gravity}
\label{gravChapter}

This chapter presents a quasilocal Hamiltonian formulation 
of gravity. I start in section \ref{gravLag}
with the standard Einstein-Hilbert
action and then in section \ref{gravHamSect} 
use a temporal foliation and
Legendre transform to define a Hamiltonian functional for general
relativity over a finite region of a spacelike surface and with respect
to an arbitrary time evolution. This Hamiltonian
and its derivation are similar to the well known ADM formalism
\cite{ADM} though here the analysis is conducted for
a finite and bounded region of a larger spacetime. I confirm that 
the proposed functional correctly generates the equations of
motion and show how its boundary terms can be used to define
quasilocal quantities such as mass, energy, and angular momentum.
These boundary terms depend only on the values of the fields at the 
boundaries.

With these quasilocal concepts defined, in section \ref{energySect} I consider conserved
quantities and calculate the time rates of change of their non-conserved
equivalents. Next section \ref{gravtransform} examines how they 
transform with respect to boosts of the time evolution 
vector field and 
shows that those transformation laws are pleasingly Lorentz-like.
From there, in section \ref{RefTerm} 
I survey a variety of proposals about
how to define the zero of the action and Hamiltonian and discuss
the specific instances in which each is useful. 
Finally, section \ref{thinshell} 
examines the close relationship between
the quasilocal formalism and the thin shell work of Israel. 
This relationship makes it possible to recast the
definition of quasilocal energy from an operational point of view 
and at the same time use quasilocal insights to shed light on the
physics of thin shells. 

Most of this work was published in 
\cite{nopaper} and parts of \cite{naked} and \cite{tidal}.
This thesis however is the first place where the variation 
of the Hamiltonian has been explicitly calculated.

\section{The gravitational Lagrangian}
\label{gravLag}

Given $M \subset \cM$ 
as described in the previous chapter and allowing for the inclusion of a 
cosmological constant, the appropriate action for general relativity
is
\bea
\label{action}
I &=& \frac{1}{2 \kappa} \int_M d^4 x \sqrt{-g} (\cR - 2 \Lambda) 
+ \frac{1}{\kappa} \int_\Sigma d^3 x 
\sqrt{h} K - \frac{1}{\kappa} \int_B d^3 x \sqrt{- \gamma} \Theta \\ 
&&+ \frac{1}{\kappa} 
\int_\Omega d^2 x \sqrt{\sigma} \sinh^{-1} \eta + \bI, \nn 
\eea
where $\int_\Sigma = \int_{\Sigma_2} - \int_{\Sigma_1}$, $\int_\Omega =
\int_{\Omega_2} - \int_{\Omega_1}$, and, choosing a system of units
where $c$, $\hbar$, and $G$ are unity, $\kappa = 8 \pi$.  
The $\sinh^{-1} \eta$ 
term is added so that the variation of the action will still be
well defined if the boundaries are not orthogonal to each other
at their intersection. It was first discussed in \cite{hayward}.
$\bI$ is any functional of the boundary metrics on $\partial M$. 

To see that this is indeed the correct action, take its variation with 
respect to the metric $g_{\az \bz}$. The result is \cite{hayward}
\bea
\label{firstVar}
\delta I &=& \frac{1}{2 \kappa} \int_M d^4 x \sqrt{-g} (G_{\az \bz} + 
\Lambda g_{\az \bz})  \delta g^{\az \bz} \\
&& + \int_\Sigma d^3 x  \left( P^{\az \bz} \delta h_{\az 
\bz} \right) + \int_B d^3 x \left( \pi^{\az \bz} \delta \gamma_{\az \bz} 
\right) \nn \\
&& +  \int_\Omega d^2 x 
\left( \frac{1}{ \kappa} \sinh^{-1}(\eta) \delta \sqrt{\sigma} \right)
+ \delta \underline{I}  , \nn
\eea
where $P^{\az \bz} \equiv \frac{\sqrt{h}}{2 \kappa} 
\left( K h^{\az \bz} - K^{\az \bz} \right)$ is the 
same tensor density defined by equation (\ref{Pab}) in the
previous chapter, and 
$\pi^{\az \bz} \equiv - \frac{\sqrt{-\gamma}}{2 \kappa} 
\left( \Theta \gamma^{\az \bz} - \Theta^{\az \bz} \right)$ is an 
equivalent tensor density defined by the surface $B$. 

For variations that leave the boundary metrics 
$h_{\az \bz}$ and  $\gamma_{\az \bz}$ fixed, the  
boundary terms, and $\delta \underline{I}$ vanish. Then
$\delta I = 0$ if and only if the Einstein's equations hold over all
of $M$. Thus with these terms fixed, 
the variation of $I$ is properly defined and 
generates general relativity as asserted.

\section{The gravitational Hamiltonian}
\label{gravHamSect}

\subsection{Form of the Hamiltonian}
\label{Hamform}

With this quasilocal action in hand, it is a relatively simple matter to
obtain the corresponding quasilocal Hamiltonian. The process is to
decompose the action with respect to the foliation and then identify
the Hamiltonian and momentum terms. Details of the calculation may
be found in appendix \ref{appGravHam}, but here I'll just present the 
results. Breaking it up with respect to the foliation the action may be
written as,
\bea
\label{MgravActdecomp}
I - \bI &=& \int dt \left\{ \int_{\Sigma_t}  d^3 x 
\left( P^{\az \bz} \pounds_T h_{\az \bz} \right)
+ \int_{\Omega_t} d^2 x \left( P_{\ssg} \pounds_T \sqrt{\sigma}
\right) - H_t \right\} 
\eea  
where $P^{\az \bz}$ retains its meaning from the previous chapter, 
$P_{\ssg} \equiv \frac{1}{\kappa} \sinh^{-1} \eta$, and
\bea
\label{gravHam}
H_t = 
\int_{\Sigma_t} d^3 x [N \cH + V^{\az} \cH_{\az}] 
+ \int_{\Omega_t} d^2 x \sqrt{\sigma} (\bN \bep - \bV^\az \bj_{\az}),
\eea
where $\cH$ and $\cH_{\az}$ are the matter free versions of the 
Einstein constraint equations (\ref{C1}) and (\ref{C2}).
$\bN$ and $\bV^\az$ are the boundary lapse and shift defined following
equation (\ref{boundaryTime}), while 
\bea
\bep 
&\equiv& \frac{1}{\kappa \lambda} k + \eta \frac{2}{\sqrt{h}} P^{ab} n_a n_b
= \frac{1}{\kappa} \bk, \mbox{ and} \label{bep}\\
\bj_\az 
&\equiv& - \frac{2}{\sqrt{h}} \sigma_{\az \cz} P^{\cz \dz} n_\dz 
- \frac{\lambda}{\kappa} \sigma^\bz_\az \partial_\bz \eta
= - \frac{1}{\kappa} \sigma_\az^\bz \bu^\cz \nabla_\bz \bn_\cz.
\label{bj}
\eea
Shortly $\bep$ and $\bj_\az$ will be identified  
as related to energy and angular momentum respectively 
but for now simply note that despite the initial appearance
of these terms, their second versions show that they 
are really defined with respect $\bu^\az$, $\bn^\az$, and $\sigma_{\az \bz}$ and as such 
are defined with respect to the foliation $\Omega_t$ 
of the boundary and the normals $\bu^\az$ and $\bn^\az$ 
rather than the foliation $\Sigma_t$ and its normals $u^\az$
and $n^\az$.

To motivate the definition of the Hamiltonian, 
recall that in elementary classical mechanics with
one degree of freedom, the action $I$ of a path $q=\Gamma(t)$
taken by a particle is given by $I = \int_\Gamma L(\Gamma(t)) dt$
where $L$ is the Lagrangian function and the integral is 
over the path. This is related to the 
Hamiltonian $H$ by the relation 
$L = p \dot{q} - H$, where $q$ is the variable giving
the configuration of the system and $p = \frac{\partial L}{\partial \dot{q}}$
is the momentum conjugate to $q$. Extending this analysis to
gravitational fields \cite{BY1} and referring back to 
equation (\ref{MgravActdecomp}), $h_{\az \bz}$ may be identified 
as a configuration variable
on the spatial $\Sigma_t$ surfaces 
and $P^{\az \bz}$ recognized as its conjugate momenta. Further 
$\sqrt{\sigma}$ is seen to be a configuration variable on $\Omega_t$ 
(albeit one that is not independent of $h_{\az \bz}$) 
and $P_{\sqrt{\sigma}}$ is its conjugate momentum. 
Finally perform an effective Legendre transform by identifying quantity $H_t$ as the required quasilocal Hamiltonian.

The bulk term integrand of $H_t$
is $N \cH + V^\az \cH_\az$ where $\cH$ and $\cH_\az$ are the
Einstein constraint equations (\ref{C1}) and (\ref{C2}). Then, in the
standard Hamiltonian way,
the lapse and shift are identified as Lagrange
multipliers rather than configuration variables. Further since the
constraints will be zero for solutions to the Einstein equations, 
the actual numerical value of $H_t$ will be
a functional of the boundary $\Omega_t$ and its normals $\bu^\az$
and $\bn^\az$ only (and recall that these normals are 
fixed by $\Omega_t$ and $T^\az$ without reference to $\Sigma_t$). 
Therefore 
the evaluation of $H_t$ doesn't require any knowledge of the
surface $\Sigma_t$ apart from the fact that is has a boundary
$\Omega_t$. This indifference to the bulk will be considered
further in section \ref{energySect}.

By contrast, the nonorthogonal Hamiltonian proposed by Hawking and 
Hunter in reference \cite{hhunter} focused on the foliation 
surfaces $\Sigma_t$ and normals $u^\az$ and $n^\az$, which meant
that their Hamiltonian was explicitly dependent on the intersection
angle parameter $\eta$. They had to resort to a clever choice of the
reference term $\bI$ to remove this dependence. 

\subsection{Variation of the Hamiltonian}
\label{HamVar}
This subsection checks that $H_t$ really does
encode the correct equations of motion for gravity.
To do this, consider $H_t$ as a functional of the surface
$\Sigma_t$, its boundary $\Omega_t$, the normal $n^a$ to that
boundary, the fields $h_{a b}$ and $\ssg$ along with their
conjugate momenta $P^{ab}$ and $P_{\ssg}$, and the Lagrange 
multipliers $N$ and $V^\az$. In the usual Hamiltonian
way the conjugate momenta are taken to be entirely 
independent of $h_{ab}$ and $\ssg$. 
Further $\bep$ and $\bj_a$ are considered to have their first 
meanings from the definitions
(\ref{bep},\ref{bj}), and $\bN$, $\bV^\az$, $\lambda$,
and $\eta$ are defined entirely with respect to $V^\az$, $N$, 
and $n_a$ as expressed by equations (\ref{v}-\ref{bNbV}).
Thus, $H_t$ is a functional on the three-space $\Sigma_t$ rather
in the four-dimensional spacetime $M$.

Now vary $H_t$ with respect to the three-metric $h_{\az \bz}$, the 
conjugate momentum $P^{\az \bz}$, and the
lapse and shift $N$ and $V^\az$. Because $\sqrt{\sigma}$
and $P_{\ssg}$ are functions of $N$, $V^\az$, and
$h_{\az \bz}$, these two secondary quantities are automatically
varied as well. This is an important calculation but
its details are not really pertinent to the main ideas of the thesis, 
so I banish them to appendix \ref{appGravVar} and go straight to the 
final result. The total variation of $H_t$ is
\bea
\label{HamVarEq}
\delta H_t &=& \int_{\Sigma_t} d^3 x \left(
 \cH \delta N + \cH_a \delta V^a + 
[h_{ab}]_T \delta P^{ab} - [P^{ab}]_T \delta h_{ab}  \right) \\
&&+ \int_{\Omega_t} d^2 x \sqrt{\sigma} \left(\bep \delta \bN - 
\bj_a \delta \bV^a - (\bN/2) \bs^{ab} \delta \sigma_{ab} \right) \nn \\
&& + \int_{\Omega_t} d^2 x \sqrt{\sigma} 
\left(\left[ \sqrt{\sigma} \right]_T 
\delta P_{\sqrt{\sigma}} - \left[P_{\sqrt{\sigma}} \right]_T \delta 
\sqrt{\sigma} \right), \nn
\eea 
where $\cH$ and $\cH_a$ retain their previous values, while
\bea
\left[ h_{ab} \right]_T &\equiv& 
\frac{4 \kappa N}{\sqrt{h}}[P_{ab} - \frac{1}{2} P h_{ab}]
+ 2 D_{(a}V_{b)}, \label{dhdt} \\
\left[ P^{ab} \right]_T & \equiv & 
- \frac{\sqrt{h}}{2 \kappa} \left( N [^{(3)}G^{ab}
+ \Lambda h^{ab}] 
- \left[ D^a D^b N - h^{ab} D_c D^c N \right] \right) 
\\
&& + \frac{N \kappa}{\sqrt{h}} \left(
[ P^{cd} P_{cd} - \frac{1}{2} P^2 ] h^{ab} 
- 4 [ P^{c(a}P_c^{\ b)} - \frac{1}{2} P P^{ab} ] \right) \nn \\
&& + \pounds_V P^{ab}, \nn \\
\label{bs}
\bar{s}^{ab} &\equiv& 
\frac{1}{\kappa \lambda} \left( k^{ab} - [k-n^d a_d]
\sigma^{ab} \right) - \frac{2}{\sqrt{h}} \eta \sigma_c^a \sigma_d^b 
P^{cd}\\
&& + \frac{1}{N} \left(
\left[ P_\ssg \right]_T
- \frac{1}{\kappa} \pounds_{\bar{V}} \eta \right) \sigma^{ab}, \nn \\
\left[ \sqrt{\sigma} \right]_T  
&\equiv& 
- \sqrt{\sigma} \left(
N \frac{2}{\lambda \sqrt{h}} 
P^{ab} n_a n_b + N \frac{\eta}{\kappa} k - 
\frac{1}{\kappa} d_b \bV^b \right),
\eea
and $ \left[P_{\sqrt{\sigma}} \right]_T $ is
an undetermined function over $\Omega_t$.
I'll interpret $\bs^{ab}$ in section \ref{gravtransform} and 
show that it is 
actually independent of $\eta$, despite initial appearances. 
If $\eta = 0$ it becomes the stress tensor 
$s^{ab}=(1/\kappa)(k^{ab} - [k-n^d a_d] \sigma^{ab})$ 
considered by Brown and York.

The Hamiltonian equations of motion can now be obtained
by calculating the full variation of the action
(\ref{MgravActdecomp}) (treating the momenta
as independent variables) and solving $\delta I = 0$.
Using the preceding result, only a little work using
the fundamental theorem of calculus\footnote{That is 
$\int_{t1}^{t2} df = f(t_2)- f(t_1)$.}
to move total time 
derivatives to the spacelike boundaries is required to show that
\bea
\delta I - \delta \bI &=& 
\int_\Sigma d^3 x P^{ab} \delta h_{ab} + 
\int_\Omega d^2 x P_{\sqrt{\sigma}} \delta
\sqrt{\sigma} \label{Act2Var}  \\
&& - \int dt \int_{\Sigma_t} d^3 x
\left\{ \cH \delta N + \cH_a \delta V^a \right\} \nn \\
&& + \int dt \int_{\Sigma_t} d^3 x
\left\{ 
 \left( 
\pounds_T h_{ab} -
\left[ 
h_{ab} \right]_T \right) 
\delta P^{ab} 
- \left( 
\pounds_T P^{ab} - 
\left[ P^{ab} \right]_T 
\right) \delta h_{ab}
\right\}
\nn \\
&& + \int dt \int_{\Omega_t} d^2 x \left\{ 
\left( \pounds_T \sqrt{\sigma} -
\left[ \sqrt{\sigma} \right]_T \right) \delta P_{\sqrt{\sigma}} 
- \left( \pounds_T P_{\sqrt{\sigma}} -
\left[ P_{\sqrt{\sigma}} \right]_T \right) \delta 
\sqrt{\sigma} \right\} \nn \\
&& - \int dt \int_{\Omega_t} d^2 x \sqrt{\sigma} \left(
\bep \delta \bN - \bj_a \delta \bV^a - \frac{\bN}{2} \bs^{ab}
\delta \sigma_{ab} \right), \nn
\eea
where again $\int_{\Sigma} \equiv \int_{\Sigma_2} - \int_{\Sigma_1}$ and
$\int_{\Omega} \equiv \int_{\Omega_2} - \int_{\Omega_1}$.

If metrics $\gamma_{\az \bz}$ (equivalently $\sigma_{ab}$, $\bN$,
$\bV^a$) and $h_{ab}$ are held constant on the
timelike and spacelike boundaries respectively, then 
$\delta \bI = 0$ and solving $\delta I = 0$ while allowing
for general variations in the bulk gives the 
following set of equations. 
\bea
\cH &=& 0, \label{Ein1}\\
\cH_a &=& 0, \label{Ein2} \\
\pounds_T P^{ab} &=& \left[ P^{ab} \right]_T, \label{Ein3} \\
\pounds_T h_{ab} &=& \left[ h_{ab} \right]_T,  \label{Geom1} \\
\pounds_T \sqrt{\sigma} &=& \left[ \sqrt{\sigma} \right]_T, 
\label{Geom2} \mbox{ and} \\
\pounds_T P_{\sqrt{\sigma}} &=& \left[ P_{\sqrt{\sigma}} \right]_T 
\label{Free1}.
\eea
Now (\ref{Ein1}), (\ref{Ein2}), and (\ref{Ein3}) are the
(matter free) projected Einstein equations (\ref{C1}),
(\ref{C2}), and (\ref{C3}) respectively, so the Hamiltonian
has recovered those correctly. 
At the same time, (\ref{Geom1}) and (\ref{Geom2})
give the correct expressions for the Lie derivatives of $h_{ab}$
and $\ssg$ as compared to direct geometric calculation.

Finally, equation (\ref{Free1}) correctly expresses the fact
that the time rate of change of $P_{\sqrt{\sigma}}$ is undetermined
by any of the other quantities -- a fact that is to be expected
since in the Lagrangian formulation $P_{\sqrt{\sigma}} 
= \frac{1}{\kappa} \sinh^{-1} \eta$, where 
$\eta = v_\vdash / \sqrt{1 - v_\vdash^2}$ and $v_\vdash = 
(V^\az n_\az)/N$ (equation (\ref{v}) and the surrounding
discussion). The lapse
and shift are Lagrange multipliers whose time evolution is not
determined by the equation of motion. Therefore the evolution
of $P_{\ssg}$ is similarly undetermined. Intuitively, 
this is to be expected
since $v_\vdash$ quantifies the ``radial'' evolution of $\Omega_t$ 
or equivalently the radial ``shape'' of $B$.
The ``shape'' of $B$ is chosen arbitrarily so one would 
certainly not expect $P_{\ssg}$ to be determined by the field
equations.

Thus, $H_t$ is a proper quasilocal Hamiltonian as supposed. 

\subsubsection{Comparison with the Lagrangian approach}
Before moving on to the next section I will compare the
variation of the Hamiltonian with the variation
of the action (and its decomposition with 
respect to the time foliation) as considered in
refs.~\cite{BY1,nopaper}. Specifically I compare
with \cite{nopaper} where we allowed for a non-orthogonal
intersection of $\Sigma_t$ with the boundary $B$. Reference
\cite{BY1}
deals with the special case where $\eta=0$. In those papers,
the variation of the action (equation (\ref{firstVar})) was decomposed
according to the foliation, the key result being that
\be
\label{Bterm}
\pi^{\az \bz} \delta \gamma_{\az \bz} = 
- \ssg \left( \bep \delta \bN - \bj_\az
\delta \bV^\az \right)
+ \frac{\bN \ssg }{2} \bs^{\az \bz} \delta \sigma_{\az \bz},
\ee
where all quantities retain their earlier definition
though with the recognition that $P^{\az \bz} = \sqrt{h}/(2\kappa)
(K h^{\az \bz} - K^{\az \bz})$ (as opposed to 
Hamiltonian calculations which treat it as 
an independent variable). Then, equation (\ref{firstVar})
becomes
\bea
\delta I - \delta \bI &=&
\frac{1}{2 \kappa} \int_M d^4 x \sqrt{-g} \left(G_{\az \bz} + \Lambda
g_{\az \bz} \right) \delta g^{\az \bz} \label{HJGravVar} \\
&& + \int_{\Sigma} d^3 x \left( P^{\az \bz} \delta h_{\az \bz} \right)
   + \int_{\Omega} d^2 x (P_{\ssg} \delta \ssg) \nn \\
&& - \int dt \int_{\Omega_t} d^2 x \sqrt{\sigma} \left( 
\bep \delta \bN - \bj_\az \bV^\az - \frac{\bN}{2} \bs^{\az \bz} 
\delta \sigma_{\az \bz} \right), \nn
\eea
where again $P^{\az \bz}$ is recognized as a function of the
metric $g_{\az \bz}$, its compatible covariant derivative $\nabla_\az$,
and the embedding of $\Sigma_t$ in $\cM$. With this viewpoint equations 
(\ref{Geom1},\ref{Geom2},\ref{Free1}) are automatically
satisfied and so the Lagrangian and Hamiltonian treatments are 
equivalent -- as of course they should be.

\section{Energy and $H_t$}
\label{energySect}
In classical mechanics the value of the Hamiltonian is identified with
the energy of the system under consideration and so by analogy 
Brown and York identified (the hypersurface orthogonal version of)
$H_t$ with the mass/energy contained by the surface 
$\Omega_t$\footnote{Or, more properly 
$H_t$ is the energy {\it associated} with the surface $\Omega_t$ 
as discussed in section \ref{QLEsect}.}.
A key point in favour of this identification is the fact that
for an asymptotically flat spacetime, $H_t$ is numerically
equivalent to the ADM and Bondi masses in the appropriate 
limits (as shown in \cite{BY1} and \cite{BYnull} respectively).

Tentatively making this association, recall that the energy of
a mechanical system is conserved if and only if it is isolated from
all outside influences. Now, a finite gravitational system can be 
considered to be isolated if the metric $\gamma_{\az \bz}$ of $B$
has a timelike Killing vector field and there is no flow of matter
across $B$ (that is $T_{\az \bz} \bn^\az \bu^\bz = 0$). 
That said consider the time rate of change of $H_t$. Because
I am considering a pure gravitational field here, there will 
no matter flows across $B$.

This is a surprisingly easy calculation to do because the 
Hamiltonian variation calculation can be easily recycled to 
do all of the work. 
Equation (\ref{HamVarEq}) showed how the Hamiltonian is changed
by general first order variations of the metrics and
their conjugate momenta. Of course, during that calculation there
was no assumption made that the metrics and momenta satisfied the
Einstein equations -- the point of that calculation
was to derive those equations. 
However, that said, the mechanics of the calculation  
equally well hold for variations that do satisfy the 
equations of motion. In particular consider a region $M$ 
of spacetime with metric $g_{\az \bz}$ that is a 
solution to the Einstein equations. Then, evaluate
$H_t$ over a spatial three-surface $\Sigma_t$ with 
two-boundary $\Omega_t$.
Lie-drag that surface forward by an infinitesimal amount of 
coordinate time, in which case 
$\delta h_{ab} = (\pounds_T h_{ab}) \delta t$,
$\delta \sqrt{\sigma} = (\pounds_T \sqrt{\sigma}) \delta t$,
$\delta P^{ab} = (\pounds_T P^{ab}) \delta t$, and
$\delta P_{\sqrt{\sigma}} = (\pounds_T P_{\sqrt{\sigma}}) \delta t$.
Combining these substitutions with the fact that the Einstein 
equations are satisfied on $\Sigma_t$, the first
and third lines of equation (\ref{HamVarEq}) go to zero and 
leave behind the time rate of change of $H_t$
\be
\label{Hdot}
{\pounds_T} H_t 
\equiv \lim_{\Delta t \rightarrow 0} \frac{\delta H}{\delta t}
= \int_{\Omega_t} d^2 x \sqrt{\sigma} \left\{
\bep {\pounds_T} \bN - \bj_\az {\pounds_T} \bV^\az - \frac{\bN}{2}
\bs^{\az \bz} \pounds_T \sigma_{\az \bz} \right\},
\ee
or alternatively using eq.\ (\ref{Bterm})
\be
\label{Hdot2}
{\pounds_T} H_t = - \int_{\Omega_t} d^2 x \left\{ \pi^{\az \bz} 
\pounds_T \gamma_{\az \bz} \right\},
\ee
which is often the most convenient form for explicit 
calculations\footnote{An alternate calculation of this result
which also allows for matter flows can be found in reference
\cite{tidal}.}.
Note that just as the Hamiltonian itself depended only on the foliation
of the boundary and its associated normals, so does its time rate of
change. What is happening in
the bulk is irrelevant.

${\pounds_T} H_t$ is zero if  ${\pounds_T} \bN = 0$, ${\pounds_T} \bV^a = 0$, and ${\pounds_T} \sigma_{ab} = 0$ or equivalently if
${\pounds_T} \gamma_{\az \bz} = 0$. This is almost the definition 
of an isolated gravitational system (in the absence of matter
flows) that was proposed a couple of paragraphs
back except that there 
only the existence of a Killing vector was required rather
than demanding that $T^\az$ be that vector.
$\pounds_T H_t$ might not be zero even if the Killing vector
exists. As an example consider Schwarzschild space with $B$ as a
surface of constant $r$. Then the lapse function $\bN$ can be 
chosen so that $H_t$ is not a constant even though the Killing vector
exists. See \cite{BY1} for a further discussion of this point.

Viewing $H_t$ as a mass, it is useful to think of $B$ as the
history/future of a set of observers as was discussed in 
section \ref{geometry}. Then, as noted there, the
foliation $\Omega_t$ defines the ``instants'' of time agreed on by
those observers and $T^\az$ defines their four-velocity. Thus, the
quasilocal Hamiltonian can be thought of as a kind of Gauss's law 
for mass, in the sense that it defines the mass contained in the 
bulk without making any reference to what is actually happening there,
just as the electromagnetic Gauss's law defines the electric charge
contained by a
surface based entirely on measurements made at that surface. It then
makes sense that the time rate of change of the Hamiltonian should
also depend only on what happens at the surface since the only way
energy can get in or out of the bulk should be through that surface. 

If the boundary is made up of observers, it is reasonable
that their notion of the energy contained by the surface should
not depend on the bulk foliation. 
There are no observers in the bulk and so there is no natural
way for the boundary observers to globally extend their notion of
simultaneity into that bulk. 
Thinking empirically these observers would say that the foliation of 
the bulk is a fiction invented by theorists that has no external 
reality. Locally the natural foliation for the observers to consider
is $\bar{\Sigma}_t$ -- the foliation that is orthogonal to their four velocity $T^\az$. That is, they would view $\bn^\az$ as the natural normal to the surface $\Omega_t$. If that foliation could be extended
throughout $M$ then the numerical
value of the Brown-York Hamiltonian is identical to its generalized
form considered here. However, whether that extension 
exists or not is irrelevant from the point of view adopted in this
thesis.

A special case of this Hamiltonian definition of energy is when
$T^\az = \bu^\az$. That is, the observers are evolved by 
the timelike boundary unit normal to $\Omega_t$ and measure proper
time. Such observers measure an energy of 
\be
E_{Geo} \equiv \int_{\Omega_t} d^2 x \sqrt{\sigma} \bep \label{Egeo}
\ee 
and it is this energy that is used in applications of the Brown-York
energy to thermodynamics (see for example
\cite{Bthermo,BY2,jolien,jolienrobb}).
Because of this identification $\sqrt{\sigma} \bep$ is usually called
the energy density. For $\eta = 0$ (that is the foliation $\Sigma_t$ is
orthogonal to $B$) it reduces to the Brown-York energy
density but in any case it will be referred to as the
quasilocal energy or QLE.

This measure of quasilocal energy has a nice geometrical 
interpretation and that is the reason for the subscript in
$E_{Geo}$. Specifically, $\ssg \bep = (\ssg/\kappa) \bk = 
- (1/\kappa) \pounds_{\bn} \ssg$ and so measures how the surface
area of $\Omega_t$ changes if it is translated ``radially'' 
outwards in the direction $\bn^\az$. Similarly, the surface area
measures how the volume of the region contained by $\Omega_t$
changes if one ``radially'' translates the surface outwards. 
Now, of course the volume in $\Omega_t$ depends on the curvature
of the space contained therein so it is not unreasonable that
its ``second radial derivative'' might tell one something about
the gravitational energy\footnote{
Contained volume here is a very hand-wavy notion since as has
been already emphasized, the volume 
contained by $\Omega_t$ is really very dependent on the behaviour
of the foliation $\Sigma_t$ and the derivatives themselves depend
on how one chooses the radial normals $\bn^\az$.
However this is a useful way to think about the energy intuitively, 
and in any case corresponds to the volume changes with respect
to the natural local orthogonal foliation $\bar{\Sigma}_t$.}.

Other symmetries of the boundary $B$ correspond to other conserved
charges. I won't consider the details here, but it is not hard 
to show (see for example \cite{BY1}) 
that if $B$ admits an angular (spatial) Killing vector
$\xi_\phi^\az$ then $J = \int_{\Omega_t} d^2 x \ssg \xi_\phi^a \bj_a$ 
is the
charge corresponding to this angular symmetry. In the appropriate limit
it agrees with the ADM definition of angular momentum \cite{BY1}
at spatial 
infinity, and so $\ssg \bj_a$ is usually called the angular momentum
density. Interestingly $\bj_a$ can also be identified with the
connection on the normal bundle to $\Omega_t$. A good  
discussion of this and its implications can be found in 
reference \cite{epp}.

\section{Transformation laws}
\label{gravtransform}
Having defined the quasilocal energy
it is natural to ask what is the relationship between the quasilocal quantities $\bep$, $\bj_\az$, $\bs^{\az \bz}$ as they are seen by
different sets of observers moving with different four-velocities.

Consider two sets of observers who instantaneously coincide on the
surface $\Omega_t$. Let the evolution of the first set of observers
be guided by the $\Sigma_t$ forward-pointing timelike unit normal 
vector $u^\az$ while the second set is evolved by the time 
vector field $T^\az$. Henceforth I'll refer to the $u^\az$ observers
as the $\Sigma_t$ ``unboosted'' observers while the $T^\az$ 
set will be the ``boosted'' observers.

The evolution of the unboosted observers is orthogonal to 
the foliation so they view $u^\az$ and $n^\az$ as the unit
normals to $\Omega_t$. Meanwhile the boosted observers regard 
$\bu^\az$ and $\bn^\az$ as the unit normals. The unboosted observers
measure the radial velocity of the boosted observers as
\be
v_\vdash \equiv - \frac{T^\az n_\az}{T^\az u_\az} = 
\frac{V^\az n_\az}{N},
\ee
and their $\Omega_t$ tangential velocity as
\be
\hat{v}_\az \equiv - \frac{\sigma_{\az \bz} T^\bz}{T^\az u_\az} = 
\frac{\sigma_{\az \bz} V^\bz}{N}.
\ee
Then, recall equation (\ref{veta}) which showed that 
$\eta = \bu^\az n_\az$ and $\lambda =
\frac{1}{\sqrt{1 + \eta^2}}$ can be rewritten as
\bea
\eta = \gamma v_\vdash  & \mbox{ and } & \frac{1}{\lambda} = \gamma,
\nn \eea
where $\gamma = 1/\sqrt{1-v_\vdash^2}$ is the Lorentz factor for
radial $v_\vdash$. With this substitution, 
equations (\ref{nutrans}) can be written as
\bea
\bn^\az = \gamma (n^\az + v_\vdash u^\az) & \mbox{and} & 
\bu^\az = \gamma (u^\az + v_\vdash n^\az).
\eea

The extrinsic curvature of $\Omega_t$ with respect to 
the timelike $u^\az$ is defined as
\be
k^{\updownarrow}_{\az \bz} \equiv - \sigma_\az^\cz
\sigma_\bz^\dz \nabla_\cz u_\dz = - \frac{1}{2}
\sigma_\az^\cz \sigma_\bz^\dz \pounds_u \sigma_{\cz \dz}
\ee 
which can be contracted to
$k^\updownarrow \equiv \sigma^{\az \bz} k^{\updownarrow}_{\az \bz}$.
The rate of change of $n^\az$ in the direction it points is $a_\az^{\updownarrow} \equiv n^\bz \nabla_\bz n_\az$. The choice
of the $\updownarrow$ superscript is meant to suggest an interchange of
$u^\az$ and $n^\az$ in these quantities (as compared to the same
expression without the superscript) and as usual the addition of 
a bar means that they are to be calculated with
respect to $\bu^\az$ and $\bn^\az$ rather than $u^\az$ and $n^\az$.
The quasilocal quantities with $u^\az$ and $n^\az$ interchanged are: 
\bea
\varepsilon^{\updownarrow} &\equiv& \frac{1}{\kappa} k^{\updownarrow}, \\ 
j_\az^{\updownarrow} &\equiv& -\frac{1}{\kappa} 
\sigma_\az^\bz n^\dz \nabla_\bz u_\dz, \mbox{ and}\\
s^{\updownarrow}_{\az \bz} &\equiv& \frac{1}{\kappa} 
\left( k^{\updownarrow}_{\az \bz} - [k^{\updownarrow} 
- u^\cz a^{\updownarrow}_\cz] \sigma_{\az \bz} \right).
\eea
Note that $j^{\updownarrow}_\az = - j_\az$.

Some of these quantities were first used in \cite{Lau} in the context of
defining quantities that are invariant with respect to boosts. The
simplest example of such an invariant is $\varepsilon^2 -
\varepsilon^{\updownarrow 2}$ which is
analogous to $m^2 c^2 = E^2 - p^2 c^2$, an invariant for a
particle with energy $E$ and momentum $p$ in special relativity. This
suggests that $\varepsilon^\updownarrow$ be viewed 
as a momentum flux through the surface $\Omega_t$. 
Support for this interpretation comes from noting that
\be
\label{dArea}
\sqrt{\sigma} \varepsilon^\updownarrow
= -\frac{\sqrt{\sigma}}{2\kappa} \sigma^{\az \bz} 
{\pounds_u} \sigma_{\az \bz}
= -\frac{1}{\kappa} \pounds_u \sqrt{\sigma}.
\ee
That is, $\varepsilon^\updownarrow$ is zero if and only if the
observers don't see the area of the surface they inhabit to be changing.
However, this means that a sphere of observers moving at 
constant radial speed in flat space will measure a momentum flux 
so this isn't entirely in
accord with intuition. Of course without reference terms such observers
will also measure a non-zero quasilocal energy so this is not entirely
unexpected. A more complete discussion of the identification of
$k^\updownarrow$ with momentum may be found in \cite{epp} which
also develops a notion of quasilocal energy from the invariant
$\sqrt{\bep^2 - \bep^{\updownarrow 2}}$ which is closely 
related to the one considered here. 

A series of straightforward calculations leads to expressions 
for the quasilocal quantities seen by the boosted observers in terms of 
quantities measured by the $u^\az$ observers. These transformation
laws are
\bea
\bep &\equiv& -\frac{1}{\kappa} \sigma^{\az \bz} 
\nabla_\az \bn_\bz  \label{epTrans} \\
&=& \frac{1}{\lambda} \varepsilon + \eta \varepsilon^{\updownarrow}
\nn \\
&=& \gamma ( \varepsilon + v_\vdash \varepsilon^{\updownarrow} ), \nn \\
\bj_\az &\equiv& - \frac{1}{\kappa} \sigma_\az^\bz \bu^\cz 
\nabla_{\bz} \bn_\cz \label{jTrans} \\ 
&=& j_\az - \frac{\lambda}{\kappa} \sigma_\az^\bz \partial_\bz 
\eta \nn \\
&=& j_\az - \frac{\gamma^2}{\kappa} \sigma_\az^\bz \partial_\bz
v_\vdash,  \nn  \mbox{   and} \\
\bs_{\az \bz} &\equiv& \frac{1}{\kappa} \left(\bk_{\az \bz} 
- [\bk-\bn^\dz \ba_\dz]\sigma_{\az \bz} \right) \label{bs2} \\
&=& \frac{1}{\lambda} s_{\az \bz} + \eta s^\updownarrow_{\az \bz} 
+ \frac{\lambda}{\kappa} \sigma_{\az \bz} \pounds_{\bu} \eta
\nn \\
&=& \gamma(  s_{\az \bz} + v_\vdash s^\updownarrow_{\az \bz} ) 
+ \frac{\gamma^2}{\kappa} 
\sigma_{\az \bz} \pounds_{\bu} v_\vdash. \nn
\eea
The reader will recall that a quantity $\bs^{\az \bz}$ has already
appeared in equations (\ref{bs}) and (\ref{Bterm}). 
Short calculations show that
$s^{\updownarrow}_{\az \bz} = - \frac{2}{\sqrt{h}} \sigma_{\az \cz}
\sigma_{\bz \dz} P^{\cz \dz} $, and $\pounds_T \eta - \pounds_{\bV} 
\eta = \bN \pounds_{\bu} \eta$ so these two quantities are the same.
Further, the first line of equation (\ref{bs2}) shows that 
$\bs^{\az \bz}$ is independent of $\eta$ and the bulk
foliation $\Sigma_t$.

If the unboosted observers and their time slice $\Sigma_t$ 
are static in the sense that $P^{\az \bz} \sigma_{\az \bz} = 0$ and
$P^{\az \bz} n_\az n_\bz = 0$,  
and the boosted observers have
a constant radial velocity over $\Omega_t$ (ie.\ $v_\vdash
= \mbox{constant}$ and $\hat{v}_\az = 0$), then 
these laws greatly simplify. Specifically,
\bea
\bep &=& \gamma \e, \label{sphEpsTrans}\\
\bj_\az &=&  j_\az, \mbox{ and}\\
\bs^{\az \bz} &=& \gamma s^{\az \bz} + \frac{\gamma^2}{\kappa}
\sigma^{\az \bz} \pounds_{\bu} v_\vdash.
\eea
So, in this case the energy density transforms as might be 
expected from special relativity. The angular momentum
density is an invariant which isn't too surprising considering
that it is perpendicular to the direction of the boost. However,
the stress tensor has a somewhat
more complicated transformation law that is dependent on the
perpendicular component of the acceleration of the boosted observers.
Breaking it up into pressure (ie.\ trace)
$p \equiv s^{\az \bz} \sigma_{\az \bz}$
and shear (ie.\ traceless) 
$\eta^{\az \bz} \equiv s^{\az \bz} - (1/2) p \sigma^{\az \bz}$
parts a little simplification results. Namely 
$\bar{\eta}^{\az \bz} = \gamma \eta^{\az \bz}$ and so it loses its 
acceleration dependence. However, $\bar{p} = \gamma p + 
(2 \gamma^2  / \kappa) \pounds_{\bu} v_\vdash$ and the 
dependence remains there. 
 
\section{The reference term}
\label{RefTerm}
If one calculates the quasilocal energy contained by a spherical
shell in Minkowski spacetime
it is immediate that the reference term $\underline{I}$ 
cannot be neglected
as I have been doing up to now. To see this, let the sphere have 
radius $R$. Then 
$\ssg \bep = - R/(4 \pi)$ and so the QLE is
$-R$. This is manifestly not zero, and what is more it actually diverges 
as $R \rightarrow \infty$, which are not properties that  
one would expect flat space to have! 
Thus, in anticipation of the upcoming 
calculations in chapter \ref{quasiExamples}, 
it is time to consider $\bI \neq 0$.

It has already been seen that
$\delta \bI = 0$ for variations that leave the 
boundary metrics unchanged, and so its exact form does not affect 
the equations of motion as derived by the Lagrangian or Hamiltonian
principles. However, it is equally clear that it does determine 
the zero of the numerical value of the action and therefore the 
zero of the evaluated Hamiltonian and all quantities derived from it
as well. In this
section I'll consider some specific choices of $\bI$ and discuss
the merits and problems of each. 

\subsubsection{Setting $\bI = 0$}
First, consider when $\bI = 0$ might be of some use. As pointed out
this means that $H_t$ will have non-zero values for finite regions of
flat space and it is not hard to see that a similar problem arises for
the action itself. However it does have the strong argument of 
simplicity in its favour, so it is worthwhile to consider circumstances
where it might be of use.

If one wishes to compare the energies contained 
by two almost identical surfaces, each embedded in the 
same space, then this may be a reasonable choice as any  
reference terms will (at least approximately) cancel each other out.
In fact, if one uses reference terms defined entirely with 
respect to the two-boundary metric $\sigma_{ab}$, instead of the
full three-boundary metric $\gamma_{\az \bz}$, and considers how
the associated energy changes as the surface is smoothly deformed,
then the terms do cancel exactly so one doesn't need to worry about
them. This is essentially because such terms can have no dependence
on the time rate of change of $\sigma_{ab}$. Examples of this class of
reference terms are the 2D into 
3D embedding reference terms considered next 
and the 2D intrinsic reference terms considered after that.

Given these facts, $\bI = 0$ is often used when one is doing thermodynamics \cite{BY2,jolien}. In section \ref{tidal}
where I examine energy flows through a quasilocal surface, 
I'll assume $\bI = 0$. My main motivation was to simplify an
already complicated calculation, but as I have just pointed
out, for a smoothly deforming surface $\Omega_t$ a wide range of
reference terms reduce to exactly this case. 

\subsubsection{Embedding $\Omega_t$ in a 3D reference space}

In \cite{BY1} Brown and York suggested that one should embed 
the two-surface ($\Omega_t,\sigma_{\az \bz})$ 
into a three-dimensional reference
space such that its intrinsic geometry is unchanged. One can
then define 
\be
\bI = \int_{B} d^3 x \ssg N \underline{\e},
\ee
where $\underline{\e}$ is calculated  
calculated for $\Omega_t$ embedded in the reference space (usually taken 
as $\R^3$  with metric $\delta_{a b} = \mbox{diag}[1,1,1]$). I omit
the $\underline{j}_a$ term since it fundamentally depends on how
$\Sigma_t$ is embedded in $M$ rather than on the geometry of $\Omega_t$
in $\Sigma_t$. 

For this reference term closed two-surfaces 
in Minkowski space have QLE
zero. What is more, for a two-sphere of constant $r$ and $t$ in
Schwarzschild space, the QLE $\rightarrow m$ as 
$r \rightarrow \infty$. Further, it is with this reference
term that the QLE was first shown to be equivalent to the ADM mass
\cite{BY1}.

There are still problems. Take a spherical set of observers in 
flat space and give them a radial boost. Then as shown in equation 
(\ref{sphEpsTrans}), $\bep = \gamma \e$ and so the QLE
is $(\gamma-1) R$. Again it is non-zero in flat space and 
actually divergent as $R \rightarrow \infty$.
That is bad enough, but there is an even more serious concern.
As Brown and York recognized in their paper, in general 
it isn't possible to embed
a two-surface in flat $\R ^3$. There are theorems that
say (see for example \cite{spivak}) that any Riemannian manifold
with two-sphere topology and everywhere positive intrinsic curvature
may be globally embedded in $\R^3$. However, most surfaces don't have 
such an intrinsic curvature and once that restriction is broken 
it is easy to find surfaces that cannot be embedded. For example, a
surface of constant $r$ and $t$ (Boyer-Lindquist coordinates)
in Kerr space cannot,
in general, be embedded in $\R^3$. For small enough $r$ (though still 
outside the horizon and even the ergosphere) 
the intrinsic curvature goes 
negative sufficiently 
close to the poles and it is not hard to show that the surface
cannot be embedded in three-dimensional flat space. 
For a further discussion of this point see \cite{martinez}.

\subsubsection{Embedding $\Omega_t$ in a 4D reference space}
The Brown-York reference term may be naturally generalized to
deal with the problem of moving observers \cite{nopaper}.
Then, instead of embedding $(\Omega_t,\sigma_{\az \bz})$ in a 
three-dimensional reference space, embed it in a four-dimensional
reference spacetime $\underline{\cM}$
and define a timelike vector field
$\underline{T}^\az$ over the embedded surface such that
\begin{description}
\item[1)] $\underline{T}^\az \underline{T}_\az = T^\az T_\az$,
\item[2)] $\bV^\az = \underline{\bV}^\az$ (in the sense
that their mappings into $\Omega_t$ are equal), and
\item[3)] $\pounds_T
\sigma_{\az \bz} = \pounds_{\underline{T}} \underline{\sigma}_{\az \bz}$
(in the sense that their mappings into $\Omega_t$ are equal).
\label{EmbedCond}
\end{description}
Together the first two conditions imply that the 
boundary lapses $\bN$ and $\underline{\bN}$ are equal
as well as the boundary shifts.
The third says that the time rate of change of the metric is the same
in the two spacetimes. Physically these conditions mean 
that an observer living in the surface $\Omega_t$ and observing 
only quantities intrinsic to
that surface (as it evolves through time) cannot tell whether she is
living in the original spacetime or in the reference spacetime.
That is, locally (in the time sense) $B$ is embedded in
$\underline{\cM}$. 
From a physical point of view the observers have calibrated their
instruments so that they will always measure the quasilocal quantities
to be zero in the reference spacetime, no matter what kind of motion
they undergo.

Then define
\be 
\bI = \int dt \int_{\Omega_t} d^2 x \sqrt{\sigma} 
[\bN \underline{\bep} - \bV^\az \underline{\bj}_\az ],
\ee
where $\underline{\bep}$ and $\underline{\bj}_\az$ are defined in the same way as before except that this time they are evaluated for 
the surface $\Omega_t$ embedded in the reference spacetime.  
Thus, the net effect of
including $\bI$ is to change $\bep \rightarrow \bep - \underline{\bep}$
and $\bj_\az \rightarrow \bj_\az - \underline{\bj}_\az$. 

With this reference term, the transformation laws for the quasilocal
quantities change. Consider unboosted observers evolved
by $u^\az$ and $\underline{u}^\az$ watching
$T^\az$ and $\underline{T}^\az$ observers. Then, in 
general $\eta = \bu^\az n_\az$ will not be equal to
$\underline{\eta} = \underline{\bu}^\az \underline{n}_\az$. 
Physically this means that in order for
$(\Omega_t,\sigma_{\az \bz})$ to evolve in the same way in the two
spacetimes, that surface will have to ``move'' at different speeds in each. Then the transformation law for the quasilocal
energy density with reference terms becomes
\be
\bep - \underline{\bep} = \left(
\frac{1}{\lambda} \varepsilon + \eta \varepsilon^{\updownarrow}
\right) 
- \left( \frac{1}{\underline{\lambda}} \underline{\varepsilon} + 
\underline{\eta} \underline{\varepsilon}^{\updownarrow} \right).
\ee

With this definition of $\bI$ the problem of observers in flat space
seeing non-zero energies is solved. Taking Minkowski space as the
reference space it is trivial that $\Omega_t$ may be embedded and
$\underline{T}^\az$ defined. Simply leave $\Omega_t$ as it is and 
define $\underline{T}^\az = T^\az$. Then observers undergoing any motion
in Minkowski space measure zero energy. Similarly the action is
zero for any region of flat space.

However, problems remain. In the first place even though it is always
possible to (locally) embed a two-surface in Minkowski space (see for
example ref.~\cite{brink}), that embedding will not be
unique \cite{epp}. Thus, the problem of existence has been replaced by
a question of uniqueness. Furthermore, there is no guarantee that
the desired vector field can even be defined so 
even the existence problem has not been fully
eliminated.


Nevertheless for the problems that are considered in this thesis this
definition of the reference terms will suffice. A good discussion 
of a closely related reference term (that combines aspects of this
approach with those reviewed in the next section)
may be found in \cite{epp}.

\subsubsection{Intrinsic reference terms}
Recently there have been several proposals
for reference terms $\bI$ that are defined with respect to 
the intrinsic geometry of $B$, rather than its extrinsic geometry  
after it has been embedded in $\underline{\cM}$.
Most but not all (for example Lau \cite{LauCT} has a different
motivation)
of these so-called counterterms have been inspired by the AdS/CFT 
correspondence and are intended to remove the divergences of the
action $I-\bI$ without having to worry about the existence or uniqueness
of embeddings or for that matter what is the proper reference space to
use -- a non-trivial issue if one is considering more complicated
spacetimes such as AdS space with 
periodic identifications \cite{dom3,dom2} or NUT black holes
\cite{NUTS3,NUTS2,NUTS1}.

Typically such terms are defined with respect to the Ricci scalar of 
$B$ or $\Omega_t$ as well as other intrinsic scalars -- their exact
form depending on the dimension of the spacetime in which they are
being defined. The original proposal \cite{AdSCFT1,AdSCFT2,AdSCFT4} 
only worked for
AdS spacetimes but later work allows for asymptotically
flat spacetimes as well \cite{LauCT,NUTS2,AdSCFT3}. 

I'll briefly consider one such proposal here. Its advantages 
and problems are typical of the wider class of intrinsic counterterms. 
For asymptotically flat space Lau and Mann \cite{LauCT,NUTS2}
suggested using 
\be
\bI \equiv \frac{1}{\kappa}
\int_{\Omega_t} d^2 x \bN \ssg \sqrt{ 2 R^{(2)}}
\ee 
where $R^{(2)}$ is the Ricci scalar for $\Omega_t$. Lau showed
that asymptotically, for a static set of observers, this 
reference term agrees with the embedding reference term of
Brown and York and so the quasilocal energy is equal to the 
ADM and Bondi energies.

Unfortunately for a finite region of flat space the quasilocal energy
defined with this reference term will not, in general, be zero. The
reason for this is easy to see. Recall from elementary differential
geometry that the mean curvature of a two-surface in flat
three-space is 
$C_m \equiv (\kappa_1 + \kappa_2)/2$ and the Gaussian  
curvature is 
$C_G \equiv \kappa_1 \kappa_2$ where $\kappa_1$ and $\kappa_2$ 
are the principal curvatures of the surface. Now, the contracted
extrinsic curvature $\underline{k} = 2 C_m$ and the Ricci scalar
$R^{(2)} = \sqrt{2 C_G}$ so $\underline{k} \geq \sqrt{ 2 R^{(2)}}$ 
simply 
because the arithmetic mean of two quantities is always greater than
or equal to their geometric mean. The equality only holds if the two 
principal curvatures are equal. 
That is, the two are only exactly equal when $\Omega_t$ is a sphere. 
Lau showed that if a rigid surface $\Omega_t$ is
blown-up to its asymptotic limit, then $\underline{k} \rightarrow
\sqrt{ 2 R^{(2)}}$. However, finite regions of flat space have
non-zero energy with this reference term 
unless their boundary is a sphere. 

Before moving on to the next section it is as well to emphasize
once again that any of these choices of reference terms are perfectly
acceptable from the point of view of the action and/or Hamiltonian
generating the correct equations of motion. The exact form is 
only important to set the zero of the quasilocal quantities.

\section{Thin shells}
\label{thinshell}

In this section I examine in some detail a correspondence
between the quasilocal formalism and the mathematics describing thin 
shells in general relativity which was developed by Lanczos and
Israel \cite{thinshell}. This was noted in passing in 
\cite{BY1} but here it will be examined in more detail and used
to reinterpret the quasilocal energy from an operational point of view.  
Following that, I'll briefly apply some results from the previous
sections to explore the physics of thin shells.

\subsection{The thin shell/QLE mathematical equivalence}
Israel considered the conditions that two spacetimes, 
each with a boundary, must satisfy so that they may be joined
along those boundaries and yet still satisfy
Einstein's equations. He showed that as an absolute minimum the
spacetimes must induce the same metric on the common boundary
hypersurface. Further the Einstein
equations will only be satisfied at the boundary if its extrinsic
curvature in each of the two spacetimes is the same. If those curvatures
are not the same then a singularity exists in the (joined) spacetime
at the hypersurface. However the singularity is sufficiently
mild that it may be accounted for by a thin shell of matter
defined on that boundary. The change in curvature may then be
interpreted as a manifestation of the thin shell of matter. 

Modifying Israel's notation and sign conventions to be compatible with
those used here, the stress-energy tensor of that matter
is defined as follows. 
Consider a spacetime $\cM$ divided into two regions
$\mathcal{V}^+$ and $\mathcal{V}^-$ by a timelike hypersurface $B$. Let
the metric on $\mathcal{V}^+$ be $g^+_{\az \bz}$ and the metric on
$\mathcal{V}^-$ be $g^-_{\az \bz}$, and assume that they induce the same
metric $\gamma_{\az \bz}$ on $B$. Further, let $\bn^+_\az$ and 
$\bn^-_\az$
be the spacelike unit normals of $B$ on each of its sides (both oriented
to point in the same direction) and define
$\Theta^+_{\az \bz}$ and $\Theta^-_{\az \bz}$ to be the extrinsic
curvature of $B$ in $\mathcal{V}^+$ and $\mathcal{V}^-$ respectively.
Then, Einstein's equation will only be satisfied if 
a thin shell of matter is present at $B$ with stress-energy
tensor $S_{\az \bz} = \frac{1}{\kappa} \left\{ ( \Theta^+_{\az \bz} -
\Theta^+ \gamma_{\az \bz}) - (\Theta^-_{\az \bz} - \Theta^- \gamma_{\az
\bz} ) \right\}$. Note that this is written as a tensor field 
in the surface $B$. 
To write it as a four-dimensional stress-energy tensor 
an appropriate Dirac delta function must be included.

Now let $\Omega_t$ be a foliation of $B$ corresponding to a timelike
vector
field $T^\az \equiv \bN \bu^\az + \bV^\az$ (which as usual lies entirely
in the tangent space to $B$). Then observers who are static with respect
to the foliation will observe the thin shell to have the
following energy, momentum, and stress densities:
\bea
{\mathcal E} = S_{\az \bz} \bu^\az \bu^\bz &=& \frac{1}{\kappa} 
\left\{ \bk^+ - \bk^-\right\}, \\
{\mathcal J}_\az =  -S_{\cz \dz} \sigma_\az^\cz \bu^\dz &=&  
 \frac{1}{\kappa} \left\{ \sigma_\az^\cz \bu^\dz \nabla_\cz \bn^+_\dz 
- \sigma_\az^\cz \bu^\dz \nabla_\cz \bn^-_\dz \right\}, \mbox{ and} \\
{\mathcal S}_{\az \bz} = S_{\cz \dz} \sigma^\cz_\az \sigma^\dz_\bz &=& 
\frac{1}{\kappa} \left\{ 
( \bk^+_{\az \bz} - (\bk^+ - \bn^{+\dz} \ba_\dz) \sigma_{\az \bz} )
\right. \\
&& \left. 
- ( \bk^-_{\az \bz} - (\bk^- - n^{-\dz} \ba_\dz) \sigma_{\az \bz} ) \right\} \nn ,
\eea
where $\bk^{\pm}_{\az \bz} = - \sigma_\az^\cz \sigma_\bz^\dz \nabla_\cz
\bn^\pm_\dz$ and $\bk^\pm = \sigma^{\az \bz} \bk^\pm_{\az \bz}$ are the extrinsic curvature of the surface
$\Omega_t$ in a (local) foliation of $\cM$ perpendicular to $B$. 
$\ba^\az$ retains its earlier meaning. 

The correspondence between the quasilocal and thin shell formalisms is now obvious. Consider the surface $(B,\gamma_{\az \bz})$ embedded in a
spacetime $(\cM,g_{\az \bz})$ and a reference spacetime $(\underline{\cM},
\underline{g}_{\az \bz})$. Further let $(\cM,g_{\az \bz})$ be isomorphic
to $({\mathcal{V}}^+ ,g^+_{\az \bz})$ (or more properly the portion of
$(\cM,g_{\az \bz})$ to one side of $B$ is isomorphic to
$({\mathcal{V}}^+,g^+_{\az \bz})$), and in the same sense let
$(\underline{\cM},\underline{g}_{\az \bz})$ be isomorphic to
$({\mathcal{V}}^-,g^-_{\az \bz})$. Then for observers living on $B$ and
defining their notion of simultaneity according to the foliation
$\Omega_t$,
\bea
\mathcal{E} &=& \bep - \underline{\bep}, \\
\mathcal{J}_\az &=& \bj_\az - \underline{\bj}_\az, \mbox{ and}\\
\mathcal{S}_{\az \bz} &=& \bs_{\az \bz} - \underline{\bs}_{\az \bz},
\eea
where $\underline{\bs}_{\az \bz}$ is defined in the obvious way
and the energy density of the matter seen by the observers is
\be
T^\az S_{\az \bz} \bu^\bz = N {\mathcal E} - V^\az {\mathcal J}_\az 
= N (\bep - \underline{\bep}) - V^\az (\bj_\az - \underline{\bj}_\az).
\ee

This mathematical identity of the formalisms can be interpreted
in  couple of ways. First, following \cite{BY1} one can note that
the quasilocal work formalism provides an alternate derivation 
of the thin
shell junction conditions and stress-energy tensor. Namely
consider two quasilocal surfaces on either side of the shell and
consider the limit as the two go to the shell. In that case any
reference terms will match and cancel leaving only the 
the stress-energy tensor defined above. This derivation is
quite different from the one used by Israel. 

From a slightly different perspective the thin shell
work can be seen as providing an operational definition of
the quasilocal energy with the two-surface-into-4D reference terms.
Given a reference spacetime $\underline{\cM}$ which is assumed
to have energy zero, then the quasilocal energy associated
with a two surface $\Omega_t$ and time vector $T^\az$ 
in a spacetime $\cM$ can be defined as the energy
of a shell of matter $\underline{\Omega}_t$ 
in $\underline{\cM}$ that has the same intrinsic geometry
as $\Omega_t$ (including the rate of change of those properties)
and a matter stress-energy tensor 
defined so that the spacetime outside of $\underline{\Omega}_t$ is
identical to that outside of $\Omega_t$ in $\cM$,
while inside it remains $\underline{\cM}$. In fact, 
the quasilocal energy 
with the embedding two-surface-into-4D reference terms 
considered in the previous section
is defined if and only if the fields outside
of $\Omega_t$ can be replicated by a shell of stress-energy
with the same intrinsic geometry embedded in $\underline{\cM}$. 
Provided that $\Omega_t$ and $T^\az$
can be embedded in the reference spacetime 
$\underline{\cM}$, the construction considered in 
this section defines the relevant stress-energy for a shell in 
$\underline{\cM}$. 

\subsection{Physics of thin shells}

Finally, there is a nice application of equation (\ref{Hdot}) to 
thin shells. Using that equation, including the reference term,
and assuming that the reference space is a solution to the Einstein
equations,
\be
{\pounds_T} H_t 
= \int_{\Omega_t} d^2 x \sqrt{\sigma} \left\{
\cE {\pounds_T} \bN 
- {\mathcal{J}}_a {\pounds_T} \bV^a 
- \frac{\bN}{2} {\mathcal{S}}^{a b}
{\pounds_T} \sigma_{ab} \right\}.
\ee
The stress tensor can be further decomposed as it was at the
end of section \ref{gravtransform}. For the stress tensor
\be
\ssg s^{ab} \pounds_T \sigma_{ab} = p \pounds_T \ssg + \sqrt{\sigma}
\eta^{ab} \pounds_T \sigma_{ab},
\ee
where $p = \sigma^{ab} s_{ab}$ is a pressure and 
$\eta^{ab} = s^{ab} - (1/2)p \sigma^{ab}$ is a shear. The 
reference space stress tensor can be broken up in the same way. 

The terms of the above can be individually interpreted. The 
${\mathcal{E}} \pounds_T N$ term records how the energy measured changes
with how the observers choose to measure their time (remember
that in a Hamiltonian approach energy is conjugate to time so if 
one measures time as going by more quickly then one also measures a
larger energy).
The ${\mathcal{J}}_a \pounds_T V^a$ evaluates the change in the energy
contribution from matter flowing around the shell -- if the observers
change their motion then they will observe different matter motions
and so see a different energy. The part of the stress tensor 
corresponding to $p \pounds_T \ssg - \underline{p} \pounds_{\underline{T}} \ssg$ term measures
energy expenditures required to rigidly shrink 
or expand the shell, while
the $\eta^{ab} \pounds_T \sigma_{ab} - \underline{\eta}^{ab} 
\pounds_{\underline{T}} \sigma_{ab}$
part records the work done to 
deform it. These are all terms that one would expect based
on an intuition on how 
classical, non-relativistic membranes under tension should
behave.

\chapter{A quasilocal Hamiltonian for matter}
\label{matterChapter}

The analysis of the previous chapter can easily be extended to include
matter fields when those fields have a Lagrangian formulation.
Such an  extension was made (in the orthogonal case) for 
dilatons and general gauge fields in ref.\ \cite{jolien, jolienrobb}
but for purposes of this work I just need the coupled Maxwell and
dilaton fields that were discussed in section \ref{appMatter}.

In this chapter I will consider a Lagrangian formulation of the
field equations from section \ref{matLag}, 
and then derive an equivalent Hamiltonian in section \ref{matHam}. 
The field equations will then be seen to follow from that Hamiltonian,
though it will be seen that the formalism itself puts restrictions on
the matter field configurations that it can be used to study.
Issues such as the transformation laws and thin shells will 
be briefly reconsidered in the light of the new matter terms in 
section \ref{MatHamProps}. 
Finally in section \ref{EMdual}, 
I'll examine all of this in the light of the duality 
that was considered in section \ref{dualreview}.

From a Hamilton-Jacobi perspective parts of this chapter were
published in \cite{naked}, but in pure Hamiltonian form they
appear here for the first time.

\section{The gravity-Maxwell-dilaton Lagrangian}
\label{matLag}
The field equations (\ref{nabF}, \ref{dil}, and \ref{Ein})
are generated by the variation of the action 
\bea
\label{matteraction}
I^m - \bI &=& 
\frac{1}{2 \kappa} \int_M d^4 x \sqrt{-g} (\cR - 2 \Lambda 
- 2 (\nabla_\az \phi)(\nabla^\az \phi) - e^{-2a\phi} F_{\az \bz}F^{\az \bz}) \\ 
&& + \frac{1}{\kappa} \int_\Sigma d^3 x 
\sqrt{h} K - \frac{1}{\kappa} \int_B d^3 x \sqrt{- \gamma} \Theta 
+ \frac{1}{\kappa} 
\int_\Omega d^2 x \sqrt{\sigma} \sinh^{-1} (\eta), \nn 
\eea
where $\phi$ is the dilaton field, $F_{\az \bz}$ is the electromagnetic
field tensor, and $a$ is the coupling constant between the two fields. 
I assume that at any point in $M$, $F_{\az \bz}$ is defined with respect
to some gauge potential one-form $A_\az$ such that 
$F_{\az \bz} \equiv \partial_\az A_\bz - \partial_\bz A_\az$. 
The existence of these vector potentials means that 
$\epsilon^{\az \bz \cz \dz} \nabla_{\bz} F_{\cz \dz} = 0$ 
(equivalently $d(dA) = 0$) so before applying the variational
principle the Maxwell equation (\ref{dF}) has been assumed.

Taking the first variation
of the metric terms with respect to the metric, gauge potential, and 
dilaton, it is straightforward to obtain
\bea
\label{firstmatvar}
&& \delta \left( -2 \sqrt{-g} (\nabla_\az \phi)(\nabla^\az \phi) 
- \sqrt{-g} e^{-2a\phi} F_{\az \bz}F^{\az \bz} \right) \\
&=& 4 \sqrt{-g} \cF_{Dil} \delta \phi + 4 \sqrt{-g} \cF_{EM}^\bz 
\delta A_\bz - \kappa \sqrt{-g} T_{\az \bz} \delta g^{\az \bz} \nn \\
&& - 4 \sqrt{-g} \nabla_\az \left( [\nabla^\az \phi] \delta \phi 
+ e^{-2 a \phi} F^{\az \bz} 
\delta A_\bz \right) \nn, 
\eea
where, 
$\cF^\bz_{EM} \equiv \nabla_\az \left[ e^{-2a \phi} F^{\az \bz}
\right]$, 
$\cF_{Dil} \equiv \nabla^\az \nabla_\az \phi + (1/2) a e^{-2a\phi}
F_{\az \bz} F^{\az \bz}$ 
and $T_{\az \bz}$ was defined in equation (\ref{Tab}). The equations
$\cF_{EM}^\bz = 0$ and $\cF_{Dil}=0$ are equivalent to 
equations (\ref{nabF}) and (\ref{dil}) respectively.

Then, assuming that there exists a {\it single} gauge
potential $A_\az$ covering the entire region $M$ 
(an assumption that I will have more to say about in 
section \ref{FormMatHam}) Stokes's theorem can be used
to move the total divergence out to the boundary 
of $M$. Combining this with the vacuum result
(\ref{firstVar}) the total variation of the action is
\bea
\label{mactvar}
\delta I^m &=& \frac{1}{2 \kappa} \int_M d^4 x \sqrt{-g} 
\left\{ (G_{\az \bz} + \Lambda g_{\az \bz} - 8 \pi T_{\az \bz} ) 
\delta g^{\az \bz} + 4 \cF_{Dil} \delta \phi + 4 \cF_{EM}^\bz \delta \tilde{A}_\bz \nn
\right\}  \\
&& + \int_\Sigma d^3 x \left\{ P^{\az \bz} \delta h_{\az \bz} + 
\wp \delta \phi + \cE^\az \delta \tilde{A}_\az \right\}  
+ \int_\Omega d^2 x \left\{ P_{\sqrt{\sigma}} 
\delta (\sqrt{\sigma}) \right\} \\
&& + \int dt \int_{\Omega_t} d^2 x \sqrt{\sigma} 
\left\{ \pi^{\az \bz} \delta \gamma_{\az \bz}
-\frac{2}{\kappa} \left( [\bn^\az \nabla_\az \phi] \delta \phi 
+ e^{-2 a \phi} \bn_\az F^{\az \bz} 
\delta A_\bz \right)
 \right\} \nn,
\eea
where
$P^{\az \bz}$ and  $P_{\sqrt{\sigma}}$ retain their 
earlier meanings and
$\cE^\az$, $\wp$, and $\tilde{A}_\az$ are the densitized
electric field, the densitized time rate of change of the dilaton, 
and the three-dimensional gauge potential as discussed in 
detail in section \ref{3dfieldeq}.

Fixing the metric, vector potential, and dilaton on
the boundaries of $M$, and solving $\delta I^{m} = 0$
the Einstein, Maxwell, and dilaton field equations must hold.
Equivalently this particular action is only 
fully differentiable if those quantities are fixed on the
boundary.

\section{The gravity-Maxwell-dilaton Hamiltonian}
\label{matHam}

\subsection{Form of the Hamiltonian}
\label{FormMatHam}
From this action it is a fairly straightforward calculation to derive
the corresponding quasilocal Hamiltonian. As in the previous chapter the 
action has to be broken up with respect to the foliation and then the
Hamiltonian and momentum terms picked out from the detritus. Details of
the calculation can be found in appendix \ref{appMatHam} but the 
foliated action is
\bea
\label{Mmactdecom}
I^m - \underline{I} &=& \int dt \left\{ 
\int_{\Sigma_t} d^3 x \left( P^{\az \bz} 
\pounds_T h_{\az \bz} + \wp \pounds_T \phi + \cE^\az 
\pounds_T \tilde{A}_\az \right) \right\} \\
&& + \int dt \left\{
\int_{\Omega_t} d^2 x \left( P_{\sqrt{\sigma}}
\pounds_T \sqrt{\sigma} \right)  - H^m_t \right\}, \nn
\eea
where
\bea
H^m_t &=& 
\int_{\Sigma_t} d^3 x [N \cH^m + V^a \cH^m_a + T^\az A_\az \cQ] \\
&& + \int_{\Omega_t} d^2 x \sqrt{\sigma} \left[ \bN (\bep + \bep^m) 
- \bV^\az (\bj_\az + \bj^m_\az) \right]. \nn
\label{HamFull}
\eea
$T^\az A_\az = - N \Phi + V^\az \tilde{A}_\az$ in terms of quantities
defined on the hypersurfaces, while 
$\cQ = - D_\bz \cE^\bz = 0$ is the free space version of
Gauss's law from electrodynamics (equation (\ref{DE})).
\bea
\bep^m &\equiv& - \frac{1}{\sqrt{h}} (n_\bz \cE^\bz) 
(\frac{1}{\lambda} \Phi - \eta \tilde{A}_\az n^\az) 
= - \frac{1}{\sqrt{h}} (\bn_\bz \bar{\cE}^\bz) \bar{\Phi} 
\label{bepm} \mbox{ and}\\
\bj^m_\az &\equiv& - \frac{1}{\sqrt{h}} (n_\bz \cE^\bz) 
\hat{A}_\az = - \frac{1}{\sqrt{h}} (\bn_\bz \bar{\cE}^\bz) 
\hat{A}_\az \label{bjm}
\eea
which can be identified with energy and angular momentum as
suggested by the notation. 
The bar retains its usual meaning, so in this case
$\bar{\Phi} = - A_\az \bu^\az$ and $\bar{\cE^\az} = -2 \sqrt{h}/\kappa
F^{\az \bz} \bu_\bz$. Note that 
$\hat{A}_\az \equiv \sigma_\az^\bz A_\bz$
and $n_\bz \cE^\bz = \bn_\bz \bar{\cE}^\bz$ and so 
are left invariant by the bar notation. 

Then, the electric field vector density $\cE^\az$ and the dilaton
rate of change $\wp$ are identified as momenta conjugate to 
$\tilde{A}_\az$ and $\phi$ respectively. Exactly what is happening
with the $T^\az A_\az$ term isn't clear at this stage, but
after calculating the variation of $H^m_t$ in the next section it
will be clear that $\Phi$ (the Coulomb potential) is a Lagrange 
multiplier. Finally, $H_t^m$ can be identified as the Hamiltonian
functional. 
As in the previous chapter, the numerical value of 
$H^m_t$ evaluated for a particular leaf of the
spacetime foliation $\Sigma_t$ depends only on the 
boundary $\Omega_t$ and how that boundary is evolving in 
time. 

In the next section I will show that the functional $H^m_t$
really does generate the correct field equations, but before
moving on 
there are a couple of points to consider regarding the 
electromagnetic gauge potential $A_\az$ and gauge invariance.

\subsubsection{No magnetic charges allowed}

In the derivation of the Hamiltonian from the action it was
assumed that there is a single vector potential $A_\az$ defined
over all of $M$. This assumption meant that total derivatives
in the bulk could be removed to the boundaries under the
auspices of Stokes's theorem. However, a corollary of this 
assumption is that there is no magnetic charge in $M$ (or contained
by any surface that is itself contained in $M$). The next few
paragraphs explore this statement from three closely related
points of view.

As a start, let $\Omega_X$ be
any closed spatial two-surface in $M$ with normals $\bu^\az$ and
$\bn^\az$. Then, the magnetic charge contained within $\Omega_X$ is $
\int_{\Omega_X} d^2 x \sqrt{\sigma} \bn^\az \bB_\az $. By equation
(\ref{Bpot}), $\bn^\az \bB_\az = - \bn^\az \bu^\bz
\epsilon_{\az \bz}^{\ \ \ \cz \dz} D_\cz \tilde{A}_\dz = 
\epsilon_\Omega^{\az \bz} d_\az \hat{A}_\bz )$ 
where $d_\az$ is the covariant derivative in the surface
$\Omega_X$, $\epsilon_\Omega^{\az \bz}$ is the Levi-Cevita
tensor on that surface, and again $\hat{A}_\bz = \sigma_\bz^\cz A_\cz$.
But this is an exact differential form and so integrated over a closed
surface it is zero\footnote{In the more efficient differential forms
notation, in the spatial slice orthogonal to $\bu^\az$, $\tilde{A}$ is
a one form and $B = d \tilde{A}$ is a two form. Then if $\hat{B}$ and
$\hat{A}$ are the forms
projected (or pulled-back) into $\Omega_X$, 
the magnetic charge contained within $\Omega_X$ is $\int_{\Omega_X} \hat{B} = \int_{\Omega_X} d\hat{A} = 0 $ since $\Omega_X$ is closed.}. 
Thus there is no
magnetic charge contained by any surface in $M$. 

Keep in mind that this is a stronger statement than just the local 
statement that $F = dA \Rightarrow dF = d(dA) = 0 \Rightarrow D_\az
\cB^\az = 0$. When working with a gauge potential, 
the manifestation of magnetic charge in the potential is  
global and topological (resulting from a twist in the 
$U(1)$ gauge bundle) rather than local as is the case for electric
charge. If one assumes that there is a single $A_\az$ that covers $M$
then the $U(1)$ gauge bundle is trivial by definition
and so there is no magnetic/topological charge. Even more strongly, as
just noted, no surface contained in $M$ can itself 
contain magnetic charge.
This means, for example, that if $M$ is the region bounded by
two concentric spheres (multiplied by a time interval), then not only is
there no charge in $M$ but also there is no charge in the region inside
the inner sphere. 

In fact, projecting into spatial slices $\Sigma_t$ of $M$, 
de Rhams theorem (see for example \cite{flanders})
says that a single vector potential is defined over all
of $\Sigma_t$ if and only if there is no magnetic charge contained
within any two-surface $\Omega_X \subset \Sigma_t$. Thus
to allow for a magnetic charge in $M$, one
must break the region of spacetime into at least two regions each
of which has its own vector potential. Then, the frequent
uses of Stokes's theorem in the derivation will remove 
total divergences to the boundaries of those regions rather
than just the boundary of $M$ itself. By definition some of those
region boundaries will actually be interior to $M$ and so
observers inhabiting $\partial M$ will not be in a 
position to measure all of the boundary terms and therefore
will not be able to fully assess what is happening in the
interior of $M$.

\subsubsection{The gauge dependence of the Hamiltonian}
Note that even though the action $I^m$ is gauge invariant
(ie.\ it depends only on $F_{\az \bz}$ and not on the exact form
of $A_\az$) the proposed
Hamiltonian doesn't necessarily inherit that invariance.
The paths by which this gauge dependence can creep in are 
quite easily found but at the same time the effect is 
important so I'll pause here to point them out in some detail. 

First, by equation (\ref{Mmactdecom}) it is clear that 
while the time integrated difference
between 
the Hamiltonian and ``kinetic energy'' terms must be gauge invariant,
that invariance can only be inherited by the Hamiltonian itself if 
part of the gauge freedom is used to ensure that 
$\pounds_T \tilde{A}_\az = 0$. If this is the case and $H_t^m$
is independent of the leaf of the foliation, then $H_t^m$
will be independent of the remaining gauge freedom. 
For stationary spacetimes that gauge and foliation are, of course, 
the natural ones to choose but it should be kept in mind that
a partial gauge fixing is required to ensure that the Hamiltonian
is invariant with respect to the remaining gauge freedom.

In the conventional usage of this work to study black holes,
there is an alternate route by which gauge dependence can find its way
into the Hamiltonian. \label{HorProb}
Namely, components of the gauge potential $A_\az$
may diverge on the (apparent) event horizon. Then, $A_\az$ has
a singularity in $M$ and so the uses
of Stokes's theorem in the derivation of the Hamiltonian aren't valid.
To avoid this problem one could cut out the horizon with a inner
timelike boundary $B'$, though in that case the region between 
$B'$ and $B$ would be under consideration rather than the full region
contained by
$B$. This problem is usually ignored however and to facilitate comparisons
between singular and non-singular spacetimes, only the outer boundary is
considered. In section \ref{HamGaugeDual}, 
this version of gauge dependence will be demonstrated for 
a Reissner-Nordstr\"{o}m spacetime, and it will also be seen that
this gauge dependence arising from neglecting the inner boundary 
amounts to little more than a choice of where to set the zero
of the electromagnetic energy.

Thus, it can be seen that gauge independence of the action doesn't
necessarily
ensure the gauge invariance of the Hamiltonian. Specializing to the case
where the lapse $N=1$ and shift $V^\az = 0$ on the boundary, and leaving
aside the issue of how such a choice affects the relative foliations of
inner and outer boundaries, note that the QLE 
will not in general be gauge independent either.

\subsection{Variation of the Hamiltonian}
\label{matterVariation}
This section will show that the proposed Hamiltonian really does 
generate the correct equations of motion. To do this 
consider $H_t$ as a functional of the surface
$\Sigma_t$, its boundary $\Omega_t$, the normal $n^a$ to that
boundary, the fields $h_{a b}$, $\ssg$, $\phi$, $\tilde{A}_a$
along with their conjugate momenta $P^{ab}$, $P_{\ssg}$, $\wp$,
$\cE^a$, and the Lagrange multipliers $N$, $V^a$ and $\Phi$. 
In the usual Hamiltonian way, the conjugate momenta are
considered to be entirely independent variables. Their
connection with $h_{ab}$, $\ssg$, $\phi$, and $\tilde{A}_a$
is forgotten. Following the lead of section \ref{HamVar}, 
$\bep$, $\bj_a$, $\bN$, $\bV^a$, $\lambda$, and $\eta$ are defined entirely with respect to $V^\az$, $N$, and $n_a$. Similarly 
$\bPhi$, $\bep$, and $\bj_a$ can be written with respect to 
$\Phi$, $\tilde{A}_a$, $\lambda$, $\eta$, and $n_a$.  

Then, the variation of $H_t^m$ with respect to the quantities
$h_{ab}$, $\ssg$, $\phi$, and $\tilde{A}_a$, their 
conjugate momenta $P^{ab}$, $P_{\ssg}$, $\wp$, and $\cE^a$, and the
Lagrange multipliers $N$, $V^a$, and $\Phi$ is 
\bea
\delta H_t^m &=& \int_{\Sigma_t} d^3 x \left(
 [\cH^m - \Phi \cQ] \delta N + [\cH^m_a + \tilde{A}_a \cQ] \delta V^a 
- N \cQ \delta \Phi \right) \\
&& + \int_{\Sigma_t} d^3 x \left( [h_{ab}]_T \delta P^{ab} 
- [P^{ab}]^m_T \delta h_{ab} \right) \nn \\
&&+ \int_{\Sigma_t} d^3 x \left( 
[\phi]_T \delta \wp - [\wp]_T \delta \phi + [\tilde{A}_a]_T \delta
\cE^a - [\cE^a]_T \delta \tilde{A}_a \right) \nn \\
&&+ \int_{\Omega_t} d^2 x \sqrt{\sigma} \left( [\bep+\bep^m] 
\delta \bN - 
[\bj_a+\bj_a^m] 
\delta \bV^a 
- (\bN/2) \bs^{ab} \delta \sigma_{ab} \right) \nn \\
&& + \int_{\Omega_t} d^2 x \sqrt{\sigma} 
\left(  \left[ \sqrt{\sigma} \right]_T 
\delta P_{\sqrt{\sigma}} - \left[P_{\sqrt{\sigma}} \right]_T \delta 
\sqrt{\sigma} \right), \nn \\
&& + \int_{\Omega_t} d^2 x \frac{\bN \ssg}{\sqrt{h}} \left(
[\cE^a n_a] 
\delta \bar{\Phi} + e^{-2a\phi}
\hat{\bar{\cB}}_a n_b \hat{\epsilon}^{abc}
\delta \hat{A}_c \right) \nn \\
&& + \int_{\Omega_t} d^2 x \frac{2 \bN \ssg}{\kappa} 
\left[ \phi \right]_{\bn} \delta \phi \nn.
\eea
Details of this calculation can be found in appendix \ref{appMatVar}
but for now note that 
$[ h_{ab} ]_T$ retains its meaning from the previous chapter
(equation (\ref{dhdt})) while
\bea
\left[ P^{ab} \right]^m_T 
& \equiv & \left[ P^{ab} \right]_T + \frac{N \sqrt{h}}{\kappa}
\left( [D^a \phi][D^b \phi]  - \frac{1}{2} [D_c \phi] [D^c \phi] h^{ab}
\right) + \frac{N \kappa}{8 \sqrt{h}} \wp^2 h^{ab} \\
&& + \frac{N \kappa}{4 \sqrt{h}}
\left( [e^{2a\phi} \cE^a \cE^b + e^{-2 a \phi} \cB^a \cB^b] - 
\frac{1}{2} [e^{2a\phi} \cE^c \cE_c + e^{-2 a \phi} \cB^c \cB_c]
h^{ab} \right) \nn,  \\
\left[ \phi \right]_T 
&\equiv& \frac{N \kappa}{2 \sqrt{h}} \wp + V^a D_a \phi, \\
\left[ \wp \right]_T 
&\equiv& \frac{2 \sqrt{h}}{\kappa} D^c [N D_c \phi] 
+ a \frac{N \kappa}{2 \sqrt{h}} [e^{-2a\phi} \cB^b \cB_b - 
e^{2a\phi} \cE^b \cE_b] + D_b [\wp V^b], \\
\left[ \tilde{A}_a \right]_T 
& \equiv & \frac{N \kappa}{2 \sqrt{h}} e^{2a\phi} \cE_a + \pounds_V 
\tilde{A}_a - D_a [N \Phi], \\
\left[ \cE^a \right]_T 
& \equiv & - \epsilon^{abc} 
D_b [N e^{2a\phi} \cB_c ] + \pounds_V \cE^a, \\
\hat{\bar{\cB}}^b & \equiv & \frac{1}{\lambda} \hat{\cB}^b
- \eta e^{2a\phi} \epsilon^{cd} \hat{\cE}^d,  \mbox{ and}\\
\left[ \phi \right]_{\bn} & \equiv & \frac{1}{\lambda} {\pounds_n} \phi
+ \frac{\kappa}{2 \sqrt{h}} \wp.
\eea
$\hat{\cE}^a \equiv \sigma^a_b \cE^b$ and $\hat{\cB}^a \equiv \sigma^a_b
\cB^b$ are the projections of the electric and magnetic vector densities
into the tangent bundle of the boundary $\Omega_t$. 

The Hamiltonian equations of motion can now be obtained
by calculating the full variation of the action
(\ref{MgravActdecomp}) (treating the momenta
as independent variables) and solving $\delta I = 0$. 
Then an application of the fundamental theorem of calculus
to remove the time derivatives to the spatial boundaries and 
a little bit of algebra shows that 
\bea
\delta I - \delta \bI &=& 
\int_\Sigma d^3 x \left( P^{ab} \delta h_{ab}
+ \cE^a \delta \tilde{A}_a + \wp \delta \phi \right)
 +  \int_\Omega d^2 x P_{\sqrt{\sigma}} \delta
\sqrt{\sigma} \label{Hmactvar} \\
&& - \int dt \int_{\Sigma_t} d^3 x \left\{
(\cH^m - \Phi \cQ) \delta N + (\cH_a^m + \tilde{A}_a \cQ)\delta V^a  - N \cQ \delta \Phi\right\} \nn \\
&& + \int dt \int_{\Sigma_t} d^3 x 
\left\{  \left( \pounds_T h_{ab} -
\left[ h_{ab} \right]_T \right) \delta P^{ab}
-\left( \pounds_T P^{ab} - 
\left[ P^{ab} \right]_T \right) \delta h_{ab} \right\} \nn \\
&& + \int dt \int_{\Sigma_t} d^3 x
\left\{ 
\left( \pounds_T \tilde{A}_a -
\left[ \tilde{A}_a \right]_T \right) \delta \cE^a 
- \left( \pounds_T \cE^a - 
\left[ \cE^a \right]_T \right) \delta \tilde{A}_a \right\} \nn \\
&& + \int dt \int_{\Sigma_t} d^3 x
\left\{ \left( \pounds_T \phi -
\left[ \phi \right]_T \right) \delta \wp 
-\left( \pounds_T \wp - 
\left[ \wp \right]_T \right) \delta \phi \right\} \nn \\
&& + \int dt \int_{\Omega_t} d^2 x
\left\{ \left( \pounds_T \sqrt{\sigma} -
\left[ \sqrt{\sigma} \right]_T \right) \delta P_{\sqrt{\sigma}} 
- \left( \pounds_T P_{\sqrt{\sigma}} -
\left[ P_{\sqrt{\sigma}} \right]_T \right) \delta 
\sqrt{\sigma} \right\} \nn \\
&& - \int dt \int_{\Omega_t} d^2 x \sqrt{\sigma} \left(
(\bep-\bep^m) \delta \bN 
- (\bj_a - \bj^m_a) \delta \bV^a 
- \frac{\bN}{2} \bs^{ab}
\delta \sigma_{ab} \right), \nn\\
&& + \int dt \int_{\Omega_t} d^2 x \frac{\ssg \bN}{\sqrt{h}} 
\left( - (2\sqrt{h}/\kappa) [\phi]_{\bn} \delta \phi 
- (\bn_a \bar{\cE}^a) \delta \bPhi
+ e^{-2a\phi}
\hat{\bar{\cB}}_a \bn_b \epsilon^{abc} \delta \tilde{A}_c \right).
\nn
\eea

Then, if $h_{ab}$ and $\tilde{A}_a = h_a^\bz A_\bz$ 
are fixed on the boundaries
$\Sigma_1$ and $\Sigma_2$, and $\gamma_{\az \bz}$
(equivalently $\sigma_{ab}$, $\bN$, $\bV^a$) and 
$\gamma_\az^\bz A_\bz$ (equivalently
$\bPhi$ and $\hat{A}_a$) are held constant on $B$,
the solution of $\delta I = 0$ gives the correct
field equations. Namely
\bea
\cH^m &=& 0, \label{mEin1}\\
\cH^m_a &=& 0, \label{mEin2} \\
\pounds_T P^{ab} &=& \left[ P^{ab} \right]^m_T, \label{mEin3} \\
\cQ &=& 0, \label{cQ}\\
\pounds_T \cE^a &=& \left[\cE^a \right]^m_T, \label{tE} \\
\pounds_T \tilde{A}_a &=& \left[\tilde{A}_a \right]^m_T, \label{tA} \\
\pounds_T \wp &=& \left[\wp \right]^m_T, \label{twp} \\
\pounds_T \phi &=& \left[\phi \right]^m_T, \label{tp}
\eea
as well as equations (\ref{Geom1}), (\ref{Geom2}), and (\ref{Free1})
from the previous chapter.

Now (\ref{mEin1}), (\ref{mEin2}), and (\ref{mEin3}) are the
projected Einstein equations (\ref{C1}),
(\ref{C2}), and (\ref{C3}) respectively, so the Hamiltonian
has recovered those correctly. Equations (\ref{cQ}) and (\ref{tE}) 
are the projected Maxwell equations (\ref{DE}) and (\ref{cE}), 
while (\ref{twp}) is dilaton equation (\ref{wpT}).
Equations (\ref{tA}) and (\ref{tp}) are just definitions of the 
respective Lie derivatives while 
(\ref{Geom1}), (\ref{Geom2}), and (\ref{Free1}) continue to
express their earlier meanings. Keep in mind that
the existence of $A_\az$ implied the remaining two 
Maxwell equations (\ref{DB}) and (\ref{cB}). 

From equations (\ref{Epot}) and (\ref{Bpot}) it is easy to see that
fixing $\bPhi$ and $\hat{A}_\az$ on the timelike boundary $B$ is
equivalent to fixing the component of 
$\bar{B}^\az$ perpendicular to $B$ (that is $\bar{B}^\az \bn_\az$) and
the components of $\bar{E}^\az$ parallel to $\Omega_t$ (that is
$\sigma^\az_\bz \bar{E}^\az$). Thus, the action is fully differentiable
only if the parameter space of spacetime studied is restricted to 
those with a specified magnetic charge. This fits in well with the
discussion at the end of the previous section that said that the 
magnetic charge is fixed (to be zero) by the existence of the 
single vector potential generating the EM fields. In contrast
there is no restriction on the electric charge. This issue will
be considered in more detail in section \ref{EMdual}.

Note too that while the value of the dilaton field $\phi$ is fixed
on $\Omega_t$, its ``radial'' rate of change $\pounds_\bn \phi$ is left
free. Therefore, the dilaton charge
\be
Q_{dil} = \int_\Omega d^2 x \ssg \pounds_{\bn} \phi,
\ee
is not fixed either.

\subsubsection{Comparison with the Lagrangian approach}
Again it is reassuring to compare this Hamiltonian analysis with a
Lagrangian analysis and in particular show that the variation of
the Hamiltonian properly fits into that of the action. 
The matter fields considered above were examined from that viewpoint in  
full nonorthogonal form in \cite{naked} and for $\eta=0$
in \cite{jolien,jolienrobb}.

Breaking up the matter term of equation (\ref{mactvar}) and bringing in
the full variation of the gravitational action (\ref{HJGravVar}) from
the last chapter, it is straightforward to show that
\bea
\delta I^m - \delta \underline{I} 
&=& \frac{1}{2 \kappa} \int_M d^4 x \sqrt{-g} 
\left\{ (G_{\az \bz} + \Lambda g_{\az \bz} - 8 \pi T_{\az \bz} ) 
\delta g^{\az \bz} + 4 \cF_{Dil} \delta \phi + 4 \cF_{EM}^\bz \delta A_\bz 
\right\}  \nn \\
&& + \int_\Sigma d^3 x \left\{ P^{\az \bz} \delta h_{\az \bz} + 
\wp \delta \phi + \cE^\az \delta \tilde{A}_\az \right\}  
+ \int_\Omega d^2 x \left\{ P_{\sqrt{\sigma}} 
\delta (\sqrt{\sigma}) \right\} \label{fullmactvar} \\
&& - \int dt \int_{\Omega_t} d^2 x \sqrt{\sigma} \left\{ 
(\bep + \bep^m ) 
\delta   \bN - 
(\bj_\az + \bj^m_\az) \delta \bV^\az 
- \frac{\bN}{2} \bs^{\az \bz} \delta \sigma_{\az \bz} \right\} \nn \\
&& + \frac{2}{\kappa}
\int dt \int_{\Omega_t} d^2 x \sqrt{\sigma} \bN \left\{ 
- {\pounds_{\bn}} \phi \delta \phi + (\bn_\bz \bE^\bz) \delta \bPhi 
- e^{-2a\phi} 
\bu_\az \bn_\bz \epsilon^{\az \bz \cz \dz} 
\bar{B}_\cz \delta \tilde{A}_\dz \right\}. \nn 
\eea
With this approach the momenta are functions of the metrics, normals,
gauge potentials, and dilaton field so equations 
(\ref{Geom1}), (\ref{Geom2}), (\ref{Free1}), (\ref{tA}), and (\ref{tp})
automatically hold. $G_{\az \bz} + \Lambda g_{\az \bz} - 8 \pi T_{\az \bz} = 0$, $\cF_{Dil}=0$, and $\cF_{EM}^\bz = 0$ are the rest of the 
field equations and so again the Lagrangian and Hamiltonian treatments
are equivalent.

\section{Properties of the Hamiltonian}
\label{MatHamProps}
In this section I'll discuss some of the issues that arose in the
previous chapter in the light of the new matter terms. As will be seen
the required changes are incremental rather than qualitative. I will
also examine the action and Hamiltonian in the light of the duality
discussed in \ref{dualreview} and show how it may be used to extend
the preceding analysis to magnetically charged spacetimes.

\subsection{$\pounds_T H^m_t$, conserved charges, and energy}
By the same arguments as used in section \ref{energySect} the 
time rate of change of the Hamiltonian functional with matter
fields included is
\bea
{\pounds_T} H_t^m &=& 
\int_{\Omega_t} d^2 x \sqrt{\sigma} \left( [{\bep}+{\bep}^m] 
{\pounds_T} \bN - [{\bj}_a+{\bj}_a^m] {\pounds_T} {\bV}^a 
- ({\bN} /2) \bs^{ab} {\pounds_T} \sigma_{ab} \right) \nn \\
&& + \int_{\Omega_t} d^2 x \frac{\bN \ssg}{\sqrt{h}} \left(
[\cE^a n_a] 
\pounds_T \bar{\Phi} + e^{-2a\phi} \hat{\bar{\cB}}_a n_b \epsilon^{abc}
\pounds_T \hat{A}_c \right) \\
&& + \int_{\Omega_t} d^2 x \frac{2 \bN \ssg}{\kappa} 
\left[ \phi \right]_{\bn} \pounds_T \phi \nn.
\eea
This is zero if the vector field $T^\az$ defines 
a symmetry of all fields on the boundary $B$ and so in that case
$H^m_t$ is a conserved charge and is conventionally identified
with the mass contained by the surface $\Omega_t$. Of course
by Noether's theorem it is to be expected
that a symmetry corresponds to a conserved charge, but once again note
that it is only symmetries of the fields at the boundary that matter.
The properties of the fields in the bulk are completely irrelevant.

Even if $H_t^m$ is not a conserved charge, I'll still label
it to be the mass contained by $\Omega_t$. Then, the discussion of section \ref{energySect}
largely applies here as well. In particular one can consider
the special case where $\bN = 1$ and $\bV^\az = 0$ and define the 
quasilocal energy
\be
E_{tot} = \int_\Omega d^2 x \ssg (\e + \e^m).
\ee
Note that a gauge choice can be made to set $\e^m = 0$ on $\Omega_t$ in 
which case this reduces to the $E_{Geo}$ defined in equation \ref{Egeo}.

\subsection{Transformation laws}
\label{mattransform}
It is easy to extend the transformation laws
to the matter terms. Again considering reference frames associated with
the normals $(u^\az, n^\az)$ and $(\bu^\az, \bn^\az)$ and reusing the
$\updownarrow$ notation of section \ref{gravtransform} define
\bea
\varepsilon^{m \updownarrow} &\equiv& 
\frac{2}{\kappa} (n^\bz E_\bz) \Phi^{\updownarrow}, \mbox{ and}\\ 
j_\az^{m \updownarrow} &\equiv& \frac{2}{\kappa} 
(n^\bz E_\bz) \hat{A}_\az,
\eea
where $\Phi^{\updownarrow} = A_\az n^\az$. Note that 
$j_\az^{m \updownarrow} = j_\az^m$.
Then, it is almost trivial to show that
\bea
\bep^m &=& \frac{1}{\lambda} \e^m + \eta \e^{m \updownarrow} 
= \gamma (\e^m + v_\vdash \e^{m \updownarrow}) \ \ \mbox{ and} \\
\bj^m_\az & = & j^m_\az.
\eea
Thus,
\bea
\bep + \bep^{m}
&=& \frac{1}{\lambda} (\varepsilon + \varepsilon^m) + 
\eta (\varepsilon^{\updownarrow} + \varepsilon^{m \updownarrow}) \\
&=& \gamma \left( ( \varepsilon + \varepsilon^m) + v_\vdash
(\varepsilon^\updownarrow + \bep^{m \updownarrow}) \right), \nn
\ \ \mbox{and} \\
\bj_\az + \bj^m_\az 
&=& j_\az + j^m_\az 
- \frac{\lambda}{\kappa} \sigma_\az^\bz \partial_\bz \eta \\
&=& j_\az + j^m_\az - 
\frac{\gamma^2}{\kappa} \sigma_\az^\bz \partial_\bz v_\vdash.
\nn
\eea

\subsection{Reference terms}
Virtually no change is required in the discussion of the reference terms
from section \ref{RefTerm}. In principle, with no implications for the
field equations, $\underline{I}$ could be allowed to 
depend on the matter field terms that are fixed on the boundary. That is
$\underline{I}$ is a functional of $h_\az^\bz A_\bz$ 
on $\Sigma_1$ and $\Sigma_2$, $\gamma_\az^\bz A^B_\bz$ on $B$,
and $\phi$ over all three of those boundaries. In practice however,
the usual use of the reference term is to calculate how different
the action of $M$ is from a similar $\underline{M}$ in an ``empty''
reference space and so this option is not generally taken up.

\subsection{Thin shells}
\label{mattershells}
The analogy between the quasilocal formalism and thin shells
can be extended to encompass the Maxwell and dilaton fields
if one allows the shells to have electric and dilaton 
charges and currents embedded in them. 
These charge and current densities are defined to account for
discontinuities in the electromagnetic/dilaton 
field just as the stress tensor is defined to account for
discontinuities in the gravitational field/geometry of spacetime. 

If one assumes that there are no electric charges/currents inside
the thin shell, then calculating the electric charge densities is an
exercise from undergraduate electromagnetism \cite{jackson}. 
Specifically for the foliation $\Omega_t$ of $B$
the electric charge density on the shell is
$-\ssg/\sqrt{h} \bn^\az \bar{\cE}_\az$. At the same time,
given an electromagnetic potential $A_\az$, 
observers on the surface of the shell 
whose evolution is guided by the vector field $T^\az$ will define a 
Coulomb potential $-T^\az A_\az = \bN \bPhi - \hat{V}^\az \hat{A}_\az$.
In the usual way the energy of the charge density in the 
electromagnetic field is then the charge times the potential. 
That is, -$\ssg/\sqrt{h} \bn^\az \bar{\cE}_\az (-T^\az A_\az) =
\bN \bep^m - \bV^\az \bj^m_\az$. As usual this component of the 
energy is gauge dependent. 

Similar reasoning gives the dilaton charge on the shell. The dilaton
charge in a given volume is given by the integral of $\bn^\az \nabla_\az
\phi$ over the surface enclosing that volume.  For black hole solutions,
the value of the dilaton charge is constrained by demanding the spacetime
has no singularities on or outside of the outermost horizon \cite{dilbh}.  
In the thin shell case, $\bn^\az \nabla_\az \phi$ then yields the
dilaton charge density on the shell $\Omega_t$.

The surface charges do not change the definition of the surface stress
energy tensor which was defined entirely by the Einstein equations. As
such they also don't change the definitions of $\mathcal{E}$,
$\mathcal{J}_\az$ and $\mathcal{S}_{\az \bz}$. Therefore, including the
stress energy with the energy of the shell in the gauge field, the total
energy density in a thin shell evolving by the vector field $T^\az$ is
$\bN (\bep + \bep^m) -\bV^\az
(\bj_\az + \bj^m_\az)$ minus the corresponding reference terms.  
This of course is exactly
the same as the Hamiltonian quasilocal energy
of the region of space on and inside of the shell as
measured by a set of observers being evolved by the same vector field, and so the correspondence between thin shells and quasilocal energies
remains.

\section{Electromagnetic duality} 
\label{EMdual} 
In section \ref{matHam} it was demonstrated that the formalism developed
so far only properly applies to spacetimes that do not contain magnetic
charge. Spacetimes with magnetic charge are often of interest however and
in particular in section \ref{nbh}, I'll want to use the formalism to
investigate to naked black holes which are magnetically charged.  Thus, it
is of interest to extend the formalism to allow for such spacetimes.

The obvious way to do this is to make use of the duality reviewed in 
section \ref{dualreview}. Applying this duality, the action becomes 
\bea 
I^{m \star} &=& \frac{1}{2 \kappa} \int_M d^4 x 
\sqrt{-g} (\cR - 2 \Lambda -
2 (\nabla_\az \phi)(\nabla^\az \phi) - e^{2a\phi} \star \! 
F_{\az \bz} \star \! F^{\az \bz}) \\ 
&& + \frac{1}{\kappa} \int_\Sigma d^3 x \sqrt{h} K -
\frac{1}{\kappa} \int_B d^3 x \sqrt{- \gamma} \Theta + \frac{1}{\kappa}
\int_\Omega d^2 x \sqrt{\sigma} \sinh^{-1} (\eta) + \bI. \nn 
\eea 
Note that $F_{\az \bz} F^{\az \bz} = - \star \! \! F_{\az \bz} 
\star \! \! F^{\az \bz}$ so this action is not numerically equal 
to $I^m$.

Breaking up this action with respect to the foliation, one must assume
that there is a single vector potential $A^\star_\az$ generating
$\star F_{\az \bz}$ so that total divergences can be removed to the
boundary. Then a corollary to this assumption is that
$d \star \! \! F = 0$ or equivalently $\nabla_\bz (e^{-2a \phi} F^{\az
\bz})=0$.
From section \ref{3dfieldeq} this relation can be rewritten in terms of
fields in $\Sigma_t$ 
as equations (\ref{DE}) and (\ref{cE}) which are
\bea
D_\az \cE^\az &=& 0 \ \ \mbox{ and} \nn \\
h^b_\bz \pounds_T \cE^\bz &=& -\ez^{bcd} D_c \left[ N e^{-2a\phi} \cB_d
\right] + \pounds_V \cE^b. \nn
\eea
Recall that assuming a single $A_\az$ implied the other pair of 
Maxwell equations. The arguments of section \ref{FormMatHam} can 
then trivially be extended to show that there are no electric
charges in a spacetime where the Maxwell field can be described
by such a single $A^\star_\az$. However, 
magnetic charge is allowed.

That said, the action may
be broken up with respect to the foliation to become,
\bea
\label{starMmactdecom}
I^{m \star} - \underline{I} &=& \int dt \left\{ 
\int_{\Sigma_t} d^3 x \left( P^{\az \bz} 
\pounds_T h_{\az \bz} + \wp \pounds_T \phi + \cB^\az 
\pounds_T \tilde{A}^\star_\az \right) \right\} \\
&& + \int dt \left\{
\int_{\Omega_t} d^2 x \left( P_{\sqrt{\sigma}}
\pounds_T \sqrt{\sigma} \right)  - H^{m \star}_t \right\}, \nn
\eea
and
\bea
H^{m \star}_t &=& \int_{\Sigma_t} d^3 x [N \cH^m + V^a H^m_a + T^\az A^\star_\az \cQ^\star ] \\ 
&& + \int_{\Omega_t} d^2 x \sqrt{\sigma} \left\{ \bN (\bep + 
\bep^{m \star})  - \bV^\az (\bj_\az + \bj^{m \star}_\az) \right\}.
\nn \eea
$T^\az A^\star_\az = - N \Phi^\star + V^\az \tilde{A}^\star_\az$
($\Phi^\star$ and $\tilde{A}^\star_\az$ defined in section
\ref{dualreview}) and
$\cQ^\star = - D_\bz \cB^\bz$ is the free space magnetic version of
Gauss's law from electrodynamics (derived from its 4D form at
equation (\ref{DB})). Further
\bea
\bep^{m \star} &\equiv& - \frac{1}{\sqrt{h}} (n_\bz \cB^\bz) 
(\frac{1}{\lambda} \Phi^\star - \eta \tilde{A}^\star_\az n^\az) 
= - \frac{1}{\sqrt{h}} (\bn_\bz \bar{\cB}^\bz) \bar{\Phi}^\star 
\label{starbepm} \mbox{ and}\\
\bj^{m \star}_\az &\equiv& - \frac{1}{\sqrt{h}} (n_\bz \cB^\bz) 
\hat{A}^\star_\az = - \frac{1}{\sqrt{h}} (\bn_\bz \bar{\cB}^\bz) 
\hat{A}^\star_\az \label{starbjm}
\eea
are the new matter energy and angular momentum terms. Note that they
are different from $\bep^m$ and $\bj^m_\az$ which depended on 
the regular vector potential $A_\az$ and electric field density
$\cE_\az$. The bar retains its usual meaning.

Sticking to the Hamiltonian perspective that momenta are independent of 
their corresponding configuration quantities, the total variation of
this action is 
\bea
\delta I^{m \star} - \delta \bI &=& 
\int_\Sigma d^3 x \left( P^{ab} \delta h_{ab}
 + \cB^a \delta \tilde{A}^\star_a + \wp\delta \phi \right)
 +  \int_\Omega d^2 x P_{\sqrt{\sigma}} \delta
\sqrt{\sigma} \label{starHmactvar} \\
&& - \int dt \int_{\Sigma_t} d^3 x \left\{
(\cH^m - \Phi^\star \cQ^\star) \delta N 
+ (\cH_a^m + \tilde{A}^\star_a \cQ^\star) \delta V^a  
-  N \cQ^\star \delta \Phi^\star \right\}
\nn \\
&& + \int dt \int_{\Sigma_t} d^3 x 
\left\{ \left( \pounds_T h_{ab} -
\left[ h_{ab} \right]_T \right) \delta P^{ab} -
\left( \pounds_T P^{ab} - 
\left[ P^{ab} \right]_T \right) \delta h_{ab} \right\} \nn \\
&& + \int dt \int_{\Sigma_t} d^3 x
\left\{ 
\left( \pounds_T \tilde{A}^\star_a -
\left[ \tilde{A}^\star_a \right]_T \right) \delta \cB^a 
-
\left( \pounds_T \cB^a - 
\left[ \cB^a \right]_T \right) \delta \tilde{A}^\star_a 
\right\} \nn \\
&& + \int dt \int_{\Sigma_t} d^3 x
\left\{\left( \pounds_T \phi -
\left[ \phi \right]_T \right) \delta \wp^\star 
-
\left( \pounds_T \wp^\star - 
\left[ \wp^\star \right]_T \right) \delta \phi 
\right\} \nn \\
&& + \int dt \int_{\Omega_t} \left\{ \left( \pounds_T \sqrt{\sigma} -
\left[ \sqrt{\sigma} \right]_T \right) \delta P_{\sqrt{\sigma}} 
- \left( \pounds_T P_{\sqrt{\sigma}} -
\left[ P_{\sqrt{\sigma}} \right]_T \right) \delta 
\sqrt{\sigma} \right\} \nn \\
&& - \int dt \int_{\Omega_t} d^2 x \sqrt{\sigma} \left\{
(\bep-\bep^{m \star}) \delta \bN 
- (\bj_a - \bj^{m \star}_a) \delta \bV^a 
- \frac{\bN}{2} \bs^{ab}
\delta \sigma_{ab} \right\}, \nn\\
&& + \int dt \int_{\Omega_t} d^2 x \frac{\ssg \bN}{\sqrt{h}} 
\left\{ - (2\sqrt{h}/\kappa) [\phi]_{\bn} \delta \phi 
- (\bn_a \bar{\cB}^a) \delta \bPhi^\star
- e^{2a\phi}
\hat{\bar{\cE}}_a \bn_b \epsilon^{abc} \delta \tilde{A}^\star_c
\right\}
\nn
\eea
where $[P^{ab}]^m_T$, $[h_{ab}]_T$, $[\wp]_T$, $[\phi]_T$ and 
$[\phi]_\bn$ retain their earlier meanings
while
\bea
\left[ \tilde{A}^\star_a \right]_T 
& \equiv & \frac{N \kappa}{2 \sqrt{h}} e^{-2a\phi} \cB_a 
+ \pounds_V \tilde{A}^\star_a - D_a [N \Phi^\star], \\
\left[ \cB^a \right]_T 
& \equiv &  \epsilon^{abc} 
D_b [N e^{-2a\phi} \cE_c ] + \pounds_V \cB^a, \mbox{ and}\\
\hat{\bar{\cB}}^b &\equiv& \frac{1}{\lambda} \hat{\cB}^b 
+ \eta e^{-2a\phi} \ez^{bc} \hat{\cE}_c.
\eea

Applying the duality relations it is easy to see that these equations of 
motion are equivalent to the earlier ones. Note however, that the 
electromagnetic quantities that must be kept constant on the 
boundaries have changed. Specifically on $\Sigma_1$ and $\Sigma_2$,
$\tilde{A}^\star_\az$ must be kept constant while on the $B$ boundary
$\gamma_\az^\bz A^\star_\bz$ (or equivalently $\Phi^\star$ and 
$\sigma_a^b \tilde{A}^\star_b$) must be held constant. This corresponds
to holding $\cE_\az n^\az$ and $\hat{\cB}^\az = \sigma^\az_\bz \cB^\bz$
constant which means that now the electric charge must be held 
constant (at zero by the earlier comments) while the magnetic charge
is not fixed.

Thus, there are now well defined formalisms which can be used to 
study spacetimes containing either electric or magnetic charges. 
What is missing is a formalism that easily handles dyonic 
spacetimes\footnote{That is those with both electric and magnetic
charges.}. A duality rotation could be used to study
a spacetime with a particular dyonic charge but even then
there would still be problems in spacetimes containing multiple
dyons with varying ratios of electric and magnetic charges.
Furthermore, there is something
fundamentally unsatisfying about having the form of the action 
depend on the charges contained in the spacetime. As it stands,
I don't have a solution for this problem 
and so will not consider dyonic spacetimes in this thesis.

\chapter{Classical applications}
\label{quasiExamples}

In this chapter I will apply the quasilocal energy derived from the 
Hamiltonians of chapters three and four to investigate a variety of 
spacetimes. First, in section \ref{sss}, I'll provide some 
orientation for the reader by examining the quasilocal energies
seen by static and moving observers in Schwarzschild and 
Reissner-Nordstr\"{o}m spacetimes. Section \ref{nbh} will then
apply the work to study naked black hole spacetimes. These
spacetimes are characterized by the fact that static and 
infalling observers experience very different tidal forces
from each other, 
and that section will demonstrate that they also measure
very different quasilocal energies and explain the connection 
between the two results. Finally, section \ref{tidal} applies the
formalism to calculate energy transfers during gravitational tidal
heating such as that seen in the Jupiter-Io system. This last section
can then be seen from two different points of view. First of all it 
can be seen as an alternate way to calculate the magnitude of these
effects from the usual Newtonian or pseudo-tensor methods or 
secondly it can be viewed as a test of the formalism 
to see if it produces the standard answers.

Note that section \ref{sss} is based on equivalent sections in  
\cite{nopaper} and \cite{naked}. The work found in section
\ref{nbh} was published in \cite{naked}, while
section \ref{tidal} formed part of \cite{tidal}.

\section{Reissner-Nordstr\"{o}m spacetimes}
\label{sss}
This section examines the quasilocal quantities measured by 
observers undergoing various motions in Reissner-Nordstr\"{o}m
(RN) spacetimes (and Schwarzschild as a special case). 
In standard form the RN metric is
\be
ds^2 = - F(r) dt^2 + \frac{dr^2}{F(r)} 
+ r^2 (d\theta^2 + \sin^2 \theta d\varphi^2),
\ee
where $F(r) \equiv 1 - \frac{2m}{r} + \frac{E_0^2 + G_0^2}{r^2}$, $m$ is
the mass, and $E_0$ and $G_0$ are respectively the electric and magnetic
charges of the hole. The accompanying electromagnetic field is described
by the two form
\be
F = -\frac{E_0}{r^2} dt \wedge dr + G_0 \sin \theta d \theta \wedge d \varphi,
\ee 
while a local vector potential generating this field is
\be
A = -\frac{E_0}{r} dt - G_0 \cos \theta d \varphi + d \chi,
\ee
where $\chi = \chi(t,r,\theta,\varphi)$ is any function defined 
over $M$. 
For $\chi = 0$, note that $A$ is not defined for all of $M$ since 
$d \varphi$ is not defined on $\theta = 0, \pi$. This is in accord
with the discussion of section
\ref{EMdual} where I showed that the Lagrangian and Hamiltonian
formalisms as constituted are not suitable for discussing dyonic
spacetimes. In the $(F_{\az \bz}, A_\az)$ form considered here,
a single $A_\az$ cannot describe the field due to a magnetic charge
and so this property of $A_\az$ is not just an annoyance that 
can be removed with a clever gauge transformation. Given this
difficulty I set $G_0=0$ and
focus on electric black holes in the following subsections.
The results for magnetic black holes are identical if one
switches $E_0$ and $G_0$ and adds in the appropriate $\star$'s and minus
signs.

I begin the study of these spacetimes by calculating
the quasilocal quantities
measured by a static, spherically symmetric set of observers. 
Their observations
will be the subject of the next subsection, while the
two that follow will compare their measurements with those of a boosted
set who instantaneously coincide with them on a surface 
$\Omega_t$, which is a surface of constant $r$ and $t$. 
The observers are evolved by the vector field 
$T^{\az} = N(r) u^\az$, where 
$N(r)$ is the lapse function while the shift
$V^\az=0$. For any choice of $N(r)$ the observers will be static in the
restricted sense that they don't observe any changes in the $\Omega_t$
metric $\sigma_{\az \bz}$, and the lapse just determines how they
choose to measure their time on the surface $B$. 
In particular, choosing $N(r) = \sqrt{F(r)}$
they measure time according to the coordinate $t$ and 
$T^\az$ corresponds to the timelike Killing vector for the full  
spacetime metric $g_{\az \bz}$, while choosing
$N(r) = 1$ the observers measure proper time (that is
$T^\az T_\az = -1$).

Then, with $u^\az = \frac{1}{\sqrt{F(r)}} \partial^\az_t$ 
and $n^\az = \sqrt{F(r)} \partial^\az_r$, where $\partial^\az_t$
and $\partial^\az_r$ are the coordinate forms of the vector fields
$\frac{\partial}{\partial r}$ and $\frac{\partial}{\partial t}$ 
respectively, a series of straightforward calculations yields
\bea
\varepsilon &=& -\frac{1}{4\pi r^2} \sqrt{ r^2 F }, \label{epSSS}\\
\varepsilon^m &=& \frac{1}{4 \pi r^2} 
\frac{E_0 (E_0 - r \partial_t \chi)}{\sqrt{ r^2 F}}, \mbox{ and}\\
\underline{\varepsilon} &=& -\frac{1}{4 \pi r}. \label{uepSSS}\\
\eea
As usual I am working in geometric units where $G=c=\hbar=1$, so
$\kappa = 8
\pi$. The reference terms are defined by embedding the sphere $\Omega_t$
statically in the obvious way in Minkowski space\footnote{ Clearly this
embedding isn't unique. Still, it is convenient and given the many issues
involved in choosing reference terms (section \ref{RefTerm}) this is all
that I will ask for!}. Since this is a spherical set of observers in a static spacetime the $j_a$ angular momentum terms vanish.

Strictly speaking, to rigorously apply the quasilocal formalism to 
black holes I should include an inner boundary $B'$ as well as
the outer boundary $B$, in which case $M$ would be homeomorphic
to $\R^2 \times S^2$ rather than $\R^4$. 
Without such a boundary to remove the
collapsing matter/singularity at the centre of the black hole
from consideration, the quasilocal formalism is not properly 
constituted since these were not accounted for in its setup.
Even for an eternal black hole foliated with 
Einstein-Rosen-bridge hypersurfaces that avoid the singularity,
there is a difficulty in that the leaves of the foliation 
all intersect
on the horizon. For a good discussion of this problem see
ref.\cite{frolov}, but in the following I will generally consider
only the outer boundary so as to facilitate comparisons with 
non-singular spacetimes (such as stars). When studying the quasilocal
energy of black holes it is conventional to proceed in this way. 
In some sense it is equivalent to using Gauss's law to calculate
the electric charge of a point particle without worrying about the
divergence of the fields at the particle itself.

\subsection{Static observers}

\subsubsection{Static geometric energy}

First I calculate the part of the 
quasilocal energy associated with the density
$\varepsilon$. Following section \ref{energySect}, I label it the 
geometric energy since it depends only on the extrinsic curvatures. It
can be thought of as the full QLE with a gauge choice made so that
$\varepsilon^m = 0$. Then,
\be
E_{Geo} = \int_{\Omega_t} d^2 x \sqrt{\sigma} \varepsilon = 
- \sqrt{r^2-2mr+E_0^2}.
\ee
In the large $r$ limit this becomes $E_{Geo} \approx -r + m + \frac{1}{2r} 
(m^2 - E_0^2)$. The $\underline{\varepsilon}$ reference term is
\be
\underline{E} = \int_{\Omega_t} d^2 x \sqrt{\sigma} 
\underline{\varepsilon} = - r,
\ee
so 
\be
E_{Geo}-\underline{E} = r - \sqrt{r^2-2mr+E_0^2} \approx m + 
\frac{1}{2r}(m^2-E_0^2),
\ee
in the large $r$ limit. Note that $E_{Geo}-\underline{E}$ monotonically
decreases as $r$ increases, starting at $2m$ on the horizon and 
reaching a minimum of $m$ at
infinity. Thus by this measure the energy contained in the fields is
negative which is to be expected for a binding energy such as gravity.

For large $r$ these are the results that would be expected from an
application of Newtonian intuition to the (equivalent) thin shell
situation.  From this viewpoint consider how much energy it would take to
construct a shell of radius $r$ with mass $m$ and charge $E_0$. First
it would cost $m$ units of energy to create the required mass
at spatial infinity 
where there would be no interactions and so no deviation
from the rest energy. Then, using Newton's and Coulomb's laws,
and assuming that mass and charge are equally distributed throughout
the matter, 
it is straightforward to show that $-\frac{1}{2r} (m^2 - E_0^2)$
units of work are required to assemble the shell out of the created 
material out at infinity. It is then very
natural to say that this energy is ``stored'' in the field, outside
radius $r$. Thus, assuming that conservation of energy holds once the
matter is created, the energy
contained on and/or inside the shell with radius $r$ is
\be
(\mbox{total energy}) - (\mbox{energy in fields outside the shell}) = m
+ \frac{1}{2r} (m^2 - E_0^2),
\ee 
as was calculated above. This limiting case was first considered 
in the original Brown and York paper \cite{BY1}. 

Note that for an extreme black hole where
$|E_0| = m$, $E_{Geo} - \underline{E} = m$ is a
constant. From the Newtonian shell point of view this 
makes sense. During the construction of the
shell out of particles which also
have equal mass and charge, equal but opposite electric and
gravitational forces would act on the particles.
Thus, no work must be done to build the shell and so no energy is
stored in the fields. Alternatively equal amounts of positive and
negative energy are stored in the electric and gravitational 
fields and cancel each other out. The only energy is then that stored
in the mass and so the energy contained by $\Omega_t$ is $m$ for 
all radii greater than $m$.

Note that even though this is the ``geometric energy'' with the matter
terms omitted, it certainly seems to include the energy contributions
from the electric as well as gravitational fields. Thus
one may think of this geometric energy as a ``configuration energy''
that arises from the spatial relationships of different parts of the
spacetime
to each other. By contrast in the next subsection 
where the gauge dependent terms are included, 
the energy also includes ``position'' terms that arise
due to the position of the different parts of the spacetime in the gauge
potential (a point of view also explored in thin shell section
\ref{mattershells}).
Of course, the form of the gauge potential is determined up to
a gauge transformation by the matter so this view of the terms as being
configurational versus positional is at best a rough way to think of them.

\subsubsection{Static total energy}

Next consider $\varepsilon + \varepsilon^m$, the full energy density
that was derived from the variational calculations (as opposed to the
geometric energy which is gauge fixed so that the $\bPhi=0$ on $B$).
Then 
\be
E_{tot} = \int_{\Omega_t} d^2 x \sqrt{\sigma} (\varepsilon + \varepsilon^m) 
= \frac{-r^2 + 2mr - E_0 r \partial_t \chi}{\sqrt{r^2 - 2mr + E_0^2}}.
\ee 
As has been emphasized before this expression is manifestly gauge
dependent. Even worse however is the fact that this energy will in general diverge at the outer horizon of a black hole.  
Before I deal with that
worry however, consider the usual $r\rightarrow \infty$ limit.

Demanding that $A_\az$ has the same spherical and time translation
symmetries as the spacetime, $\chi = - \Phi_\infty t + f(r)$ where
$\Phi_\infty \equiv \lim_{r \rightarrow \infty} \Phi$ is a constant and
$f(r)$ is an arbitrary function of $r$. Then,
\be
E_{tot} - \underline{E} = r - \frac{r^2 - 2mr - E_0 r \partial_t
\chi}{\sqrt{r^2 - 2mr + E_0^2}} \approx (m + E_0 \Phi_\infty) +
\frac{1}{2r}(m^2+E_0^2 + 2mE_0 \Phi_\infty).
\ee
Since the total energy is the sum of the geometric energy and the 
gauge dependent term, it isn't surprising that this Newtonian limit
is the sum of the Newtonian limit of the geometric energy and 
the ``positional'' potential term. One can think of $\Phi_\infty$ as the
zero level of the potential throughout space (it remains even if 
$E_0 \rightarrow 0$) and so by the thin shell analogy 
$E_0 \Phi_\infty$ is the energy cost to create matter with
charge $E_0$ at infinity (apart from the energy costs associated with
the mass).
For extreme black holes recall that $E_{Geo} - \underline{E} = m$ so
$E_{tot} - \underline{E} = m + \int_{\Omega_t} d^2 x \sqrt{\sigma} 
\varepsilon^m$ and the only energy is the mass $m$ plus the
energy of the charge with respect to the potential. 

In most situations the exact choice of gauge is just a matter of 
convenience. For black hole spacetimes however, 
$E_{tot}$ will diverge on the horizon with most gauge choices.
This divergence can be directly traced to the fact that the 
Coulomb potential $\Phi = - u^\az A_\az = \frac{1}{\sqrt{F}} 
(\frac{E_0}{r} - \partial_t \chi)$ also diverges at the horizon. To
remove both divergences choose $\chi$ such that
$\partial_t \chi \rightarrow 
\frac{E_0}{r_+}$ as $r \rightarrow r_+$, where $r_+$ is the outer
black hole horizon. That is, set the Coulomb potential
to zero on the black hole horizon. Then, assuming that $A_\az$ has the 
symmetries discussed above $\Phi_\infty = - \frac{E_0}{r_+}$. 
Making that choice, a little algebra leads to
\be
E_{tot} = -r \sqrt{ \frac{r-r_+}{r-r_-}}
\ee
where $r_\pm = m \pm \sqrt{m^2 - E_0^2}$ are the radial positions of
inner and outer horizons. This gauge will also
be used for the naked black holes.
For extreme black holes $r_+ = r_- = |E_0| = m$ and so 
$E_{tot}= -r - \underline{E} = 0$ 
everywhere. 
Physically the gauge has been chosen so that the electromagnetic
potential energy is a constant and everywhere equal $-m$. The
(in this case negative) 
electric potential energy cancels the mass-energy while at the same time
the positive energy of the electric field cancels the negative 
binding energy of the gravitational field. 

So, as suggested at the end of the last section, the total energy 
may be split into two parts. The geometric
part depends only on the configuration of the spacetime and 
examining the Newtonian limit one can see that it appears to include
not only the gravitational but also the
electromagnetic ``configurational'' energies. 
By contrast this section showed that 
the gauge dependent part exclusively deals with the 
potential of the matter relative to the gauge field. 
As has been seen,
for a given solution to the Einstein-Maxwell equations the
total QLE for a given surface
may take any value (including zero) depending
on the exact gauge choice. As such it is clear that this gauge 
dependent part of the energy should not be reflected in the 
geometry of the spacetime as indeed it isn't, since the 
stress-energy tensor $T_{\az \bz}$ doesn't depend on the 
gauge potential. 
On the other hand it should not be concluded that this
gauge dependent part is meaningless. It certainly plays
a role equal to the geometric energy in both thermodynamics 
\cite{BY2, jolien, jolienrobb} and black hole pair creation (chapter
\ref{PC}).

\subsubsection{Value of the Hamiltonian}

\label{HamGaugeDual}

Next consider the value of the Hamiltonian as calculated by 
the same static sets of observers who now
measure time according to the (Killing) time coordinate
$t$  (that is lapse $N = \sqrt{F}$). Then
\bea
H_{Geo} &=& N E_{Geo} = - r F, \\  
H_{Geo} - \underline{H} &=& N(E_{Geo} - \underline{E})
= \sqrt{r^2 F}(1-\sqrt{F}),    \\ 
H_{tot} &=& N E_{tot} = -r + 2m + E_0 \partial_t \chi, \ \ \ \ \ \
 \mbox{ and }
\\
H_{tot} - \underline{H} &=& N (E_{tot} - \underline{E}) 
= 2m + E_0 \partial_t \chi + \sqrt{r^2 F} -r
\eea
In the large $r$ limit, $H_{Geo} - \underline{H} \approx m -
\frac{m^2+E_0^2}{2r}$ and $H_{tot} - \underline{H} \approx 
(m + E_0 \Phi_\infty) - 
\frac{m^2 - E_0^2}{2r}$. Thus as is usual for asymptotically flat 
spacetimes the Hamiltonian corresponds to the QLE in 
the $r \rightarrow \infty$ limit. Note too though that
even in the large $r$ limit, away from infinity 
these Hamiltonians don't agree with
the Newtonian limits discussed earlier. In 
particular the contribution from the gravitational field doesn't
have the right sign in either case and the electric
contribution is also wrong for the total Hamiltonian.

Finally consider the earlier comments on the gauge dependence
of the Hamiltonian (clearly seen above by the $\chi$ dependence) 
from section \ref{FormMatHam}. 
To avoid the complications of singularities
in the gauge potential redefine $M$ as the 
region of $\cM$ contained by the two timelike hypersurfaces
$r=r_1$ and $r=r_2$ where $r_+<r_1 < r_2$. Again
foliate that region according to the standard time coordinate $t$.
Since I am considering the gauge dependence of the Hamiltonian
the reference terms are ignored since they are gauge invariant.

Then the total $H^m_t$ for a spacelike slice $\Sigma_t$ with boundary
$\Omega_t$ is
\be
H^m_t[\Sigma_t] = \Sigma H_{tot} = (r_1 - r_2)  
+ \frac{E_0}{2} [\partial_t \chi]^{r2}_{r1}
\ee
where $[\partial_t \chi]^{r_2}_{r_1} = \partial_t \chi |_{r=r_2} - 
\partial_t \chi |_{r=r_1}$ and the sum is over the two boundary 
components.  Thus, at first glance 
the Hamiltonian appears to be gauge dependent. 
Recall however that
section \ref{FormMatHam} showed that it could only be expected
to be (partially) gauge independent if $M$ was
a region containing no singularities and 
${\pounds_T} \tilde{A}_\az = 0$.
Well, there are no singularities in $M$ and a quick calculation 
shows that ${\pounds_T} \tilde{A}_\az = 0$ 
implies that $\partial_t \chi$
is constant over $\Sigma_t$. If this is true then 
$[\partial_t \chi]^{r_2}_{r_1} = 0$ and the Hamiltonian is
(partially) gauge independent as expected.

\subsection{Radially boosted observers}
\label{infall}

This section considers the energies measured by spherically symmetric
sets of observers who are moving radially towards or away from 
the gravitational source in the RN spacetime. 
As before,
$\Omega_t$ is a surface of constant $r$ and $t$
but this time set the time vector $\bT^\az = \bN \bu^\az$ where 
$\bu^{\az} = \frac{1}{\lambda} u^\az + \eta n^\az = 
\gamma ( u^\az + v_\vdash n^\az)$. As in section \ref{geometry},
$v_\vdash$ is the speed of the $\bT^\az$ observers in the 
$n^\az = \sqrt{F} \partial_r^\az$ 
direction as measured by the static set of observers that I have been
working with up until now.

Then, a straightforward calculation shows $\varepsilon^\updownarrow = 0$
so
\be
\label{EGF}
\bar{E}_{Geo} = \gamma E_{Geo} = - \gamma r \sqrt{F}.
\ee 
Unfortunately, from the point of view of simplicity, within the
gauge freedom $\varepsilon^{m \updownarrow}$ is not necessarily zero.
Even if one considers only gauge choices that give
$A_\az$ the same symmetries as the spacetime 
$\chi = - \Phi_\infty t + f(r)$, where as noted earlier
$\Phi_\infty$ is a constant and $f$ is any function of $r$. Then  
\be
\varepsilon^{m \updownarrow} = -\frac{E_0 }{4 \pi} \sqrt{F} 
\partial_r f.
\ee 
In the interests of simplicity however, I make the standard gauge
choice for electrostatics and let $\partial_r f = 0$. Then
the Lorentz-type transformation laws apply and
\be
\label{Etot}
\bar{E}_{tot} = \gamma E_{tot} = -\gamma r \left( \sqrt{F}  
- \frac{E_0}{r \sqrt{F}} \left[ \Phi_\infty + \frac{E_0}{r} \right] \right).
\ee
As before, I choose
$\Phi_\infty = - \frac{E_0}{r_+}$ so that this quantity doesn't 
diverge at the horizon.

To include the reference terms, it is necessary to find a time 
vector $\underline{\bT}^\az$ for the reference spacetime such that
$\underline{\bT}^{\az}
\underline{\bT}_\az = \bT^\az \bT_\az$
and $\pounds_{\underline{\bT}} \underline{\sigma}_{\az \bz} = 
\pounds_{\bT} \sigma_{\az \bz}$ (the conditions from page
\pageref{EmbedCond}). Such a vector field is given by
\be
\underline{\bT}^{\az} = \gamma \left( \sqrt{1 - (1-F)v_\vdash^2}
\underline{u}^\az +
 v_\vdash \sqrt{F} \underline{n}^\az \right),
\ee
where $\underline{u}^\az = \partial_{\underline{t}}^\az$,
$\underline{n}^\az = \partial_{\underline{r}}^\az$ and
$\underline{t}$
and $\underline{r}$ are 
the usual time and radial coordinates for Minkowski space.

Then
\be
\underline{v}_\vdash \equiv -\frac{ \underline{\bT}^\az 
\underline{n}_\az }{ \underline{\bT}^\az \underline{u}_\az}
= \frac{ v_\vdash \sqrt{F} }{ \sqrt{ 1 - (1-F)v_\vdash^2 }}
\ \ \mbox{ and} \ \ 
\underline{\gamma} = \gamma \sqrt{1 -  (1-F)v_\vdash^2},
\ee
which implies
\be
\label{Eb}
\underline{\bar{E}} = \underline{\gamma} \underline{E} 
= -r \gamma \sqrt{1 - (1-F) v_\vdash^2},
\ee
and thence
\be 
\bar{E}_{Geo} - \underline{\bar{E}} = r \gamma \left( 
\sqrt{1 - (1-F)v_\vdash^2}  - \sqrt{F} \right) ,
\ee
and
\be
\bar{E}_{tot} - \underline{\bar{E}} = r \gamma 
\left( \sqrt{1 - (1-F) v_\vdash^2} 
  - \left(\sqrt{F} - \frac{E_0}{r \sqrt{F} } 
\left[ \Phi_\infty + \frac{E_0}{r} \right] \right) \right). 
\ee

As they stand these expressions are quite complicated and their
physical interpretation isn't at all obvious. To simplify things
a little consider the large-$r$/small-$v_\vdash$ limit. 
Then, to first order 
in $\frac{1}{r}$ and first order in $v_\vdash^2$
\be
\bar{E}_{Geo} - \underline{\bar{E}} \approx  
m - \frac{1}{2}mv_\vdash^2 + \frac{1}{2r} \left(m^2 - E_0^2\right),
\ee
and 
\be
\bar{E}_{tot} - \underline{\bar{E}} \approx  (m + E_0 \Phi_\infty) 
- \frac{1}{2} (m + E_0 \Phi_\infty) v_\vdash^2 
+ \frac{1}{2r} (E_0^2 + m^2 + 2m E_0 \Phi_\infty).
\ee
These results are interesting but unfortunately confound Newtonian 
intuition. Radial motion of the observers serves to 
decrease the quasilocal energy measured. Specifically
the boosted quasilocal energies are equal to their
unboosted counterparts minus a kinetic term equal to the 
$\frac{1}{2} (\mbox{Total Energy of Fields}) v_\vdash^2$. 
The thin shell equivalence and Newtonian intuition would lead one
to expect the opposite sign for this kinetic energy term so this
is a bit disturbing. By contrast the no-reference-term quantities
increase with motion in the expected way. Some discussion of
why this happens may be found in section \ref{nbh}
where the equivalent effect is considered for naked black holes,
but briefly the decrease can be thought of as occurring because
the relativistic effects of the boost compete with those
of gravity. Thus, $\bep$ and $\underline{\bep}$ begin to 
converge even as they are both boosted to larger values by 
the motion.

\subsubsection{Infalling observers}

Next consider the special case where the radially moving observers are 
falling along timelike geodesics towards the gravitational source (be
it a black hole or any other spherically symmetric matter distribution).
Let these observers have started with velocity 
zero ``close to infinity'' and then have 
fallen along radial timelike geodesics inwards. Rigorously, the
geodesic is the one that, with respect to the standard time foliation,
has radial velocity zero at infinity and $-1$ (ie.\ the speed of light) 
at the outer horizon (if the source is a black hole).
Now, a test particle starting with velocity zero at radial coordinate
$r_0$ and then allowed to fall towards a black hole on a 
radial geodesic will have coordinate velocity
\be
\label{velocity}
\frac{dr}{d\tau} = -\sqrt{F(r_0) - F(r)},
\ee
as a function of $r$, where $\tau$ is the proper time. Thus
an observer infalling on a geodesic that was static at 
infinity will have coordinate velocity 
$\frac{dr}{d\tau} = -\sqrt{1 - F(r)}$.

Let these observers measure time in
the natural way (that is $\bN = 1$), so 
$\bT^\az = \frac{1}{\sqrt{F}} u^\az - \sqrt{\frac{1-F}{F}}
n^\az.$ Then the instantaneous radial velocity of the 
$\bT^\az$
observers as measured in the static $u^\az$ frame is
\be
v_\vdash \equiv - \frac{\bT^\az n_\az}
{\bT^\bz u_\bz}
= - \sqrt{1 - F},
\ee
and so the Lorentz factor is $\gamma = \frac{1}{\sqrt{F}}$.

Substituting this value for $\gamma$ into equations 
(\ref{EGF},\ref{Etot},\ref{Eb}) and making the gauge choice
$\Phi_\infty = - \frac{E_0}{r_+}$ so that $\bar{E}_{tot}$ doesn't diverge
at the horizon, 
\bea
\bar{E}_{Geo} &=&  - r, \\
\bar{E}_{tot} &=&  - \frac{r^2}{r-r_-}, \mbox{ and}\\
\underline{\bar{E}} &=& -r \sqrt{2-F}.
\eea
Note that as $r \rightarrow r_+$ all of these take non-zero values.
By contrast $E_{Geo}$ and $E_{tot}$ both are zero at $r_+$. Also, keep
in mind that for a near extreme black hole, $r_+ \approx r_-$.
Therefore for a black hole that is very close to being extreme, the
observers measure
$\bar{E}_{tot}$ to have a very large negative value as they approach the
horizon. 

Including the reference terms,
\be 
\bar{E}_{Geo} - \underline{\bar{E}} = r (\sqrt{2-F} - 1),
\ee
and
\be
\bar{E}_{tot} - \underline{\bar{E}} = r \left( \sqrt{2-F} 
- \frac{r}{r-r_-} \right)
\ee
So near the horizon the infalling gravitational energy (including the
reference term) goes to $(\sqrt{2}-1)r_+$ compared to $r_+$ for the static
gravitational energy. By contrast, the infalling total energy (including
reference term) attains arbitrarily large negative values as the observers
approach the horizon for black holes that are arbitrarily close to being
extreme. Static observers however, will measure $E_{tot} - \underline{E}
= r_+$ as they hover around the horizon. The difference is essentially 
due to the hugely boosted matter terms. The boosting of 
the geometric terms has a comparatively minor effect. 

\subsection{Z-boosted observers}

Finally consider the slightly more complicated example of a spherical
set of observers in Schwarzschild space 
who are boosted to travel ``in the z-direction''
with ``constant'' velocity $v_z$. In this case, ``constant'' 
means with respect to the usual set of static and spherically
symmetric observers whose four-velocity is $u^\az$.

Then the four-velocity of the boosted observers is 
$\bT^\az = \bN \bu^\az + \bV^\az$ in the usual way where
\bea
\bN &=& \sqrt{ \frac{1 - v_z^2 \cos^2 \theta}{1-v_z^2} }, \\ 
\bu^\az &=& \frac{1}{\sqrt{1 - v_z^2 \cos^2 \theta}} 
\left( \frac{1}{\sqrt{F}} \partial_t^\az + v_z \sqrt{F} \cos 
\theta \partial_r^\az \right), \mbox{ and}\\
\bV^\az &=& \frac{v_z \sin \theta}{r \sqrt{1-v_z^2}}
\partial_\theta^\az,
\eea
where $\partial_r^\az$ and $\partial_\theta^\az$ 
are the coordinate forms of the 
vector fields $\frac{\partial}{\partial r}$ and
$\frac{\partial}{\partial \theta}$ respectively. 
Note that $\bT^\az$ has been normalized so that the boosted observers
measure proper time. Then, the static observers see the boosted
observers as having velocity $v_\vdash = v_z \cos \theta$ in the radial
direction and $v_\theta = v_z \sin \theta$ in the $\theta$ direction. 

Now, taking $\Omega_t$ as a spacelike surface of constant 
$r$ and $t$, equation
(\ref{epSSS}) along with the transformation laws 
(\ref{epTrans}) and (\ref{jTrans}) gives
\bea
\bep &=& \gamma \varepsilon = 
- \frac{2}{\kappa r} \sqrt{ \frac{F}{1-v_z^2 \cos^2 \theta} } 
\ \ \ \mbox{ and }\\
\bj_\az &=& \frac{\gamma^2}{\kappa} \sigma_\az^\bz \partial_\bz v_\vdash
= - \frac{v_z \sin \theta}{\kappa (1-v_z^2 \cos^2 \theta) }
[d \theta]_\az,
\eea
where $[d \theta]_\az$ is $d \theta$ in coordinate form.

Next, I calculate flat-space-embedding reference terms. By 
embedding conditions (1-3) on page \pageref{EmbedCond} 
and taking
$\underline{\Omega}_t$ as an $r,t$ constant surface
in the reference space (same coordinate values as in 
the Schwarzschild space) the time vector in the reference
space is $\underline{\bT}^\az = \underline{\bN} 
\underline{\bu}^\az + \underline{\bV}^\az$ where
the lapse is the same as above while
\bea 
\underline{\bu}^\az &=& \frac{1}{\sqrt{1 - v_z^2 \cos^2 \theta}} 
\left( \partial_{\underline{t}}^\az 
+ v_z \cos \theta \partial_{\underline{r}}^\az \right), 
\mbox{ and}\\
\underline{\bV}^\az &=& \frac{v_z \sin \theta}{r \sqrt{1-v_z^2}}
\partial_{\underline{\theta}}^\az.
\eea
The underlined coordinates are of course in the reference space. 
Then
\bea
\underline{v}_\vdash = 
\frac{v_z \sqrt{F} \cos \theta}{\sqrt{1+v_z^2 (F-1)
\cos^2 \theta}}, \\
\underline{\gamma} = \sqrt{ \frac{1 + v_z^2 (F-1) \cos^2 \theta}{1-v^2
\cos^2 \theta} },
\eea
and equation (\ref{uepSSS}) along with the transformation laws 
(\ref{epTrans}) and (\ref{jTrans}) gives
\bea
\underline{\bep} &=& \underline{\gamma} \underline{\varepsilon} = 
- \frac{2}{\kappa r} \sqrt{ \frac{1 + v^2 (F-1) \cos^2 \theta}{1-v^2
\cos^2 \theta} } \ \ \ \
\mbox{ and }\\
\underline{\bj}_\az &=& \frac{\underline{\gamma}^2}{\kappa}
\sigma_\az^\bz \partial_\bz \underline{v}_\vdash
= - \frac{v \sin \theta}{\kappa (1-v_z^2 \cos^2 \theta)} 
\sqrt{ \frac{F}{1 + v_z^2 (F-1) \cos^2 \theta} }
[ d \underline{\theta} ]_\az
\eea

Not unexpectedly these results are quite a bit messier than 
the corresponding ones for radially boosted observers, and in 
particular they don't integrate over $\Omega_t$ into nice
tidy forms. To clear things up a little, I consider 
a limiting case. For, $r \rightarrow \infty$,
\be
\left( \bep - \underline{\bep} \right)_{r \rightarrow \infty}  
= \frac{m}{4 \pi r^2} \sqrt{1 - v_z^2\cos^2 \theta  }
\ \mbox{ and } \
\left( \bj_\az - \underline{\bj}_\az \right)_{r \rightarrow \infty} 
= \frac{m v_z \sin \theta}{4 \pi} [d\theta]_\az.
\ee
Then, integrating over the two-surface of constant $r$ and $t$ 
the result is,
\bea
E_\infty &=& \int_\Omega d^2 x \sqrt{\sigma} (\bep - \underline{\bep}) \nn \\
&& = \frac{m}{2} \left( \sqrt{1-v_z^2} + \frac{\arcsin v_z}{v_z}
\right), 
\mbox{ and } \\
H_\infty &=& \int_\Omega d^2 x \sqrt{\sigma} 
\left( \bN (\bep - \underline{\bep})
- \bV^\az (\bj_\az - \underline{\bj}_\az ) \right) \nn \\ 
&& = \sqrt{1-v_z^2} m.
\eea
Thus the quasilocal Hamiltonian decreases in the same way that the 
quasilocal geometric energy did in the radial boost case. The decrease
can again be thought of as occurring because of a competition between the 
relativistic effects of the boost versus that of the gravity. Thus
$\bep$ and $\underline{\bep}$ converge even as they are boosted. 
Another interesting interpretation of this result can be found in 
the non-orthogonal paper by Hawking and Hunter \cite{hhunter}
who considered this case using their Hamiltonian method. They
interpret the decrease as occurring because some of the energy
has been transformed into a non-zero gravitational momentum
by the boost.

\section{Naked black holes}
\label{nbh}

An interesting application of the quasilocal energy formalism is 
found in the study of the so-called naked black holes.
These are low-energy-limit solutions to string theory and are characterized by the fact that static
observers hovering close to their horizons feel only very small
transverse tidal forces while  
infalling observers are crushed by arbitrarily large
tidal forces. Thus they are
naked in the sense that even though they are not 
Planck scale themselves, Planck scale curvatures may 
still be experienced outside their horizons by those infalling
observers. Several classes of these holes were studied in a couple of
papers by Horowitz and Ross \cite{naked1,naked2} 
but here I will consider only those satisfying the
equations of motion (\ref{dF}--\ref{Ein}). The naked black holes are then a subset of the following class of Maxwell-dilaton
black hole solutions. The metric is
given by
\be
ds^2 = -F(r) dt^2 + \frac{dr^2}{F(r)} + R(r)^2 ( d \theta^2 + \sin^2 \theta 
d \varphi^2 ),
\ee
where
\be
F(r) = \frac{(r-r_+)(r-r_-)}{R^2} \ \ \mbox{ and} \ \ R(r) = r 
\left( 1 - \frac{r_-}{r} \right)^{a^2/(1+a^2)}.
\ee
In the above, 
$r_+$ is the radial coordinate of the black hole horizon and 
$r_-$ is that of its central singularity. 
The accompanying dilaton and electromagnetic fields are
defined by
\be
e^{-2\phi} = \left(1 - \frac{r_-}{r} \right)^{2a/(1+a^2)}
\ee
and
\be
\star F = \frac{G_0}{r^2} dt \wedge dr.
\ee
These solutions are all magnetic black holes so as discussed
earlier the dual form of the quasilocal Hamiltonian must be used. 
The ADM  mass and magnetic charge are
\bea
M &=& \frac{r_+}{2} + \frac{1-a^2}{1+a^2} \frac{r_-}{2} \mbox{ and} \\
G_0 &=& \left( \frac{r_+ r_-}{1+a^2} \right)^{1/2}.
\eea
Solving this pair of equations in terms of $r_+$ and $r_-$ one finds
that $r_\pm = \frac{1 \mp a^2}{1-a^2} 
( M \pm \sqrt{ M^2 - (1-a^2) G_0^2} )$
for $a \neq 1$ or $r_+ = 2M$ and $r_- = G_0^2/M$ for $a = 1$.
Note that for $a=0$ these spacetimes reduce to magnetically
charged RN black holes.

Massive near-extreme members of this class of solutions are dubbed
``naked''. To see the reason for this nomenclature 
note that in terms of the orthonormal tetrad $\{u^\az,
n^\az, \hat{\theta}^\az, \hat{\phi}^\az \}$ where $u^\az =
1/\sqrt{F} \partial^\az_t$, $n^\az = \sqrt{F} \partial^\az_r$,
$\hat{\theta}^\az = 1/R \partial^\az_\theta$ 
and $\hat{\varphi}^\az = 1/(R \sin \theta) \partial^\az_\varphi$ 
the non-zero components of the Riemann tensor are
\bea
\cR_{u n u n} &=& \frac{\ddot{F}}{2}, \\
\cR_{\hat{\varphi} \hat{\theta} \hat{\varphi} \hat{\theta}} &=& \frac{1 - F 
\dot{R}^2}{R^2}, \\
\cR_{u \hat{\theta} u \hat{\theta}} &=& \cR_{u \hat{\varphi} u \hat{\varphi}} 
= \frac{\dot{F} \dot{R}}{2R}, 
\mbox{ and}\\
\cR_{n \hat{\theta} n \hat{\theta}} &=& 
\cR_{n \hat{\varphi} n \hat{\varphi}} 
= - \frac{\dot{F} \dot{R}}{2R} - \frac{F \ddot{R}}{R}.
\eea
In this section overdots indicate partial derivatives with
respect to $r$ (as opposed to the time derivatives that they
represent elsewhere in this thesis).

In the alternate infalling tetrad $\{\bu^\az, \bn^\az,
\hat{\theta}^\az,
\hat{\varphi}^\az \}$, where as usual $\bu^\az =
(1/\lambda) u^\az + \eta n^\az$ 
and $\bn^\az = (1/\lambda) n^\az + \eta u^\az$, 
the non-zero components of the
Riemann tensor are (in terms of the non-moving components) 
\bea
\cR_{\bu \bn \bu \bn}  &=& \cR_{u n u n} = 
\frac{\ddot{F}}{2} , \\
\cR_{\bu \hat{\varphi} \bu \hat{\varphi}} &=& 
\cR_{u \hat{\varphi} u \hat{\varphi}} + \eta^2  \left( \cR_{u \hat{\varphi} 
u \hat{\varphi}} + \cR_{n \hat{\varphi} n \hat{\varphi}} 
\right)
= \frac{\dot{F} \dot{R}}{2R} - \eta^2 \frac{F \ddot{R}}{R}, \mbox{ and}\\
\cR_{\bn \hat{\varphi} \bn \hat{\varphi}} 
&=& \cR_{n \hat{\varphi} n \hat{\varphi}} + \eta^2 \left( \cR_{u \hat{\varphi} u 
\hat{\varphi}} + \cR_{n \hat{\varphi} n \hat{\varphi}} \right)
= - \frac{\dot{F} \dot{R}}{2R} - \frac{F \ddot{R}}{R} - \eta^2 \frac{F 
\ddot{R}}{R}.
\eea
$\cR_{\hat{\varphi} \hat{\theta} \hat{\varphi} \hat{\theta}}$ is unchanged, 
$\cR_{\bu \hat{\theta} \bu \hat{\theta}} = 
\cR_{\bu \hat{\varphi} \bu \hat{\varphi}}$, and 
$\cR_{\bn \hat{\theta} \bn \hat{\theta}} = \cR_{\bn \hat{\varphi} \bn\hat{\varphi}}$.
Clearly if $a=0$ then $R(r) = r$ and all of the components are the same 
as for the unboosted frame.

If $a \neq 0$ and $\delta \equiv \left(1 - r_-/r_+
\right)^{1/(1+a^2)}$, then the naked black holes are the subset of the
above solutions 
whose parameters satisfy the conditions $\frac{\delta^2}{a^2}
\ll \frac{1}{R_+^2} \ll 1$, where $R_+ = R(r_+)$. That is
$\frac{a}{\delta} \gg R_+$ which in turn is much larger than the
Planck length. Note that if 
$\delta \ll 1$ then $r_- \approx r_+$ and if 
$R_+ \gg 1$ then $M, G_0 \gg 1$. Thus
naked holes are near-extreme as well as being very large
(relative to the Planck length).  

In the static frame as $r \rightarrow r_+$, 
\bea
\left| \cR_{u n u n} \right| &\rightarrow& \frac{1}{R_+^2} 
\left( 1 - \frac{2 r_-}{(1+a^2)r_+}  \right) \ll 1 , \\ 
\cR_{\hat{\varphi} \hat{\theta} \hat{\varphi} \hat{\theta}} &\rightarrow& 
\frac{1}{R_+^2} \ll 1, \\
\cR_{u \hat{\varphi} u \hat{\varphi}} & \rightarrow & \frac{1}{2 R_+^2 } 
\left( 1 - \frac{r_-}{(1+a^2)r_+} \right) \ll 1 , \mbox{ and}\\
\left| \cR_{n \hat{\varphi} n \hat{\varphi}} \right| 
&\rightarrow & \frac{1}{2 R_+^2 } \left( 1 - \frac{r_-}{(1+a^2)r_+} \right) 
\ll 1.
\eea 
Thus, all of the 
curvature components (and consequently the curvature invariants
calculated from them) are small compared to the Planck scale. 

By contrast, choosing the    
tetrad to be that carried by the infalling observers, 
$\eta^2 = \gamma^2 v_\vdash^2 
= \frac{1-F}{F}$  and as $r \rightarrow r_+$,
\bea
\left| \cR_{\bu \hat{\varphi} \bu \hat{\varphi}} \right| 
&\rightarrow& \frac{a^2}{(1+a^2)^2} \frac{r_-^2}{r_+^2} 
\frac{1}{R_+^2 \delta^2} \gg 1 \mbox{ and} \\
\left| \cR_{\bn \hat{\varphi} \bn \hat{\varphi}} \right| &\rightarrow& 
\frac{a^2}{(1+a^2)^2} \frac{r_-^2}{r_+^2} \frac{1}{R_+^2 \delta^2} 
\gg 1.
\eea
Thus these infalling observers see Planck scale curvatures. 
Interpreting these components in terms of the relative
acceleration of neighbouring geodesics it is easily seen that
these observers are laterally crushed by huge tidal forces. 

\subsection{QLE of naked black holes}
Now,
consider a spherical shell of observers
falling into a naked black hole. It is to be expected that the
huge transverse tidal forces will cause the area of the shell to 
shrink at
a very rapid rate. Such rates of change of area are an important
factor in evaluating the quasilocal energy defined in this thesis.
In particular  $\varepsilon^\updownarrow$ is (up to a 
normalization factor) exactly the (local) rate of change of the area
of an infalling surface of observers. As such it is of interest to 
calculate the quasilocal energies measured by static versus infalling
observers and to see how they compare to the observed curvatures. 
As the first step in calculating these energies one finds that
\bea
\varepsilon &=& - \frac{\dot{R} }{4 \pi R^2} \sqrt{ (r-r_+)(r-r_-) }, \\
\varepsilon + \varepsilon^m &=& -\frac{1}{4 \pi R} 
\sqrt{ \frac{r-r_+}{r-r_-}}, \\
\varepsilon^\updownarrow = \varepsilon^{m \updownarrow} &=& 0, \mbox{ and} \\
\underline{\varepsilon} &=& - \frac{1}{4 \pi R}.
\eea
The gauge choice for the matter term is the same one that was used
in the previous section. That is, I choose the gauge so that
$A^\star_\az \parallel u_\az$, as
well as being static, spherically symmetric, and non-diverging on the
black hole horizon. 
Though this is a long list of requirements, as noted earlier 
they amount to little more than deciding to make the standard gauge
choice of electrostatics (or in this case magnetostatics).

For the type of infalling observers that were 
considered in the last section,
$\bT^\az = \frac{1}{\sqrt{F}} u^\az - \sqrt{\frac{1-F}{F}}
\tilde{n}^\az$ which implies that 
$v_\vdash = - \sqrt{1-F}$ and $\gamma = 1 /\sqrt{F}$. 
By contrast the joint
requirements that $\underline{\bT}^\az \underline{\bT}_\az 
= \bT^\az \bT^\az$ and
$\pounds_{\underline{\bT}} \underline{\sigma}_{\az \bz} = 
\pounds_{\bT} \sigma_{\az \bz}$ imply that
\bea
\underline{\bT}^\az &=& 
\sqrt{ 1 + \dot{R}^2 (1-F) } \underline{u}^\az 
- \dot{R} \sqrt{{1-F}} \underline{n}^\az, \\
\underline{v}_\vdash &=&  
- \frac{\dot{R} \sqrt{1-F}}{\sqrt{1 + \dot{R}^2 (1-F) }}, 
\mbox{ and} \label{uv} \\
\underline{\gamma} &=& \sqrt{1 + \dot{R}^2 (1-F) }.
\eea 
Then,
\bea
E_{Geo} &=& -\sqrt{ (r-r_+)(r-r_-) } \dot{R}, \\
\bar{E}_{Geo} &=& - R \dot{R} , \\
E_{tot} &=& -R \sqrt{ \frac{r-r_+}{r-r_-} }, \\
\bar{E}_{tot} &=& - \frac{R^2}{r-r_-}  \\
\underline{E} &=&  -R, \mbox{ and} \\
\underline{\bar{E}} &=& - R \sqrt{1 + \dot{R}^2 (1-F) } .
\eea
Evaluating these expressions at $r=r_+$ is straightforward with the
only complication being
\be
\dot{R}_+ \equiv \dot{R}(r_+) = \frac{1}{1+a^2} 
\left( \delta^{a^2} + \frac{a^2}{\delta} \right).
\ee
If $a^2 \ll \delta$ then the square of the coupling constant is
extremely small even relative to $\delta$, and $\dot{R}_+ \approx 1$. 
In fact even if $a^2 \approx \delta$ then $\dot{R}_+$ is of the same
order as 1. By contrast for $a^2 \gg \delta$, $\dot{R}_+ \approx 
\frac{1}{1+a^2} \frac{a^2}{\delta} \gg 1$. Thus, it is simplest
to calculate the quasilocal energies for the cases 
$a^2 < \approx \delta$
(which includes the magnetic Reissner-Nordstr\"{o}m case for $a=0$)
and $a^2 \gg \delta$ separately. The results along with those for
$r \rightarrow \infty$ are displayed in table \ref{Chart}. Note that if 
$a^2 < \approx \delta < 1 $ then $R_+ \approx r_+$

From table \ref{Chart}
static observers outside a naked black hole measure $ E_{Geo} , E_{tot}
\rightarrow 0$ near to the horizon while the infalling observers
measure those same quantities to be very large. This effect
occurs for both $\delta \ll a^2$ and the $a^2 < \approx \delta$ (which
include the RN holes) and so cannot be attributed to the ``nakedness'' 
of the holes. Of course since the reference terms have been omitted,
both of these expressions blow up if the quasilocal surface is taken
out to infinity.

\begin{table}
\centering
\begin{tabular}{|c|c|c|c|}
\hline
         & $\delta > \approx a^2  $ & $ \delta \ll a^2 $ &   \\
Quantity & $r \rightarrow r_+$ & $r\rightarrow r_+$ & 
$r \rightarrow \infty$ \\
\hline
$-E_{Geo}$ & 0 & 0 & $r$ \\ 
$-\bar{E}_{Geo}$ & $R_+ \gg 1 $ & $\frac{a^2}{1+a^2} \frac{R_+}{\delta} 
\gg \gg 1 $ & $r$ \\ 
$E_{Geo} - \underline{E}$ & $R_+ \gg 1 $ & $R_+ \gg 1$ & $M$ \\ 
$\bar{E}_{Geo} - \underline{\bar{E}}$ & $C_1 R_+ \gg 1$ & $ 
\frac{1+a^2}{2a^2} R_+ \delta \ll 1$ & $M$ \\ 
$-E_{tot}$ & 0 & 0 & $r$ \\
$-\bar{E}_{tot}$ & $ \frac{R_+}{\delta} \gg \gg 1 $ & $\frac{R_+}{\delta} 
\gg \gg 1 $ & $r$ \\
$E_{tot} - \underline{E}$ & $R_+ \gg 1$ & $R_+ \gg 1 $ &  $0 < R_+  
\delta \ll 1 $\\
$-(\bar{E}_{tot} - \underline{\bar{E}})$ & $ \frac{R_+}{\delta} 
\gg \gg 1 $ & $ \frac{R_+}{\delta} \gg \gg 1$ & $-1 \ll R_+ \delta < 0 $\\
\hline
\end{tabular}
\caption[Quasilocal energies of near-extreme dilaton-Maxwell black holes]{Asymptotic and
near horizon values of the
quasilocal energies for near-extreme dilaton-Maxwell black holes. 
$\delta = (1-r_-/r_+)^{1/(1+a^2)} \ll 1$, 
$R_+ = r_+ \delta^{a^2} \gg 1$
and $R_+^2 \delta^2 \ll 1$, where $R_+ = R(r _+)$. $C_1$
is a constant of the same order as $1$.} 
\label{Chart}
\end{table}

Including the reference terms,
$E_{tot} - \underline{E}$ is very large for static
observers near to the horizon, where it is $R_+$. 
It is even larger in the absolute
sense for infalling observers who measure it as $-R_+/\delta$. 
Again however,
those effects are seen by observers surrounding both naked and 
near-extreme RN holes
and so cannot really be attributed to the extreme curvatures. As $r
\rightarrow \infty$ the two expressions agree which is not surprising
since as $r \rightarrow \infty$ the velocity of the infalling observers
goes to zero. Note however that this is not the ADM mass.

More interesting are the measurements of $E_{Geo}- \underline{E}$.
If $a \approx 1$ and the holes are large ($R_+^2 \gg 1$) then 
while static observers near to the horizon measure large 
values, sets of observers falling into naked black holes actually
measure very small values for this quasilocal energy. By contrast 
observers falling into an RN hole will measure large values. In fact
one can see that if $a \approx 1$ and $R_+^2 \gg 1$ then
these infalling observers will measure 
$\bar{E}_{Geo}- \underline{\bar{E}} \ll 1$, 
if and only if the black hole
is naked. Thus this is an alternate characterizing feature of naked
black holes when the coupling constant is of a reasonable size.
The equivalence is broken if $a^2 < \approx \delta$ in which
case the static and infalling observers both measure large energies.
Consider for example the case where $a^2 = \delta$. Then the black hole
can still be naked if $\delta$ (and therefore $a^2$) is 
small enough that $\delta^2 R_+^2 \ll 1$. 

\subsection{Why do naked holes behave this way?}
\label{nakedexp}

At the beginning of the previous subsection it was suggested that the
curvature results could be understood in terms of the rates of change 
of the surface area of shells of infalling observers. In this section
the idea is explored in more detail and used to provide an
explanation of the $E_{Geo}- \underline{E}$ result.

First I quantify the expectation that the surface area of a shell
of infalling observers will be changing extremely quickly as they cross
the horizon of a naked black hole. Recall that naked black holes are 
near extreme and so the singularity sits ``just behind'' the 
horizon ($r_- \approx r_+$). More rigorously,
Horowitz and Ross \cite{naked1} noted that an observer passing
through the horizon after falling from $r_0$ (the situation described
by equation (\ref{velocity})),
will hit the singularity at $r_-$ after a proper
time of $\Delta \tau < \approx \frac{r_+ - r_-}{\sqrt{F(r_0)}} = 
\frac{R_+ \delta}{\sqrt{F(r_0)}}$. 
Thus a set of observers infalling on geodesics that were
stationary at infinity ($F(r_0)=1$) will only have a very short
time before they reach $r_-$.  
At $r_-$, $R(r) \rightarrow 0$ and
so the area of the shell goes to zero. However, by assumption 
$R_+ \gg 1$ and 
so at the horizon itself, that same area is very large. For
the area to go from very large to zero in such a small time, one would
naively expect it to be decreasing very quickly as the observers pass
the horizon. This expectation can be quantified by using (\ref{dArea}) 
to show that the
fractional rate of change of the area of the surface $\Omega_t$
as measured by the observers who inhabit that surface is
\be
\frac{{A'}}{A} = \frac{8 \pi \int_{\Omega_t} d^2 x \sqrt{\sigma} 
\bep^\updownarrow}{\int_{\Omega_t} d^2 x \sqrt{\sigma}} =
- \frac{2 \dot{R}_+}{R_+}
= -\frac{2}{(1+a^2) R_+}  \left( \delta^{a^2} + \frac{a^2}{\delta^2}
\right),
\ee
where for the rest of this section primes indicate proper time
derivatives.
If $a^2 \gg \delta$ (that is, it isn't pathologically small), 
$\frac{{A'}}{A} \approx \frac{1}{R_+ \delta^2} \gg 1$
as expected. By contrast for the RN case ($a=0$), 
$\frac{{A'}}{A} \approx \frac{1}{R_+} \ll 1$.  
However, the expectation is confounded if
$\delta > a^2 \neq 0$ in which case the hole remains naked
even while the rate of change is more along the lines of the RN values.
In that case the extremely small value of $a$ suppresses the rapid
decrease in area until the observers get even closer to the singularity
(basically $r-r_- \ll a^2$).  

These rates of change of the area also nicely explain why 
$E_{Geo} - \underline{E}$ is small while the observed
curvature components are large. Recall that to define
the reference term $\underline{E}$,  
$\Omega_t$ had to be embedded into flat space along 
with a vector field
$\underline{T}^\az$ defined so that if $\Omega_t$ was evolved
by that vector field and only intrinsic observations were made
in the resulting timelike three-surface, those observations are 
identical whether they were in the original or reference spacetimes. 
In particular, the area of $\Omega_t$ should change at the same
rate. Thus, if the area decreases extremely rapidly, the embedded
shell of observers in the reference spacetime would have to be
moving at a correspondingly fast speed. Equation
(\ref{uv}) quantifies this saying that
$\underline{v}_\vdash = \dot{R} / \sqrt{1+ \dot{R}^2}$ at the horizon.
Then for $a^2 \approx 1$, $\dot{R} \gg 1 \Rightarrow 
\underline{v}_\vdash \approx 1$ and 
the observers would 
have to move at close to the speed of light in the reference
time to match the rate of change of the area.
By contrast, 
for $a^2 \approx 0$, $\dot{R} \approx 1 \Rightarrow 
\underline{v}_\vdash \approx \frac{1}{2}$.
The area is changing at a relatively leisurely rate so the observers 
would not need to move so fast in the reference time. 

For observers moving at extremely rapid velocities 
there is a sense in which the relativistic effects of their speed
become more important than those due to gravity. To see
this recall equation (\ref{mattransform}) where it was shown that 
$\varepsilon^2 - \varepsilon^{\updownarrow 2}$ is a constant
independent of the speed of the observers. Now, by construction
$\bep^\updownarrow$ is the same in both the reference and
original spacetime and so the geometric QLE can be rewritten as,
\be
\bar{E}_{Geo} - \underline{\bar{E}} 
= \int_{\Omega_t} d^2 x \sqrt{\sigma} \left( 
\sqrt{\underline{\varepsilon}^2 + \bep^{\updownarrow 2}} 
- \sqrt{\varepsilon^2 + \bep^{\updownarrow 2}}
\right).
\ee
If $\bep^\updownarrow$ is much larger than 
$\varepsilon$ and $\underline{\varepsilon}$, and so in a sense the
relativistic effect of speed
dominates over that of curvature, then at the horizon
\be
\bar{E}_{Geo} - \underline{\bar{E}}  
\approx \frac{1}{2} \int_{\Omega_t} d^2 x \sqrt{\sigma}
\left( \frac{\underline{\varepsilon}^2 - 
\varepsilon^2}{\bep^\updownarrow} \right),
\ee
and so as $\varepsilon^\updownarrow$ becomes larger and larger the observed quasilocal energy becomes smaller and smaller. Physically, though $\bep$ and 
$\underline{\bep}$ are boosted to be very large,
the difference between them simultaneously becomes
very small. In particular for naked black holes
\be
\bar{E}_{Geo} - \underline{\bar{E}} \approx 2 \pi R_+^2 \left( 
\frac{\underline{\varepsilon}^2 }{\bep^\updownarrow}
\right) = \frac{R_+}{2 \dot{R}_+} = - \frac{A}{A'},
\ee
and it can be seen that in this case the geometric quasilocal energy 
is actually the inverse of the (normalized) rate of change of the area. 
As noted in section \ref{infall} these general ideas explain the 
much less dramatic decrease of the quasilocal energy for boosted 
observers in the Reissner-Nordstr\"{o}m spacetime as well. 

By contrast $\bar{E}_{tot} - \underline{\bar{E}}$ includes matter terms
which are also boosted to be very large. There is
no corresponding term in the reference spacetime to cancel these
large terms out. The result is that
the matter terms dominate over the geometrical terms in 
$\bar{E}_{tot} - \underline{\bar{E}}$ and so this
total quasilocal energy is very large.

\section{Tidal heating}
\label{tidal}

As a final classical calculation I use the quasilocal formalism to 
calculate the amount of work done by an external gravitational field
when it deforms a self-gravitating body. The canonical example of this
effect in our own solar system is found in the gravitational 
interactions between Jupiter and its moon Io. In that
instance, the gradient of Jupiter's gravitational field distorts 
the shape of Io away from being a perfect sphere
and then tidally locks it in its orbit so that it
always presents the same face to Jupiter. That orbit is
strongly perturbed by the other Galilean moons and so
its radial distance from Jupiter varies with time. 
With this variation comes
a corresponding one in the gradient of the field and so Io is gradually
stretched and then allowed to relax. 
The energy transferred by this pumping is largely dispersed as heat 
and it is this heat that produces the volcanic activity on Io.  
The same type of process occurs in principle 
for any two bodies in non-circular orbits about each other.

To calculate the gravitational energy transferred to Io during this
process using the quasilocal formalism, 
I'll need a metric describing the situation. 
To this end, first consider the situation 
from a Newtonian perspective. Assume that the
self-gravitating body is far enough away from the source 
of the external field that that field is nearly uniform close to the
body. Then in a rectangular coordinate system that orbits with the body
(with origin fixed at the center of mass), 
the Newtonian potential of the
external field may be written as $\Phi_{ext}=\cE_{ab} x^a x^b$
where $\cE_{ab}$ is the (time-dependent but symmetric and trace-free)
quadrupole moment of the field and $x^a$ is the position vector based at
the body's centre of mass. At the same time, to quadrupolar order the
Newtonian potential of the body is 
$\Phi_{o}=-M/r-(3/2)r^{-3}\cI_{ab }n^a n^b$, where $M$ is the mass
of the body, $r$ is the radial distance from the centre of mass, $\cI_{ab}$ is its (time-dependent but symmetric and trace-free)
quadrupole moment, and $n^a = x^a/r$ is the unit normal radial vector. 

From this description one can use the techniques of Thorne and Hartle 
\cite{thorne:1985} to construct a metric that 
describes these situations in the slow moving, nearly Newtonian limit.
First, define an annulus surrounding the self-gravitating body whose
inner boundary is
chosen so that its gravitational field is weak throughout
and whose outer boundary is chosen close enough 
so that the external field
is nearly uniform. This region is termed the buffer zone. 
The rectangular coordinate system from the Newtonian limit
is replaced with one that is chosen
so that the metric is as close to Minkowskian as possible over the
buffer zone. Then to first order in perturbations
from Minkowski and first order in time derivatives the metric can be
written as \cite{purdue:1999}
\begin{eqnarray}
  ds^2 &=& - (1+2 \Phi) dt^2 + 2(A_b + \partial_t \xi_b) dx^b dt \nonumber\\ 
  && + [(1-2\Phi) \delta_{ab} + \partial_a \xi_b + \partial_b\ \xi_a]
dx^a dx^b
\end{eqnarray}
where the indices run from one to three and 
$\delta_{ab} = {\mathrm{diag}}[1,1,1]$ is the Cartesian metric on a spacelike slice.  The Newtonian potential is still
$\Phi=-M/r-(1/2)(3r^{-3}\cI_{ab}-r^2\cE_{ab})n^a n^b$ while
\be
A_b \equiv -\frac{2}{r^2} n^c \frac{d \cI_{bc}}{dt}
- \frac{2}{21} r^3 (5n_b n^c - 2 \delta^c_b) n^d
 \frac{d \cE_{cd}}{dt}
\ee 
is a vector potential that must be included so that the metric is
a solution to the first order Einstein equations.  
Here, $n^a$ is the radial
normal with respect to the flat spatial metric $\delta_{ab}$ and
$r^2=x^2+y^2+z^2$.  The diffeomorphism generating vector field $\xi_b$
represents the gauge ambiguity in setting up a nearly Minkowski coordinate system. 
In order that the metric be slowly evolving and nearly Minkowski,
$\xi_b$ must be of the form
\be
\xi_b= \frac{\alpha}{r^2}
\cI_{bc}n^c+\beta r^3 \cE_{bc}n^c+\gamma r^3\cE_{cd}n^c n^d n_b,
\ee
where  $\alpha$, $\beta$, and $\gamma$ are free constants of order one. 

To measure the flow of quasilocal energy, I define $B$ as a 
surface of constant $r$ surface in the buffer zone, foliate it 
with constant $t$ spacelike two-surface $\Omega_t$, and define 
the time vector $T^a$ as $\partial/\partial t$. Then I can 
calculate $\dot{H}_t$ from equations (\ref{Hdot}) and (\ref{Hdot2}).
As I mentioned in section \ref{RefTerm}, I will neglect reference
terms here because for a wide range of choices of how to define
them, they don't contribute in a situation such as this
where I am calculating rates of change. Of course this also serves 
to simplify the already messy calculations. 

In calculating the time rate of change it is most convenient
to switch to spherical coordinates. Making 
the standard transformation 
$x^a=r[\sin\theta\cos\phi,\sin\theta\sin\phi,\cos\theta]$,
the metric becomes
\begin{eqnarray}
ds^2 &=& -(1-2\Phi) dt^2 + 2 \bar{A}_r dr dt + 2 r \bar{A}_\theta 
d\theta dt \nonumber\\
&&+ 2 r \sin \theta \bar{A}_\phi d\phi dt
+ (1+2\Phi+ H_{rr}) dr^2 \nonumber \\
&&+ r^2(1+2\Phi+H_{\theta \theta}) d\theta^2
+ r^2 \sin^2 \theta (1+2\Phi+H_{\phi \phi}) d\phi^2 \nonumber \\
&&+ rH_{r \theta} dr d\theta + r\sin\theta H_{r \phi}drd\phi + r^2
\sin\theta H_{\theta \phi} d\theta d\phi,
\end{eqnarray}
where 
\bea
H_{rr} &=& -\frac{4\alpha}{r^3} 
\cI_{rr} +6 (\beta+\gamma) r^2\cE_{rr}, \\
H_{\theta\theta} &=& \frac{2\alpha}{r^3}\cI_{\theta\theta}+2\beta r^2\cE_{\theta\theta} +2\gamma r^2\cE_{rr}, \\
H_{\phi\phi} &=& 
\frac{2\alpha}{r^3}\cI_{\phi\phi}+2\beta r^2\cE_{\phi\phi}
+2\gamma r^2\cE_{rr}, \\
H_{r\theta} &=& 
-\frac{\alpha}{r^3}\cI_{r\theta}+(4\beta+2\gamma)r^2\cE_{r\theta}, \\
H_{r\phi} &=& 
-\frac{\alpha}{r^3}\cI_{r\phi}+(4\beta+2\gamma)r^2\cE_{r\phi}, \mbox{and}
\\
H_{\theta\phi}&=& \frac{2\alpha}{r^3}\cI_{\theta\phi}+2\beta r^2\cE_{\theta\phi}.
\eea  
In these expressions $\cE_{rr}=\cE_{ab} e^a_r e^b_r$,
$\cE_{r\theta}=\cE_{ab}e^a_r e^b_\theta$, etc., with $e^a_r=n^a$,
$e_\theta^a=\partial_\theta e^a_r$ and
$e_\phi^a=(1/\sin\theta)\partial_\phi e^a_r$.  Also,
$\bar{A}_r=(A_b+\partial_t\xi_b)e^b_r$,
etc., but their expanded forms are not needed since only time
derivatives of them show up in later calculations and the calculation is
only been done up to first order in time derivatives.

As might be expected the subsequent calculations are quite involved
and I did them with a lot of help from the GRTensor~\cite{grtensor}
package for Maple. Ultimately though after a huge amount of 
algebra, equation (\ref{Hdot2}) works out to become
\begin{eqnarray}
\label{e:power1}
\dot{H} &=& -\frac{1}{2}\int_\Omega d^2x \sqrt{-\gamma}\,
  \pi^{ab} \pounds_T \gamma_{ab} \\
&=& \frac{1}{2} \cE_{ab} \dot{I}_{ab} \nn \\ 
&&+ \frac{d}{dt} \left\{
\frac{r^5}{30} (-3 - 2\beta - 2\beta^2 + 4\gamma + 4\gamma^2
+ 8 \beta \gamma ) \cE_{ab} \cE_{ab} \right\} \nonumber \\
&&+ \frac{d}{dt} \left\{ \frac{1}{30} (3 - 2\alpha + 6\beta 
- 12\gamma +8\alpha\gamma) \cE_{ab} I_{ab} \right\} 
\nonumber \\
&&- \frac{d}{dt} \left\{ 
\frac{1}{60 r^5} (-9 + 12 \alpha + 4 \alpha^2) I_{ab} I_{ab}
\right\}. \nonumber
\end{eqnarray}
Note that repeated indices continue to indicate summation.
Since $\cE_{ab}$ and $I_{ab}$ are Cartesian tensors, the index
position doesn't matter. These calculations used the identities
\bea 
&& \int d\theta d\phi \sin \theta A_{rr} B_{rr} 
= (8 \pi/15) A_{ab} B_{ab}\ \ \  \mbox{and} \\
&&\int d\theta d\phi \sin \theta (2A_{\theta \phi} B_{\theta \phi}
- A_{\theta \theta} B_{\phi \phi} - A_{\phi \phi} B_{\theta \theta} )
= (4 \pi/3) A_{ab} B_{ab},
\eea 
where the integrations are over the unit sphere.

This result requires some interpretation. As the external
field changes with time and thereby forces the self-gravitating
body to change configuration, the work done by the external
field can be split into time reversible and irreversible parts
(as seen in equation \ref{e:power1}). 
The reversible part represents work being done
to increase the potential energy of the system and is recoverable.
On the other hand the irreversible part represents work 
being done
to deform and/or heat up the system. This is the tidal work
that I am interested in and by the above it is
$(1/2) {\mathcal E}_{ab} I_{ab}$,
which is the same leading term obtained when one
does the corresponding calculation in Newtonian gravity or with 
pseudo-tensors \cite{purdue:1999}. It is
completely independent of the diffeomorphisms generated by 
$\xi_b$ which correspond to fluctuations of the
quasilocal surface. 

Note that a gauge ambiguity similar in form to (though not identical 
with) that found in the time
reversible term is also found in the corresponding 
results obtained by the Newtonian and pseudo-tensor methods. 
What is much more clear in this calculation however, is that the 
ambiguity is a result of fluctuations of the quasilocal 
surface through the fields as generated by the $\xi_t$
diffeomorphisms. Keep in mind that those other methods also give
answers with time reversible and time irreversible parts 
so that is not unique to the quasilocal procedure but instead
is a physical property of the system as I argued in the 
previous paragraph. 

Finally for completeness consider how the energy flow
splits up into its component parts as considered in 
equation~(\ref{Hdot}). In the approximation 
in which I am working, the angular
momentum term is zero and what is left are the two terms
$\dot{H}_N = - \int d\theta d \phi \sqrt{\sigma} 
\varepsilon \pounds_t N$ and $\dot{H}_\sigma = \int d\theta d \phi
\sqrt{\sigma} \frac{N}{2} s^{ab} \pounds_t \sigma_{ab}$. It can be
shown that
\begin{eqnarray}
\dot{H}_N
&=& 
\frac{1}{2} \cE_{ab} \dot{I}_{ab} + \frac{\alpha}{15} \dot{\cE}_{ab}
I_{ab} - \frac{\beta}{5} \cE_{ab} \dot{I}_{ab} - \frac{4 \gamma}{5}
\cE_{ab} \dot{I}_{ab} \\
&+& \frac{d}{dt} \left\{ 
\frac{4 \gamma + \beta - 2}{30} r^5 \cE_{ab}\cE_{ab}
- \frac{1}{10} \cE_{ab} I_{ab} - \frac{2 \alpha - 3}{20 r^5}
I_{ab}I_{ab} \right\}. \nonumber
\end{eqnarray}
The second term is a bit more complicated. It is
\begin{eqnarray}
\dot{H}_\sigma &=& 
- \frac{\alpha}{15} \dot{\cE}_{ab} I_{ab} + \frac{\beta}{5}
 \cE_{ab} \dot{I}_{ab} + \frac{4 \gamma}{5} \cE_{ab} \dot{I}_{ab}\\
&& + \frac{d}{dt} \left\{ \frac{r^5}{30} (-1 - 3 \beta
- 2 \beta^2 + 4 \gamma^2 + 8 \beta \gamma ) \cE_{ab} \cE_{ab} 
\right\} \nonumber \\
&& + \frac{d}{dt} \left\{ \frac{1}{15} (3 - \alpha + 3 \beta
- 6 \gamma + 4 \alpha \gamma) \cE_{ab} I_{ab} \right\} \nonumber  \\
&& - \frac{d}{dt} \left\{
\frac{1}{30 r^5} (2\alpha^2 - 9 \alpha + 9) I_{ab} I_{ab} \right\}
\nonumber.
\end{eqnarray}
Thus part of the work done is measured by deformations of the 
surface and part is measured by changes in how observers choose
to measure the rate of passage of time. Note that individually
the time irreversible sections of the two parts are gauge 
dependent but when one combines them 
equation (\ref{e:power1}) returns 
and the gauge dependence vanishes back
into the reversible part where it would be expected.

There are two ways to look at this calculation of tidal heating.
The first is to see
it as an astrophysical application of the quasilocal energy and so an
alternate way to calculate the tidal heating effects. As I have 
argued above, it has an advantage over previous methods of calculating
the magnitude of the effects in
that the source of the gauge ambiguity
in the final result can be clearly identified.  
It is also somewhat tidier
than the corresponding pseudo-tensor methods since the
integrals are defined in terms of tensor quantities and so are 
covariant. On the other hand, the second way to look at the result is as
a check on the physical relevance of the Brown-York energy. That it
can reproduce the results produced by other methods is a good 
argument for its physicality.

On the down side, I haven't
shown that this result is independent of the
exact choice of the form of the reference term. For example, it would
be good to show that the final results would be the same with the
two-surface embedded in 4D reference term.  
Further, from the work of section
\ref{sss}, one is led to think that it is the 
geometric quasilocal energy
that is the physically relevant quantity. Here I have calculated
the Hamiltonian based on a physically arbitrary coordinate time vector.
However, to resolve either of these questions would require extensive
calculations so for now I let the result rest in its computationally
simplest form that I have considered here.

\chapter{Quantum creation of black hole pairs}
\label{PC}
While the previous chapter considered applications of the quasilocal
Hamiltonian in classical general relativity the current chapter will
consider its application to semi-classical quantum gravity.
Specifically, I combine it with the path integral formulation of
quantum gravity to calculate the probability that a 
pure deSitter spacetime will transform itself into
a pair of charged and rotating black holes in a deSitter background via
a quantum tunneling process. This work was published in  
\cite{pairprl, pairprod}.

As a short outline, the section \ref{formalism}
reviews path integrals as applied
to quantum gravity and then the following sections flesh out that 
introduction as applied to the case of black hole pair creation in 
a deSitter background. Section \ref{KNdSsect} 
examines the classical description
of spacetimes containing pairs of black holes. Section 
\ref{InsCons} constructs the instantons used to mediate the creation 
of such spacetimes while section \ref{ActPick} uses the Brown-York
formalism to decide which is the correct action to use in the
path integral calculations. Finally section \ref{evAct} evaluates
those integrals to lowest order and section \ref{CompExt}
looks back on some questions that arose during the calculations.

\section{The idea}
\label{formalism}

A standard problem of quantum mechanics is to calculate the probability 
that a system passes from an initial state $X_1$ to a final state $X_2$.
If the classical equations of motion for that system can be derived from
a Lagrangian action $I$ then the path integral formulation of quantum
mechanics provides a prescription for calculating the probability 
amplitude that that transition occurs. Basically it says that one
should consider all conceivable ``paths'' $\Gamma$ that the system could 
follow to evolve between $X_1$ and $X_2$ (and not just those
that satisfy the classical equations of motion). If one
calculates the action $I[\Gamma]$ for each of those paths then 
the probability amplitude that the system will move from state
$X_1$ to state $X_2$ is hypothesized to be given by the path 
integral
\be
\Psi_{12} = \int d[\Gamma] e^{-i I [\Gamma]},
\ee
where the integral is over all possible paths. Note that I use the word
``hypothesized'' above because in general, this integral is not well
defined and so the path integral methods are sometimes more of a way 
thinking about these 
problems rather than actually calculating exact amplitudes. A 
more complete
description of the approach can be found in \cite{feynman}.

Despite problems of definition, the procedure was generalized to 
a formulation of quantum gravity in the 1970's (see for example 
\cite{OrigPathInt}). The philosophy behind the approach remains
the same but the details change quite a bit.

In the first place, it is no longer a trivial matter to define
an instantaneous configuration of a system if that system
is a general relativistic one. 
For a system with gravitational and electromagnetic
fields (the case in which I'll be interested in this chapter)
an ``instant'' will be defined as it was in chapter \ref{Setup}.
Namely it will consist of a three-manifold $\Sigma$ 
with Riemannian metric $h_{ab}$, conjugate momentum density $P^{ab}$ (or 
equivalently extrinsic curvature $K_{ab}$) describing how the
system is evolving at that ``instant'', and vector field densities 
$\cE^a$ and $\cB^a$ defining the electric and magnetic
fields on $\Sigma$. These four fields must satisfy 
the constraint equations (\ref{DB}), (\ref{DE}), (\ref{C1}), and
(\ref{C2}) and if they do, the ``instant'' can be embedded in a larger
four-dimensional solution to the Einstein-Maxwell equations. In fact
if $\Sigma$ is a Cauchy surface then it uniquely determines that 
solution via the evolution equations (\ref{cB}), (\ref{cE}), and
(\ref{C3}). 
 
Then, using the path integral approach one must consider  
{\it all} possible interpolations (or ``paths'') between the
states (not just those that would be allowed by 
the classical evolution of the system). This  
means considering four-manifolds (with boundaries) 
$M_{12}$, along with metric fields 
$g_{\az \bz}$ and electromagnetic field tensors $F_{\az \bz}$ on 
those manifolds such that the surfaces $\Sigma_1$ and $\Sigma_2$ and 
their accompanying fields, may be embedded in $M_{12}$ 
and its accompanying 
fields \footnote{In this context, a three manifold $\Sigma$ and its 
accompanying fields $\{h_{ab}, P^{ab}, 
\cE^a, \cB^a\}$ is said to be embeddable in the spacetimes 
$(M_{12}, g_{\az \bz}, F_{\az  \bz})$ if there 
exists an embedding (in the differential topology sense), $\Phi:\Sigma 
\rightarrow M_{12}$ such that $\Phi^*(h_{ab}) = \left. h_{\az \bz} 
\right|_{\Sigma}$, $\Phi^*(P^{ab}) = \left. P^{\az \bz} \right|_{\Sigma}$,  $\Phi^*(\cE_a) = \left. \cE_\az \right|_\Sigma= \left. - 2 \sqrt{h} / \kappa F_{\az \bz} u{^\bz} 
\right|_\Sigma$, and $\Phi^*(\cB_a) 
= \left. \cB_\az \right|_\Sigma = \left. - 2 \sqrt{h}/\kappa \frac{1}{2} \varepsilon_{\az  
\bz}^{\ \ \ 
\cz \dz} F_{\cz \dz} u^\bz \right|_\Sigma$. In the preceding $\Phi^*$ 
represents the appropriate mapping as   
derived from $\Phi$ for each quantity.}. 
I reiterate  that the spacetime paths 
$(M_{12},g_{\az \bz}, F_{\az \bz})$ are
not, in general, solutions to the Einstein-Maxwell equations.  

Next, the action  
\be
\label{actIM} 
I[M_{12},g_{\az \bz},F_{\az \bz}] 
= \int_{M_{12}} 
d^4 x \sqrt{-g} (\cR - 2 \Lambda - F_{\az \bz} F^{\az \bz}) 
+ \mbox{ (boundary terms) }, 
\ee 
for each path must be calculated,  
where the integration is over all of $M_{12}$ between the two embedded 
surfaces $\Sigma_1$ and $\Sigma_2$, and the boundary terms are 
calculated on the boundaries of $M_{12}$ that are consistent with the
boundaries of $\Sigma_1$ and $\Sigma_2$. How an appropriate 
action functional can be picked
will be discussed in section \ref{ActPick}.

Finally, the value of the action for each path is used 
to assign a probability amplitude for that path. 
The amplitudes are summed over all 
of the possible paths to give a net probability amplitude that 
the system passes from $X_1$ to $X_2$. This summation is represented as 
a functional integral over all of the possible manifold topologies, metrics, and vector 
potentials $A_\alpha$ (generating the field strength $F_{\alpha \beta}$)  
interpolating between the two surfaces. That is, 
\be 
\label{psi}  
\Psi_{12} = \int d[M_{12}]d[g]d[A] e^{-i I[M_{12}, g, F]}.  
\ee 
 
Thus at least in principle, the probability that a spacetime initially
in a state $(\Sigma, h_{ab}, P^{ab},\cE^a,\cB^a)_1$ 
passes to a state $(\Sigma, h_{ab}, P^{ab},\cE^a,\cB^a)_2$ 
is proportional to $|\Psi_{12}|^2$ (the wave function
hasn't been normalized). Unfortunately the integral 
(\ref{psi}) cannot be directly calculated. In the first place, 
there is no known way to define a measure for the 
integral. Second, even if such a measure were known, it seems 
quite likely that calculation of the integral would be impractical, 
considering that the parameter space of paths from 
$X_1$ to $X_2$ has an uncountably infinite number of dimensions.  
 
Fortunately there is a well-motivated simplifying assumption available. 
In analogy with flat-space calculations, it is argued \cite{OrigPathInt} 
that to lowest order in $\hbar$, the probability amplitude may be  
approximated (up to a normalization factor) by 
\be  
\label{pairc} 
\Psi_{12} \approx e^{-I_c},  
\ee  
where $I_c$ is the real action of a (not necessarily real) Riemannian
solution to the Einstein-Maxwell equations that interpolates
between the given initial and final conditions. Essentially, it is
assumed that such a solution is a saddle point  
of the path integral. This solution (if it exists) is referred to as an 
instanton. The probability that such a tunnelling occurs is then
proportional to $|\Psi_{12}|^2 \approx e^{-2I_c}$.  Note that this 
interpretation requires that the action $I_c$ be real and positive, and 
ideally that all of the fields on its boundary match those in the 
Lorentzian solution ``instants'' so that it can smoothly
match onto that solution. As will be seen in section
\ref{InsCons} this is sometimes a bit much to ask for, but if
one only requires a match, rather than a smooth match, it can be 
done.

As an alternative to paths and instantons interpolating between two 
spatial slices $\Sigma_1$ and $\Sigma_2$, one can consider those
with a single spacelike boundary 
that match onto a single slice labelled $\Sigma_2$. In that case
one can interpret the resultant path integral as calculating the
probability for the creation of the three-space $\Sigma_2$ from 
nothing and the initial boundary condition is the 
no-boundary condition of cosmology \cite{nobound}. One can then
compare the relative creation rates for different spacetimes
(eliminating the need to calculate a normalization factor) and
even interpret those probabilities as giving the chance that
the different spacetimes tunnel into each other \cite{boussochamblin}.
This is the approach that will be taken here.

Finally, before passing on to consider the classical solutions that
describe the spacetimes that I want to create, note that 
path integrals (especially in the single boundary case) 
can be interpreted as sums over all the possible histories
of the system being considered
\cite{feynman} and in particular this interpretation is often carried
over into gravity \cite{OrigPathInt}. Then the path integral can be
interpreted as a thermodynamic partition function and so this formalism
naturally lends itself to the study of gravitational thermodynamics.
As was discussed in \cite{BY2} and I will consider to some
extent in section \ref{ActPick}, the choice of the action $I$ 
will determine the exact partition function being considered --
that is the canonical, microcanonical, or grand canonical partition
functions. Given this correspondence the terminology of thermodynamics
will sometimes be used in the following. Ultimately I will also
use the connection to extract some conclusions about black hole
entropy from my calculations.

\section{Accelerating and rotating pairs of black holes}
\label{KNdSsect}  
Since I am interested in calculating the creation rate for a 
pair of black holes accelerating away from each other in a cosmological
background, the first step in the path integral calculation discussed
above is to find a solution to the Einstein-Maxwell equations that
describes such a physical situation. Such solutions are the subject
of this section. 

\subsection{The generalized C-metric and KNdS spacetime}
\label{CtoK}  
The well-known C-metric solution to the Einstein equations (first  
interpreted in \cite{KW})  describes  
a pair of uncharged and non-rotating black holes that are uniformly  
accelerating away from each other. In \cite{PlebDem} this metric was  
generalized to allow the holes to be charged and rotating, as well as to  
allow the inclusion of a cosmological   
constant and NUT parameter.  
  
In general, spacetimes of this type contain conical singularities.  
Physically these arise if the rate of acceleration of the black holes does not match the energy source available to accelerate them. 
Thus, in the  
cosmological case, if the black holes are accelerating faster or more  
slowly than the rest of the universe, conical singularities will exist.   
Physically, these may be interpreted as cosmic strings or ``rods'' that are pulling or pushing the black holes apart (or together)
to make them accelerate faster (or slower) than the rate of expansion 
of the universe as a whole. 
The singularities are eliminated if the acceleration
of the holes is matched to the amount of energy that is 
available to accelerate them. In that case no extra acceleration 
is required and so the cosmic strings or rods aren't needed to 
provide the extra energy.
 
The generalized C-metric takes the form
\begin{equation}  
\label{PlebDemMetricM}  
ds^2  = \frac{1}{(p-q)^2} \left\{
\begin{array}{llll}
& \frac{1+p^2q^2}{P} dp^2   
& + & \frac{P}{1+p^2q^2} \left( d \sigma - q^2 d \tau \right)^2 \\
  
- & \frac{1+p^2q^2}{Q} d q^2 &   
+ & \frac{Q}{1+p^2q^2} \left(p^2 d \sigma + d \tau \right)^2 
\end{array} 
\right\},  
\end{equation} 
with accompanying electromagnetic field defined by the vector potential 
\be 
A = - \frac{e_0 q (d \tau + p^2 d\sigma)}{1 + p^2 q^2} + \frac{g_0 p (d 
\sigma -  
q^2 d\tau)}{1 + p^2 q^2}, 
\ee 
where $p,q,\tau$, and $\sigma$ are coordinates,
\begin{equation}  
\label{Pp1m}  
P(p) = (-\frac{\Lambda}{6} - g_0^2 + \gamma) + 2 n p - \epsilon   
p^2   
+ 2 m p^3 + (- \frac{\Lambda}{6} - e_0^2 - \gamma) p^4,  
\end{equation}  
and $Q(q) = P(q) + \frac{\Lambda}{3} (1 + q^4)$. $\Lambda$ is the   
cosmological constant,  $\gamma$ and $\epsilon$ are constants connected   
in a non-trivial way with rotation and acceleration, $e_0$ and $g_0$   
are linear multiples of electric and magnetic charge, and $m$ and   
$n$ are the respectively mass and the NUT parameter (up to a linear 
factor). This solution can be analytically extended across the
coordinate singularity at  $p=q$, so that on the other side of $p=q$ 
there is a mirror image of the initial solution (though with opposite
electric/magnetic charge and direction of spin). 
Thus, if one views it as describing a pair of black holes, the 
two holes will be on opposite sides of that $p=q$ hypersurface. 

In general this metric has a conical singularity in the $(p,\sigma)$ 
hypersurface which corresponds to the above mentioned string or rod. 
There are a few limiting processes that can be used to remove this
singularity, but on setting the NUT charge to zero, 
at least one of them reduces the metric to the
Kerr-Newman-deSitter metric. Details of that process can be found in 
appendix \ref{AppCtoK}. In Boyer-Lindquist type coordinates, 
the KNdS metric takes the form \cite{MM}  
\bea  
\label{KNdS}  
ds^2  &=&  -\frac{\cQ}{\cG \chi^4} \left(  dt - a \sin^2\theta d\phi  
\right)^2 + \frac{\cG}{\cQ} dr^2  \\
  && + \frac{\cG}{\cH} d\theta^2 + \frac{\cH \sin^2 \theta}{\cG \chi^4}  
\left( a dt - \left[ r^2 + a^2 \right] d\phi \right)^2, \nn 
\eea  
where
$\cG \equiv r^2 + a^2 \cos^2 \theta$, $\cH \equiv 1 +  
(\Lambda/3) a^2 \cos^2 \theta$, $\chi^2 \equiv 
1 + (\Lambda/3) a^2$, and
\be 
\cQ \equiv -\frac{\Lambda}{3} r^4 + \left( 1 - \frac{\Lambda}{3} a^2  
\right)  
r^2 - 2Mr + \left( a^2 + E_0^2 + G_0^2 \right).  \nn
\label{Qpoly}  
\ee  
The individual solutions are defined by the values of the   
parameters $\Lambda$, $a$, $M$, $E_0$, and $G_0$ which are   
respectively  
the cosmological constant (since I'm interested  
in deSitter type spacetimes, assume that it is positive), the   
rotation parameter, the mass, and   
the effective electric and magnetic charge of the solution.  
Along with the electromagnetic field  
\be  
\label{EMfield}   
F = -\frac{1}{\cG^2 \chi^2} \left\{ X dr \wedge (dt - a \sin^2 \theta  
d\phi) + Y \sin  
\theta d\theta   
\wedge ( a dt - (r^2+a^2)d\phi) \right\},  
\ee  
where $X = E_0 \Gamma + 2aG_0 r \cos \theta $, $Y = G_0 \Gamma -   
2 a E_0 r   
\cos \theta$, and   
$ \Gamma = r^2 - a^2 \cos^2 \theta$, this metric is a solution to the  
Einstein-Maxwell equations. For reference note that a vector  
potential generating this field is
\be  
\label{EMpot}  
A = \frac{E_0 r}{\cG \chi^2} \left(dt  
 - a \sin^2 \theta d\phi \right) + \frac{G_0  
\cos \theta }{\cG \chi^2} \left( a dt - \left(  
r^2 + a^2 \right) d\phi \right).  
\ee  
Keep in mind however the restrictions against dyonic spacetimes
that were discussed in previous chapters. Thus, even though this
is a dyonic solution I'll only be able consider the creation of
spacetimes where either $E_0=0$ or $G_0=0$. 
  
The roots of the polynomial $\cQ$ correspond to horizons of the   
metric.   
As a quartic with real coefficients, $\cQ$ may have zero, two, or  
four real roots. I will be interested in cases where there are four
real roots, three of which are positive. In ascending order,
these positive roots correspond to   
the inner black hole horizon, the outer black hole horizon, and the cosmological horizon.    
  
\begin{figure}[h!]
\begin{center}
\rotatebox{270}{\resizebox{10cm}{!}{\includegraphics{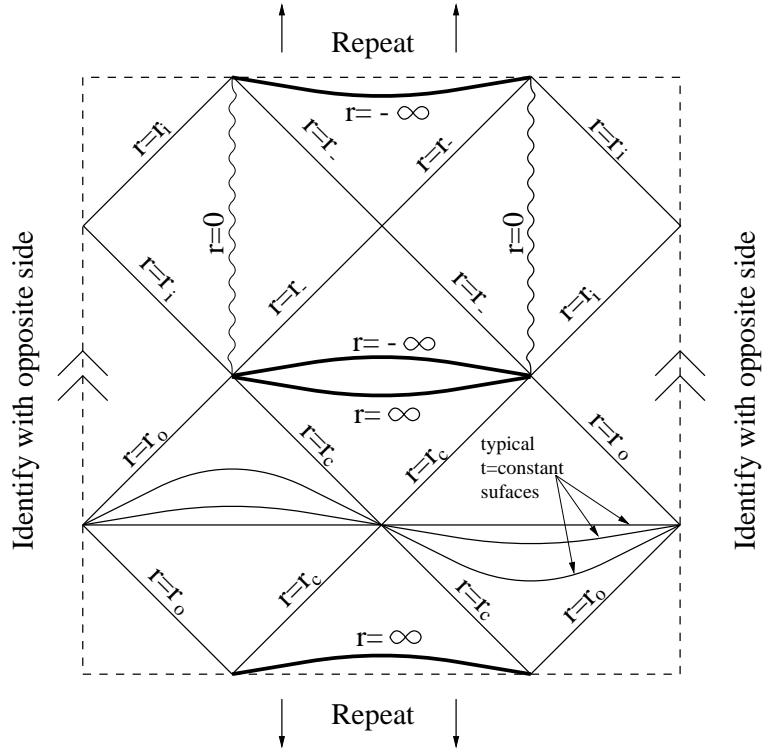}}}
\end{center}
\caption[Global structure of general KNdS solutions]{The global structure of the KNdS solutions with periodic 
identifications so that $t=\mbox{constant}$ hypersurfaces contain 
only two black holes.}   
\label{PenReg}  
\end{figure}

If all of the roots of $\cQ$ are distinct, then by the standard 
Kruskal techniques the metric may be analytically continued through 
the horizons to 
obtain the maximal extension of the spacetime \cite{cosmo}. Though this 
maximal
extension is infinite in extent, a variety of other global 
structures are possible if periodic identifications are made.
In particular, demanding that there be no closed 
timelike curves in the spacetime and also that there are two  
black holes in spatial cross-sections of constant time coordinate
$t$, the global
structure is uniquely determined and is shown in figure \ref{PenReg}  
(for a two-dimensional constant $\phi$, 
$\theta = \frac{\pi}{2}$ cross section). As indicated the figure is repeated vertically and periodically identified horizontally. $r=r_c$ is the
cosmological horizon, $r=r_o$ is the outer black hole horizon, and
$r=r_i$ is the inner black hole horizon. The wavy lines at $r=0$
represent the ring singularity found there for $a \neq 0$.  
If $a = 0$ then this singularity may not be avoided and the spacetime  
cuts off at $r=0$. Otherwise the singularity may be bypassed and one may  
proceed to negative values of $r$. $r=r_-$ is the (negative) fourth root of $\cQ$. The constant $t$ spatial hypersurfaces are closed and
span the two black hole regions, cutting through the intersections of
both the $r=r_c$ and $r=r_o$ lines.     
The matching conditions are such that, in the  
spatial hypersurfaces,  the two holes have opposite  
spins as well as opposite charges. Thus, the net charge and net spin of  
the system are both zero. Note that it is not possible to periodically
identify the spacetime so that the spatial sections contain only a
single black hole.  
    
\subsection{The allowed range of the KNdS solutions}  
The allowed ranges of the parameters so that $\cQ$ has three 
non-negative roots are calculated in appendix \ref{KNdSRange}
and shown in figure \ref{allowedRange}. The parameter space
is the region bounded by the two solid sheets plus the 
$a^2=0$, $M=0$, and $E_0^2 + G_0^2 = 0$ sheets. The darkest
sheet corresponds to the extreme black hole for which the 
inner and outer black hole event horizons are degenerate and
the lighter gray sheet is the case where the outer black hole
horizon is degenerate with the cosmological horizon. Taking
nomenclature from the non-rotating instantons discussed in 
\cite{robbross} I'll call the extreme black hole case the cold
KNdS spacetime while the second will be the Nariai KNdS
spacetime. I'll denote their intersection the ultracold
KNdS spacetime. The transparent sheet represents a 
special case of solutions corresponding to
lukewarm spacetimes, which will be discussed in subsection 
\ref{equil}.  
\begin{figure}  
\begin{center}
\resizebox{13cm}{!}{\includegraphics{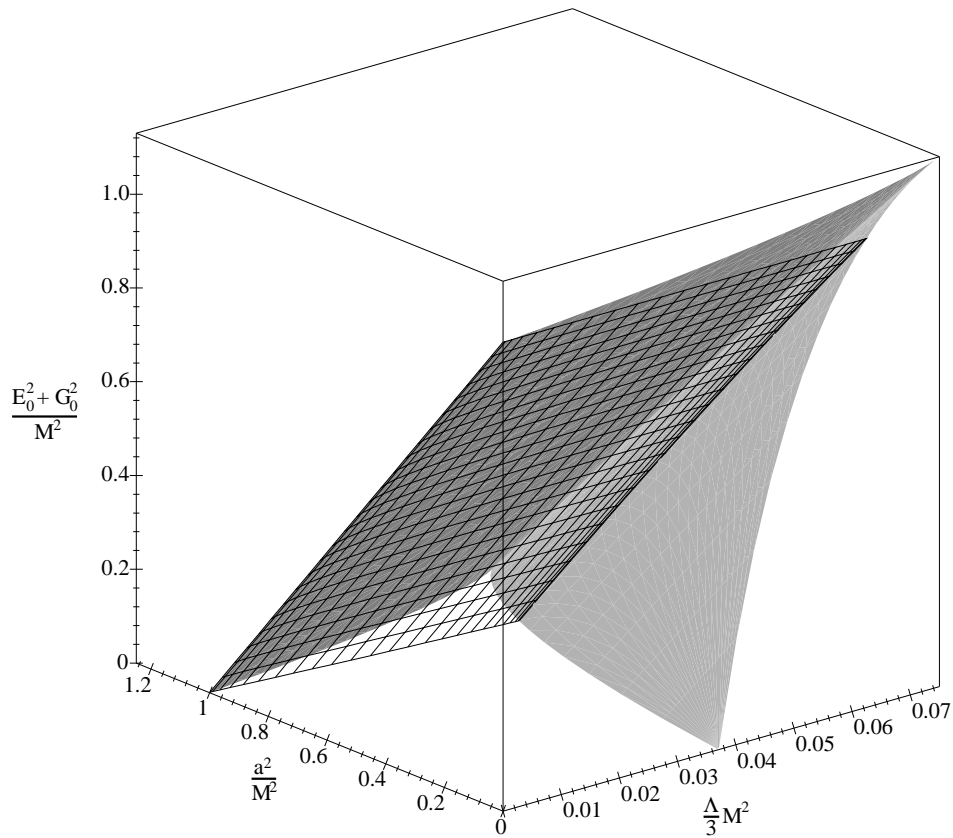}}
\end{center}
\caption[Allowed range of KNdS parameters]{The allowed range of the KNdS parameters. The range is bounded 
by the planes $M=0$, $a^2 = 0$, $E_0^2 + G_0^2 = 0$, the cold solutions 
(the darkest sheet) and the rotating Nariai solutions (the lighter gray sheet).
Also shown as a meshed sheet are the lukewarm solutions.}
\label{allowedRange}   
\end{figure}  

Note that the extreme cases, though limits of the KNdS metric,
have different global topological structures. In fact the 
Nariai and ultracold spacetimes do not even contain black holes. 
Their metrics in coordinate form may be found in appendix \ref{KNdSRange}, but here I'll just comment briefly on some of 
their properties. 

\begin{figure}[h]  
\begin{center}
\rotatebox{270}{\resizebox{6cm}{!}{\includegraphics{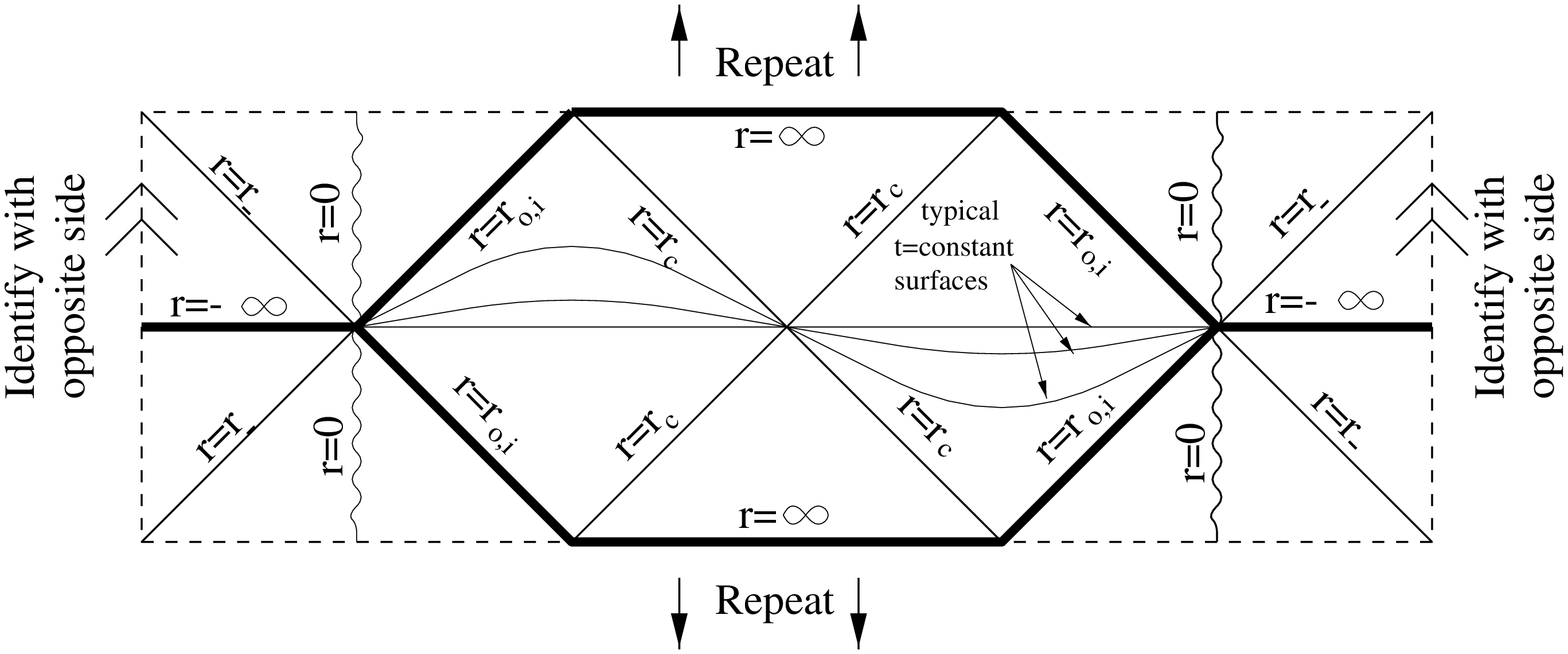}}}
\end{center}
\caption[Global structure of cold KNdS spacetime]{The Penrose-Carter diagram for a two hole cold KNdS spacetime.}  
\label{PenCold}  
\end{figure}

In the cold case  
the double horizon of the black hole recedes to an  
infinite proper distance from all other parts of the spacetime (as 
measured in a spacelike surface of constant $t$).  
Thus, the global structure of the spacetime changes. In particular, 
the region inside the  
black hole is cut off from the rest of the spacetime. Making 
appropriate periodic identifications of the global structure 
so that the spacetime contains two (in this case   
extreme) black holes, the structure is shown in figure \ref{PenCold}.
In that figure opposite sides of the rectangle are identified.
$r=r_c$  is the cosmological horizon and $r=r_{o,i}$ is the double 
black hole horizon. 
If $a=0$, then the spacetime cuts off at the singularity at   
$r=0$. Otherwise, one may pass through the ring singularity to the  
negative values of $r$, including $r_-$, the fourth root of $\cQ$.
Note that in this case, the $t=\mbox{constant}$ hypersurfaces contain 
two extreme black holes, and so are not closed as they are in the 
regular KNdS spacetime. The metric for this case is given in appendix
\ref{coldsect}.
  
\begin{figure}[t]  
\begin{center}
\rotatebox{270}{\resizebox{4cm}{!}{\includegraphics{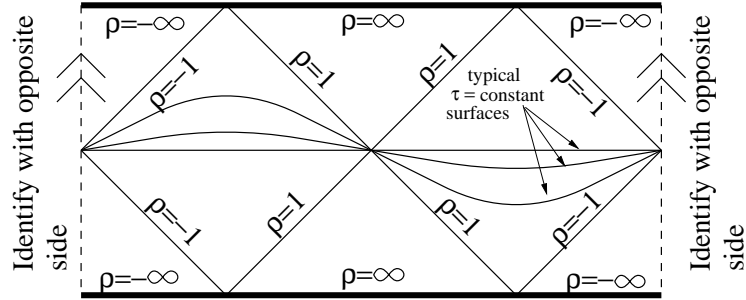}}}
\end{center}
\caption[Global structure of rotating Nariai spacetime]{The Penrose-Carter diagram for the Nariai limit spacetime.}  
\label{PenNar}  
\end{figure}

As noted the Nariai solution shown in \ref{PenNar}
is no longer a black hole solution, and there is no 
longer a singularity at finite distance beyond either of the horizons
at $\rho = \pm 1$. In fact, the diagram is the same as that for 
two-dimensional deSitter space. If there were no rotation ($a=0$), 
then this
spacetime would just be the direct product of two-dimensional deSitter  
space, and a two-sphere of fixed radius. With rotation, of course the  
situation is not so simple but
if $a=0$, it reduces to the non-rotating charged Nariai solution
considered in \cite{robbross}. The metric may be found in 
appendix \ref{narsect}.

Even though the Nariai solution is not a black hole solution itself,  
it was shown in \cite{ginsper} that an uncharged, non-rotating Nariai  
solution is unstable with respect to quantum tunnelling into 
an almost-Nariai 
Schwarzschild-deSitter spacetime. It is usually argued     
\cite{bousso} that this tunnelling carries over analogously with the  
inclusion of charge and rotation, in which case Nariai solutions  
decay into near Nariai KNdS spacetimes. Thus, in the future sections   
where I study black hole  
pair creation this solution will remain of interest, 
as a route to black  
hole pair creation will be to create a Nariai spacetime and then let it  
decay into a black hole pair.  

A similar argument can be made \cite{robbross}
for the ultracold spacetimes found
at the intersection of the parameter spaces of the cold and Nariai
solutions. There are two possible spacetimes (appendix \ref{ucsect}),
one with one horizon and the structure of Rindler space and the other 
which is
conformally Minkowski and has no horizons. Neither contain black holes.

\subsection{Issues of equilibrium}  
\label{equil}  
Before passing on to the next section where  
instantons to create the above spacetimes will be constructed, 
I'll pause to examine whether these solutions to the Einstein-Maxwell  
equations are stable with respect to semi-classical effects. This
is relevant because traditionally one only considered the quantum 
creation of black hole pairs in thermodynamic equilibrium, as it 
was thought that these
were the only cases where regular instantons could be constructed.
It is not so clear today (see for example
\cite{wuPairBase, ConsInst}) that that requirement must be 
enforced, but since I'll be more-or-less using the traditional
methods here and also try to draw some conclusions about the 
thermodynamics of the spacetimes, 
it is an issue that must be considered. 
To check for this equilibrium, one must consider three
phenomena: thermodynamically driven particle   
exchange between the horizons, electromagnetic discharge of the holes (due to emission of charged particles), and spin-down of the holes 
(due to emission of spinning particles and super-radiance).   
  
It is well known that a black hole emits particles in a black body   
thermal spectrum and thus may be viewed as having   
a definite temperature \cite{hawkNat}. In the same way, it has   
been shown that deSitter horizons may also be viewed as black bodies   
and have a definite temperature \cite{cosmo}. For a spacetime with  
non-degenerate horizons, these  
temperatures may be most easily calculated by the conical singularity 
procedure  
\cite{OrigPathInt} (which will show up again in the next
section during the instanton construction). 
First, corotate the coordinate system with the horizon whose temperature is being calculated. Second, analytically continue  
the time coordinate to imaginary values. For definiteness  
label the imaginary time  
coordinate $\cT$, the radial coordinate $\cR$, and let the horizon be  
located at $\cR = \cR_h$. Next,  
consider a curve in the $\cT-\cR$ plane with constant radial coordinate  
$\cR = \cR_0$.  
Periodically identify the imaginary time coordinate with some period
$P_0$ so that  this  
curve  becomes a coordinate ``circle'' and may be assigned a radius  
$R_0$ and circumference $C_0$ according to the integrals
\be  
R_0 \equiv \left. \left( \int_{\cR_h}^{\cR_0} \sqrt{g_{\cR \cR}} d\cR  
\right) \right|_{\cT=0},  
\;\;\;\;\;   
\mbox{and} \;\;\;\;\;   
C_0 \equiv \left. \left( \int_0^{P_0} \sqrt{g_{\cT \cT}} d\cT  
\right) \right|_{\cR=\cR_0}.  
\ee  
Finally, calculate $\lim_{\cR_0 \rightarrow \cR_h}   
\frac{C_0}{R_0}$.  
Pick the value of $P_0$ so that the limit has value $2\pi$. Then, the  
horizon has temperature $T_h = 1/P_0$, and surface gravity $\kappa_h =  
2\pi/P_0$. 
 
If there is a degenerate horizon as is the case for a cold black hole,  
then that horizon is an infinite proper distance from all non-horizon   
points of the spacetime. In such a situation there is no restriction 
on the period  
with which the degenerate horizon can be identified, and it  
has been argued \cite{hhr} 
that the black hole can therefore be in  
equilibrium with thermal radiation of any temperature.   
  
Now consider which of the spacetimes are in thermodynamic equilibrium.  
First, consider the general non-extreme KNdS solutions. 
The temperature  
of the outer black hole horizon and the cosmological horizon are  
respectively,   
\be  
T_{bh} = \left. \left( \frac{1}{4 \pi \chi^2 (r^2 + a^2)} \frac{d \cQ}{d  
r} \right) \right|_{r=r_{bh}} \;\;\; \mbox{and} \; \; \; T_{ch} = \left.  
\left(\frac{-1}{4 \pi \chi^2 (r^2 + a^2)} \frac{d \cQ}{d    
r} \right) \right|_{r=r_{ch}}  \label{temperatures}
\ee   
There are two ways that these two temperatures may be equal. The first
is if $r_{bh} = r_{ch}$ which actually is the Nariai spacetime. 
However there is
also a non-extreme solution labelled the lukewarm case. Its 
parameterization is considered in appendix \ref{luke}. 
  
The cold limit is in thermodynamic equilibrium at the temperature of  
the cosmological horizon, for as has been noted an  
extreme black hole may be in equilibrium with thermal radiation of  
any temperature. As noted, the Nariai limit too is in thermodynamic   
equilibrium, with both horizons having the same temperature  
\be  
T_{Nar} = \frac{\lt (4e^2-\delta^2)}{4 \pi},
\ee   
where $e$ and $\delta$ are defined in appendix \ref{narsect}.
The first ultracold case has only one horizon with temperature  
\be  
T_{UCI} = \frac{1}{2\pi},  
\ee  
and so with no other horizon to balance this one off, it is not in thermal equilibrium. The second ultracold case has no horizons, and so
is trivially in equilibrium.
  
Next consider discharge of the black holes. Even if the black hole   
and cosmological horizon are in equilibrium with respect to thermal
particle exchange between them, there can still
be a net exchange of charge between the horizons. 
The mechanism is that even  
though both may create the same number and masses of particles,  
an excess of charged particles will be created at the black hole horizon,  
and so it will discharge \cite{discharge}. This effect
can be quite rapid and so in most cases a charged black hole
cannot be said to be truly in equilibrium. 
However, there are a couple of ways to avoid the discharge.   
If there are no particles of the appropriate charge
that are also lighter than the black hole then discharge
cannot occur. Thus, if magnetic monopoles
do not exist then the magnetic holes will be stable with respect to
discharge. Further, even if the appropriate light charged particles
exist, the discharge effects will be small if the temperature of the
black hole is small relative to the mass of those particles. That is,
the more massive the black hole, the slower the discharge.

Finally consider the spin-down of the black holes. If the black  
holes and cosmological horizons are at the same temperature, then there  
will be no net energy exchange between the horizons, but the particles  
created at the black hole horizons may still have an excess of angular  
momentum relative to those created at the cosmological horizons. 
Unfortunately this effect is not as well studied for cosmological
spacetimes as is the equivalent discharge case. Still, from
the extensive calculations in asymptotically flat 
space \cite{page1,page2,chambers} one can say the following. 
In flat space, the direct spin-down by particle creation is a relatively 
slow process but it is greatly amplified by super-radiance. 
In combination the two processes cause angular momentum to be
radiated relatively more quickly than mass is radiated 
unless there are a
truly ridiculous number of scalar fields in the spacetime
\cite{chambers}. Preliminary indications \cite{maeda} are that
spin-down occurs at least as quickly and possibly more quickly
in asymptotically deSitter spacetimes which are not in thermal
equilibrium. There aren't any corresponding calculations for
black holes which are in thermal equilibrium
with cosmological horizons, but that said, in the other cases
spin-down is a relatively quick effect
which means that a rotating black hole in deSitter space 
probably cannot be thought of as being in full equilibrium.
Possibly the presence of thermal equilibrium might cause 
something miraculous to happen, but that is unlikely
and in any case a matter to be resolved by future calculations.
However, even in the absence of such a miracle,  
the physically intuitive notion
that a black hole that is rotating slowly relative to its
mass will discharge slowly is supported by the existing
results, and so it seems likely that at least a class
of these holes may be considered quasi-static in a thermodynamic
sense.

That said, in the following section I'll show that only
thermal rather than full thermodynamic equilibrium appears
to be required for the construction of smooth instantons. Of course
if the created system is not in full equilibrium one cannot really
draw conclusions about its thermodynamics. Thus, the reader
who is uncomfortable with the quantum 
creation of spacetimes that are not in full thermodynamic
equilibrium can consider all of the following to apply only
to the subset of spacetimes that are at least quasi-static. 

\section{Instanton assembly}     
\label{InsCons}     
In this section I construct the instantons that will be used to study
the creation of the spacetimes considered in the previous section. 
As discussed in     
the review of the path integral formalism, these instantons must both be     
solutions to the Einstein-Maxwell equations and also should match
as smoothly as possible along a spacelike hypersurface onto     
the spacetime that they create. The instantons constructed here will
satisfy the cosmological no boundary condition, and so I will not need
to worry about matching to initial conditions.  
  
\subsection{Analytic continuation}     
For static spacetimes, the first step of instanton construction 
is usually to analytically continue $t \rightarrow i \tau$. 
For a static spacetime expressed in appropriate coordinates, this
gives a real Euclidean solution to the equations
of motion but for a spacetime that is only stationary it will
usually produce a complex solution to the equations of motion. For now
I accept this complex solution but at the end of this section I'll
consider its relative merits
compared to the more standard approach where other metric parameters are
also analytically continued in order to obtain a real Euclidean metric.
That said, I proceed in the following manner  
(which is equivalent to continuing $t \rightarrow i \tau$). 
 
If a spacetime is foliated by a set of space-like hypersurfaces     
$\Sigma_t$ labelled by a time coordinate $t$, the most general
Lorentzian metric can be written as    
\bea     
\label{lapshif}     
ds^2 &=& -N^2 dt^2 + h_{ab}(dx^a + V^a dt)(dx^b + V^b dt) \\     
     &=& (-N^2 + h_{ab}V^a V^b ) dt^2 + 2 h_{ab}V^b dx^a dt + h_{ab}dx^a     
dx^b, \nn     
\eea     
where as usual 
$h_{ab}$ is the induced metric on the hypersurfaces, $N$ is the     
lapse function, and $V^a$ is the shift vector field.     
Using the prescription of \cite{brownprl}, the analytic continuation
can be made by making all of the Lagrange multipliers from the
Hamiltonian
purely imaginary. To wit, I start by changing the lapse and 
shift so that 
$N \rightarrow i N$ and $V^a \rightarrow i V^\az$. 
The spacetime metric for the proto-instanton then becomes
\be   
\label{lapshifInst}     
ds^2 = (N^2 - h_{ab}V^aV^b) dt^2 + 2 i h_{ab} V^b dx^a dt +     
h_{ab}dx^a dx^b.     
\ee   
If  $V^i=0$ then this metric has a Euclidean signature, whereas if 
$V^i \neq 0$ then the metric is complex and its signature is not so
easily defined. There is a sense however in which it is still Euclidean.
At any point $x_0^\az$ one can make a  complex coordinate transformation
$x^a = \tilde{x}^a - i t \left. V^a \right|_{x_0} $ 
(or equivalently add a complex constant to the shift),  
to obtain the metric 
\be     
\left. ds^2 \right|_{x^a_0} = N^2 dt^2 + h_{ij} dx^i dx^j,     
\ee     
at $x^a_0$. Thus the signature is Euclidean at any point modulo a  
complex coordinate transformation. Following the Lagrange multiplier
prescription, the electromagnetic field is made complex by rotating 
the Coulomb potential 
$\Phi \rightarrow i \Phi$ which changes the 
Maxwell field tensor as 
\be     
F_{ta} \rightarrow i F_{ta}, \; \; F_{at} \rightarrow i F_{at}, 
\ \  \mbox{and} \ F_{ab} \rightarrow F_{ab},
\ee   
where as usual the Latin indices indicate a restriction to the 
spatial slices. If the original Lorentzian metric and electromagnetic
field were solutions to the Einstein-Maxwell equations, then so are this
complex metric and electromagnetic field.

I now show that this complex solution can be 
matched onto the real solution from which it was derived.

\subsection{Matching the complex to the Lorentzian}     
 
The obvious hypersurface along which to match the Lorentzian solution to     
its complex ``Euclidean'' counterpart described above, is a   
hypersurface of constant $t$. 
I specialize the general metric (\ref{lapshif}) to the     
stationary, axisymmetric case where $x^1 = \phi$, $x^2 = \theta$, and $x^3     
= r$. Then, $V^a = [V^\phi(r,\theta), 0, 0]$, $N = N(r,\theta)$,     
and     
$h_{ab} = \mbox{diag} [h_{\phi \phi} (r,\theta), h_{\theta     
\theta}(r,\theta), h_{r r} (r,\theta)]$. 
This specialization will remain general enough to     
cover the cases of interest in this thesis.     

Now, consider how the complexified solution does or does not
match onto the Lorentzian solutions across a 
surface of constant $t$. First, the unit 
normal to $\Sigma_t$ is $u_\az = \pm N [dt]_\az$ where 
$[dt]_\az$ is the coordinate version of $dt$. Choosing
it to be forward pointing on the Lorentzian side and consistently
oriented on the ``Euclidean'' side $u_\az = - N [dt]_\az$ in 
each case. Then, on the 
Lorentzian side the induced metric is $h_{ab} = g_{ab}
+ u_a u_b$ while on the ``Euclidean'' side it is
$\tilde{h}_{ab} = g_{ab} - u_a u_b$ which are both
equal to $\mbox{diag} [h_{\phi \phi}, h_{\theta     
\theta}, h_{r r}]$. Thus, the induced
hypersurface metrics match and so a geometrical 
matching is possible.
In the same way, the same vector potential $\tilde{A}_a$ is 
induced from both sides, so from a purely Hamiltonian 
point of view, the configuration variables match.

Of course for the matching to be smooth, both sides should
also induce the same extrinsic curvature on the surface 
(as discussed by Israel in \cite{thinshell} and 
already discussed for
timelike surfaces in section \ref{thinshell} of this 
thesis). Unfortunately with $u_\az$ as defined above,
the extrinsic curvatures $K_{ab} = 
e_a^\az e_b^\bz \nabla_\az u_\bz$ are not the same\footnote{In \cite{pairprl,pairprod} Robert Mann
and I took a slightly different view of this by letting $u_\az$
become imaginary over the instanton. Then the same extrinsic
curvatures are induced on the surface. Here though I choose not
to take this view.}. Namely on the
Lorentzian side, 
\be     
K_{ab} \equiv e_a^\az e_b^\bz u_{\az;\bz} = \left[      
         \begin{array}{ccc}     
    0 & \frac{h_{\phi \phi} \partial_\theta V^\phi}{2N} &     
        \frac{h_{\phi \phi} \partial_r V^\phi}{2N} \\     
    \frac{h_{\phi \phi} \partial_\theta V^\phi}{2N} & 0 & 0 \\     
    \frac{h_{\phi \phi} \partial_r V^\phi}{2N} & 0 & 0     
         \end{array}     
         \right],     
\ee   
while on the ``Euclidean'' side the extrinsic curvature is
$i K_{ab}$. 
In a similar way, the induced electric field on the
``Euclidean'' side is $i E_a$ where $E_a$ is the Lorentzian
field.

Then, 
from the Hamiltonian perspective adopted in this thesis the situation
is as follows. Configuration variables $h_{ab}$ and $\tilde{A}_a$ 
remain real under the complex transformation, while their
conjugate momenta $P^{ab} = \sqrt{h}/(2 \kappa) (K h^{ab} - K^{ab})$ 
and $\cE^a  = -2 \sqrt{h}/\kappa E^a$ become purely imaginary along with
the Lagrange multipliers. Thus a matching is possible, though it
isn't smooth.

The conclusions of section \ref{thinshell} for spacetimes where there
is an extrinsic curvature discontinuity across a timelike hypersurface
apply equally well in this section where the discontinuity is
across a spacelike surface. That is, the discontinuity corresponds to 
a thin shell of matter. In this case the stress-energy tensor 
representing the matter is imaginary and since it is spacelike
exists only instantaneously. This is unusual to say the 
very least, but then again the surface $\Sigma_t$ separates regions
with different metric signature so perhaps it isn't surprising that 
something strange might occur at that surface. What is more of a concern 
however is that the presence of this strange matter at the borders of
the instanton might shift the action of the solution away from
extremality. If that is the case then the instanton cannot be used 
to approximate the full path integral. This matter deserves
further investigation, though I will not do that 
here\footnote{From the point of view adopted in 
\cite{pairprl, pairprod}, as discussed in the previous footnote,
this problem doesn't arise because the extrinsic curvatures match
exactly.}.

In section \ref{ActPick}, where I select an
appropriate action for evaluating the path integral, it will
be seen that this situation of real and complex fields actually
integrates quite nicely into the path integral formalism, but at
first glance the discontinuities are a bit disturbing. 
It is clear however, that
the complications have arisen from the inclusion of rotation. 
In earlier pair creation studies (such as \cite{em1,em2,em3,hhr,robbross,dom1,dom2,dom3,bousso}) 
there was no rotation which meant 
$K_{ab}=0$ and the geometric matching was smooth. The
discontinuity in the electric field remained, though
it wasn't usually considered.
     
Before moving on, I'll point out that by the 
traditional methods of instanton construction
such as those used in \cite{MM,LP,wu} the situation would be
even worse. The standard method would require that I analytically  
continue as many parameters of the metric as necessary to arrive at a
real and Euclidean solution to the Einstein-Maxwell equations. For  
example, with the Kerr-Newman-deSitter solutions, which will soon
be under consideration, the rotation and electric charge parameters
would be made complex ($a \rightarrow i a$, 
$E_0 \rightarrow  i E_0$) so that the   
Euclidean metric and electric field would be real.  
Although this approach avoids dealing with complex metrics, it  
incurs several serious problems of its own. Specifically, 
sending $a \rightarrow i a$ and   
$E_0 \rightarrow i E_0$ means that 
the surface metric $h_{ab}$ itself is affected by the
transformation. In detail, the polynomial $\cQ$,
and functions $\cG$ and $\cH$ (defined in 
and before equation (\ref{Qpoly})) are all changed and so 
$h_{rr} \neq \tilde{h}_{rr}$, 
$h_{\theta \theta} \neq \tilde{h}_{\theta \theta}$,
and $h_{\phi \phi} \neq \tilde{h}_{\phi \phi}$. That
this is not just a problem of coordinates
is made clear most dramatically by the fact that the change in 
$\cQ \rightarrow -\frac{\Lambda}{3} r^4  
+ (1 + \frac{\Lambda}{3} \tilde{a}^2)r^2 - 2 M r - \tilde{E_0}^2 +  
G_0^2$ will certainly shift, and often entirely change its 
number of roots, which means that the ``horizon'' structure of the 
spatial surface will be different on each side of $\Sigma_t$.
Thus, by the traditional method the ``Euclidean'' solution emphatically
would not match onto the Lorentzian solution. 
 
Given that the  
matching conditions are the only existing prescription that 
definitively link instantons with physical Lorentzian solutions, 
I choose to keep what matching conditions I can,
abandon the requirement that the full spacetime metric be real, and 
proceed with the calculation.
     
\subsection{Putting the parts together}     
\label{ppt}     
 
With these general steps taken, I'm now ready to finish off the
instantons. They will come in three classes: 
i) those creating spacetimes with two     
non-degenerate horizons bounding the primary Lorentzian sector (this case     
will create Nariai and lukewarm spacetimes), ii) those creating      
spacetimes with only a single non-degenerate horizon bounding the     
Lorentzian sector,      
(this case will create cold spacetimes and ultracold I spacetimes),     
and iii) and those creating zero horizon spacetimes (here, the ultracold II spacetime).     
     
\subsubsection{Spacetimes with two nondegenerate horizons}     
\label{space2hor}
\begin{figure}[t]
\begin{center}     
\rotatebox{270}{\resizebox{8cm}{!}{\includegraphics{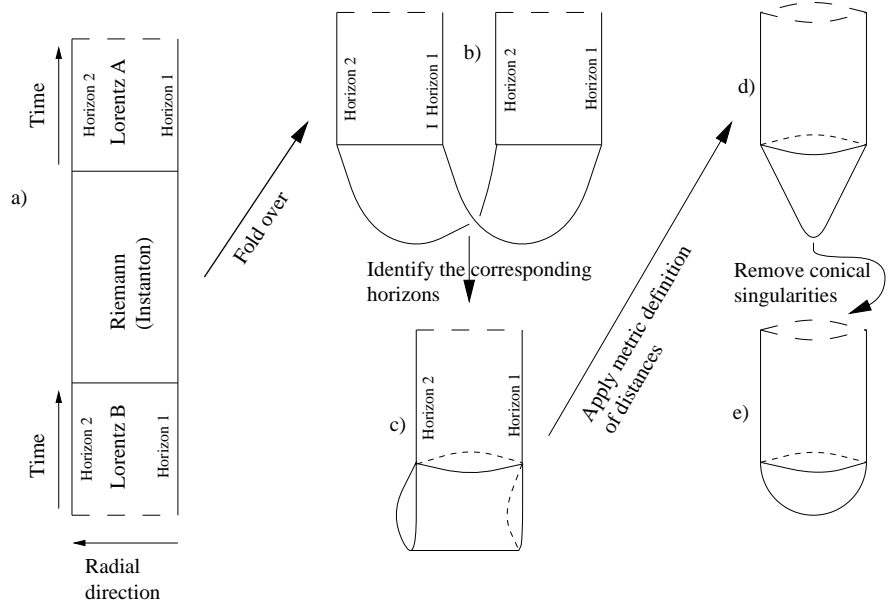}}}
\end{center}
\caption[Construction of a two-horizon instanton]{Construction of a 
two-horizon instanton. The radial/time     
sector is shown. The heavily dashed lines indicate that the solution     
continues in that direction.}     
\label{H2hor}     
\end{figure}

By the procedure described above I have found a complex solution     
that may be joined to the Lorentzian 
solution from which it was generated. However a subtlety arises 
in that the constant $t$ spatial hypersurfaces of the
nondegenerate KNdS and     
Nariai spacetimes both consist of two Lorentzian regions that are     
connected to each other across their corresponding horizons, while the     
constant $t$ hypersurfaces of the complex solution consist of  
only one such region.
The complex solution may be connected to both sections simultaneously 
by the following procedure (that is illustrated in     
figure \ref{H2hor}).     

First,   
connect half of a full Lorentzian solution (the region bounded by the  
outer black hole and cosmological horizons) to each of the $t=0$ and $t=\frac{P_0}{2}$  hypersurfaces of the ``Euclidean'' solution 
(as in figure \ref{H2hor}a). Next (figure \ref{H2hor}b) fold the construction  
over,  
and identify outer horizon to outer horizon, and inner horizon to inner  
horizon  
(figure \ref{H2hor}c). The $t=\mbox{constant}$ hypersurfaces of the  
Lorentzian part of     
the construction now consist of two regions with opposite spin and charge,     
and are the complete $t=\mbox{constant}$ hypersurfaces of the maximally 
extended but periodically identified KNdS solutions that I considered 
earlier.
     
Next note that the metric at any point of the Riemannian part of the     
construction is      
\be     
ds^2 = N^2 dt^2 + h_{ij}dx^i dx^j 
\ee     
under the coordinate transformation that eliminates the shift at that point. 
At the horizons $N^2 \rightarrow 0$     
for these solutions. Therefore it is reasonable to     
identify the entire time coordinate along the horizons as a single time     
(figure \ref{H2hor}d). The instanton is nearly complete.  
The ``Euclidean'' part is smooth everywhere except at the points
where I made the identification and probably 
introduced conical singularities.
 
Now, 
for a given horizon at $r=r_h$, I can find a period 
$P_0$ such that $\lim_{r     
\rightarrow r_h} \frac{ P_0 \partial_r \tilde{N}}{\sqrt{h_{rr}}} =
2\pi$ which in turn implies that the conical singularity has been
eliminated. This 
is the same condition used in calculating the temperature of the     
horizons in section \ref{equil}, and so those results may be reused
here. Hence the only double-horizon cases where the conical 
singularities at the two horizons may be simultaneously eliminated  (figure \ref{H2hor}e)
and so the only cases where the
instanton will everywhere be a solution to     
the Einstein equations, are the lukewarm and Nariai instantons,     
for which     
\be     
P_0^{lw} = \frac{4 \pi \chi^2 (r_{bh}^2 + a^2)}{Q'(r_{h})} \ \mbox{and} 
\ \     
P_0^{Nar} = \frac{4 \pi}{\frac{\Lambda}{3} (4e^2 - \delta^2)}     
\ee     
respectively.
$Q' = \frac{d Q}{d r}$, and $r_{bh}$ is the radius of the outer black 
hole horizon in the lukewarm solution. 
Then, the full construction of Lorentzian and Euclidean parts is
smooth everywhere, except on the $\Sigma_t$ transition 
surface where there will be a mild jump discontinuity in the 
extrinsic curvatures. 

Next consider the single-horizon spacetimes.

\subsubsection{Spacetimes with one non-degenerate horizon}     
 
\begin{figure}
\begin{center}     
\rotatebox{270}{\resizebox{8cm}{!}{\includegraphics{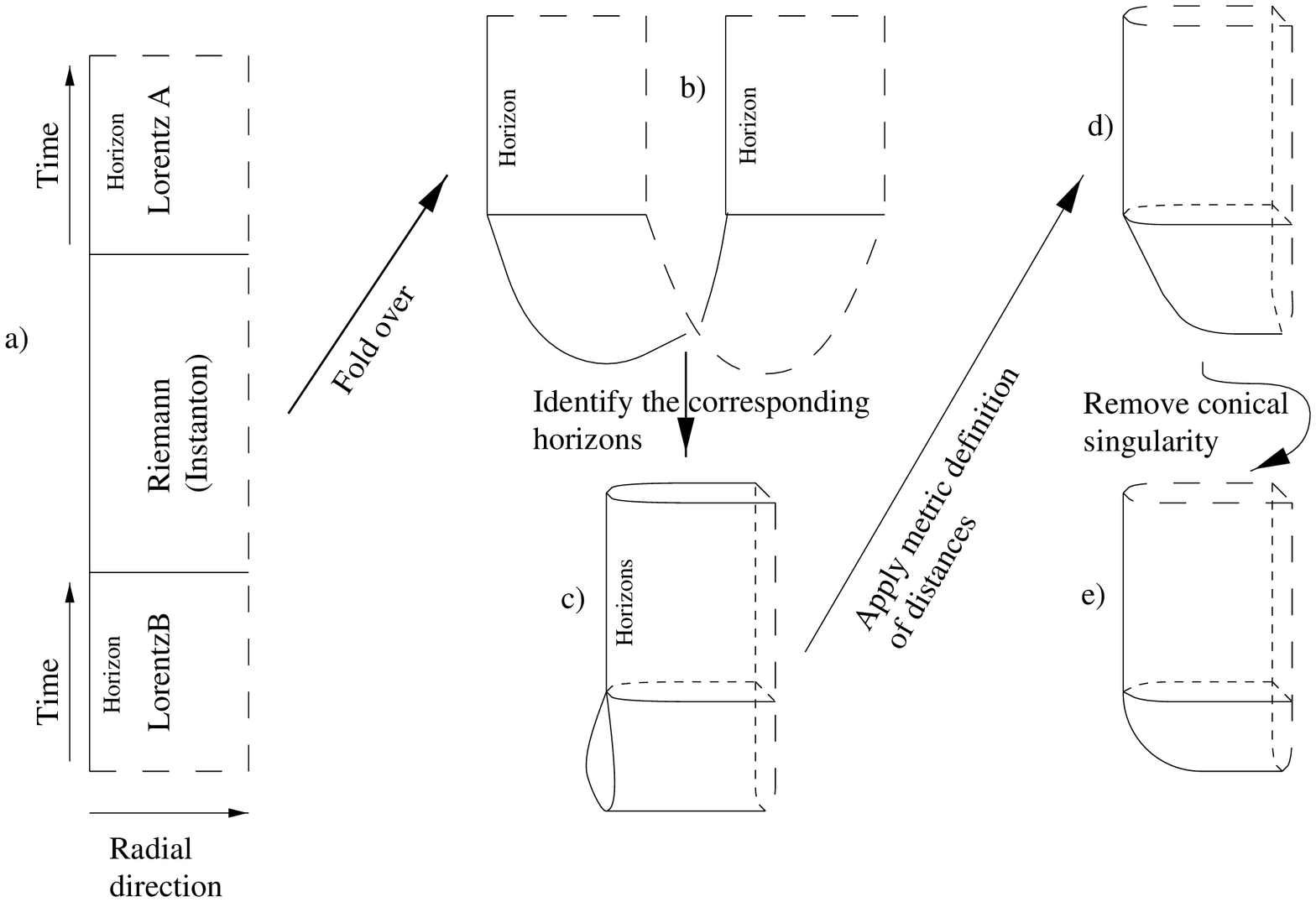}}}
\end{center}
\caption[Construction of a one horizon instanton]{Construction of a 
one-horizon instanton. The radial/time      
sector is shown. The heavily dashed lines indicate that the solution      
continues in that direction.}     
\label{H1hor}     
\end{figure}

It is now fairly easy to build the single     
non-degenerate horizon instantons for the cold  
and ultracold I spacetimes (even though the cold  
spacetime has two horizons, the inner horizon is a degenerate, 
double horizon).     
For these spacetimes, attach     
half-copies of the Lorentzian spacetime at the $t=0$ and     
$t=\frac{P_0}{2}$ hypersurfaces of the complex Riemannian section  
(figure      
\ref{H1hor}a).     
Then fold and identify the cosmological horizons to reconstruct     
the full Lorentzian $t=\mbox{constant}$ hypersurfaces (figure  
\ref{H1hor}b and c).     
Next, identify the time coordinate along the cosmological horizon     
(figure \ref{H1hor}d). Finally, with just one horizon choose     
\be     
P_0^{cold} = -\frac{4 \pi \chi^2 (r_{ch}^2 + a^2)}{Q'(r_{ch})}     
\qquad \mbox{ and} \qquad     
P_0^{UCII} = 2\pi,     
\ee     
where $Q'(r_{ch}) = \left. \frac{d Q}{d r} \right|_{r=r_{ch}}$ and
$r_{ch}$ is the radius of the cosmological horizon. Then the instanton will have
no conical singularities (figure \ref{H1hor}e). 
 
\subsubsection{No-horizon spacetimes} 
     
\begin{figure}
\begin{center}     
\rotatebox{270}{\resizebox{6cm}{!}{\includegraphics{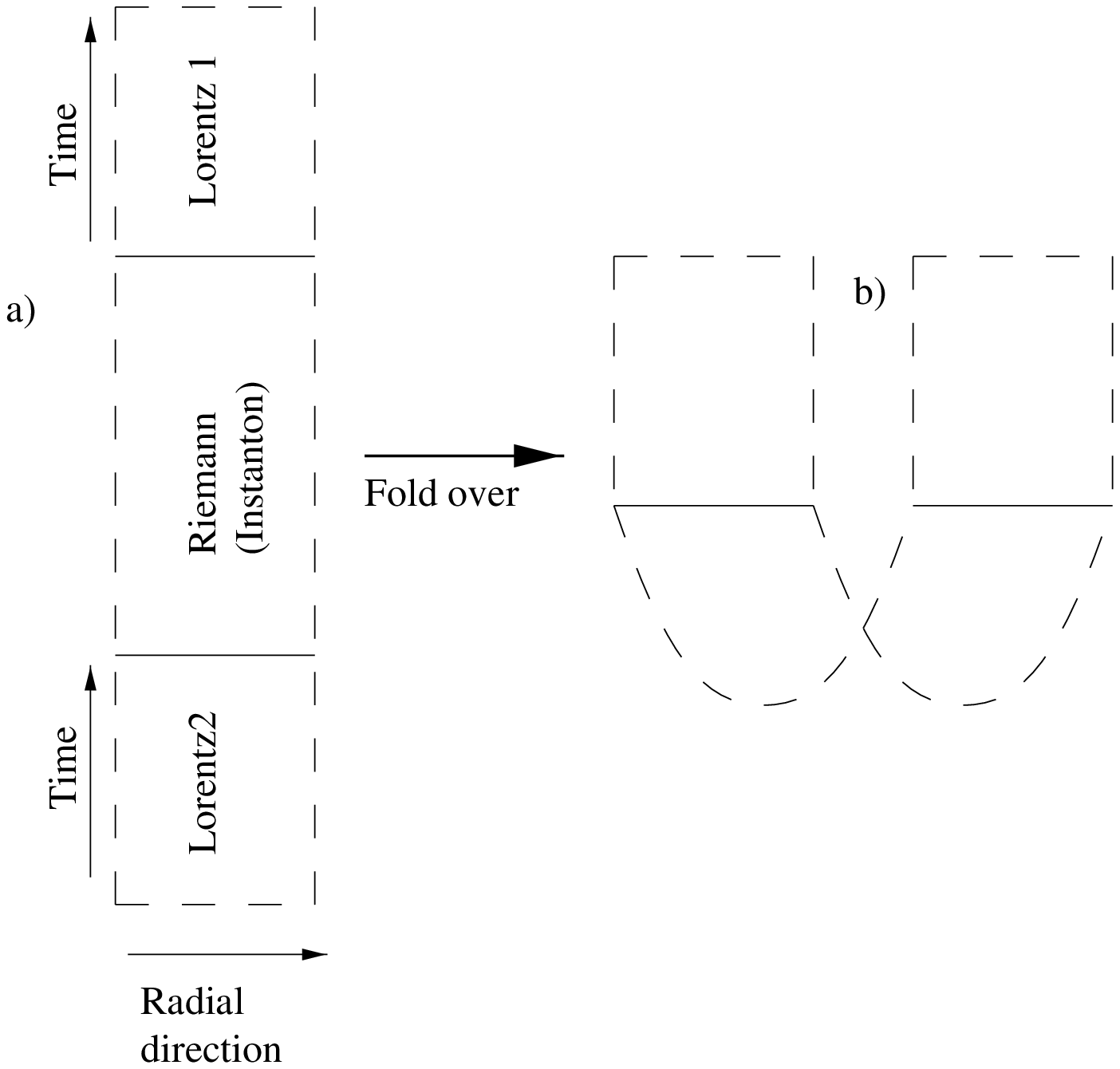}}}
\end{center}
\caption[Construction of a no-horizon instanton]{Construction of a 
no-horizon instanton. The radial/time      
sector is shown. The heavily dashed lines indicate that the solution      
continues in that direction.}        
\label{H0hor}     
\end{figure}

This time the construction is less definite. With no identifications being     
made, and no horizons to define a period, the instanton has     
indefinite period creating two disjoint spacetimes (figure \ref{H0hor}).     
This corresponds to the ultracold II case.     
         
\section{Choosing an appropriate action}
\label{ActPick}
As was discussed in chapters \ref{gravChapter} and \ref{matterChapter},
if one chooses a
Lagrangian action $I$, takes its first variation $\delta I$ over
a finite region $M$, and solves $\delta I = 0$, then the
solution includes
not only field equations in the bulk, but also boundary 
conditions on the fields over $\partial M$. Thus in choosing 
an action that is appropriate to a particular situation, one
must keep in mind the implied boundary conditions that are
attendant upon it. 

In particular, in the path integral formulation of gravity for
finite regions of spacetime, the action choice also fixes
boundary conditions that the possible paths must satisfy, and 
therefore restricts the allowed parameter space of those paths.
There are two ways of approaching the choice of how the parameter 
space should be restricted. The first is from a geometrical/topological
point of view. There one considers what geometrical 
properties the paths should have so that they will properly 
match onto the Lorentzian solutions. The second way is much 
more physical and considers what physical restrictions should
be placed on the paths so that they will produce the types of 
spacetime that one is interested in. That is, one demands that
the created spacetime has certain physical characteristics such
as horizons, temperatures of those horizons, and a particular angular
momentum or electric/magnetic charge and then enforces those same restrictions on the ``paths'' so that they will create the correct
spacetimes. Happily, as will be seen below, these apparently disparate
approaches complement each other and produce compatible lists of
restrictions. 

First from a geometrical point of view, it is essential that the
``paths'' match, onto the Lorentzian solution along the interface
surface $\Sigma_2$. That is, they should all induce the correct
surface metric $h_{ab}$ and vector potential $\tilde{A}_a$ 
on $\Sigma_2$. Examining the Hamilton-Jacobi
variation (\ref{HJGravVar}) (the orthogonal version is 
sufficient in this case) it is clear that the standard action 
functional has this property. Note however that the formalism
does not guarantee that the conjugate momenta $P^{ab}$ 
(or 
equivalently the extrinsic curvatures) and $\cE^a$ will match 
as well.
In an ideal world both would be fixed but since they are
conjugate to each other this is not possible. Given this
and the fact that the instanton work showed that
for that solution 
the conjugate momenta don't in fact match across
the transition surface, I'll fix the configuration
variables and leave the other two free.

Continuing with the geometry recall the conditions that were
placed on the instantons. Namely I required that they have only one
boundary ($\Sigma_2$ that matches onto the Lorentzian solutions)
and further that they be smooth and without conical singularities. 
That is I demanded that $N=0$ (because the foliation of the spacetime
is othogonal to the boundary in this case, I'll drop the 
bar notation) at the coordinates values of $r$
corresponding
to non-degenerate horizons in the Lorentzian solution and further that 
\bea    
\lim_{r \rightarrow r_h} \frac{\int_0^{P_0/2} dt N}{\int_{r_h}^r    
dr \sqrt{h_{rr}}} 
= \lim_{r\rightarrow r_h} \frac{P_0 \partial r N}{2 \sqrt{h_{rr}}} = 
\pi,    
\label{ConCond}
\eea  
where $r_h$ is the coordinate of the horizon. At first glance that
second condition appears to be awkward and abstruse but in fact it 
is quite straightforward to show that if $N = 0$ at $r_h$ then
\be
\lim_{r \rightarrow r_h} N p = \frac{2}{\kappa} n^a \partial_a N,
\label{PressTerm}
\ee
where $p$ is the pressure defined at the end of section
\ref{gravtransform}. Now $n^a = \frac{1}{\sqrt{h}} \partial^a_r$ 
and so to avoid conical singularities one must fix $Np$.
To ensure this, it is
more than sufficient to fix $N s^{ab}$. Turning
again to the action variation (\ref{HJGravVar}) the reader will
note that $N$ is already fixed while 
$N s^{ab}$ has been left free.
Adding
\bea
\Delta I_{p} &\equiv& -\frac{1}{2} \int dt \int_{\Omega_t} d^2 x    
\sqrt{\sigma} N p,
\eea
to the action the situation is corrected and one can force the paths to
be closed and smooth at the points corresponding to non-degenerate
horizons. 

There is also a nice physical interpretation of these conditions. Namely
fixing the lapse $N$ to be zero at $r_h$ means that there will be an 
(apparent) horizon at that point while putting the restrictions on $Np$
fixes the temperature of those horizons (see section \ref{equil}).
Therefore enforcing these conditions at $r_h$ means that there will be a
horizon of predetermined temperature there. If there are two
non-degenerate horizons then each will have a temperature.
By the no-conical-singularity requirement of geometrical smoothness,
they must have the same
temperature and so geometrical smoothness is equivalent to the
thermal equilibrium of the final state.

Next consider what should be done at a degenerate horizon such as 
that found in a cold spacetime and what restrictions should be
placed on the ``paths'' that might create 
it. To match onto the Lorentzian  solution all paths must have the    
``tapered  horn" shape characterized by $N \rightarrow 0$ at 
the degenerate horizon.
Since the horizon is an infinite proper distance from the rest of the spacetime, there is no need to worry  
about conical singularities, and therefore no need to fix the pressure.    
Instead leave $\sigma_{ab}$ fixed to ensure that the metric will
have the correct asymptotic behaviour. Thus at degenerate horizons
do not add $\Delta I_p$ to the action. Note that
this geometrical behaviour is 
the basis of the claim \cite{hhr} that an extreme black hole doesn't
have a fixed temperature but instead can be in thermodynamic
equilibrium with any background.

Having set the boundary conditions to ensure that the spacetimes
contain black holes it is natural to fix the angular momentum
and electromagnetic charge of those black holes. Afterall the
ultimate intention is to calculate the pair creation rates for
pairs of black holes of specified mass, angular momentum, and 
charge so these quantities must be fixed in advance or else
the path integral will calculate the creation rate for some other
situation. At first it might seem natural to fix $\e$ as well
so that one could specify the mass of the holes being created
but as was noted above one can fix $N$ or $\e$ but not both.
$N$ must be fixed so that the black holes can be guaranteed to 
exist, so $\e$ has to be left free. That said, to fix the 
angular momentum one must add
\be
\Delta I_j \equiv - \int dt \int_{\Omega_t} d^2 x \sqrt{\sigma} 
V^a j_a,
\ee
to the action.

Next consider fixing the electromagnetic charges. First recall
from chapter \ref{matterChapter}, that by choosing to work with $I_m$,
$F_{\az \bz}$, and $A_\az$ I have automatically excluded magnetically
charged solutions from consideration. At the same time however, the 
electric charge has been left free (see the variation 
(\ref{fullmactvar})). 
By the previous paragraph it should be fixed and so add  
\be
\Delta I_{el} \equiv \frac{1}{\kappa} \int dt \int_{\Omega_t} d^2 x
N \sqrt{\sigma} n_\az F^{\az \bz} A_\bz,
\ee
to the action. There is a choice of whether to
fix $\cE^a$ or $\tilde{A}_a$ on the boundary $\Sigma_2$. As has
already been noted $\tilde{A}_a$ is the appropriate quantity to fix
and that doesn't require an extra boundary term on $\Sigma_2$.

By contrast if I want to consider magnetic black holes I use
$I_m^\star$, $\star F_{\az \bz}$, and $A^\star_\az$. Then electric
charges are automatically eliminated from consideration and 
\be
\Delta I^\star_{mg} 
\equiv \frac{1}{\kappa} \int dt \int_{\Omega_t} d^2 x
N \sqrt{\sigma} n_\az \star \! F^{\az \bz} A^\star_\bz,
\ee
should be added to the action to fix the created magnetic charges.
No additional boundary term is required to fix $\tilde{A}^\star_a$
on $\Sigma_2$.

Switching to the thermodynamic interpretation of the path integrals
it is immediate that what is being considered here is a canonical
partition function. That is, extensive variables (angular momentum and 
electric/magnetic charge) are fixed except for the energy which is
left free in favour of holding the temperature constant. This is then
in accord with the standard approach to pair creation calculations which
uses that partition function \cite{hawkingross}. This choice
then ensures that created spacetimes are in thermal equilibrium, that 
there is no discontinuity in physical properties such as
electromagnetic charge and angular momenta at the juncture of the paths
and the Lorentzian solution, and
from a geometric point of view that the paths are smooth and
match onto the Lorentzian solution.    
    
\section{Evaluating the actions - pair creation rates and entropy}
\label{evAct}

As noted earlier, creation rates for these spacetimes are    
proportional to $e^{-2I_{inst}}$, where $I_{inst}$ is the 
numerical value of the action of the appropriate instanton.
Now, as was laid out in section \ref{formalism}, 
those rates are calculated only up to a normalization factor.
Evaluating this normalization factor would involve fully
evaluating another ill-defined path integral so I will 
side-step that issue by calculating the probability of creation
of these spacetimes relative to deSitter space. Then the normalization
factors cancel each other out and the relative probability of 
creation of the black holes in a deSitter background is 
\be 
P = \exp{(2I_{dS}-2I)},    
\ee    
where $I$ is the action of the instanton, and $I_{dS}$ is the    
action of an instanton mediating the creation of deSitter space with
the same cosmological constant. 
Conventionally, this probability is also interpreted as    
the probability that deSitter space will tunnel into a given black hole 
spacetime \cite{boussochamblin}.    
    
A further link to thermodynamics is found by the following argument.
The spacelike hypersurfaces of the spacetimes that I 
have considered are all topologically closed and with finite volume.
Then, the energy
is trapped in the hypersurfaces and so they can be interpreted as
having constant energy even though that condition has not been 
enforced by boundary conditions \cite{robbross,hawkingross}.
By that reasoning the canonical partition function is equivalent
to the microcanonical partition function in this case,
and so as is standard for a microcanonical partition function, 
the entropy of the created spacetime is
\be   
\label{entropy}   
S = \ln{\Psi^2} = - 2I.    
\ee    
Thus, at least in the case where the created spacetime is 
quasi-static,
there is a close connection between pair creation rates
and the entropy of the spacetimes and in particular it is
consistent with the idea that
the entropy is the logarithm of the number of 
quantum states. With all of this in mind I evaluate the appropriate
action for each spacetime.

Momentarily leaving aside the matter terms, the appropriate action 
by all of the above considerations is
\bea   
I_{(N,p,j)} &=& I + \Sigma_{SH} \Delta I_p 
                  +  \Sigma_{AH} \Delta I_j \nn \\   
                  &=& \left( \mbox{ terms that vanish for stationary solutions }    
\right) \nn \\    
        &&  + \Sigma_{AH} \int_{B_h} d^3 x \sqrt{\sigma} N \varepsilon
 -\frac{1}{2} \Sigma_{SH} \int_{B_h} d^3 x  \sqrt{\sigma} N p, 
\eea   
where the subscript $SH$ indicates a sum over all single, non-degenerate
horizons and $AH$ means a sum over all horizons regardless of their 
degeneracy. Keeping in mind that $N = 0$ on all of the 
horizons, it is clear that the $N \varepsilon$ terms are zero. Further,
recall equation (\ref{PressTerm}) which says $N p = (2/\kappa) n^a
\partial_a N$ and equation (\ref{ConCond}) which implies that
$n^a \partial_a N = (2 \pi)/P_0$ on a non-degenerate horizon. Then
\be
I_{(N,p,j)} = - \Sigma_{SH} \frac{\cA_H}{8},
\ee
where $\cA_H$ is the surface area of the event horizons in the spatial
surface $\Sigma_2$.
 
Next, consider the matter terms. For electric solutions it is a trivial 
use of Stokes's theorem to show that,
\bea
&& -\frac{1}{2 \kappa} \int_M d^4 x \sqrt{-g} (F_{\az \bz} F^{\az \bz}) 
+ \Delta I_{el} \nn \\
&=& \int_M d^4 x \sqrt{-g}( A_\bz \nabla_\az F^{\az \bz} )  
- \frac{1}{\kappa} \int_\Sigma d^3 x \sqrt{h} E^a \tilde{A}_a. 
\eea
Of course the first term includes 
the constraint equation (\ref{nabF}) and
so is zero for solutions to the Maxwell equations. Thus all that is left
is the second term, but it too is zero for the solutions in which 
I'm interested, and so the total electric term is also zero.
Thus, the value of the action that keeps $N$, $p$ (if appropriate), $j_a$, and $E_0$ fixed (and $G_0 = 0$) is
\be
I_{(N,p,j,E_0)} = - \Sigma_{SH} \frac{\cA_H}{8}.
\ee
The same line of reasoning shows that the value of the action that keeps
$N$, $p$ (if appropriate), $j_a$, $G_0$ fixed (and $E_0=0$) is 
\be
I^\star_{(N,p,j,G_0)} = - \Sigma_{SH} \frac{\cA_H}{8}.
\ee

Assuming that the spacetimes are at least quasi-static, then equation    
(\ref{entropy}) says that the entropy of these spacetimes is equal to
one-quarter of the sum of the areas of non-degenerate horizons 
bounding the Lorentzian region of the spacetime.   
Consistent with references \cite{hhr} and \cite{robbross}, the degenerate horizon in the cold case does not contribute to the entropy of the cold spacetime.

Using these general formulae for the pair creation rates and
entropy of the spacetimes, I now consider each of the specific 
spacetimes separately.

\underline{The lukewarm action:}     
In this case, there are non-degenerate cosmological and outer black hole    
horizons. Therefore the numerical value of the action of the 
electric/magnetic instantons is   
\be   
I_{LW} = - \frac{\cA_c + \cA_h}{8} = - \frac{\pi (r_c^2 +  
a^2)}{2\chi^2}    
- \frac{\pi(r_h^2 + a^2)}{2 \chi^2},   
\ee   
where $\cA_c$ and $\cA_h$ are respectively the areas of the cosmological    
and outer black hole horizons at $r_c$ and $r_h$
in the Lorentzian solution.   
   
\underline{The Nariai action:}   
Again  there are two non-degenerate horizons, this time at $\rho = \pm 1$. Therefore the total action of the electric/magnetic 
Nariai instantons is   
\be   
I_{N} = - \frac{\cA_{\rho=-1} + \cA_{\rho=1}}{8} =  
- \frac{ \pi (e^2+a^2)}{\chi^2},   
\ee   
where $\cA_{\rho = \pm 1}$ is the area of the horizon at $\rho = \pm 1$.    
Note that for the Nariai solutions $\cA_{\rho=1} = \cA_{\rho=-1}$.   
   
\underline{The cold action:}   
Here there is only one non-degenerate horizon, and so 
\be   
I_{C} = - \frac{\cA_c}{8} = - \frac{\pi (r_c^2 + a^2)}{2 \chi^2},   
\ee   
where $\cA_c$ is again the area of the cosmological horizon at
$r_c$.  
   
\underline{The ultracold I actions:}   
Again there is only a single nondegenerate horizon, this time at $R=0$.    
The action of the magnetic ultracold I instanton is then 
\be   
I_{UCI} = - \frac{ \cA_{0}}{8} = -\frac{\pi (e^2+a^2)}{2 \chi^2}.   
\ee   
where $\cA_0$ is the area of the Rindler horizon as $R=0$.

\underline{Ultracold II Actions:}   
There are no horizons whatsoever for this case, and so 
\be   
I_{UCII} = 0,   
\ee   
irrespective of the chosen period $P_0$ of the time coordinate,
which is good since as was seen in section \ref{InsCons}, 
that period is not specified by the formalism!   
   
\begin{figure}[t]    
\begin{center}
\resizebox{13cm}{!}{\includegraphics{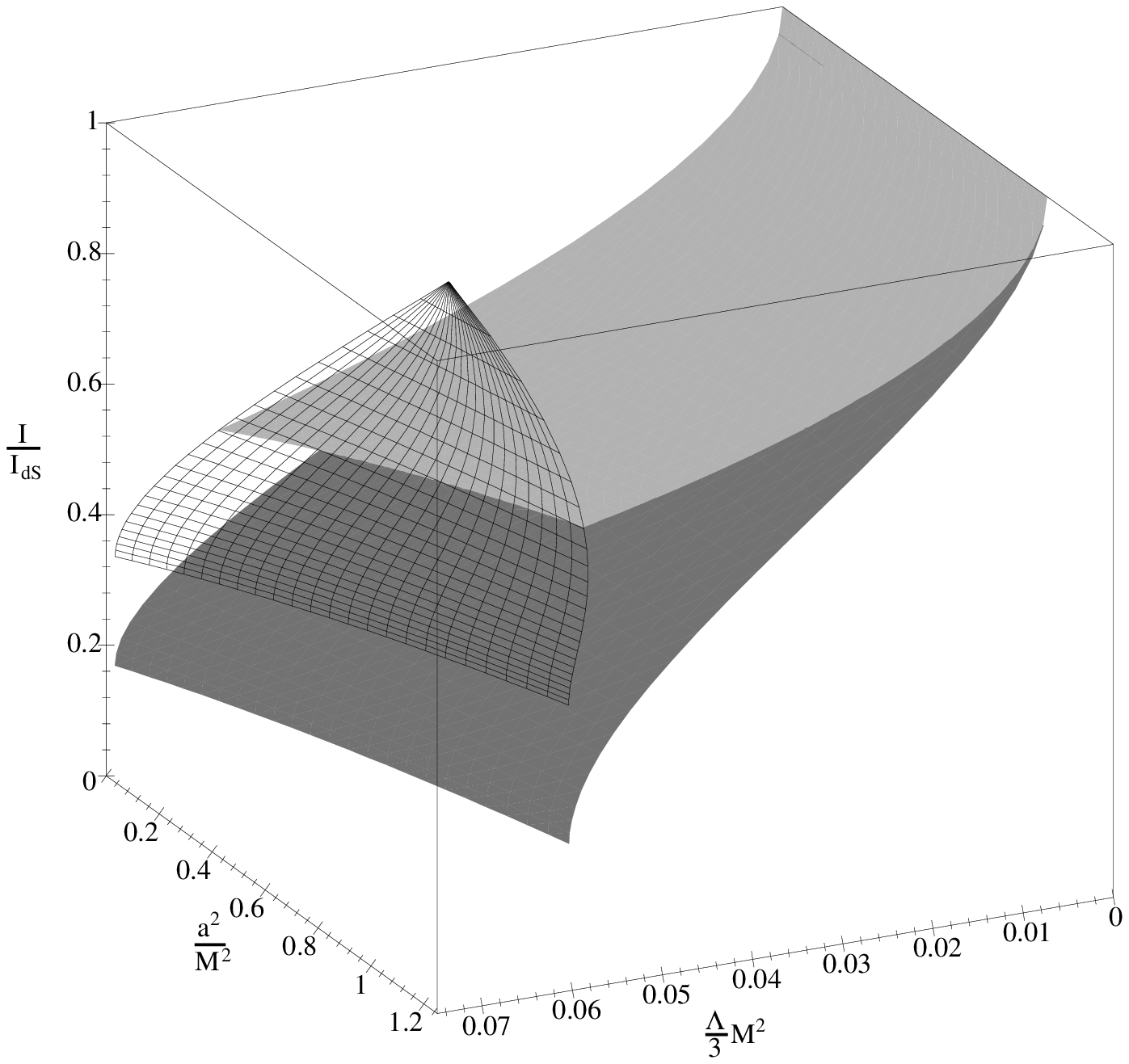}}
\end{center}
\caption[Charged and rotating instantons actions]{The actions for the charged and rotating lukewarm, cold, and
Nariai instantons plotted as a fraction of the action of deSitter
space with the same cosmological constant.  
The Nariai instantons are the meshed sheet, the
lukewarm instantons are the lighter grey sheet, and the cold instantons
are the darker grey sheet.The ultracold I instanton actions may be found
at the ``bottom'' end of the cold sheet, while the ultracold II
instanton actions are zero.}    
\label{creationrates}    
\end{figure}

In figure \ref{creationrates}, I plot these actions as a fraction of    
the action of the instanton creating deSitter space with the same    
cosmological constant. The instantons/created spacetimes are
parameterized by $\frac{a^2}{M^2}$ and $\frac{\Lambda}{3}M^2$. 
For all cases $I, I_{dS} < 0$ and from the diagram it is clear that    
$|I| < |I_{dS}|$. Then $I_{dS} - I < 0$ and so each of the spacetimes
considered above is less likely to be created than pure deSitter space
with the same cosmological constant.   
Note that the Nariai spacetime is the most likely to be created  
provided the parameter values are such that the instanton exists, 
while the cold spacetime is the least likely to be created.    
As might expected on physical grounds, smaller and more slowly rotating
holes are more likely to be created than larger and more 
quickly rotating ones. 
As $\frac{a}{M} \rightarrow 0$ and $M \rightarrow 0$,    
the creation rates approach those of deSitter space which again
is physically reasonable.  
    
\section{Reflections on the calculation}
\label{CompExt}
The approach to pair creation taken here is a bit different from 
that taken in most of the literature and because rotation has
been included new issues have arisen that were not present in 
those papers. Thus in this section I will compare the methods
and examine those issues a little further.

First I will compare the way I calculated actions 
here with the way it was done in reference \cite{robbross}
(which is representative of the more traditional 
calculations done for non-rotating black holes). 
There, the fact that the instantons are closed 
and smooth at the points    
corresponding to the non-degenerate horizons was taken to mean that no    
boundary terms need be considered there, implying that  
the basic action used for the lukewarm and Nariai instanton should be 
\bea    
I_{old} = - \frac{1}{2 \kappa} \int_M d^4 x \sqrt{-g} \left( \cR -    
2\Lambda - F^2\right)    
- \frac{1}{\kappa} \int_{\Sigma} d^3x \sqrt{h} K,    
\label{oldact}    
\eea    
which from my point of view is the Lagrangian action 
(\ref{matteraction}) with the boundary term    
\bea    
\frac{1}{\kappa} \int_{B} d^3x \sqrt{-\gamma} \Theta   
\eea    
added on.  
It is easy to see that this term is equivalent to the pressure term   
\bea    
-\int_{B} d^3x N \sqrt{\sigma} \frac{p}{2} = -\int_{B} d^3x N    
\sqrt{\sigma}    
\left[ \frac{n^a \partial_a N}{N} - \frac{k}{2} \right],    
\eea    
evaluated on the equivalent horizons. To see this, note that    
$\Theta = k - \frac{n^a \partial_a N}{N}$,    
$k = - \frac{1}{2 \sqrt{h_{rr}}} \partial_r \ln\sigma$,     
and $\frac{1}{\sqrt{h_{rr}}} \rightarrow 0$ at each horizon. So    
on those horizons $\Theta = -\frac{p}{2}$ and in the absence of 
rotation my approach is equivalent to that of \cite{robbross}.   

For the cold case, $k$ still vanishes on the boundary and so the inclusion of    
the $\Theta$ term in \cite{robbross} is  equivalent to 
the omission of the    
pressure term in my calculation. Finally, in the ultracold cases $k=0$    
everywhere and so once more the omissions/inclusions are equivalent.   
   
For electric instantons in both calculations, electromagnetic boundary terms are added to the action to fix the electric charge for all paths
considered in the path integral. Further, in both calculations for
solutions to the Maxwell equations, these boundary terms may be
converted into the $F^2$ bulk term that was used in this work.    
For electric instantons that earlier work added a boundary term
\be
\frac{2}{\kappa} \int_{\Sigma2} d^3 x \sqrt{h} E^a \tilde{A}_a
\ee    
on the $\Sigma_2$ boundary (the only boundary for those instantons).
For solutions to the Maxwell equation this is then equivalent to
adding a bulk term
\be
\frac{1}{\kappa} \int_M d^4 x \sqrt{-g} F_{\az \bz} F^{\az \bz}
\ee
to the action which then makes it numerically equivalent to the 
action $I^m + \Delta I_{el}$. For the magnetic case it was 
argued that nothing needed to be added since the magnetic
charge was already fixed on the boundary.
Note however that while that
approach works out numerically it not quite right because, as was noted
in chapter \ref{matterChapter}, if one assumes that a single 
$A_\az$ covers $M$ then no magnetic charge can exist.

Although for non-rotating instantons the approach here     
is equivalent to earlier ones, differences arise when rotation is
included. In earlier approaches 
\cite{em1,em2,em3,hhr,robbross,othercosmo,str1,str2,str3,dom1,dom2,dom3} there was no provision made for fixing the angular momentum and so 
the action differs by the term $\Delta I_j$ and its omission 
is tantamount to working with an incorrect thermodynamic ensemble.
Evaluating the action of rotating  
instantons with (\ref{oldact}) will not yield the preceding relationships 
linking surface areas, actions, and entropies.  
Indeed, using (\ref{oldact}) the creation rate of    
rotating black holes is enhanced relative to that of    
non-rotating black holes and with an appropriate choice of physical    
parameters may be made arbitrarily large.  
    
Second, 
around the same time that this work was originally published, 
Wu Zhong Chao published a series of papers on the 
creation of a single black hole (see for example \cite{wu,wuPairBase}) 
using a slightly different set of instantons to create spacetimes that
are not in any kind 
of equilibrium (thermodynamic or otherwise)\footnote{This type
of instantons have also been proposed
for use in cosmology by Hawking and Turok in \cite{ConsInst}.}. 
He recognized that
the angular momentum needed to be fixed but used an ad hoc approach
to work out what the angular momentum fixing term should be. 
For the cases considered however, that term was equivalent to the
one used here. 

However, despite the results being similar, his
approach was quite different. In the first place he asserted
that his approach could create a single black hole. From a physical
point of view, this would violate conservation of angular momentum 
and electric/magnetic charge. Even apart from this, 
the instantons that he considered do not properly  
match to real Lorentzian solutions for two reasons. 
In the first place there  
are no periodic identifications of the universal covering space of the  
basic KNdS solution that can be made such that hypersurfaces of
constant $t$ will contain only a single black hole. The smallest number  
of black holes that may be contained are     
the two discussed here. Second, as argued     
earlier, an analytic continuation of $a$ to $ia$ and $E_0$ to $iE_0$ will mean in general that an instanton generated from a classical
solution will not properly match onto that classical     
solution. In general, there will not even be the correct number of    
horizons available in the instanton to match onto the Lorentzian solution. In later
papers (for example \cite{wuADS1,wuADS2}) 
he considered the creation of pairs of black
holes instead, but the other differences remain.

\chapter{Discussion}
\label{discuss}
In this thesis I have taken the quasilocal energy formalism of Brown and
York and generalized it in several different directions. First, for
the finite region of spacetime $M$, I dropped the requirement that
the foliation hypersurfaces $\Sigma_t$ and spacelike boundaries
$\Sigma_1$ and $\Sigma_2$ be orthogonal to the timelike boundaries
$B$. From a theoretical perspective this obviously allows one to 
consider much more general regions $M$ and further does not 
restrict the allowed variations of the metric when the 
variational principle is applied. The ensuing calculations
then make it clear that the numerical value of the 
quasilocal Hamiltonian (and thus the quasilocal energies 
derived from it) is a function of the  foliation of $B$ and time evolution vector field $T^\az$. It does
not care about the foliation of the bulk $M$ as a whole. As was
repeatedly emphasized throughout the thesis, this is a very 
desirable characteristic for a quasilocal energy to have
since the correspondence between foliations of the 
bulk and foliations of the boundary is many-to-one. 
Further from a practical, computational
point of view, focusing only on the foliation of the boundary 
(as opposed to the bulk) makes it much easier to 
calculate the quasilocal energies seen by moving observers. 

Second, I shifted the calculations from the usual Lagrangian 
framework into a pure Hamiltonian form. Of course the two are 
equivalent but this thesis was the first place where that
was shown explicitly for a quasilocal region of spacetime. 
A side benefit of this shift
of emphasis was that the variational calculations could
easily be adapted to calculate rates of change of the
quasilocal quantities and so give a slightly different
outlook on conserved quantities than that discussed in 
the original Brown-York work. 

Third, I examined the reference terms that set the zero of 
the quasilocal energy. In the process of reviewing some of the 
extant proposals I showed that all had problems dealing with 
moving observers in Minkowski space -- namely such observers
measure non-zero energies in flat space. To deal with this 
problem I proposed a new definition that embeds
the two-surface of observers into a four-dimensional reference 
space. While this new reference term is by no means perfect
at least it ensures that the action and quasilocal energy 
of flat space is zero in all cases. Unfortunately, the inclusion
of this reference term (or any other reference term for that matter)
complicates the Lorentz-like transformation laws derived for the 
referenceless quasilocal energy. Specifically, the reference
terms must be transformed with respect to a different boost
velocity than the referenceless terms.

Fourth, with the reference term discussed above, 
I showed that it is possible 
to recast the (generalized) Brown-York QLE in an 
operational form. Roughly speaking I showed that the 
QLE contained by a closed two-surface $\Omega_t$ is
exactly equal to the total stress-energy of a 
particular thin shell of matter. That thin shell
must be embedded in the reference space such that: 
i) it has the same intrinsic geometry as $\Omega_t$, and 
ii) outside of that surface the spacetime
geometry is identical to that found outside of
$\Omega_t$ in the original spacetime. 

Finally, I added Maxwell and dilaton matter fields into the 
mix. These have previously been considered for orthogonal
foliations, but my work was the first to examine them 
in the non-orthogonal case. Their integration into the 
purely gravitational scheme of things proceeded smoothly
resulting in small, but usually non-qualitative, changes.
The only significant problem arose because
the quasilocal formalism as constituted doesn't allow
magnetically charged configurations of the Maxwell field.
They can be accommodated by switching to a dual formalism
but in doing so electric charges were excluded. Thus, the
formalism was shown to allow either 
electric or magnetic charges but not both. 

Having developed this formalism, I applied it to a variety of
situations, both classical and quantum, to investigate whether 
or not the QLE can reasonably be thought of as defining a
physical energy. To get some orientation, I started
by examining Schwarzschild and Reissner-Nordstr\"{o}m
spacetimes. For a spherical set of observers far from an RN 
source, I showed that the geometric QLE matches a Newtonian
intuition of how energy should be distributed in the spacetime
including the contributions from electromagnetic forces. 
I then showed that the total QLE included a contribution
to the energy from the base Coulomb potential for the spacetime
(by which I mean the Coulomb potential that would exist even if
there was no charged matter in the spacetime). However, the
Killing-vector-adapted QLE didn't behave in such an intuitive 
way. 

From there I considered the quasilocal energy measured by 
boosted observers. I started with the easy case of radial
boosts and showed that while the referenceless QLE increases
in the expected Lorentzian way with the boost, the numerical
value of the referenced QLE actually decreases! In particular
I showed that for observers far from the hole, the 
boosted and referenced QLE equals $M/\gamma$ where $\gamma$
is the usual Lorentz factor. Later on 
in the naked black hole section, I showed that this is 
due to the competing relativistic effects of the motion of
the observers and the gravitational field. I then examined
the trickier case of z-boosted observers. The results were much
more complicated, but far from the hole I found that 
the boosted and referenced Hamiltonian again measures $M/\gamma$.

Next, I examined the quasilocal energies measured by spherical
observers who are either hovering around or falling into a 
class of naked black holes. Such observers respectively
feel either negligible or Planck scale transverse tidal forces. 
In contrast,
I demonstrated that the static observers measure a large
geometric QLE while the infalling set measure a very small geometric QLE. This can be explained because the extremely strong tidal force
corresponds to a massive Lorentz boost of the reference terms which in 
turn means that the relativistic effects of the motion completely 
overwhelm those of the gravity. Thus, even though the unreferenced QLE
and reference terms are both hugely boosted, at the same time they
converge towards the same value 
so the difference between them goes to zero.

As a final classical example, I applied the formalism to investigate
energy flows that arise during gravitational tidal heating. I 
successfully used it to reproduce the standard Newtonian
and pseudo-tensor result and explain their gauge
ambiguities in terms of fluctuations of the quasilocal two-surface. 
Thus I demonstrated the utility of the formalism in an astrophysical
context which also helps to boost its claims to physical relevance.

The thesis finishes up in the last, rather long chapter,
by applying the quasilocal formalism to study pair 
production of rotating black holes in deSitter space.
It was seen that the results for non-rotating black 
holes can be qualitatively extended to the rotating case.
That is, created spacetimes can be classified as 
lukewarm (regular KNdS solutions where the horizons are
in thermal equilibrium), cold (extreme KNdS solutions), 
and Nariai (a limiting case where the outer 
black hole horizon approaches the cosmological
horizon). The entropy of such spacetimes
continues to be proportional to the surface areas
of the non-degenerate horizons and the pair creation
creation rates continue to be proportional to the negative
exponential of those entropies and suppressed relative to 
the creation of a pure deSitter space. 

To obtain these results I was forced to make a choice between 
the real instantons that are usually employed to evaluate the path 
integrals and the standard 
Lorentzian solution/``Euclidean'' instanton 
matching conditions. Since the 
matching conditions are the only way that I know of to 
associate an instanton with a given Lorentzian solution,
I opted for the matching conditions and allowed complex 
instantons.

Using the quasilocal formalism to fix the ensemble of paths
considered in the path integral, I showed that the standard
Einstein-Hilbert action is not the appropriate action to 
use for rotating pair creation. In particular it does not
fix the angular momentum of the ensemble and therefore
does not guarantee the creation of a black hole pair with 
a prespecified angular momentum. A careful application of the
quasilocal formalism allowed me to identify the correct action
and so obtain physically reasonable results. 

\subsubsection{Possible future work}

I see two main directions in which to continue work started
in this thesis. First, I have concentrated almost entirely
on the Brown-York definition of QLE. However, there are many other 
Hamiltonian based QLE's and it would be of interest to examine
them closely in the same way that I have dealt with the Brown-York
QLE here. For example, it would be
interesting to examine how the various definitions measure the 
energy flow in the tidal heating example. More generally one could
compare them in the limit of weak gravitational fields where
they could also be compared with the perturbative treatment of 
gravity as a spin-2 field propagating in a flat background. There
is a gauge ambiguity in such a treatment and it seems possible
that each measure of QLE might correspond to a different gauge
choice. Even more generally, a close examination and comparison of their
mathematical formalisms might help to shed light on each.

A second project can be found in the pair creation calculation. There
I noted that there was a difference between thermal and thermodynamic
equilibrium in two horizon spacetimes. 
The first is defined by the temperatures of the horizons, 
which in turn are most easily calculated using Euclidean quantum 
gravity techniques. Thus such a spacetime is in thermal equilibrium if 
a regular instanton can be constructed from it. At the same time
such spacetimes do not have to be in full thermodynamic equilibrium 
since angular momentum and/or electromagnetic charge could still be
exchanged between the horizons. This dichotomy deserves a fuller 
investigation. To this end it would be profitable to investigate
the evolution of black holes in deSitter space using the 
techniques of quantum field theory in curved spacetime to calculate
rates of particle emission, and the mass, charge, and angular momentum
carried off by those particles. Much work has been done in this area for
asymptotically flat spacetimes, but cosmological spacetimes are not
as well studied. In particular no one has studied them when 
they are in thermal equilibrium.
Apart from understanding the difference between
the notions of equilibrium, this issue is quite topical with the
recent interest in non-zero cosmological constant
spacetimes, that has arisen from the astronomical measurements
which indicate that our universe may have a positive cosmological constant and the string theory inspired AdS/CFT correspondence.

\appendix

\chapter{Hamiltonian Calculations}
\label{HamCalc}

This appendix presents the calculations behind the results of chapters
\ref{gravChapter} and \ref{matterChapter}.

\section{Foliating the gravitational action}
\label{appGravHam}
First I decompose the action (\ref{action})
\bea
I - \bI  &=& \frac{1}{2 \kappa} \int_M d^4 x \sqrt{-g} (\cR - 2 \Lambda) 
+ \frac{1}{\kappa} \int_\Sigma d^3 x 
\sqrt{h} K - \frac{1}{\kappa} \int_B d^3 x \sqrt{- \gamma} \Theta 
\nn \\ 
&&+ \frac{1}{\kappa} 
\int_\Omega d^2 x \sqrt{\sigma} \sinh^{-1} (\eta), \nn 
\eea
as discussed in chapter \ref{gravChapter}
into three-fields and time derivatives of those fields 
defined on the foliation hypersurfaces $\Sigma_t$ and
$\Omega_t$. 
To start, with the help of the Gauss-Codacci relations one 
can rewrite
\bea
(\cR -2 \Lambda) = 
R -2 \Lambda - K^2 + K_{\alpha \beta} K^{\alpha \beta} 
- 2 \nabla_\az (K u^\az + a^\az),  
\eea
where $a_\az \equiv u^\bz \nabla_\bz u_\az = \frac{1}{N} D_\az N$ 
is the acceleration of the foliation's unit normal vector field 
along its length. Then using Stokes's theorem
to move the total derivative 
out to the boundary, it is trivial to show that
\bea
\int_M d^4 x \sqrt{-g} (\cR - 2 \Lambda) &=& 
\int_M d^4 x \sqrt{-g} \left( R -2 \Lambda - K^2 
+ K_{\alpha \beta} K^{\alpha \beta} \right) \\
&& - 2 \int_\Sigma d^3 x \sqrt{h} K - 2 \int_B d^3 
x \sqrt{-\gamma} \left( K \eta + \bn_\az a^\az \right) \nn.
\eea

Next, referring back to the expressions for $n^\az$ and $\bu^\az$ 
given in equations (\ref{nou}) it is a simple matter to show that
\be
\Theta = \bk - \bn_\bz a^\bz =  
\frac{1}{\lambda} k + \lambda \bu^\az \nabla_\az \eta
- \eta K - \bn_\bz a^\bz.
\ee

Then, these two results can be combined to rewrite the Lagrangian
as 
\bea
I-\bI &=& \frac{1}{2 \kappa} \int_M d^4 x \sqrt{-g} 
\left( R - 2\Lambda - K^2 + K_{\az \bz} K^{\az \bz} \right) \\ 
&& - \frac{1}{\kappa} \int_B d^3 x \sqrt{-\gamma} \left( \frac{k}{\lambda} + 
\lambda \bu^\az \nabla_\az \eta \right) 
+ \frac{1}{\kappa} \int_\Omega d^2 x \sqrt{\sigma} \sinh^{-1} (\eta) \nn.
\eea

The next step in the process combines (the matter-free versions of) 
the Einstein constraint equations (\ref{C1}) and (\ref{C2})
with the extrinsic curvature of $\Sigma_t$ in $\cM$ written as
$K_{\az \bz} = - \frac{1}{2} {\pounds_{u}} h_{\az \bz} =
- \frac{1}{2 N} \left( {\pounds_T} h_{\az \bz} - 2 D_{(\az} V_{\bz)}
\right)$, to rewrite the integrand of the 
remaining bulk term of the Lagrangian as
\bea
\label{inter1}
&& R - 2 \Lambda - K^2 + K_{\az \bz} K^{\az \bz} \\
&=& 
\frac{2 \kappa}{\sqrt{-g}} P^{\az \bz} 
{\pounds_T h_{\az \bz}}  - \frac{2\kappa}{\sqrt{h}} \cH - 
\frac{2 \kappa}{\sqrt{-g}} 
V^\az \cH_\az - \frac{4 \kappa}{N} 
D_\az \left[ \frac{1}{\sqrt{h}} P^{\az \bz} V_\bz 
\right], \nn
\eea
where $P^{\az \bz} \equiv \frac{\sqrt{h}}{2 \kappa} 
\left(K h^{\az \bz} - K^{\az \bz} \right)$. Recalling 
that $\sqrt{-g} = N \sqrt{h}$ (eq.\ (\ref{dets}))
and once again using Stokes's theorem, this time on the 
$\Sigma_t$ hypersurfaces to move the total 
divergence term out to the boundary surfaces $\Omega_t$,
I can rewrite the action as
\bea
I - \bI &=& \int_M d^4 x \left( P^{\az \bz} \pounds_T h_{\az \bz} 
- N \cH - V^\az H_\az \right)\\ 
  && -\frac{1}{\kappa} \int dt \int_{\Omega_t} d^2 x \sqrt{\sigma} 
\left( Nk - V^\az [K_{\az \bz} - K h_{\az \bz} ] n^\bz - 
\bN \lambda \bu^\az \nabla_\az \eta \right)  \nn \\
  && + \frac{1}{\kappa} \int_{\Omega} d^2 x \sqrt{\sigma} \sinh^{-1} (\eta) \nn.
\eea

Up to this point I have been working with the foliation of $M$ and
therefore with the lapse $N$, shift $V^\az$, and normal vectors $u^\az$
and $n^\az$. On the term evaluated on $B$, I now switch to work with the
foliation of $B$ and therefore the boundary lapse $\bN$, the boundary shift $\bV^\az$, and normal vectors $\bu^\az$ and $\bn^\az$. Then,
\be
N k = \frac{1}{\lambda^2} \bN \bk - \eta N \sigma^{\az \bz} 
\nabla_\az \bu_\bz,
\ee
and 
\be
-V^\az \left( K_{\az \bz} - K h_{\az \bz} \right) n^\bz = N \eta
 \sigma^{\az \bz} \nabla_\az \bu_\bz - \bN \eta^2 \bk + \bn^\az \bV^\bz 
\nabla_\bz \bu_\az + \lambda \bV^\bz \nabla_\bz \eta.
\ee
Writing the timelike vector $T^\az$ in terms of the 
boundary quantities (eq.\ (\ref{boundaryTime})) it is
easy to see that
\bea
&& \int_\Omega d^2 x \sqrt{\sigma} \sinh^{-1}(\eta) - \int dt 
\int_{\Omega_t} d^2 x \sqrt{\sigma} \lambda \bN \bu^\az \nabla_\az \eta
\\
&=& \int dt \int_{\Omega_t} d^2 x \left( ({\pounds_T} \sqrt{\sigma}) 
\sinh^{-1}(\eta) + \sqrt{\sigma} \lambda \bV^\az \nabla_\az \eta \right) \nn.
\eea
Thus, the action takes its final form given in 
eq.\ (\ref{MgravActdecomp}). That is
\bea
\label{gravActdecomp}
I - \bI &=& \int_M d^4 x \left
( P^{\az \bz} {\pounds_T} h_{\az \bz} 
- N \cH - V^\az H_\az \right) \\
  &&  + \int dt \int_{\Omega_t} d^2 x 
P_\ssg ( {\pounds_T} \sqrt{\sigma})
  - \int dt \int_{\Omega_t} d^2 x \sqrt{\sigma} 
\left( \bN \bep - \bV^\az \bj_\az \right) 
\nn, 
\eea  
where $
\bep \equiv \frac{1}{\kappa} \bk = - \frac{1}{\kappa} 
\sigma^{\az \bz} \nabla_\az \bn_\bz$ and 
$\bj^\bz \equiv
\frac{1}{\kappa} \sigma_\az^\bz \bu^\cz \nabla_\bz
\bn_\cz$, and $P_{\ssg} \equiv \frac{1}{\kappa} \sinh^{-1} \eta$.  

\section{Gravitational Hamiltonian variation}
\label{appGravVar}
Next, I calculate the variation of the Hamiltonian with respect
to the Lagrange multipliers $N$ and $V^a$, surface metric $h_{ab}$, and
its conjugate momentum $P^{ab}$. Because $\ssg$ and $P_\ssg$ are 
functions of these other quantities they are automatically varied
as well. From eq.\ (\ref{gravHam}), the Hamiltonian is
\be
H_t = \int_{\Sigma_t} d^3 x \left( N \cH + V^a H_a \right)
+ \int_{\Omega_t} d^2 x \sqrt{\sigma} \left( \bN \bep - \bV^a \bj_a
\right) \nn  
\ee
where
\bea
\cH &=& - \frac{\sqrt{h}}{2 \kappa} \left( R - 2 \Lambda \right)
      + \frac{2 \kappa}{\sqrt{h}} \left(P^{ab} P_{ab} 
      - \frac{1}{2} P^2 \right), \nn \\
\cH_a &=& - 2 D_b P^b_{ \ a}, \nn \\
\bN &=&  \lambda N, \nn \\
\bV^a &=& V^a - (V^b n_b) n^a \nn, \\
\bep &=& \frac{1}{\kappa \lambda} k 
+ \frac{2}{\sqrt{h}} \eta P^{ab} n_a n_b, \ \ \ \mbox{ and}  \nn \\
\bj_a &=& - \frac{2}{\sqrt{h}} \sigma_{ab} P^{bc} n_c -
\frac{1}{\kappa} \sigma_a^b \partial_b (\sinh^{-1} \eta). \nn
\eea
The calculation is quite lengthy and so is tackled in parts. 

\subsubsection{Variation of the bulk term}
I start with the bulk term
\be
H_{blk} \equiv \int_{\Sigma_t} d^3 x \sqrt{h} (N \cH + V^a \cH_a),
\ee
and calculate its variation with respect to each quantity.

First, the variation with respect to the hypersurface momentum
$P^{ab}$ is easily calculated as
\bea
\label{blk1}
\delta_{P^{ab}} H_{blk} &=& \int_{\Sigma_t} d^3 x \left(
\frac{4 \kappa N}{\sqrt{h}} \left[ P_{ab} - \frac{1}{2} P h_{ab} \right]
+ 2 D_{(a} V_{b)} \right) \delta P^{ab} \\
&& - 2 \int_{\Omega_t} d^2 x \frac{\sqrt{\sigma}}{\sqrt{h}}
n_a V_b \delta P^{ab}. \nn
\eea
The boundary term arises from using Stokes's theorem to remove a total 
divergence to the boundary.

Next and more challenging is the variation with respect to the metric 
$h_{ab}$. To this end note that
\bea
\delta_{h_{ab}} \left\{ \sqrt{h} R \right\} 
&=& \sqrt{h} \left(R_{ab} - \frac{1}{2} R h_{ab} \right) \delta h^{ab}
\\
&& + \sqrt{h} D_a \left(h^{ad} h^{bc} \left[ D_c \delta h_{bd} - D_d 
\delta h_{bc} \right] \right). \nn 
\eea
This may be calculated from first principles, but the easiest way
to do it is to simply adapt the variation of the four dimensional 
Ricci scalar with respect to the four metric $g_{\az \bz}$. Such a
calculation may be found in any text book that deals with 
Lagrangian formulations of general relativity (for example Wald 
\cite{wald}). Then, recalling from the previous section that
the acceleration vector can be written in terms of the lapse 
as $a_b = \frac{1}{N} D_b N$, a not too lengthy computation obtains
\bea
\label{blk2}
&& \delta_{h_{ab}} \left\{ \int_{\Sigma_t} d^3 x \left( 
- \frac{N \sqrt{h}}{2 \kappa} [R-2\Lambda] 
\right) \right\} \\
&=&  \int_{\Sigma_t} d^3 x
\frac{\sqrt{h}}{2 \kappa} \left( N \left[ ^{(3)} G^{ab} + 
\Lambda h^{ab} \right] + h^{ab} D_c D^c N - D^a D^b N \right) 
\delta h_{ab} \nn \\
&& + \int_{\Omega_t} d^2 x \frac{N \sqrt{\sigma}}{2 \kappa} 
\left( - n^a h^{bd} D_d \delta h_{ab} + n^d h^{ab} D_d \delta h_{ab}
+a^b n^a \delta h_{ab} - [a^d n_d] h^{ab} \delta h_{ab} \right), \nn
\eea
where $^{(3)} G^{ab} = R^{ab} - \frac{1}{2} R h^{ab}$.
The $h_{ab}$ variation of the rest of the $N \cH$ is quickly found to be
\bea
&&\delta_{h_{ab}} \left\{ \int_{\Sigma_t} d^3 x 
\left( \frac{2 \kappa}{\sqrt{h}} \left[P^{ab} P_{ab} -
\frac{1}{2} P^2 \right] \right) \right\} \\
&=& \int_{\Omega_t} d^2 x 
\frac{\kappa}{\sqrt{h}} \left( - \left[ 
P^{cd} P_{cd} - \frac{1}{2} P^2 \right] h^{ab} 
+ 4 \left[ P^{ac} P_c^{\ b} - \frac{1}{2} P P^{ab} \right] 
\right) \delta h_{ab}. \nn 
\eea
Slightly more difficult is the variation of the $V^a \cH_a$ term. 
For that I use the relation 
\be
\delta_{h_{ab}} \{ \Gamma^c_{ab} \}= \frac{1}{2}
h^{cd} ( D_a \delta h_{bd} + D_b \delta h_{da} - D_d \delta h_{ab} )
\ee
where $\Gamma^c_{ab}$ is the Levi-Cevita connection for $h_{ab}$.
Then, keeping in mind that $P^{bc}$ is a tensor density (a relative 
tensor of weight one) and so 
$D_c P^{bc} = \partial_c P^{bc} + \Gamma^b_{cd} P^{cd}$, it can 
be shown that
\bea 
\label{blk3}
&& \delta_{h_{ab}} \left\{ 
\int_{\Sigma_t} d^3 x \left( -2 V^a D_b P_a^{\ b} \right) \right\} \\
&=& 2  \int_{\Sigma_t} d^3 x \left( P^{c(b} D_c V^{a)} - \frac{1}{2}
D_c [ P^{ab} V^c ] \right) \delta h_{ab} \nn \\
&& - 2  \int_{\Sigma_t} d^3 x \frac{\sqrt{\sigma}}{\sqrt{h}} \left(
V^a P^{bc} n_c \delta h_{ab} - \frac{1}{2} [V^a n_a] P^{ab} \delta 
h_{ab} \right) \nn. 
\eea

Finally, it is trivial to calculate the variation of the bulk
term with respect to the lapse and shift. To wit,
\be
\delta_{N,V^a} H_{blk} = \int_{\Sigma_t} d^3 x \left( 
\cH \delta N + \cH_a \delta V^a \right).
\ee

Focusing back on the boundary terms of 
(\ref{blk1},\ref{blk2},\ref{blk3}) one can break
up the variation of $h_{ab}$ in terms of the variation of $\sigma_{ab}$ 
and $n_a$, where $\delta \sigma_{ab} \equiv \sigma_a^c \sigma_b^d \delta
h_{ab}$. Further, $n_a$ is defined as the unit covariant vector 
normal to the
surface $\Omega_t$ in $\Sigma_t$ and so it is defined up to 
a normalization constant without reference to the metric (crudely, 
if $\Omega_t$ is defined to be a surface of constant $r$, then
$n_a \| dr$). Therefore $\delta n_a = \alpha n_a$ and 
$\delta n^a = -\alpha n^a + \beta^a$
where $\alpha \equiv \frac{1}{2} n^a n^b \delta h_{ab}$ and
$\beta^a \equiv -\sigma^{ac}n^d \delta h_{cd}$.
Thus the total variation of $H_{blk}$ can be written as
\bea
\delta H_{blk} &=& \int_{\Sigma_t} d^3 x \left( 
\cH \delta N + \cH_a \delta V^a - [P^{ab}]_T \delta h_{ab} + 
[h_{ab}]_T \delta P^{ab} \right) \\
&& - 2 \int_{\Omega_t} d^2 x \frac{\sqrt{\sigma}}{\sqrt{h}}
\left(n_a \bV_b + V^c n_c n_a n_b \right) \delta P^{ab} \nn \\
&& + \int_{\Omega_t} d^2 x \frac{N \sqrt{\sigma}}{2 \kappa} 
\left( \sigma^{bc} D_b \beta_c - k^{ab} \delta \sigma_{ab}
+ 2 \alpha k \right) \nn \\
&& + \int_{\Omega_t} d^2 x \frac{N \sqrt{\sigma}}{2 \kappa} 
\left(\sigma^{ab} n^c D_c \delta \sigma_{ab} - a^b
\beta_b - [a^c n_c] \sigma^{ab} \delta \sigma_{ab} \nn \right)\\
&& + \int_{\Omega_t} d^2 x \frac{\sqrt{\sigma}}{\sqrt{h}}
\left( - 2 \bV^a P^{bc} n_c \delta \sigma_{ab} - 3 \alpha V^c n_c
P^{ab} n_a n_b + V^c n_c P^{de} \sigma_d^a \sigma_e^b \delta 
\sigma_{ab} \right), \nn
\eea
where
\bea
\left[ P^{ab} \right]_T & \equiv & 
- \frac{\sqrt{h}}{2 \kappa} \left( N ^{(3)}G^{ab} 
- \left[ D^a D^b N - h^{ab} D_c D^c N \right] \right) 
+ \pounds_V P^{ab}\\
&& + \frac{N \kappa}{\sqrt{h}} \left(
[ P^{cd} P_{cd} - \frac{1}{2} P^2 ] h^{ab} 
- 4 [ P^{c(a}P_c^{\ b)} - \frac{1}{2} P P^{ab} ] \right), \nn
\eea
and 
\be
[h_{ab}]_T = \frac{4 \kappa N}{\sqrt{h}}[P_{ab} - \frac{1}{2} 
P h_{ab}]
+ 2 D_{(a}V_{b)}.
\ee 
As the notation suggests (and is discussed in section \ref{HamVar})
the above equations define time derivatives.

Though this expression is quite a mess, it will be substantially
improved once the variation of the boundary terms of 
$H_t$ is added on. Thus, I now calculate that variation.

\subsubsection{Variation of the boundary term}

It is simplest to calculate the total variation of 
\be
H_{bnd} = \int_{\Omega_t} d^2 x \sqrt{\sigma} \left( \bN \bep
- \bV^a \bj_a \right),
\ee
with
respect to all of the variables simultaneously. Then,
\be
\delta H_{bnd} = \int_{\Omega_t} d^2 x
\left( [\bN \bep - \bV^a \bj_a] \delta \sqrt{\sigma}
+ \sqrt{\sigma} [\bep \delta \bN - \bj_a \delta \bV^a 
+ \bN \delta \bep - \bV^a \delta \bj_a]
\right).
\ee
Tackling the individual terms one at a time, first note that 
\be
\delta \sqrt{\sigma} = \frac{1}{2} \sqrt{\sigma} \sigma^{ab} \delta
\sigma_{ab},
\ee
just as $\delta \sqrt{h} = \frac{1}{2} \sqrt{h} h^{ab} \delta h_{ab}$.
The $\delta \bN$  and $\delta \bV^a$ 
terms are left as they are while the
$\delta \bep$ and $\delta \bj_a$ terms become,
\bea
\bN \delta \bep &=& \frac{2 \bN}{\sqrt{h}} P^{ab} n_a n_b \delta [ 
\sinh^{-1} \eta ] - \alpha N \varepsilon 
+ \frac{1}{\kappa} N \sigma^{ab} D_a \beta_b \\
&& - \frac{1}{\kappa} N 
\sigma^{ab} n^c D_c \delta \sigma_{ab}
- \frac{1}{\sqrt{h}} V^c n_c P^{de}n_d n_e \sigma^{ab} 
\delta \sigma_{ab} \nn \\
&& + 3 \alpha V^c n_c P^{ab} n_a n_b 
+ \frac{2}{\sqrt{h}} V^c n_c n_a n_b \delta P^{ab}, \nn
\eea
and 
\bea
- \bV^a \delta \bj_a &=& - \frac{1}{\sqrt{h}} P^{cd} n_c \bV_d 
\sigma^{ab} \delta \sigma_{ab} + \frac{2}{\sqrt{h}} \bV^a P^{bc} n_c
\delta \sigma_{ab} \\
& &+ \frac{2}{\sqrt{h}} n_a \bV_b \delta P^{ab} 
+ \frac{1}{\kappa} \bV^a \partial_a ( \delta [\sinh^{-1} \eta]).
\nn 
\eea 

Again the result is a bit of a mess. 
Luckily, however the unpleasant terms cancel
each other out once this is combined with $H_{blk}$. 

\subsubsection{The complete Hamiltonian}

Putting the two variations together there is significant simplification.
Apart from cancellations, the only other computational trick 
required for the recombination is to keep in mind that $\Omega_t$ is a
closed surface, so 
\be
\int_{\Omega_t} d^2 x \sqrt{\sigma} d_a \eta^a = 0 ,
\ee
for any smooth vector field $\eta^a \in T \Omega_t$. Then,
the total variation of $H_t$ is
\bea
\delta H_t &=& \int_{\Sigma_t} d^3 x \left(
 \cH \delta N + \cH_a \delta V^a - [P^{ab}]_T \delta h_{ab} + 
[h_{ab}]_T \delta P^{ab} \right) \\
&&+ \int_{\Omega_t} d^2 x \sqrt{\sigma} \left(\bep \delta \bN - 
\bj_a \delta \bV^a - (\bN/2) \bs^{ab} \delta \sigma_{ab} \right) \nn \\
&& + \int_{\Omega_t} d^2 x \sqrt{\sigma} \left( 
[ \sqrt{\sigma} ]_T 
\delta P_{\ssg} -  \left[ P_{\ssg} \right]_T \delta 
\sqrt{\sigma} \right), \nn
\eea 
where
\bea
\bar{s}^{ab} &\equiv& 
\frac{1}{\kappa \lambda} \left( k^{ab} - [k-n^d a_d]
\sigma^{ab} \right) - \frac{2}{\sqrt{h}} \eta \sigma_c^a \sigma_d^b 
P^{cd} \\
&& + \frac{1}{N} \left( 
\left[ P_\ssg \right]_T
- \frac{1}{\kappa} \pounds_{\bar{V}} \eta \right) \sigma^{ab}, \nn \\
\left[ \sqrt{\sigma} \right]_T 
&\equiv& 
- \sqrt{\sigma} \left(
N \frac{2}{\lambda \sqrt{h}} 
P^{ab} n_a n_b + N \frac{\eta}{\kappa} k - 
\frac{1}{\kappa} d_b \bV^b \right),
\eea
and
$[P_{\ssg}]_T $ is an undetermined function over $\Omega_t$.

\section{Foliating the matter action}
\label{appMatHam}
This section decomposes the matter action
\bea
I^m - \bI &=& 
\frac{1}{2 \kappa} \int_M d^4 x \sqrt{-g} (\cR - 2 \Lambda 
- 2 (\nabla_\az \phi)(\nabla^\az \phi) - e^{-2a\phi} F_{\az \bz}F^{\az \bz}) \nn \\ 
&& + \frac{1}{\kappa} \int_\Sigma d^3 x 
\sqrt{h} K - \frac{1}{\kappa} \int_B d^3 x \sqrt{- \gamma} \Theta 
+ \frac{1}{\kappa} 
\int_\Omega d^2 x \sqrt{\sigma} \sinh^{-1} (\eta), \nn 
\eea
from chapter \ref{matteraction}
into three-fields and their
time derivatives as defined on the hypersurfaces of 
the foliations $\Sigma_t$ and $\Omega_t$.

First, after breaking up the purely gravitational terms as 
before, the bulk term integrand is
\be
\label{bulkterm}
P^{\az \bz} {\pounds_T} h_{\az \bz} - N \cH - V^\az H_\az - 
\frac{\sqrt{-g}}{\kappa} \nabla_\az \phi \nabla^\az \phi - 
\frac{\sqrt{-g}}{2 \kappa} e^{-2 a \phi}
F_{\az \bz}F^{\az \bz}.
\ee
Then bringing in the Einstein constraint equations (\ref{C1})
and (\ref{C2}) this may be rewritten as
\bea
&& P^{\az \bz} {\pounds_T} h_{\az \bz} - N \cH^m - V^\az H^m_\az 
+ ( \frac{N \kappa}{2 \sqrt{h}} \wp^2 + \wp V^\az D_\az \phi ) \\
&& + \frac{\kappa}{2 \sqrt{h}} \left( e^{2a\phi} N \cE_\az \cE^\az 
- u^\az V^\bz \epsilon_{\az \bz \cz \dz} \cE^\cz \cB^\dz \right). \nn
\eea
Next, from eq.\ (\ref{Et}) it is not hard to rewrite $\cE_\az$ 
as 
\be
\cE_\az = \frac{e^{-2a\phi}}{N} \left( \frac{\sqrt{h}}{2\kappa}
D_\az [N \Phi - V^\bz \tilde{A}_\bz] 
+ \frac{\sqrt{h}}{2\kappa} h_\az^\bz \pounds_T \tilde{A}_\bz 
+ u^\dz V^\bz \epsilon_{\dz \az \bz \cz} 
B^\cz \right). \label{Erewrite}
\ee  
Using this relation and the trivial $\pounds_T \phi = N \wp 
+ V^\az D_\az \phi$, the bulk integrand (\ref{bulkterm}) may be 
written entirely with respect to fields on the hypersurface, time
derivatives of those fields, constraints, and a total derivative.
It becomes
\bea
&& P^{\az \bz} \pounds_T h_{\az \bz} + \wp \pounds_T \phi + 
\cE^\az \pounds_T \tilde{A} - N \cH^m - V^\az \cH^m_\az \\
&&- T^\az A_\az \cQ + D_\bz (\cE^\bz T^\az  A_\az). \nn
\eea 
$\cQ \equiv - D_\az \cE^{\az}$ is the constraint
equation (\ref{DE}) for the electric field with no sources.
Thus the action can be written as shown in equation (\ref{Mmactdecom}).
That is
\bea
\label{mactdecom}
I^m &=& \int dt \int_{\Sigma_t} d^3 x \left\{ P^{\az \bz} 
\pounds_T h_{\az \bz} + \wp \pounds_T \phi + \cE^\az 
\pounds_T \tilde{A}_\az \right\} \\
&& + \int dt \int_{\Omega_t} d^2 x \left\{ P_{\sqrt{\sigma}}
(\pounds_T \sqrt{\sigma}) \right\} \nn \\ 
&& - \int dt \int_{\Sigma_t} d^3 x \left\{ N \cH^m + V^\az H^m_\az 
+ T^\az A_\az \cQ \right\} \nn \\
&& - \int dt \int_{\Omega_t} d^2 x \sqrt{\sigma} 
\left\{ \bN (\bep + \bep^m)  - \bV^\az (\bj_\az + \bj_\az^m) \right\}, 
\nn 
\eea
where 
\bea
\bep^m &\equiv& - \frac{1}{\sqrt{h}} (n_\bz \cE^\bz) 
(\frac{1}{\lambda} \Phi - \eta \tilde{A}_\az n^\az) 
= - \frac{1}{\sqrt{h}} (\bn_\bz \bar{\cE}^\bz) \bar{\Phi} 
 \mbox{ and}\\
\bj^m_\az &\equiv& - \frac{1}{\sqrt{h}} (n_\bz \cE^\bz) 
\hat{A}_\az = - \frac{1}{\sqrt{h}} (\bn_\bz \bar{\cE}^\bz) 
\hat{A}_\az,
\eea
and
$\bar{\Phi} = - A_\az \bu^\az$, $\bar{\cE^\az} = -2/\sqrt{h}
F^{\az \bz} \bu_\bz$, and $\hat{A}_\az = \sigma_\az^\bz A_\bz$.
Here, I have once again used Stokes's theorem and therefore the 
assumption that there exists a single $A_\az$ defined over all
of $M$. 

\section{Matter Hamiltonian variation}
\label{appMatVar}
Next I calculate the first variation of the matter Hamiltonian 
with respect to 
$h_{ab}$, $\ssg$, $\tilde{A}_a$, and $\phi$, their
conjugate momenta $P^{ab}$, $P_{\ssg}$, $\cE^a$, and $\wp$, and 
the Lagrange multipliers $N$, $V^a$, and $\Phi$.
From eq.\ (\ref{HamFull}) the Hamiltonian is
\bea
H^m_t &=& 
\int_{\Sigma_t} d^3 x [N (\cH^m - \Phi \cQ) 
+ V^a (\cH^m_a + \tilde{A}_\az \cQ) ] \\
&& + \int_{\Omega_t} d^2 x \sqrt{\sigma} \left[ \bN (\bep + \bep^m) 
- \bV^\az (\bj_\az + \bj^m_\az) \right]. \nn
\eea
where $\cH^m=0$, $\cH^m_a=0$, and $\cQ=0$ 
are the constraint equations 
(\ref{C1}),(\ref{C2}), and (\ref{DE}) respectively. 

Now, the variations of the purely geometric terms were calculated in
section \ref{appGravVar}, so only those of the matter terms need to be
considered separately here. This time, it is easiest to calculate the
first variations of the full expression with respect to each quantity 
separately. First for the dilaton, it is trivial to show that
\bea
\delta_{\wp} H_t^m = \int_{\Sigma_t} d^3 x \left( \left[ \phi \right]_T
\delta \wp \right),
\eea
where
\bea
\left[ \phi \right]_T \equiv \frac{N \kappa}{2 \sqrt{h}} \wp + \pounds_V \phi.
\eea
It is only a little more difficult to calculate the variation with 
respect to dilaton $\phi$ as
\bea
\delta_{\phi} H_t^m = - \int_{\Sigma_t} d^3 x \left(
\left[ \wp \right]_T \delta \phi \right)
+ \int_{\Omega_t} d^2 x \ssg \frac{2 \bN}{\kappa} 
\left[ \phi \right]_\bn \delta \phi,
\eea
where
\bea
\left[ \wp \right]_T &\equiv& \frac{2 \sqrt{h}}{\kappa} D^b (N D_b \phi)
+ D_a ( \wp V^a ) \\
& & + a \frac{N \kappa}{2 \sqrt{h}} (e^{-2a\phi} \cB^b \cB_b 
- e^{2a\phi} \cE^b \cE_b) \nn \mbox{ and}\\
\frac{2 \bN}{\kappa} \left[ \phi \right]_\bn & \equiv &  
\left( 
\frac{2 N}{\kappa} \pounds_{n} \phi 
- \frac{V^a n_a}{\sqrt{h}}\wp \right). 
\eea
Changing to the four-dimensional perspective it is easy to show
that $\left[ \phi \right]_{\bn} = \pounds_\bn \phi$. 

Variations with respect to the EM terms are a little more difficult but
still not too bad. A few lines of calculation are 
required to show that
\bea
\delta_{\cE} H_t^m = \int_{\Sigma_t} d^3 x 
\left( [\tilde{A}_b]_T \delta \cE^b \right),
\eea
where
\bea
[\tilde{A}_b]_T \equiv \frac{N \kappa}{2 \sqrt{h}} e^{2a\phi} \cE_b
+ \pounds_V \tilde{A}_b - D_b[N \phi],
\eea
and a few more give the variation with respect to $\Phi$ and
$\tilde{A}_a$ as 
\bea
\delta_{\Phi, \tilde{A}} H_t^m &=&
\int_{\Sigma_t} d^3 x \left( - \left[ \cE^b \right]_T \delta \cE^b 
- N \cQ \delta \Phi \right) \label{Pvar} \\
&& + \int_{\Omega_t} d^2 x \frac{\bN \ssg}{\sqrt{h}} 
(\cE^c n_c) \left( - \frac{1}{\lambda} \delta \Phi + \eta
n^c \delta \tilde{A}_c \right) \nn \\
&& + \int_{\Omega_t} d^2 x \frac{\bN \ssg}{\sqrt{h}} 
\left( - \frac{1}{\lambda} e^{-2a\phi} 
\epsilon^{bcd} n_b \cB_c \delta \tilde{A}_d + \eta
\sigma^b_c \cE^c \delta A_b  \right), \nn
\eea
where
\bea
\left[ \cE^b \right]_T \equiv - \epsilon^{bcd} D_c [N e^{2a\phi} \cB_d]
+ \pounds_V \cE^b.
\eea
Switching again to a four dimensional perspective, it is 
only a little more involved to show that
\be
\hat{\bar{\cB}}_b \equiv \sigma_b^c \bar{\cB}_c = 
\frac{1}{\lambda} \hat{\cB}_b + \eta e^{2a\phi} \epsilon_{bcd}
n^c \cE^d,
\ee
which is
a generalization of one of the standard Lorentz transform
laws of electrodynamics. Then the term in the brackets of the
third integral of (\ref{Pvar}) is equal to 
$\bar{\cB}_b n_c \epsilon^{bcd} \delta \tilde{A}_d$. 

Next consider the variation with respect to $P^{ab}$ and the 
lapse $N$, shift $V^\az$, and metric $h_{ab}$. The variation with 
respect to $P^{ab}$ is unchanged from the pure gravitational case
considered in section \ref{appGravVar}.
At the same time the variation with respect
to the lapse $N$, shift $V^a$, and metric $h_{ab}$ is fairly easily
shown to be
\bea
\delta_{g} H_t^m &=& \int_{\Sigma_t} d^3 x 
\left( [\cH^m - \Phi \cQ] \delta N + [\cH^m_a + \cQ \tilde{A}_a ] \delta V^a - [P^{ab}]_T^m \delta h_{ab} \right) \\
&& + \int_{\Omega_t} d^2 x \ssg \left(
[\bep + \bep^m] \delta \bN - [\bj_a + \bj^m_a] \delta \bV^a \right),
\nn \eea
where
\bea
\left[ P^{ab} \right]^m_T 
& \equiv & \left[ P^{ab} \right]_T + \frac{N \sqrt{h}}{\kappa}
\left( [D^a \phi][D^b \phi]  - \frac{1}{2} [D_c \phi] [D^c \phi] h^{ab}
\right) + \frac{N \kappa}{8 \sqrt{h}} \wp^2 h^{ab} \\
&& - \frac{N \kappa}{4 \sqrt{h}}
\left( [e^{2a\phi} \cE^a \cE^b + e^{-2 a \phi} \cB^a \cB^b] - 
\frac{1}{2} [e^{2a\phi} \cE^c \cE_c + e^{-2 a \phi} \cB^c \cB_c]
h^{ab} \right) \nn.
\eea

Thus, the total variation of $H^m_t$ is
\bea
\delta H_t^m &=& \int_{\Sigma_t} d^3 x \left(
 [\cH^m - \Phi \cQ] \delta N + [\cH^m_a + \tilde{A}_a \cQ] \delta V^a 
- N \cQ \delta \Phi \right) \\
&& + \int_{\Sigma_t} d^3 x \left( 
[h_{ab}]_T \delta P^{ab}
- [P^{ab}]^m_T \delta h_{ab} 
\right) \nn \\
&&+ \int_{\Sigma_t} d^3 x \left( 
[\phi]_T \delta \wp - [\wp]_T \delta \phi + [\tilde{A}_a]_T \delta
\cE^a - [\cE^a]_T \delta \tilde{A}_a \right) \nn \\
&&+ \int_{\Omega_t} d^2 x \sqrt{\sigma} \left( [\bep+\bep^m] 
\delta \bN - 
[\bj_a+\bj_a^m] 
\delta \bV^a 
- (\bN/2) \bs^{ab} \delta \sigma_{ab} \right) \nn \\
&& + \int_{\Omega_t} d^2 x \sqrt{\sigma} 
\left(\left[ \sqrt{\sigma} \right]_T 
\delta P_{\sqrt{\sigma}} - \left[P_{\sqrt{\sigma}} \right]_T \delta 
\sqrt{\sigma} \right), \nn \\
&& + \int_{\Omega_t} d^2 x \frac{\bN \ssg}{\sqrt{h}} \left(
[\cE^a n_a] 
\delta \bar{\Phi} + e^{-2a\phi} 
\hat{\bar{\cB}}_a n_b \hat{\epsilon}^{abc}
\delta \hat{A}_c \right) \nn \\
&& + \int_{\Omega_t} d^2 x \frac{2 \bN \ssg}{\kappa} 
\left[ \phi \right]_{\bn} \delta \phi \nn.
\eea

\chapter{Pair creation calculations}

\section{Reducing the generalized C-metric to KNdS}
\label{AppCtoK}

As noted in section \ref{CtoK}, the general Plebanski-Demianski 
metric \cite{PlebDem} contains conical singularities that correspond to 
cosmic strings or rods that supply the energy 
necessary to give black holes their extra acceleration 
above or below the rate of the rest of the universe. In this 
section I show that one way of eliminating the conical 
singularities corresponding to those strings/rods reduces the 
Plebanski-Demianski metric to the Kerr-Newmann-deSitter metric. 
This serves to emphasize that the global KNdS metric contains
at least two black holes (see section \ref{CtoK} 
for more on this point).

The generalized C-metric takes the form
\be 
\label{PlebDemMetric}  
ds^2  = \frac{1}{(p-q)^2} \left\{
\begin{array}{llll}
& \frac{1+p^2q^2}{P} dp^2 & + & \frac{P}{1+p^2q^2} \left( d \sigma - q^2 d \tau \right)^2 \\   
 - & \frac{1+p^2q^2}{Q} d q^2 & + & 
\frac{Q}{1+p^2q^2} \left(p^2 d \sigma + d \tau \right)^2  
\end{array}
\right\}, 
\ee 
with accompanying electromagnetic field defined by the vector potential 
\be 
A = - \frac{e_0 q (d \tau + p^2 d\sigma)}{1 + p^2 q^2} + \frac{g_0 p (d 
\sigma -  
q^2 d\tau)}{1 + p^2 q^2}, 
\ee 
where $p,q,\tau$, and $\sigma$ are coordinate functions,
\begin{equation}  
\label{Pp1}  
P(p) = (-\frac{\Lambda}{6} - g_0^2 + \gamma) + 2 n p - \epsilon   
p^2   
+ 2 m p^3 + (- \frac{\Lambda}{6} - e_0^2 - \gamma) p^4,  
\end{equation}  
and $Q(q) = P(q) + \frac{\Lambda}{3} (1 + q^4)$. $\Lambda$ is the   
cosmological constant,  $\gamma$ and $\epsilon$ are constants connected   
in a non-trivial way with rotation and acceleration, $e_0$ and $g_0$   
are linear multiples of electric and magnetic charge, and $m$ and   
$n$ are respectively the mass and NUT parameters (up to a linear 
factor).   
This solution can be analytically extended across the coordinate singularity at $p=q$, so that on the other side of $p=q$ there is
a mirror image of the initial solution. Thus, it can be seen
as describing a pair of black holes on opposite sides of that
$p=q$ hypersurface. 

To apply this metric to more specific physical situations, the   
coordinate functions are best converted to spherical-type 
spacetime  coordinates as $q \leftrightarrow \frac{1}{r}$, 
$p \leftrightarrow \pa + \alpha \cos \theta$ for some constants 
$\alpha$ and $\pa$, $\sigma \leftrightarrow \phi$ and 
$\tau \leftrightarrow t$. Now in general, a periodic identification of
$\sigma$ will introduce conical singularities at the roots of $P$. To
avoid such singularities, restrictions must be placed on the constants
defining $P$. Defining $\pa$, $\pb$, $\alpha$, and $\beta$ so that the
roots of $P(p)$ are at $\pa+\alpha$, $\pa-\alpha$, $\pb+i\beta$, and
$\pb-i\beta$, one may write $P$ as  
\be 
\label{Pp2} 
P(p) = - C ([p-\pa]^2 - \alpha^2)([p-\pb]^2 + \beta^2),  
\ee  
where $C = -\frac{\Lambda}{6} - e_0^2 - \gamma$. 
Specialize this by assuming that only $\pa - \alpha$ and $\pa +   
\alpha$ are real roots, $\pa - \alpha < \pa + \alpha$ and $\pb,\beta 
\in \R$, which means that
there are only two real roots. Restricting $p$ to lie
between these two roots, I reparameterize by setting $p = \pa + \alpha
\cos \theta$, where as usual $\theta \in [0, \pi]$. Then if $\pb=\pa$
(that is, $P(p)$ has an axis of symmetry along the line $p=\pa$),
potential conical singularities at $\pa-\alpha$ or $\pa+\alpha$ may be
simultaneously eliminated by identifying $\sigma$ with period $T =   
\frac{4 \pi}{P'(\pa - \alpha)}$ where $P' = \frac{d P}{d p}$.  
  
Next, I make the following extended series of coordinate transformations and definitions:  
\bea  
q &=& \frac{1}{\sqrt{\frac{\Lambda}{3}} \beta r}, \\   
\pa &=& \sqrt{\frac{\Lambda}{3}} \beta \tilde{\pa}, \\  
\pb &=& \sqrt{\frac{\Lambda}{3}} \beta \tilde{\pb}, \\ 
\alpha &=& \sqrt{\frac{\Lambda}{3}} \beta \tal, \\  
\chi^2 &=& 1 + \frac{\Lambda}{3} \tilde{\alpha}^2, \\  
\sigma &=& \frac{\phi}{\sqrt{\frac{\Lambda}{3}} C \beta^3   
\tilde{\alpha} \chi^2}, \\  
\tau &=& \frac{ t - \tilde{\alpha} \phi}{\sqrt{\frac{\Lambda}{3}} C   
\beta \chi^2}, \\   
\cH &=&  1 + \frac{\Lambda}{3} \tilde{\alpha}^2 \cos ^2 \theta,\\  
\cG &=& r^2 + (\tilde{\pa} + \tal \cos\theta)^2, \mbox{ and} \\  
\cQ(r)  &=& - \frac{\Lambda}{3 C} r^4 Q(q).  
\eea  
Equating (\ref{Pp1}) and (\ref{Pp2}) obtains the   
following three equalities relating the two forms of $P$:  
\bea  
m &=& 2C\pa \\  
n &=& C\pa(2\pa^2 - \alpha^2 + \beta^2) ,\mbox{ and} \\  
g_0^2 + e_0^2 &=& C(1 + [\pa^2-\alpha^2][\pa^2 + \beta^2]) -   
\frac{\Lambda}{3}.  
\eea  
Then, after a significant amount of algebra, these transformations and 
equations modify the metric (\ref{PlebDemMetric}) so that it becomes
\begin{equation}  
ds^2 = \mathcal{A} \left\{  
\begin{array}{l}  
\frac{\cG}{\cH} d \theta^2 + \frac{\cH \sin^2 \theta}{\cG \chi^4}   
\left( \tal d t 
+ [ r^2 + \tal^2] d \phi \right)^2  
+ \frac{\cG}{\cQ} dr^2 \\
- \frac{\cQ}{\cG \chi^4} \left( d t + \left[   
(\frac{\tilde{\pa}^2}{\tal} + 2 \tilde{\pa} \cos \theta ) -   
\tilde{\alpha} \sin^2 \theta \right] d \phi \right) ^2   
\end{array}  
\right\},
\ee  
where 
\be
{\mathcal A} =  \frac{\Lambda/(3C)}{(1 - (\Lambda/3) \beta^2 r 
[\tilde{\pa} + \tilde{\alpha} \cos \theta ])^2}.
\ee
Setting $e_0 = \sqrt{\frac{\Lambda}{3}} E_0 \beta^2$, $g_0 =  
\sqrt{\frac{\Lambda}{3}} G_0 \beta^2$, and $\tilde{\pa} = M \beta^2$,  
$\cQ$ becomes,  
\bea  
\cQ(r) &=& - \frac{\Lambda}{3} \left( \frac{1 - (E_0^2 + G_0^2) (M^2   
\beta^4 - \tal^2) ( 1 + \frac{\Lambda}{3} M^2 \beta^4) \beta^8}{1 -   
(E_0^2 + G_0^2) \beta^4} \right) r^4 \\ 
&&- 2 \frac{\Lambda}{3}M \left( 1  +   
\frac{\Lambda}{3} (2 M^2 \beta^4 - \tal^2) \right)  \beta^2 r^3   
 + (1 + \frac{\Lambda}{3} (6M^2 \beta^4 - \alpha^2) )r^2 \nn
\\ &&- 2Mr +   
\frac{E_0^2 + G_0^2 + (\tal^2 - M^2 \beta^4)(1 + \frac{\Lambda}{3}   
M^2 \beta^4)}{1 + (E_0^2 + G_0^2) \beta^4}.  \nn
\eea  
 
The $r^3$ term of the above is identified with the NUT parameter. 
To set this equal to zero while keeping the mass parameter 
$M$ non-zero, one of $\beta$ or $1 + 
\frac{\Lambda}{3} (M^2 \beta^4 - \tal^2)$ must be set to zero. 
Here I choose to take the limit as $\beta \rightarrow 0$ 
(choosing $1 + \frac{\Lambda}{3} (M^2 \beta^4 - \tal^2) = 0$ 
results in a metric that is similar to but not quite the KNdS metric --
most notably it retains a leading conformal factor).
Then, replacing 
$\tal$ with the more traditional symbol $a$ the metric becomes the 
standard KNdS metric, and similarly the vector 
potential $A$ becomes a vector potential that generates the 
associated electromagnetic field.  
Thus, the KNdS metric describes two black holes in deSitter space that 
are accelerating away from each other due to the cosmological 
expansion of the universe.  

Note that there are other ways to eliminate the conical singularities in 
(\ref{PlebDemMetric}).  Although most yield the KNDS metric, some
will give rise to other spacetimes. They will not be considered in
this thesis.

\section{Range of KNdS spacetimes}
\label{KNdSRange}

This section explores the allowed parameter range of KNdS spacetimes by
analyzing the root structure of the polynomial $\cQ$. 

If $\cQ$ has three positive real roots then they may be written in 
increasing order as $d-\delta$, $d+\delta$,  
$e-\varepsilon $, and $e+\varepsilon$, where $e$ and $d$ are reals and  
$\varepsilon$ and $\delta$ are non-negative reals. The absence of a   
cubic term in $\cQ$ forces $d=-e$. Two additional conditions
\be  
\label{rootorder}  
\begin{array}{ccccl}  
0 & \leq & \varepsilon & < & e,  \ \ \ \ \ \ \ \ \mbox{  and}\\  
e & < & \delta & \leq & 2e - \varepsilon \nn  
\end{array} 
\ee
ensure that the roots are ordered as proposed.  
Then without loss of generality
\be  
\label{Qpoly2}  
\cQ = -\frac{\Lambda}{3} \left( (r-e)^2 - \varepsilon^2 \right) \left(  
(r+e)^2 - \delta^2 \right).  
\ee  
   
Now, the requirement that $\cQ$ has three positive real roots enforces
restrictions on the allowed values of the physical parameters $a$, $M$,  
$E_0$, and $G_0$. $\cQ$ is a quartic and therefore can be    
solved algebraically, so in principle it is possible to directly discover
under what circumstances it has four real roots. In practice however,
the exact solution to a quartic is too complicated to work with. Thus, I
tackle the problem in reverse instead. First I determine the  
allowed ranges of the $\cQ$ structure parameters $e$, $\delta$, and  
$\varepsilon$, and then use these to parameterize the allowed range of
the physically meaningful parameters $a$, $M$, $E_0$, and $G_0$.  
  
Matching (\ref{Qpoly}) with (\ref{Qpoly2}) it is easy to obtain
expressions for the physical parameters in terms of the structure
parameters. Namely  
\bea   
\label{a2}   
a^2 & = & \frac{3}{\Lambda} - \delta^2 - \varepsilon^2 - 2e^2, \\   
\label{M}   
M & = & \frac{\Lambda}{3} (\delta^2 - \varepsilon^2) e, \ \ \ \ \ 
\mbox{ and} \\  
\label{Q2}   
E_0^2 + G_0^2 & = & \lt (\delta^2 - e^2)(e^2 - \e^2) + (\delta^2 +   
\e^2 +  
2e^2) - \frac{3}{\Lambda}.  
\eea   
  
Requiring that each of these parameters be non-negative imposes  
further restrictions (beyond the root ordering conditions  
(\ref{rootorder}))  
on the allowed ranges of $e$, $\varepsilon$, and $\delta$.   
If $a^2 \geq 0$ then   
\be   
\label{pc1} \frac{3}{\Lambda} - \delta^2 - \varepsilon^2 -  
2e^2 \geq 0.  
\ee   
$M$ will automatically be non-negative because of the root-ordering 
conditions while $E_0^2+G_0^2 \geq 0$ implies that  
\be   
\label{pc2}  
\frac{\Lambda}{3} (\delta^2 - e^2)(e^2 - \varepsilon^2) + (\delta^2 +  
\varepsilon^2 + 2e^2) - \frac{3}{\Lambda} \geq 0.    
\ee  
In order to disentangle these structure parameters, I rescale them   
as follows. $\Lambda$ and $e$ are non-zero so one can define $\Delta$,
$E$, and $X$ such that $\delta = \Delta e$, $\varepsilon = E e$, and
$e = \sqrt{3/\Lambda} X$.   
Then, the conditions (\ref{pc1}) and (\ref{pc2}) respectively become,  
\bea  
\label{c1}  
&& 1 - (\Delta^2 + E^2 + 2) X^2 \geq 0, \mbox{ and } \\  
\label{c2}   
&& (\Delta^2 - 1)(1 - E^2) X^4 + (\Delta^2 + E^2 + 2) X^2 - 1 \geq 0  
\eea  
The first of these provides an upper bound on the allowed range $X$  
for given values of $\Delta$ and $E$. $a^2 \geq 0$ if and only if  
\be  
X \leq X_U \equiv \frac{1}{\sqrt{2 + \Delta^2 + E^2}}.  
\ee 
  
In the meantime, (\ref{c2}) is quadratic in $X^2$ and so may be easily   
solved. It turns  
out that over the allowed ranges of $\Delta$ and $E$, it has only  
one positive real root. Further, it is upward  
opening, and therefore the positive real root provides a lower bound   
for the allowed values of $X$. $E_0^2 + G_0^2 \geq 0$ if and only if  
\be  
X \geq X_L \equiv \sqrt{  \frac{  -(\Delta^2 + E^2 + 2) +  
\sqrt{8(E^2+\Delta^2)+(E^2-\Delta^2)^2 }}{2 (\Delta^2 - 1)(1 - E^2)   
}}.  
\ee  
  
On plotting $X_U$ and $X_L$ one finds that for 
$0 \leq E \leq 1$ and $1 \leq \Delta \leq 2$, 
$X_L \leq X_U$ and so there exists a non-zero range for $X$   
for all the possible values of $E$ and $\Delta$. With this range of  
allowed values for $X$ in hand, the possible KNdS solutions have
been fully parameterized. This   
parameterization is given by the restrictions  
\be  
1 < \Delta \leq 2, \mbox{ } 0 \leq E < 2 - \Delta, \mbox{ and } X_L  
\leq X \leq X_U.  
\ee  
  
These ranges are shown in figure \ref{allowedRange}. In that figure 
the allowed parameter range of KNdS spacetimes is the region bounded by 
the five sheets defined by $a = 0$, $M = 0$,  
$E_0^2+G_0^2 =0$, $E=0$, and $E = 2 - \Delta$. 
The last two conditions respectively correspond to an extreme (cold)
black hole spacetime    
where the inner and outer black hole horizons coincide 
and a Nariai-type spacetime where the outer black hole horizon 
coincides with the cosmological horizon (though it will soon 
be seen that this apparent degeneracy  of the metric is an artifact 
of the coordinate system and that the distance between the two 
horizons remains finite and non-zero throughout  
the limiting process). The intersection of the Nariai and cold   
sheets is referred to as the ultracold solution. This nomenclature is taken from the corresponding non-rotating instantons discussed in
\cite{robbross}. 

Having established the range of KNdS solutions allowed by the  
structure of the polynomial $\cQ$, it remains to be demonstrated that   
the extreme limits of the range are realizable as a set of well defined
metrics. In particular the current coordinate representation of the
metric breaks down in the Nariai  
($\varepsilon \rightarrow 0$, $\delta \neq 0$)  and ultracold  
($\varepsilon  
\rightarrow 0$, $\delta \rightarrow 2e - \varepsilon$) cases. The  
following three subsections show how these various limits may be  
achieved while the fourth provides some details of the lukewarm 
KNdS solution discussed in section \ref{equil}.
     
\subsection{The cold limit}  
\label{coldsect}   
This limit can be taken within the current coordinate system.
Therefore, the metric keeps the form  
(\ref{KNdS}) and the electromagnetic field and potential remain as  
(\ref{EMfield}) and (\ref{EMpot}) respectively. The physical   
parameters  
are given by:   
\bea   
a^2 &=& \frac{3}{\Lambda} - 2(3e^2 - 2 \varepsilon e +  
\varepsilon^2) \\   
M &=& \frac{4 \Lambda}{3} e^2 (e-\varepsilon), \mbox{  
and} \\   
E_0^2 + G_0^2 &=& \lt (3e - \e)(e - \e)^2 (e+\e) + 2(3e^2 - 2e\e +   
\e^2)  
- \frac{3}{\Lambda},   
\eea   
where the range  
of the parameters is limited by the relations   
\bea   
0 <& E &< 1, \mbox{ and}\\   
\sqrt{\frac{-3 + 2E - E^2 + 2 \sqrt{3 - 4E + 2E^2}}{(3-E)(1+E)(1-  E)^2}}  
\leq & X & \leq \frac{1}{\sqrt{2(E^2 - 2E +3)}}.   
\eea   
As before, $e =  
\sqrt{ \frac{3}{\Lambda} } X$, and $\varepsilon = E e$.   
  
In this spacetime, the double horizon of the black hole recedes to an  
infinite proper distance from all other parts of the spacetime (as 
measured in the $\Sigma_t$ hypersurfaces).  Thus, the global structure of the spacetime changes -- in particular, the
region inside the black hole is cut off from the rest of the spacetime.
Choosing the global structure so that the spacetime contains two (in this case   
extreme) black holes, this structure is shown in figure \ref{PenCold}. Note that in this case, the hypersurfaces of constant $t$ consist of
two extreme black holes, and so are not closed as they were in the
lukewarm case (the horizons recede to infinite proper distance from all
other points in the spacetime).
  
Finally, note that 
for the cases where $a=0$, this solution reduces to the  
cold solutions discussed in \cite{robbross}.   
  
\subsection{The Nariai limit}   
\label{narsect}   
The coordinate  
system breaks down in the $\varepsilon = 0$ limit. Specifically, for  
$\varepsilon = 0$, $r=e$ (becomes a constant), and $\cQ=0$, so the  
coordinate system becomes degenerate, and the metric ill-defined.  
These problems may easily be avoided however, if one makes the  
transformations
\bea   
r &=& e + \varepsilon \rho, \\   
\phi &=& \varphi + \frac{a}{e^2 + a^2} t, \mbox{ and}\\   
t &=& \frac{(e^2+a^2)\chi^2}{\varepsilon} \tau.    
\eea   
Then, the $\varepsilon \rightarrow 0$ limit may  
be taken without hindrance, and the metric becomes   
\be ds^2 = - \tQ \cG d \tau^2 + \frac{\cG}{\tQ} d \rho^2 +   
\frac{\cG}{\cH}  
d\theta^2  
     + \frac{\cH \sin^2\theta}{\cG} \left( 2 a e \rho  d \tau +  
\frac{e^2+a^2}{\chi^2} d\varphi\right)^2,  
\ee     
while the electromagnetic  
field becomes,  
\be  
F = \frac{-X}{\cG} d\rho \wedge d\tau + \frac{Y \sin \theta}{\cG^2}   
d\theta \wedge \left( 2ae\rho   
d\tau + \frac{e^2 + a^2}{\chi^2} d\varphi \right).
\ee  
An electromagnetic potential generating this is
\be  
A = - E_0 \frac{(e^2 - a^2)}{e^2 + a^2} \rho d\tau - \frac{a E_0 e  
\sin^2 \theta + G_0 (e^2+a^2) \cos \theta}{\cG (e^2+a^2) }  \left( 2  
a e \rho d\tau + \frac{e^2 + a^2}{\chi^2} d \varphi \right).   
\ee
In the above, $\tQ = \frac{\Lambda}{3} (2e-\delta)(1-  
\rho^2)(2e+\delta)$, $\cG   
= e^2 + a^2 \cos^2 \theta$, $\Gamma = e^2 - a^2 \cos^2 \theta$, $X =   
E_0 \Gamma + 2a G_0 e   
\cos \theta$, and $Y = G_0 \Gamma - 2 a E_0 e \cos \theta$.  
Note that the above potential is not the simply ($\ref{EMpot}$) under
the coordinate transformation as the $A$ generated in that way diverges
when $\varepsilon \rightarrow 0$. The divergence is removed (and the
above result obtained) if one makes the gauge transformation 
$A \rightarrow A - \frac{E_0 e}{\varepsilon} d \tau$ before the coordinate transformation and limit.
  
The physical parameters are given in terms of $e$ and $\delta$ as  
\bea  
a^2 &=& \frac{3}{\Lambda} - 2e^2 - \delta^2, \\  
M &=& \frac{\Lambda}{3} \delta^2 e, \mbox{ and}\\  
E_0^2 + G_0^2 &=& \lt(\delta^2 - e^2)e^2 + (2e^2 + \delta^2) -  
\frac{3}{\Lambda},  
\eea  
and the allowed ranges of $e = \sqrt{\frac{3}{\Lambda}} X$ and  
$\delta = \Delta e$ are given by
\bea  
1 <& \Delta &\leq 2 \mbox{ and} \\  
\sqrt{\frac{-(\Delta^2+2)+\Delta \sqrt{\Delta^2 + 8}}{2(\Delta^2-1)}}   
\leq   
& X & \leq \frac{1}{\sqrt{ 2 + \Delta^2}}.  
\eea  
  
Note that the Nariai solution is no longer a black hole solution.  
Extending the metric through the horizons by the standard Kruskal  
techniques, its Penrose diagram appears as in figure \ref{PenNar} (for   
the ($\tau$,$\rho$) sector). 
There is no longer a singularity at finite distance beyond either of the
horizons. In fact, the diagram is the same as that for 
two-dimensional deSitter space. If there were no rotation ($a=0$), 
then this  
spacetime would just be the direct product of two-dimensional deSitter  
space, and a two-sphere of fixed radius. With rotation, of course the  
situation is not so simple.  

If $a=0$, and one makes the coordinate transformation $\rho = \cos   
\chi$, then this solution reduces to the non-rotating charged Nariai solution considered in \cite{robbross}.  
  
\subsection{The ultracold limits}  
\label{ucsect}  
Finally consider the ultracold limits where both $\varepsilon  
\rightarrow 0$ and $\delta \rightarrow 2e-\varepsilon$. It turns out
that there are two such limits which I label the ultracold I and
II limits. In this subsection I only demonstrate how they may  
be reached from the Nariai limit. Similar coordinate  
transformations (which sometimes must be iterated two or three times)  
allow one to reach the same two limits both from the cold limit, and,  
taking $\delta \rightarrow 2e-\varepsilon$ and $\varepsilon \rightarrow   
0$ simultaneously, straight from the non-extreme standard KNdS form of
the metric. I deal with the two cases separately.  
 
\underline{Ultracold I:} 
Making the transformations, 
\bea 
\rho &=& \eta - \eta k (2e-\delta) R, \\ 
\varphi &=& \Phi - 2 \eta \frac{ae \chi^2 \tau}{e^2+a^2}, \mbox{ 
and}\\ 
\tau &=& \frac{ \eta T}{k (2e-\delta)}, 
\eea   
where $\eta = \pm 1$, and $k = 8 \lt e$, and taking the 
limit as $\delta \rightarrow 2e$ one obtains the metric, 
\be 
ds^2 = -\cG R dT^2 + \frac{\cG}{R} dR^2 + \frac{\cG}{\cH} 
d\theta^2 + 
\frac{\cH}{\cG}\sin^2\theta \left( 2 a e R dT + 
\frac{e^2+a^2}{\chi^2} 
d \Phi\right)^2. 
\ee 
The electromagnetic field and potential become, 
\be    
\label{UCF} 
F = \frac{-X}{\cG} dR \wedge dT + \frac{Y \sin \theta}{\cG^2} 
d\theta \wedge \left( 2 a e R dT + \frac{e^2 + a^2}{\chi^2} d\Phi 
\right), 
\ee 
and, 
\be 
\label{UCA} 
A = -E_0 \frac{e^2-a^2}{e^2+a^2} R dT - \frac{a E_0 e \sin^2 \theta 
+    
G_0 (e^2+a^2) \cos \theta}{\cG (e^2+a^2)} \left( 2 a e R dT + 
\frac{e^2+a^2}{\chi^2} d \Phi \right). 
\ee 
$R \in (0, \infty)$, $T \in (-\infty, \infty)$, 
$\theta \in [0, \pi]$, and $\Phi$ inherits a $2\pi$ periodicity from its 
predecessors. 
$\cG$, $\cH$, $\chi^2$, $X$, and $Y$ all retain their old 
definitions. Note that the EM potential and field have retained their Nariai form. 
 
The ($R$,$T$) sector of the spacetime is conformally the 
same as the Rindler spacetime (which of course is actually a sector 
of two-dimensional Minkowski space). The Rindler horizon is at 
$R=0$ and as this is the only horizon, the space does not contain black 
holes. Before giving the parameterization of this solution, 
consider the transformations leading to the ultracold II case. 
 
\underline{Ultracold II:}  
Making the transformations,  
\bea  
\rho &=& b + k \sqrt{2e-\delta} R, \\  
\varphi &=& \Phi - 2 \frac{aeb \chi^2 \tau}{e^2+a^2}, \mbox{ and}\\  
\tau &=& \frac{T}{k \sqrt{2e-\delta}},  
\eea  
where $b \neq \pm 1$, and $k = 2\sqrt{ \lt (1-b^2) e }$ and taking the  
limit as $\delta \rightarrow 2e$, one obtains  
\be  
ds^2 = -\cG dT^2 + \cG dR^2 + \frac{\cG}{\cH} d\theta^2 +  
\frac{\cH}{\cG}\sin^2\theta \left( 2 a e R dT +   
\frac{e^2+a^2}{\chi^2}  
d \Phi \right)^2.  
\ee   
The electromagnetic field and potential again take the forms (\ref{UCF}) 
and
(\ref{UCA}).  
$R, T \in (-\infty, \infty)$, $\theta \in [0,\pi]$, and  
$\Phi$ inherits a period of $2 \pi$ from its predecessors.   
$\cG$, $\cH$, $X$, and $Y$ again retain their meanings from the Nariai 
case. 
 
Clearly the ($R$,$T$) sector of this spacetime is conformally the same   
as two dimensional Minkowski flat space. There is no  
horizon structure and therefore no black hole.  
  
The physical parameters in both of these cases are given by  
\bea  
a^2 &=& \frac{3}{\Lambda} - 6e^2, \\  
M &=& 4\frac{\Lambda}{3} e^3,\\  
E_0^2 + G_0^2 &=& \Lambda e^4 + 6e^2 - \frac{3}{\Lambda},  
\eea  
and the allowed range of $e = \sqrt{\frac{3}{\Lambda}} X$ is given   
by,  
\bea  
\sqrt{-1 + \frac{2}{\sqrt{3}}} \leq  
& X & \leq \frac{1}{\sqrt{6}}.  
\eea  
  
Once more note that when $a=0$ these
ultracold cases reduce to the two non-rotating ultra-cold solutions
considered in \cite{robbross}. However, neither of these
spacetimes contains black holes. Perhaps one can make an argument
for them decaying like the Nariai metric into black hole spacetimes,
but in any case for completeness I shall
include them in my considerations throughout the thesis.

\subsection{The lukewarm solution}  
\label{luke}  
  
As discussed in section \ref{equil}, the lukewarm solution is
defined as a KNdS solution where
the black hole and cosmological horizons are
in thermal equilibrium. Their temperatures are given by equations (\ref{temperatures}) and a little algebra shows that they are 
equal (and not degenerate) when 
$2e^2 - 2a^2 - \varepsilon^2 - \delta^2=0$.  
This relation can be used to eliminate $\delta$ from the
parameterizations of the physical parameters. Then  
\bea  
a^2 &=& 4e^2 - \frac{3}{\Lambda}, \\  
M   &=& 2e(1 - \lt(3 e^2 + \varepsilon^2)),  \mbox{ and}\\  
E_0^2 + G_0^2 &=& - \lt (7e^2+\e^2)(e^2-\e^2)-2(e^2-\e^2)+  
\frac{3}{\Lambda}.  
\eea  
The expression for the charge may also be written  
as $E_0^2 + G_0^2 = \frac{M^2}{\chi^2} - a^2 \chi^2$.   
  
The range of the parameters is limited by the relations:  
\bea  
0 &\leq E <& 1 \\  
\frac{1}{\sqrt{5-2E-E^2}} &\leq X <& \sqrt{\frac{2}{E^2+7}}   
\label{lw2}\\   
\frac{1}{2} &\leq X \leq& \sqrt{ \frac{2 \sqrt{2-E^2} - 1 -  
E^2}{(E^2+7)(1-E^2)}} \label{lw3},  
\eea  
where as earlier $\e = E e$ and $e = \sqrt{\frac{3}{\Lambda}} X$.   
The second condition above is the $1 < \Delta < 2-E$ inequality
for this case, while the third is the $a^2 \geq 0$, 
$E_0^2+G_0^2 \geq 0$ condition. Plotting the two conditions over
the allowed range of $E$ one finds that (\ref{lw2}) is redundant, 
and so the lukewarm range is given by the first and third conditions.
   
These spacetimes are non-extreme KNdS spacetimes, and so have the
global structure displayed in figure \ref{PenReg}. This spacetime
was first discussed in \cite{MM}. Just as for the other special KNdS 
spacetimes that I considered, in the absence of rotation the lukewarm
case reduces to its non-rotating counterpart discussed \cite{robbross}.

\bibliography{ThesisBib}

\bibliographystyle{plain}

\end{document}